**Preprint**

# Sulfur on Venus: Atmospheric, Surface, and Interior Processes

**Mikhail Zolotov**

School of Earth and Space Exploration, Arizona State University, Tempe, AZ, USA e-mail: zolotov@asu.edu

Sulfur-bearing species play crucial roles in atmospheric physical-chemical processes, atmosphere-surface interactions, and the geological evolution of Venus. This chapter provides a comprehensive overview of (1) instrumental data on the abundance and speciation of sulfur in atmospheric and crustal materials, (2) the behavior of sulfur-bearing species in the mesosphere, clouds, and lower atmosphere, (3) chemical and mineralogical aspects of atmosphere-surface interactions, and (4) the fate of sulfur during the formation, differentiation, and geological evolution of Venus, including volcanic degassing, gas-solid reactions at the surface, a proposed aqueous period, and subsequent evolution. The chapter also outlines key questions and discusses further exploration of Venus in the context of sulfur-relevant investigations.

## Outline

**Abstract.**

Sulfur-bearing species play critical roles in atmospheric physical-chemical processes, atmosphere-surface interactions, and the geological evolution of Venus. Most data on atmospheric sulfur have been obtained through telescopic and spacecraft spectroscopic investigations, along with limited in situ measurements using entry probes. The atmospheric composition below ≈20 km is poorly understood and estimated by extrapolating measured contents toward the surface and using thermochemical and mixing models. The bulk sulfur content was measured in solid surface samples collected from three landing sites, while the geochemistry of sulfur in the interior remains poorly constrained. Sulfur dioxide is the most abundant sulfur-bearing gas in the middle and lower atmosphere. Photochemical dissociation of $CO_2$ above the clouds into CO and atomic oxygen, along with the subsequent oxidation of $SO_2$ to $H_2SO_4$, sustains thick global clouds rich in sulfuric acid. Interaction of $SO_2$ with basaltic glasses and Ca-bearing minerals leads to the formation of sulfates, consistent with the elevated sulfur content in surface probes. Carbonyl sulfide, OCS, is the most abundant reduced atmospheric gas. It forms in the lower atmosphere from CO and sulfur gases and is oxidized by $SO_3$, which forms via pyrolysis of $H_2SO_4$ gas below the clouds. OCS can be formed and consumed by altering and forming sulfide minerals. In Hadley cell-type atmospheric circulation, low-latitude upwelling delivers OCS to the clouds, while high-altitude downwelling brings OCS-depleted and CO-enriched gas to the deep atmosphere. The occurrence and fate of $S_{1–8}$ gases and condensates ($S_8$) must be more constrained. Some models suggest the formation of $S_n$ gases below the clouds, while others imply that they are formed via photochemical reactions in the upper clouds. $S_2$ is consumed through OCS production in the deep atmosphere and released via sulfatization of silicates. Chemical disequilibria drive reactions in the atmosphere, but some gases may reach equilibrium at the surface. Fe sulfide-oxide and/or sulfate-silicate phase equilibria in a permeable surface layer could control the abundance and speciation of sulfur-bearing gases in the deep atmosphere. Pyritization and sulfatization of surface materials through gas-solid reactions are thermodynamically more favorable in the highlands but require confirmation. Past volcanic degassing mainly constituted the current inventory of atmospheric sulfur. Although cometary dust is a minor net source of atmospheric sulfur, it may account for the elevated mixing ratios of sulfur-bearing gases in the upper mesosphere. Cosmic sources could account for possible metal sulfates in the clouds. Aside from sequestered atmospheric sulfur, the chemical composition and radar-based morphology of immense volcanic formations suggest an abundance of sulfur and sulfide mineralogy (pyrrhotite, pentlandite) typical of tholeiitic and alkaline basalts. The morphology and thermal emissivity of highly tectonically deformed tessera terrain do not rule out the possibility of exposure to rocks formed in $H_2O$-rich exogenic and/or endogenic environments. The presence of sulfates in those rocks depends on the $H_2O$ history, which may or may not involve oxidizing aqueous and/ or magmatic environments. Hydrogen escaping from early Venus with a putative water ocean could have led to sulfate-rich seawater and subsequent sulfate-rich rock formations. In contrast to Earth's marine sediments, abundant pyrite may not have precipitated in Venus' counterparts due to the absence of biological sulfate reduction and sufficient organic matter. The formation of a Fe sulfide-oxide assemblage in surface materials could result from the coupled physical-chemical evolution of the atmosphere-surface system following global volcanic resurfacing. It is essential to prioritize the examination of sulfur-bearing species and isotopes in future remote and in situ studies of atmospheric gases, aerosols, and both chemically altered and pristine rocks. The upcoming

Venus Orbital Mission, DAVINCI, VERITAS, and EnVision missions will explore atmospheric and surface compositions, leading to a better understanding of the behavior of sulfur on Earth's sister planet.

## 16.1 Introduction

### *16.1.1* Fates of sulfur on Earth and Venus

Venus and Earth possess differentiated interior structures characterized by metal-rich cores, silicate mantles, and basalt-rich crusts that cover most of their surfaces. These planets have secondary atmospheres and likely share significant formation processes and geological histories, including accretion from solid materials, the formation of magma oceans during the later stages of accretion, the establishment of Fe metal-rich cores, the generation of secondary carbon- and nitrogen-rich atmospheres through magmatic degassing, and the coupled physical-chemical evolution of deep interiors, silicate crusts, and atmospheres. The mantles of both planets have oxidized beyond the iron-wüstite (IW) $fO_2$ buffer conditions. On Venus and Earth, sulfur compounds are present among atmospheric gases. In the atmosphere, sulfur exists in multiple redox states; however, the oxidized S(VI), S(IV), and S(II) species ($SO_4^{2-}$, $SO_2$, and SO) are predominant. Both sulfate (S(VI)) and sulfide (S(-II) and S(-I)) redox forms are most common in surface and crustal materials, although sulfides dominate in deeper interiors. Sulfur is abundant in exogenic surface materials. Exogenic sulfur-bearing minerals are mainly represented by sulfates of rock-forming elements (Ca, Na, etc.) and pyrite. Pyrrhotite ($Fe_{1-x}S$) and troilite (FeS) likely constitute the most sulfur-bearing compounds found in crustal magmatic and mantle/core materials.

Sulfur is active in processes that have shaped and continue to affect deep interiors, crusts, surfaces, and atmospheres. It participates in magma ocean processes, global differentiation and sequestration of Fe metal and FeS to cores, the formation of silicate mantles depleted in sulfur and chalcophile elements, the separation of sulfide liquids in magmatic processes, magmatic assimilation of crustal materials, volcanic degassing, sequestration of volcanic gases in crustal materials, atmospheric thermochemical and photochemical reactions, the formation of atmospheric acid sulfate aerosols, the supply and transformation of cosmic materials in atmospheres, geochemical cycles in coupled atmospheric-crustal systems, and isotopic fractionation during these processes.

Crustal sulfur is a typical 'excess' volatile on Earth after Rubey (1951). The term 'excess' indicates that the chemical weathering of endogenic sulfides cannot account for the mass of sulfur in the crust and the hydrosphere. This suggests a significant contribution of volcanic degassing to sulfur inventory in crustal materials. This concept implies net sulfur accumulation in planetary envelopes through volcanic degassing. Although the last period of global volcanic activity on Venus ceased by ≈0.3 to 0.9 Ga, the abundance of sulfur in the atmosphere and surface materials suggests that sulfur released through degassing continues to be sequestered.

On both planets, the abundance, speciation, and isotopic composition of sulfur serve as indicators of physical-chemical processes throughout history and in the present. These compositional data inform atmospheric chemical and photochemical processes, atmospheric circulation, redox conditions, volcanic degassing, magmatic and metamorphic processes, and the chemical weathering of surface materials. The atmospheric sulfur and sulfur mineralogy of surface and crustal materials are interconnected on both planets, although the sulfur cycles differ.

Differences in size, bulk composition, scale, specifics, and timing of geological events on Earth and Venus are reflected in the fate of sulfur. Significant discrepancies include the presence of the hydrosphere throughout much of Earth's history, plate tectonics, and organic matter in crustal materials, life, and the $O_2$-rich atmosphere. This chapter examines the abundances of sulfur and sulfur-bearing compounds. It discusses the fate of sulfur in Venus' atmosphere, surface, and endogenic processes, considering current conditions, geological history, and the planet's future. Poorly constrained processes in the interior and throughout the planet's history are judged considering data from Earth and other solar system bodies.

### *16.1.2* Venus: General Knowledge

General knowledge of Venus can be found in books (Hunten et al. 1983; Bougher et al. 1997; Esposito et al. 2007; Taylor 2014) and review chapters (Fegley 2014; Taylor et al. 2018; Gillmann et al. 2025). Venus' distance from the Sun is ≈0.72 of the Earth's distance. Venus' orbital period is ≈225 Earth days, and its slow retrograde rotation period (sidereal day) is 243 days. Venus has a slightly lower modal radius (6051.4 km) and mean density (5.243 g $cm^{-3}$) than Earth. Venus' dimensionless moment of inertia factor of 0.33 to 0.34 (Margot et al. 2021) indicates a differentiated structure, comprising a metal-rich core, a silicate mantle, and a crust. In contrast to Earth, Venus' unimodal hypsometric curve indicates a lack of Earth-like compositional dichotomy in crustal materials with thicker felsic and thinner mafic crusts. There is no evidence of plate tectonics, and Venus likely has a stagnant- or squishy-lid lithosphere (Rolf et al. 2022). The apparent lack of an intrinsic magnetic field on Venus indicates suppressed convection in the core, which could be entirely solid or liquid.

Venus has a dense atmosphere with global sulfuric acid clouds located at ≈43 to 73 km, depending on latitude, which restricts observations of the surface from space. The mesosphere, or upper atmosphere, extends from the cloud top to ≈120 km above. Clouds characterize the middle atmosphere, while the lower atmosphere spans from the cloud base down to the surface. The atmosphere mainly consists of $CO_2$ and $N_2$ with volume mixing ratios ($x$) of ≈0.965 and ≈0.035, respectively. A gas volume mixing ratio is equivalent to a mole fraction and represents the ratio of the gas's partial pressure ($p$) to the total pressure. $SO_2$ is the most abundant sulfur-bearing gas, with a mixing ratio of ≈ $(1-2)\times10^{-4}$, or 100 to 200 parts per million by volume (ppmv). The strong greenhouse effect accounts for a surface temperature of ≈660 to 760 K, depending on elevation. At the modal planetary radius, the temperature is 740 K, and the pressure is 95.6 bars.

Lowlands are primarily characterized by volcanic plains created by low-viscosity lava flows. The highlands feature plateaus, volcanic rises, and large shallow circular volcanic-tectonic structures (coronas) that are unique to Venus. About 8% of the surface consists of highly tectonically deformed plateaus known as tessera. The density of about one thousand impact craters indicates a surface age of ≈0.3 to 0.9 Ga, and there is no statistically significant difference in the relative ages among surface formations based on crater density. One reasonable interpretation of the crater statistics suggests rapid global volcanic resurfacing on a geological timescale. Interpretations of remote sensing data on recent and ongoing volcanism require confirmation. However, a correlation between gravity and topography data suggests active mantle plumes contributing to the upwelling beneath some volcanic structures. Most surface materials consist of volcanic products with mafic (basaltic) composition. There are no signs of explosive volcanic activity that would lead to the formation of stratovolcanoes associated with the subduction of lithospheric plates. The composition of tessera terrains remains unknown, but they may also consist of mafic igneous materials.

Low-velocity surface winds permit only minimal erosion, migration, and aeolian deposition of fine-grained materials. However, surface materials are affected by gas-solid type chemical weathering, which involves the sequestration of atmospheric sulfur-bearing species and reactions with carbon-, hydrogen-, Cl-, and F-bearing gases. Secondary sulfates, sulfides, oxides, and other compounds are suggested in altered surface materials. Low emissivity in the near-infrared (near-IR) and microwave spectral ranges suggests diverse compositions of surface materials across highlands.

There is no geomorphological information on Venus' geological history before ≈0.3 to 0.9 Ga, and it remains unclear whether the planet even had a plate tectonics regime. Although the high D/H ratio in the atmosphere may indicate an early $H_2O$-rich Venus, substantial evidence for past surface water and related processes is lacking.

### 16.1.3 Data Sources and Methods to Assess Sulfur and Sulfur-Bearing Species

The lack of returned samples from Venus restricts our understanding of the chemical and phase composition of sulfur-bearing materials. The data on atmospheric sulfur-bearing gases ($SO_2$, SO, $SO_3$, OCS, $H_2S$, $H_2SO_4$, $S_n$, CS, and $CS_2$) and cloud aerosols (sulfuric acid and possible native sulfur and metal sulfates) have been obtained from ground-based and orbital telescopic observations in the ultraviolet (UV) to millimeter ranges, from Venus' spacecraft flyby and orbital data, and entry probes, as reviewed by von Zahn et al. (1983), Esposito et al. (1983, 1997), de Bergh et al. (2006), Mills et al. (2007), Marcq et al. (2018), Titov et al. (2018), Vandaele (2020), and Dai et al. (2025). A significant portion of the data is obtained through remote spectroscopic observations. Although a wealth of atmospheric data is available, sulfur was measured in only three surface samples, and there is no apparent Venus-sourced specimen in the meteorite collection. Insights can be gained from chondritic and cometary materials, terrestrial data and processes, as well as experimental and numerical modeling. This section outlines the methods and instruments employed to constrain sulfur bulk abundances and the phase composition of sulfur-bearing compounds.

#### 16.1.3.1 Ground-Based and Spacecraft Data on the Atmosphere

Telescopic observations assess the speciation and concentration of sulfur-bearing compounds from the top of the mesosphere to the lower atmosphere down to ≈30 km (Table 16.1). These measurements enable the monitoring of atmospheric physical properties and composition, as well as the observation of temporal and spatial variations between spacecraft observations. Since the initial detections of $SO_2$ at the cloud tops with UV observations (Barker 1979; Conway et al. 1979), multiple telescopic data have revealed spatial, short-term, and long-term variations in the concentrations of $SO_2$ and SO. International Ultraviolet Explorer (IUE) satellite observations (Na et al. 1990) and rocket-borne UV telescopes (McClintock et al. 1994; Na et al. 1994) were used to assess the abundances of $SO_2$ and SO at and above the cloud top from Earth. Hubble Space Telescope (HST) UV spectroscopy was used to constrain the concentrations of $SO_2$ and SO at ≈70 to 81 km (Na and Exposito 1995; Jessup et al. 2015). Telescopic IR observations at 7.4 μm, 8.6 μm, and 19 μm provided data on $SO_2$ variability in upper clouds and cloud top at 57 to 67 km (Encrenaz et al. 2012, 2016, 2019, 2020, 2023). Telescopic near-IR observations assessed $SO_2$ and OCS contents at cloud top at 65 to 72 km (Krasnopolsky 2008, 2010b). Microwave telescopic data constrained the abundances of $SO_2$ and SO (Sandor et al. 2010, 2012; Jessup et al. 2015; Encrenaz et al. 2015) and $H_2SO_4$ gas (Sandor et al. 2012) above clouds. Microwave radioastronomy data enabled the evaluation of the concentration of $H_2SO_4$(g) down to an altitude of ≈30 km (Butler et al. 2001; Jenkins et al. 2002). Ground-based polarimetry measurements constrained the refractive index and corresponding concentration of liquid sulfuric acid in the cloud aerosol (Hansen and Hovenier 1974).

Allen and Crawford (1984) discovered narrow spectral windows in the $CO_2$ spectrum near 1.7 and 2.3 μm using near-IR ground-based telescopic images of Venus' night side. This discovery enabled observations of $SO_2$, OCS, and other gases ($H_2O$, CO, HCl, and HF) below the clouds (Taylor et al. 1997). Night sky observations through near-IR spectral windows (Bézard et al. 1990, 1993; Pollack et al. 1993; Marcq et al. 2005, 2006, 2021; Arney et al. 2014) provided data on $SO_2$ and OCS from the cloud deck to ~30 km and constrained the concentration of $H_2SO_4$ in cloud aerosol.

Orbital observations from Pioneer Venus 1 (Stewart et al. 1979; Esposito 1984; Esposito et al. 1979, 1988), Venus Express (Belyaev et al. 2008; Marcq et al. 2011, 2013, 2020), and Akatsuki (Yamazaki et al. 2018) in the UV spectral range provided data on the spatial, temporal, and long-term (since 1978) concentrations of $SO_2$ and SO at and above the cloud top. IR measurements with a Fourier spectrometer from the Venera 15 orbiter (Schaefer et al. 1990) constrained the $SO_2$ content in clouds at different latitudes (Moroz et al. 1990; Zasova et al. 1993). Solar and stellar occultation data obtained with Venus Express IR and UV spectrometers provided information on the abundances, variability, and gradients of $SO_2$, SO, $SO_3$, and OCS at the cloud top and in the mesosphere (Belyaev et al. 2008, 2012, 2017; Mahieux et al. 2023). Observations during the second MESSENGER Venus flyby in the UV range were utilized to constrain the composition of a species that absorbs solar irradiation in the blue and UV spectral ranges in the upper clouds (Pérez-Hoyos et al. 2018).

Within and below clouds, Venus Express orbital observations in the near-IR spectral windows constrained concentrations of $SO_2$ and OCS down to ≈30 km (e.g., Marcq et al.

2008, 2023). Akatsuki UV data were used to assess $SO_2$ abundances and variability at the cloud top (Yamazaki et al. 2018). Gaseous $H_2SO_4$ is a strong absorber of radio waves, and radio occultation data from the Mariner 10 flyby in 1974 (Lipa and Tyler 1979; Kolodner and Steffes 1998) and subsequent Pioneer Venus 1 (Jenkins and Steffes 1991), Magellan (Jenkins et al. 1994; Kolodner and Steffes 1998), Venus Express (Oschlisniok et al. 2012, 2021), and Akatsuki (Imamura et al. 2017) orbiters provided data to constrain $xH_2SO_4$(g) within and below clouds down to ~35 km. Venus Express polarimetry data were used to constrain the sulfuric acid concentration in the cloud aerosol (Barstow et al. 2012).

In situ atmospheric measurements of sulfur-bearing gases ($SO_2$, OCS, $H_2S$, $S_n$) were performed with the Pioneer Venus Large Probe, PVLP (1978), Venera 11 and 12 (1978), Venera 13 and 14 (1982), and Vega 1 and Vega 2 (1985) descent space-craft, using gas chromatography (GC), mass spectrometry (MS), and gas absorption spectroscopy in the UV and visible ranges (Table 16.1). Most data were obtained for the sub-cloud atmosphere above 20 km. The GC measurements onboard the PVLP (Oyama et al. 1980) and Venera 12 probe (Gel'man et al. 1980) led to the detection of $SO_2$ at 22 km and below 42 km, respectively. The UV absorption spectra at 220 to 400 nm obtained with Vega 1 and Vega 2 were used to estimate the $SO_2$ profile for 0–60 km (Bertaux et al. 1996). The GC data on $H_2S$ and OCS obtained with the Venera 13 and Venera 14 probes (Mukhin et al. 1982, 1983) exhibit inconsistencies compared to other observations. These discrepancies are often attributed to issues related to the carrier gas, which compromised peak identification (Krasnopolsky 1986). The blue absorption (<0.5 μm) of solar light observed at several Venera probes in the sub-cloud atmosphere was used to estimate mixing ratios of $S_3$ and $S_4$ in the deep sub-cloud atmosphere (San'ko 1981; Golovin et al. 1982; Maiorov et al. 2005; Krasnopolsky 1987, 2013).

PVLP MS data suggested that concentrated sulfuric acid solution is a major component of the cloud aerosol (Hoffman et al. 1980), which aligns with in situ Pioneer Venus polarization and nephelometry data regarding the refractive index of aerosol particles (Knollenberg and Hunten 1980; Ragent and Blamont 1980; Ragent et al. 1985). Vega reaction gas chromatography measurements confirmed the domi-nance of $H_2SO_4$ in the aerosol and were used to estimate the aerosol mass load at altitudes of 48 to 62 km (Gel'man et al. 1986; Porshnev et al. 1987). The X-ray fluorescent (XRF) method has been employed to evaluate the bulk sulfur abundance in cloud aerosols collected during the descents of Venera 12 (Surkov et al. 1982a), Venera 14 (Surkov et al. 1982b), and Vega 1 and Vega 2 (Andreichikov et al. 1987) entry probes/landers. Mass spectrometric analysis of collected and pyrolyzed cloud particles (Surkov et al. 1986a, 1987a) on Vega 1 constrained the mass load of sulfuric acid aerosol.

#### 16.1.3.2 Exploration of Surface and Crustal Materials

The bulk sulfur content in surface materials has been measured using the XRF method at the landing sites of Venera 13 (Fig. 16.1), Venera 14, and Vega 2 (Surkov et al. 1984, 1986b); however, the mineralogy of the solids has not been determined. Some constraints on the original sulfur content and speciation in these samples can be inferred from the FeO/MnO ratio as an indicator of the magma's $fO_2$ (Wänke et al. 1973; Schaefer and Fegley 2017).

Indirect constraints on the phase composition and abundance of sulfides in surface and near-surface materials can be obtained from the specific electrical resistance of surface materials measured at Venera 13 and 14 (Kemurdzhian et al. 1983) and the Vega 2 landing sites. Dielectric properties of surface materials inferred from remote microwave observations are also informative. Microwave emissivity data of surface materials were obtained from ground-based telescopic investigations (e. g., Campbell et al. 1997, 1999; Pettengill et al. 1997) as well as from the Pioneer Venus 1 (Ford and Pettengill 1983; Garvin et al. 1985; Pettengill et al. 1982, 1988), Magellan (Pettengill et al. 1992, 1996, 1997; Klose et al. 1992; Arvidson et al. 1994; Campbell et al. 1992, 1997), Venus Express (Simpson et al. 2009), and Venera 15 and 16 orbiters. The emissivity is a function of the material's dielectric constant and may inform the composition and the structure of a decimeter-scale surface layer. The correlation of inferred dielectric properties with topography or geological features provides insights into the origins of phases with an elevated dielectric constant that could be represented by metal sulfides (Pettengill et al. 1988, 1997; Klose et al. 1992; Schaefer and Fegley 2004). Venus' emission at decimeter wavelengths, measured through ground-based interferometric observations, provides indirect information on dielectric properties and the composition of the upper meter of the surface materials, such as the occurrence of Fe sulfides and oxides (Antony et al. 2022).

Sulfur content in widespread volcanic rocks can be inferred from magma composition, suggested by the morphology of lava flows and numerous volcanic centers observed in radar images obtained from Venera 15 and Venera 16 (Barsukov et al. 1986a) and Magellan orbiters (Head et al. 1992; Crumpler et al. 1997; Hahn and Byrne 2023; Ghail et al. 2024). Petrological types of rocks can be constrained from K, U, and Th concentrations obtained from in situ γ-ray spectroscopy of surface materials at the landing sites of Venera 8, 9, 10, Vega 1, and Vega 2 (Surkov et al. 1987b). The rough petrological type of the surface materials (e.g., mafic vs. felsic) and/or the degree of their physical and chemical weathering could be estimated from the nightside thermal emissivity observed through spectral windows in the $CO_2$ spectrum at ≈1 μm from Galileo (Hashimoto et al. 2008), Cassini (Baines et al. 2000), and Venus Express data (Mueller et al. 2008; Helbert et al. 2008; Basilevsky et al. 2012; Gilmore et al. 2015). Nightside surface emission data at ~0.65–0.8 μm obtained from the Parker Solar Probe Venus flybys could inform compositional differences between large regions (Wood et al. 2021; Lustig-Yaeger et al. 2023). Other assessments have been derived from considering mineral stability under surface conditions through calculations of the chemical equilibria of selected reactions and in multicomponent gas-solid systems (e.g., Mueller 1964, 1965; Lewis 1968, 1970; Nozette and Lewis 1982; Khodakovsky 1982; Barsukov et al. 1980b, 1982c, 1986d; Fegley and Treiman 1992a, b; Klose et al. 1992; Fegley et al. 1992, 1997a; Schaefer and Fegley 2004; Zolotov 2018; Semprich et al. 2020) or through experimental approaches to chemical weathering of minerals and glasses by sulfur-bearing gases (e.g., Fegley and Prinn 1989; Radoman-Shaw et al. 2022; Santos et al. 2023; Ried et al. 2024).

## 16.2 Abundance and Speciation of Sulfur

### *16.2.1* Atmospheric Compounds

This section presents data on the speciation of sulfur in atmospheric gases and aerosols obtained through in situ spacecraft and remote measurements (Sect. 16.1.3.1), while atmospheric processes are described in Sect. 16.3.1. Nine sulfur-bearing gases ($SO_2$, SO, $SO_3$, $H_2SO_4$, OCS, $S_3$, $S_4$, CS, and $CS_2$) have been identified below an altitude of 100 km (Esposito et al. 1997; Mills et al. 2007; Marcq et al. 2018; Johnson and Oliviera 2019; Vandaele 2020; Mahieux et al. 2023). The most abundant sulfur-bearing gas is $SO_2$ below ≈70 km, OCS being the second most abundant gas below ≈30 km, and $H_2SO_4$ vapor is prevalent in clouds. $SO_3$, OCS, $SO_2$, SO, CS, and $CS_2$ are notable gases in the mesosphere (Table 16.1). Concentrations of these and other gases ($S_n$, $H_2S$, $S_2O$, $ClSO_2$, etc.) have been evaluated through chemical kinetic modeling (0–112 km), while chemical equilibrium models are most relevant to the near-surface atmosphere (e.g., Krasnopolsky and Parshev 1981a, 1983; Yung and DeMore 1982; Mills 1998; Zhang et al. 2012; Fegley et al. 1997b; Krasnopolsky and Pollack 1994; Krasnopolsky 2007, 2012, 2013; Bierson and Zhang 2020). Liquid sulfuric acid solution, $H_2SO_4 \cdot nH_2O$, is the primary component of cloud aerosols, and other sulfur-bearing compounds may also be present, such as condensed native sulfur, metal sulfates, and cosmogenic acid-insoluble organic matter. Some of these sulfur-bearing compounds could account for absorption in the blue and UV spectral ranges of upper clouds. In this section, the speciation and concentration of sulfur-bearing species are described in the mesosphere above cloud top, within clouds (≈47–73 km at low latitudes and 43–64 km at high latitudes), and in the sub-cloud atmosphere. Although ≈75% of the atmospheric mass is located below an altitude of ≈20 km, most compositional data are obtained above that altitude.

#### 16.2.1.1 Sulfur Oxides Sulfur Dioxide

Sulfur dioxide is the second most chemically active gas after $CO_2$ below ≈70 km. At the cloud top and in the mesosphere, $SO_2$ has been observed using ground-based, Earth's orbital, and rocket telescopes, as well as from Venus' orbiters in the UV, IR, and microwave spectral ranges (Table 16.1). Ground-based sub-mm spectroscopy data (Sandor et al. 2010, 2012) suggest high temporal variability of $SO_2$ at 85 to 100 km, with higher $SO_2$ concentrations than at lower altitudes. Venus Express solar occultation data in the near-IR and UV ranges (Belyaev et al. 2012) also reveal greater mixing ratios at 85 to 100 km than above the clouds at 70 to 85 km. Venus Express UV data from stellar occultation for 85 to 105 km shows an increase in $xSO_2$ with altitude, with values 5 to 4 times higher on the nightside than on the terminator (Belyaev et al. 2017). Microwave observations with the ALMA telescope suggest significant spatial and diurnal variations by a factor of four and a short-scale patchy pattern of $xSO_2$ at ≈88 km (Encrenaz et al. 2015). HST observations in the UV range by Jessup et al. (2015) demonstrate variable abundances of $SO_2$ and SO at 74 to 81 km. Venus Express UV data also revealed spatial and temporal variability of $SO_2$ in the mesosphere and confirmed increasing $xSO_2$ with altitude at 70 to 100 km (Vandaele et al. 2017a, 2017b). A significant increase in $xSO_2$ from ≈0.02 ppmv below 90 km to ≈5 ppmv at 100 km is reported from the Venus Express infrared spectrometer (SOIR) data obtained via solar occultation (Mahieux et al. 2023). These data reveal variabilities over time and/or latitude of at least one order (typically, 2–3) of magnitude in the mesosphere. Ground-based IR mapping by Encrenaz et al. (2012) revealed $xSO_2$ variations between 50 and 175 ppbv over a 24-hour timescale at altitudes of 60 to 80 km. Similar observations indicate temporal and spatial variability ranging from 30 to 700 ppmv at the cloud top (Encrenaz et al. 2016, 2019, 2020, 2023).

Overall, $xSO_2$ at the cloud top (≈64–73 km) is much smaller than at the cloud base and below (Bertaux et al. 1996; Esposito et al. 1988, 1997; Marcq et al. 2018, 2020). The cloud top $xSO_2$ generally ranges from 10 to 400 ppbv, exhibiting variations across space, short-term (hourly to daily), and long-term (annual) scales, sometimes fluctuating by a factor of ten or more. Higher abundances are reported for the night side, and elevated $xSO_2$ values are observed at the morning terminator. $SO_2$ features associated with plumes at the cloud top have a lifetime ranging from a few hours to several tens of hours (Marcq et al. 2013; Encrenaz et al. 2016, 2019). Several observations reveal generally lower $xSO_2$ levels at higher latitudes and near the sub-solar point (Esposito et al. 1979; Marcq et al. 2011, 2013, 2020, 2021; Belyaev et al. 2012; Encrenaz et al. 2019, 2023; Jessup et al. 2015) (Fig. 16.2). In contrast to other data, orbital Venera 15 IR data obtained around 1983 (Schäfer et al. 1990; Zasova et al. 1993) and sounding rocket UV observations from 1988 and 1991 (McClintock et al. 1994; Na et al. 1994) indicated elevated cloud top$xSO_2$ levels at high latitudes. These observations suggest variable $xSO_2$ typically within 10–200 ppbv at 69 km, tens of ppbv $SO_2$ at latitudes below ≈45 °, and 100 to 200 ppbv (reaching one ppmv in places) in the polar regions.

Local and temporal low-latitude $SO_2$ enrichments at the cloud top are often associated with significantly more prominent plumes than those at higher latitudes (Encrenaz et al. 2019, 2023). In addition to plumes, localized $SO_2$ enrichments are observed above the western slopes of elevated Aphrodite Terra (Marcq et al. 2020), and concentration maps at the cloud top show elevated $xSO_2$ above other highlands (Encrenaz et al. 2023). The topography-related patterns may reflect orographic gravity waves reaching the clouds (Sect. 16.3.1.2).

Global cloud top $xSO_2$ systematically observed since 1978 indicates a decline by a factor of 8 ± 4 from 1978 to 1995 (Pioneer Venus and Venera 15 orbital UV data), followed by an increase from 1995 to 2006 (no observations), and a subsequent decline seen in the Venus Express UV and IR data (Fig. 16.3). For the period from 2006 to 2014, a moderate decrease in averaged low latitude cloud top $xSO_2$ values is observed in the HST UV (Jessup et al. 2015) and IR telescopic data (Encrenaz et al. 2016), as summarized by Marcq et al. (2020). Mid-IR ground-based data from Encrenaz et al. (2020) for 64 km revealed a long-term anti-correlation of $SO_2$ and $H_2O$ cloud top abundances. However, no correlation is observed in the data from 2019 to 2020 (Encrenaz et al. 2023).

Compared to the mesosphere and cloud top, $SO_2$ is significantly more abundant in the cloud layer (Table 16.1). For upper clouds at 62 km, Zasova et al. (1993) reported concentrations of 2 to 20 ppmv from Venera 15 IR data. For lower clouds at 51 to 54 km, Venus Express radio occultation data from 2006 to 2014 suggest a mean $xSO_2$ of 90 ± 60 ppmv at low latitudes and 150 ± 50 ppmv at polar latitudes (Oschlisniok et al. 2021). These data reveal a potential increase in high latitude $xSO_2$ (and $H_2SO_4$ vapor) starting from low values in 2006, a trend that contradicts the overall decline (some spikes counteracted the downward trend, Fig. 16.3) of cloud top $SO_2$ from 2006 to 2012 as shown in Venus Express UV data (Marcq et al. 2013). According to Oschlisniok et al. (2021), a mean $xSO_2$ of 180 ± 50 ppmv in lower clouds was observed in 2008–2011 at high northern latitudes, while an enhanced value of 140 ± 40 ppmv was reported for southern polar latitudes from 2009 to 2012. A gradual decrease in polar $xSO_2$ is observed after 2011/2012, with lower values noted in 2013. These patterns suggest long-term fluctuations in global circulation (Sect. 16.3.1.2). Below the clouds, $SO_2$ is much more abundant than in the middle atmosphere.

Although $SO_2$ variability decreases at lower altitudes, both altitudinal and latitudinal gradients are reported there. Pioneer Venus Large Probe GC data for 22 km unveil 185 ± 43 ppmv (Oyama et al. 1980), and Venera 12 GC data suggest 130 ± 35 ppmv $SO_2$ below 42 km (Gel'man et al. 1980). These measurements are comparable to 100 to 200 ppmv estimated from ground-based nightside near-IR data for 30 to 52 km (Bézard et al. 1993; Pollack et al. 1993; Marcq et al. 2021; Arney et al. 2014) and with analogous Venus Express data (Marcq et al. 2023). Ground-based nightside near-IR data from Marcq et al. (2021) reveal a latitudinal dependence of $xSO_2$ at ≈33 km: ≈130 ppmv at 10º S and 190–220 ppmv at 30–60º N (Fig. 16.4), with a latitudinally averaged $xSO_2$ of ≈180 ppmv. Vega in situ UV spectra at 10–60 km suggested a strong vertical $SO_2$ gradient (Bertaux et al. 1996). Their 120 ppmv and 150 ppmv $SO_2$ at 43 km and 52 km, respectively, are comparable with other data (Table 16.1). However, decreasing $xSO_2$ toward the surface to 25 to 40 ppmv below 20 km inferred by Bertaux et al. (1996) is inconsistent with in situ GC data. It may not be explained by atmospheric models (Sect. 16.3.1.2). These in situ and remote data disagree with the much higher and more uncertain $xSO_2$ values assessed for 42 km from PVLP mass spectrometry (up to 500 ppmv, Hoffman et al. 1980) and gas chromatography (Oyama et al. 1980), which could characterize thermal decomposition of captured aerosols (Hoffman et al. 1980; Donahue et al. 1982; Oyama et al. 1980; Mogul et al. 2025) rather than atmospheric $SO_2$.

**Sulfur Monoxide**

Sulfur monoxide is detected at the cloud top and in the mesosphere (Table 16.1). At altitudes of 64 to 96 km, xSO is typically within 1 to 200 ppbv. Sulfur monoxide was initially observed at the cloud top through Earth's satellite UV measurements (Na et al. 1990). Rocket UV observations by Na et al. (1994) suggested 7 to 17 ppbv SO at 64–96 km, which aligns with data from Na et al. (1990). Sandor et al. (2010, 2012) reported sub-millimeter telescopic observations that revealed diurnal variations and other SO variability by an order of magnitude at altitudes of 70 to 100 km. These data show a decrease in the $SO_2$/SO mixing ratio with altitude. The ratio is lower on the dayside (1.9 ± 1.2) than on the nightside (37 ± 15). Venus Express UV observations of the solar occultation of Belyaev et al. (2012) indicated an increasing $x$SO from ≈1 to 20 ppbv at ≈85 km to ≈50 to 200 ppbv at 95 km, along with a decrease in the $SO_2$/ SO ratio with altitude. A constant $x$SO of ~150 ± 50 ppbv is inferred for 95 to 103 km. Further analysis of these data by Belyaev et al. (2017) suggests decreasing the $SO_2$/SO ratio from 7 ± 1 at 85 km to 1.5 ± 0.5 at 100 km. The HST UV data for 80 km corresponds to $SO_2$/SO of ≈10 (Jessup et al. 2015). These HST measurements, coordinated with ground-based sub-mm data, suggest $x$SO within 1 to 30 ppbv at 74 to 81 km. Ground-based mm-wave data from Encrenaz et al. (2015) indicated significant local and diurnal SO variations at ≈88 km and suggested a uniform mixing ratio at higher altitudes. No SO is detected on the night side, consistent with its short lifetime and exclusive dayside photochemical source from $SO_2$ dissociation (Sect. 16.3.1.3). Belyaev et al. (2017) and Jessup et al. (2015) reported a long-term correlation between $xSO_2$ and the $SO_2$/SO ratio at altitudes of 70 to 95 km. However, the observations indicate episodic $SO_2$ increases occurring without a simultaneous rise in SO (Jessup et al. 2015; Belyaev et al. 2017). Similarly, Encrenaz et al. (2015) reported only a partial correlation between SO and $SO_2$ compositional maps.

**Sulfur Trioxide**

$SO_3$ is believed to be a transient gas that forms via the oxidation of sulfur-bearing gases at and above cloud top, converts to sulfuric acid in upper clouds, and is released through the thermal decomposition of $H_2SO_4$(g) and possibly ferric sulfates in the lower atmosphere (Sect. 16.3.1.4). Based on the Venus Express near-IR data (Mahieux et al. 2023), the average $xSO_3$ increases from 0.1 ppmv at cloud top to 10 ppmv at 95 km. No observational information on $SO_3$ is available within and below clouds.

**Sulfur Isotopes**

Sulfur isotope ratios in atmospheric gases were not reported in the initial analysis of LNMS data (Hoffman et al. 1980). Mogul et al. (2021) re-examined the initial LNMS mass counts and reported $^{34}S/^{32}S$ as $(5.8 \pm 0.7) \times 10^{-2}$ and $^{33}S/^{32}S$ as $(1.4 \pm 0.9) \times 10{-}2$ at altitudes of 26 to 39 km. These ratios were obtained from counts supposedly caused by the $SO^+$ ion, formed in the mass spectrometer as a fragment of sulfur-bearing cloud aerosol material that clogged the instrument's inlet. The $^{34}S/^{32}S$ ratio corresponds to $\delta^{34}S$ of 289 ± 156 ‰ relative to the Vienna Canyon Diablo troilite (V-CDT) standard value ($^{34}S/^{32}S$ = 0.045) and is significantly higher than known terrestrial materials, which have a $\delta^{34}S$ ranging from −50 ‰ to 40 ‰. The significant discrepancy with terrestrial values raises concerns about interpreting LNMS data regarding isotopically heavy sulfur. Elevated counts at 49 u attributed to $^{33}SO$ may be due to $H^{32}SO$, and some counts at 48 u may result from $H_2{}^{32}SO$ fragment rather than $^{34}SO$. The $^{33}S/^{32}S$ ratio is even less specific, and the current data analysis suggests a value of $(7 \pm 4) \times 10^{-3}$ (R. Mogul, personal communication).

The LNMS $\delta^{34}S$ value from Mogul et al. (2021) falls within the data range reported for cometary dust and gases (−400 ‰ to +600 ‰, Bergin et al. 2024). Venus' $^{34}S/^{32}S$ ratio matches that in $H_2S$ in comet Hale-Bopp at $(6.1 \pm 0.3) \times 10^{-2}$ (Crovisier et al. 2004). The ratio also pairs with that in the atmosphere of Io, $(5.95 \pm 0.38) \times 10^{-2}$, which could indicate preferential $^{32}S$ loss throughout the history of volcanic degassing of $SO_2$, $S_2$, and SO (De Kleer et al. 2024) and may also reflect an outer solar origin on the moon. If the interpretation of LNMS mass counts is confirmed to be accurate, it paves the way for conjectures about the contribution of cometary material to atmospheric sulfur. However, mass balance assessments indicate that space sources are not significant contributors to the atmospheric sulfur inventory (Sect. 16.4.3).

#### 16.2.1.2 Sulfuric Acid Vapor

Sulfuric acid vapor is abundant within clouds and in the sub-cloud ≈10 km thick layer. The concentration of $H_2SO_4$(g) varies and is typically within several ppmv, revealing altitudinal and latitudinal patterns (Table 16.1). In clouds and down to ≈30 km, $H_2SO_4$(g) is detected through ground-based microwave and sub-mm observations (Butler et al. 2001; Jenkins et al. 2002; Sandor et al. 2012) and radio occultations obtained with Mariner 10, Pioneer Venus, Magellan, Venus Express, and Akatsuki orbiters (Jenkins and Steffer 1991; Jenkins et al. 1994; Kolodner and Steffes 1998; Oschlisniok et al. 2012, 2021; Imamura et al. 2017; Ando et al. 2024). The only upper limit of 3 ppbv for $H_2SO_4$(g) is obtained at 85–100 km from ground-based microwave observations (Sandor et al. 2012). Jenkins et al. (2002) used microwave observations to retrieve $H_2SO_4$(g) profiles at altitudes below 100 km. They reported $xH_2SO_4$(g) at approximately 0.1 to 10 ppmv at altitudes of 30 to 55 km, with peak concentrations at 46 km, and indicated a substantially higher abundance at latitudes above 45°. Mariner 10, Magellan, and Akatsuki radio occultations suggest an increase in $xH_2SO_4$(g) at lower altitudes that roughly follows the saturation curve with sulfuric acid solution in middle and lower clouds at 47 to 56 km (Kolodner and Steffes 1998; Imamura et al. 2017; Ando et al. 2024), as shown in Fig. 16.5. Ando et al. (2024) noted an annual correlation of $xH_2SO_4$(g) with temperature at 46 to 50 km, supporting the effect of saturation in lower clouds. Venus Express radio occultation data from Oschlisniok et al. (2012) indicate ≈1 to 2 ppmv $H_2SO_4$(g) within clouds (50–55 km), and values of <1 ppmv are reported for southern polar latitudes. At higher latitudes, the highest concentrations are observed at lower altitudes, consistent with the results from Kolodner and Steffes (1998). A subsequent analysis of VeRa data by Oschlisniok et al. (2021) confirmed these findings. It showed that the topside of a sub-cloud $H_2SO_4$(g)-rich layer is located ≈4 km higher at equatorial latitudes than at polar latitudes. In the equatorial region, a maximum of 12 ppmv is reported at 47 km, while polar latitudes have a maximum of 9 to 12 ppmv at about 43 km. These VeRa data show significant variability in the equatorial and polar regions. In contrast, the middle latitudes are characterized by lower and less variable mixing ratios (5–7 ppmv) at altitudes of 43 to 47 km. These data indicate an increased $H_2SO_4$(g) column density at 42 to 47 km in the northern polar latitudes from 2006 to 2011. No notable distinction was found in the global distribution of $H_2SO_4$(g) between the dayside and nightside regions in these VeRa data. For altitudes between 47 and 38 km, Akatsuki sub-cloud radio oscillations indicate an uneven decrease in $xH_2SO_4$(g) from 10 ± 2 ppmv at the cloud deck to 3.5 ± 1.5 ppmv at 38 km (Imamura et al. 2017) and to 6 ± 1 ppmv at 42 ± 2 km (Ando et al. 2024). The behavior of $H_2SO_4$(g) gas within and below clouds is governed by saturation with respect to aerosol and thermal dissociation, respectively (Sect. 16.3.1.5).

#### 16.2.1.3 Reduced Gases (OCS, CO, $S_n$, and $H_2S$) Carbonyl Sulfide

None of the in situ measurements provided reliable data on OCS. Krasnopolsky (2008, 2010b) used ground-based IR spectroscopy to detect OCS at 64 to 72 km and reported a highly variable content at the cloud top. Higher abundances occur at lower latitudes, peaking at 0–15° N. Venus Express near-IR solar occultation data initially provided the upper limit for OCS at 70–90 km (Vandaele et al. 2008). Based on this dataset, Mahieux et al. (2023) reported an average $x$OCS of 1 ppbv at 65 km and a significantly higher concentration of 1 ppmv at 100 km, with variability spanning up to four orders of magnitude. No dependence on latitude, time, or the side of the terminator has been observed; however, spatial decreases in detections were reported in the 0 to 30° latitude region. The data from Mahieux et al. (2023) and Krasnopolsky (2008, 2010b) are consistent in the altitude range of 64 to 72 km. OCS was observed from the ground through near-IR nightside spectral windows (Bézard et al. 1990; Pollack et al. 1993; Marcq et al. 2005, 2006; Taylor et al. 1997; Arney et al. 2014) and from the Venus Express orbiter (Marcq et al. 2008, 2018, 2023). Three observations assessed *x*OCS between the cloud deck and 30 km. The nightside near-IR data for 33 km (Pollack et al. 1993) and 36 km (Marcq et al. 2006, Arney et al. 2014) suggest OCS gradients of $-1.6 \pm 0.3$ ppmv km$^{-1}$ and $-0.4 \pm 0.2$ ppmv km$^{-1}$, respectively. These data indicate a decrease in *x*OCS by a factor of ten from 30 to 36 km. Extrapolating OCS gradients from Pollack et al. (1993) and Marcq et al. (2006) results in *x*OCS values ranging from 8 to 36 ppmv at 29 km (Krasnopolsky 2007). Marcq et al. (2018) discussed the applicability of a constant logarithmic gradient ($d \log x\mathrm{OCS}/d \log P$) over the 30 to 37 km interval. There are indications of more abundant sub-cloud OCS at low latitudes (Marcq et al. 2005, 2006, 2008, 2023). For ≈30 to 40 km altitude, Marcq et al. (2023) reported higher *x*OCS at lower latitudes and the lowest values at high northern polar latitudes. No data is available for lower altitudes.

### Carbon Monoxide

The mixing ratio of CO in the lower atmosphere (Table 16.1) is comparable to that of OCS at an altitude of ≈30 km, though the compositional gradients are opposite. The negative correlation between OCS and CO in the lower atmosphere with altitude (Pollack et al. 1993; Marcq et al. 2005, 2006, 2008; Arney et al. 2014; Marcq et al.2018) and latitude (but not longitude, Marcq et al. 2023) indicates a significant role of CO in the atmospheric chemistry of OCS and other sulfur-bearing gases (Yung et al. 2009; Sect. 16.3.1). Gas chromatography data from the PVLP (Oyama et al. 1980) and Venera 12 (Gel'man et al. 1980; Krasnopolsky 2007) suggest *x*CO between 12 and 37 ppmv in the altitude range of 12 to 42 km. These measurements indicate a decrease in *x*CO towards the surface, possibly reaching ≈9 ppmv at the modal radius, as Fegley et al. (1997a, 1997b) estimated. Ground-based near-IR nightside observations (Pollack et al. 1993; Taylor et al. 1997; Marcq et al. 2005, 2006; Arney et al. 2014) and Venus Express (Marcq et al. 2008, 2018, 2023) confirmed these in situ data and the increase in xCO with altitude at 33 to 36 km. Galileo, ground-based, and Venus Express near-IR observations indicate higher *x*CO at high latitudes for 33–36 km (Collard et al. 1993; Arney et al. 2014; Marcq et al. 2005, 2006, 2008, 2018, 2021, 2023), Fig. 16.4. Based on the comprehensive data from Venus Express, a notable difference of ≈15 ppmv has been assessed between the equatorial and polar regions (Tsang and McGouldrick 2017). Interestingly, observations by Marcq et al. (2023) for lower latitudes suggest that zonal CO variability (and possibly OCS) is correlated with the longitudinally shifted average surface elevation, indicating that topography affects circulation. Models suggest the formation of CO via $CO_2$ photolysis above clouds and oxidation of OCS below clouds, as well as its consumption through conversion to OCS closer to the surface (Sect. 16.3.1).

### $S_n$ Gases

The strong absorption observed in low-resolution spectra at 0.45 to 0.6 μm, obtained in situ with Venera 11, 12, 13, and 14 probes below ≈30 km, has been interpreted in terms of light absorption by $S_3$ and $S_4$ gases (Moroz et al. 1981, 1983; Golovin et al. 1982; San'ko 1981; Maiorov et al. 2005; Krasnopolsky 1987, 2013). Moroz et al. (1981) evaluated $xS_3$ at 3 to 15 km (Table 16.1). The interpretation of absorption spectra in terms of concentration required knowledge of sulfur speciation. Calculations of chemical equilibria in the Sn system by San'ko (1981), based on thermodynamic data from Mills (1974), suggested $xS_3/xS_4$ of $\approx 10^4$ at 0 to 40 km and implied a significant contribution of $S_3$ to absorption. Zolotov (1985) calculated the speciation of the $S_n$ ($n = 1 - 8$) system, based on Gurvich et al. (1989–1994), and found significantly lower $xS_3/xS_4$ ratios alongside comparable concentrations of these gases above ≈25 km. Krasnopolsky (1987) assessed the absorption cross-section of $S_3$ to constrain its abundance at 5 to 25 km by considering $S_n$ speciation based on Mills (1974). Maiorov et al. (2005) separated the effect of Rayleigh scattering from true sulfur absorption and inferred $xS_3$ and $xS_4$ based on Venera 11 spectra. Krasnopolsky (2013) noted that using the thermodynamic data of Mills (1974) results in extremely low $S_4$ densities and argued for a higher $xS_4/xS_3$ ratio at chemical equilibrium. He used a new experimental absorption cross-section of $S_n$ gases and sulfur absorption spectra from Maiorov et al. (2005) to evaluate $xS_3$ and $xS_4$ at 3 to 19 km. All these evaluations suggest that $xS_3$ and $xS_4$ are within $10^{-10}$ to $10^{-12}$.

Although $S_2$ gas is not detected anywhere in the atmosphere, it is the most abundant $S_n$ gas in the near-surface atmosphere according to chemical equilibrium calculations (Table 16.2; e.g., San'ko 1981; Zolotov 1985; Fegley et al. 1997b). If $xS_2$ is derived from $xS_3$, then $xS_2 =$ (0.01–0.02) ppmv (Golovin et al. 1982; Moroz et al. 1981, 1983; Zolotov 1985; Model 10 of Table 16.2); if it is calculated based on measured $SO_2$, CO, and $CO_2$ abundances, then $xS_2 =$ (0.1–0.4) ppmv (Table 16.2).

### Hydrogen Sulfide

The concentration of $H_2S$ is insufficiently constrained from remote and in situ measurements. Venus Express solar occultation data provide highly uncertain upper limits at altitudes of 60 to 100 km (Mahieux et al. 2023), while Krasnopolsky (2008) assessed an upper limit for cloud top based on ground near-IR observations (Table 16.1). The interpretation of the Pioneer Venus Large Probe LNMS data initially suggested an $H_2S$ (if counts at 34 u correspond to $H_2S$) gradient in the sub-cloud atmosphere, an average of 3 ± 2 ppmv from 24 km to the surface, and about one-third of this value in clouds (Hoffman et al. 1980). The subsequent LNMS report mentions only a marginal detection of ≈1 ppmv $H_2S$ in clouds (von Zahn et al. 1983). PVLP GC data provided an upper limit of 2 ppmv at 22 km (Oyama et al. 1980), and $H_2S$ is absent from Venera 12 CG data (Gel'man et al. 1980). The reported 80 ± 40 ppmv $H_2S$, along with abundant $O_2$ and $H_2$ based on Venera 13 and Venera 14 GC data at 29 to 37 km (Mukhin et al. 1982, 1983), is commonly considered incorrect (Krasnopolsky 1986) and has not been mentioned in comprehensive reviews by Esposito et al. (1997), Mills et al. (2007), and Marcq et al. (2018). All reported $xH_2S$ values at or above the ppmv level have not been independently confirmed and are inconsistent with atmospheric models (Sect. 16.3.1.9). Mogul et al. (2021) reported possible signs of $H_2S$ in the LNMS mass counts at 34 u related to middle cloud samples, particularly if $PH_3$ (phosphine) is absent. As with $SO_2$, the suspected presence of $H_2S$ in gases analyzed below clouds could reflect the thermal decomposition of captured non-$H_2SO_4$ cloud aerosols rather than atmospheric composition (Zolotov et al. 2023). In addition to $H_2S$, the interpretation by Mogul et al. (2021) does not exclude a variety of reduced gases that are not expected to be abundant in the clouds but could be released from heated aerosols. All in situ data on $H_2S$ may reflect the thermal decomposition of captured aerosols, consistent with Vega GC data (Porshnev et al. 1987) on possible $H_2S$ release from heated aerosols (Sect. 16.2.1.4). Chemical equilibrium models for the near-surface atmosphere (Table 16.2) suggest an order of magnitude less $x$$H_2S$ than Hoffman et al. (1980) reported.

#### 16.2.1.4 Aerosols in Clouds and Hazes

Multiple remote and mission data provide information on the physics of clouds and hazes (refractive index, particle size, mass loads, layering, plumes, winds, etc.), as reviewed by von Zahn et al. (1983), Esposito et al. (1983, 1997), Ragent et al. (1985), Krasnopolsky (1989, 2006), Titov et al. (2018), and Diai et al. (2025). In the 1970s, information on the cloud vertical profile was obtained via nephelometry aboard Venera 9, 10, and 11 (Marov et al. 1980), as well as from the four Pioneer Venus entry probes (Ragent and Blamont 1980; Ragent et al. 1985), and the PVPL cloud particle size spectrometer (Knollenberg and Hunten 1979). The analysis of the Pioneer Venus data by Knollenberg and Hunten (1980) revealed a primary cloud with a thickness of 20 km (47.5–70 km), surrounded by lower density upper (70–90 km) and lower hazes (31–47.5 km) (Fig. 16.6). According to their data, the primary cloud layers consist of upper (56.5–70 km), middle (50.5–56.5 km), and lower (47.5–50.5 km) clouds. Particles have distinct sizes that may reflect different compositions, 0.3 to 0.4 μm (Mode 1), 2 to 2.5 μm (Mode 2), and 7 to 8 μm particles that contain the bulk of cloud mass loading (Mode 3). All size Modes are suggested in the middle and lower clouds, Modes 1 and 2 are inferred in the upper clouds, and Mode 1 particles are the presumed constituents of the upper and lower hazes. Based on optical properties, Knollenberg and Hunten (1980) noted that Mode 1 particles could be rich in sulfur compounds, Mode 2 likely consists of spherical droplets of sulfuric acid solution, and Mode 3 may consist of non-spherical solid grains.

The cloud structure inferred from Venera 13 and 14 probes' spectrometer data is reported by Grieger et al. (2004). The cloud structure obtained with the particle size spectrometer and nephelometer onboard the Vega probes (Moshkin et al. 1986; Gnedykh et al. 1987) was generally consistent with the Pioneer Venus data and confirmed a sharp boundary for the primary cloud layer and the presence of hazes down to ≈30 km. Two particle size modes (0.25–2.5 μm and 1–2.5 μm) were reported in the middle and lower cloud. The Mode 2 particles were significantly less numerous than in the Pioneer Venus data, and Gnedykh et al. (1987) suggested the non-sphericity of small particles. Vega data suggest a dense (0.1–2 mg $m^{-3}$) sub-cloud haze of small (<0.25 μm) particles down to ≈35 km. The increasing aerosol mass load from the upper to lower cloud regions was inferred from a collection of aerosols on filters onboard the Venera 12 (Surkov et al. 1982a) and Vega (Andreichikov et al. 1987) probes. Vega in situ measurements of UV light absorption

from 63 km to the surface, reported by Bertaux et al. (1996), confirmed the layered cloud structure consistent with the Pioneer Venus data (Fig. 16.6). Other investigations indicated a latitudinal dependence of the cloud structure and altitudinal boundaries (e.g., Cottini et al. 2012). Venera 15 IR observations suggest Mode 3 particles are only present at low altitudes (Zasova et al. 2007). All the listed in situ cloud physics data imply spatial and temporal variability. In addition to hazes above and below clouds, Grieger et al. (2004) suspected a near-surface haze layer at 1 to 2 km altitude based on Venera 13 and 14 in situ spectrophotometry. Further interpretation of Venera 13 data by Kulkarni et al. (2025) suggests the presence of sub-micron basaltic dust particles located 3.5 to 5 km above the surface.

#### Bulk Sulfur Content, Other Elements, and Aerosol Decomposition Products

The presence of condensed sulfur-bearing compounds in clouds is supported by the detection of bulk sulfur by the XRF method in aerosols collected on filters by Venera 12, Venera 14, Vega 1, and Vega 2. The interpretation of Venera 12 data suggests the presence of Cl (Petryanov et al. 1981a) and Fe (Petryanov et al. 1981b). Based on this experiment, Surkov et al. (1982a) did not report any Fe but estimated a sulfur mass load of 0.1 mg $m^{-3}$ and a Cl/S mass ratio of ≈21 in aerosol collected at 47 to 54 km. In a comparative study, Surkov et al. (1982b) estimated sulfur mass loads to be $1.1 \pm 0.13$ mg $m^{-3}$ and determined the Cl/S mass ratio of ≈0.09 in the Venera 14 samples collected at 47 to 63 km. These data indicated the highest aerosol density at 47 to 56 km. Vega XRF data reported by Andreichikov et al. (1987) suggested that abundant sulfur exists in the primary cloud aerosol at altitudes of 47 to 61 km, while the presence of Cl, Fe (possible), and P was also indicated. Their data analysis implied an alternating dominance of sulfur, Cl, and P in narrow cloud layers, though Krasnopolsky (1989) questioned the validity of such a model-based interpretation. The supremacy of P over sulfur at 47 to 51 km, as implied by Andreichikov et al. (1987), and Venera 12 data on the high Cl/S ratio (Surkov et al. 1982a) disagree with $xH_2SO_4(g)$ values (Sect. 16.2.1.2) at equilibrium with sulfuric acid in the middle and lower cloud regions (Fig. 16.5) and with other measurements described below.

Mass spectrometry and gas chromatography data from PVLP and Vega probes indicate the presence of condensed sulfur-bearing compounds in cloud aerosols. Although the LNMS instrument was not designed to analyze aerosols, aerosols entered the inlet in the primary cloud and clogged the inlet at ≈51 km (Hoffman et al. 1980; Mogul et al. 2023, 2025). These interpretations suggest that sulfur-bearing fragments identified in the mass counts below ≈51 km represent thermally decomposed aerosols captured and temporarily stored in the LNMS inlet assembly. The mass counts indicate the release of $SO_2$, $H_2O$, and trace sulfur-bearing gases (e.g., $H_2S$, $S_n$) from the aerosols (Hoffman et al. 1980; Donahue et al. 1982; Mogul et al. 2021, 2023, 2025; Zolotov et al. 2023). Significantly higher than the atmospheric concentrations of $SO_2$ and $H_2O$ reported for the 42 km sample by PVPL GC measurements could also reflect the decomposition of aerosol captured in the primary cloud (Oyama et al. 1980; Mogul et al. 2025). Vega 1 mass spectrometric analysis of collected and pyrolyzed (up to 673 K) aerosols suggested the release of sulfur-oxygen and Cl-bearing gases, with bulk estimates of 2–10 mg $m^{-3}$ for sulfur and >0.3 mg $m^{-3}$ for Cl in the lower cloud aerosol (Surkov et al. 1986a, 1987a). In another investigation, aerosols from Vega 1 and Vega 2 were collected in two separate cells, heated to 353 K and 573 K, and the

released water vapor and $SO_2$ were detected in the Vega 2 samples using gas chromatography (Gel'man et al. 1986). In addition to $H_2O$ and $SO_2$, Porshnev et al. (1987) reported traces of $H_2S$ and OCS ($H_2S/SO_2 = OCS/SO_2$ of $\approx 10^{-3}$) possibly released from heated Vega 2 aerosols.

## Sulfuric Acid Solution

The presence of liquid sulfuric acid solution as a major cloud aerosol component was inferred from ground-based polarimetry and near-IR spectroscopy, the refractive index of clouds, mass spectrometry, and gas chromatography data on captured aerosols. Observations commonly suggest >75% sulfuric acid ($H_2SO_4$ wt% in liquid $H_2SO_4 \cdot nH_2O$) and exhibit altitudinal, longitudinal, spatial, and temporal variations in composition. Young (1973) first noted that the IR spectra of Venus are consistent with the presence of sulfuric acid in the upper clouds. Hansen and Hovenier (1974) interpreted ground-based polarimetric data at visible and near-IR wavelengths regarding scattering from 1 μm particles with a refractive index of 75% sulfuric acid. The spectroscopic ground-based near-IR data from Pollack et al. (1978) and Krasnopolsky (2008) inferred a refractive index of 1 μm practices corresponding to 85% $H_2SO_4$ at cloud top. Polarimetry data from the Pioneer Venus atmospheric probes on the refractive index suggested an 85% sulfuric acid concentration. The nightside Venus Express near-IR observations suggested 75 to 85% acid in optically thin upper clouds and 90 to 100 wt% in optically thick lower clouds (Barstow et al. 2012). These observations also indicate an increase in concentration from 80 wt% at low latitudes to 90 wt% poleward of 60° S. The interpretation of this dataset by Cottini et al. (2012) for ≈1 μm cloud top droplets suggested that the acid concentration ranges from 75 to 85%, with a narrower range of 80 to 83% at low latitudes (±40°). The analysis of these data by McGouldrick et al. (2021) suggests >60% acid in upper clouds and 85 to 90% in lower clouds, showing a latitudinal dependence. The nightside ground-based near-IR observations of Arney et al. (2014) were used to map $H_2SO_4$ concentration in the aerosol and reported an average sulfuric acid concentration of 79 ± 4%. These observations indicate a correlation between $H_2SO_4$ concentration and the highly variable cloud opacity, revealing slightly more systematic changes with altitude and latitude. More concentrated acid is tentatively suggested for the northern hemisphere and low latitudes. A positive correlation between $xSO_2$ and altitude revealed by Venera 15 orbital IR observations from Zasova et al. (1993) is consistent with sulfuric acid aerosol (75–85%) being a significant component of upper clouds (Mills et al. 2007).

An 85% sulfuric acid concentration in lower clouds was inferred from the amount of sulfur-oxygen-hydrogen species released from the clogged LNMS inlet assembly at the Pioneer Venus Large Probe (Hoffman et al. 1980). Recent analysis of LNMS data suggests 86 ± 17 wt% $H_2SO_4$ and 14 ± 4 wt% $H_2O$ in the liquid phase of aerosol collected at altitudes between 48 and 51 km (Mogul et al. 2025). This study also suggests a mass load of $H_2SO_4$ of $1.5 \pm 0.2$ mg m$^{-3}$ in that aerosol. Results of Vega 1 reaction gas chromatography analysis confirmed the dominance of $H_2SO_4$ in collected and heated aerosols, with an average $H_2SO_4$ mass load of $0.6 \pm 0.1$ mg m$^{-3}$ at altitudes of 48 to 61.5 km, though significantly lower loads were inferred from analogues Vega 2 data (Gel'man et al. 1986; Porshnev et al.

1987). The mass spectrometric detection of abundant $SO_2$ (64 u) released from pyrolyzed Vega 1 aerosol aligned with the sulfuric acid composition and indicated a lower limit of aerosol $H_2SO_4$ mass load of 2 mg $m^{-3}$ (Surkov et al. 1986a, 1987a).

The presence of sulfuric acid in the primary cloud aerosols aligns with the physical chemistry of the $H_2SO_4$-$H_2O$ system that is consistent with measured $xH_2SO_4(g)$ (Sect. 16.2.1.2, Fig. 16.5) and the modeled vaporization of liquid aerosol particles under temperature-pressure conditions at ≈48 km (Krasnopolsky and Pollack 1994; Krasnopolsky 2015; Dai et al. 2022) (Sect. 16.3.1.5). The composition of hazes above and below clouds remains unknown; however, metastable $H_2SO_4$ aerosols are suspected to be the main component of the upper hazes at 70 to 90 km (Luginin et al. 2016). However, higher concentrations of $SO_3$, SO, $SO_2$, OCS, CS, and $CS_2$ in the upper mesosphere (Table 16.1) align more closely with the cosmogenic origins of mesospheric sulfur-bearing aerosols (Sect. 16.3.1).

#### Other Sulfur-Bearing Species

Although only sulfuric acid is detected in the clouds, several physical and compositional data suggest the presence of other sulfur-bearing species in aerosols, particularly in Mode 1 and Mode 3 particles. A long list of sulfur-bearing compounds has been considered to account for a blue-UV absorber in upper clouds (Sect. 16.3.1.8). Condensed native sulfur, often referred to as polyatomic sulfur or polysulfur ($S_8$, $S_x$), is widely recognized as a component of cloud aerosol in chemical, photochemical, and physical models (Sect. 16.3.1.7). Analysis of captured aerosols onboard Vega suggested that condensed polysulfur exists in clouds, based on reaction gas chromatography, which tentatively indicated a $H_2SO_4/S_n$ mass ratio of ≈7 to 10 in the aerosol (Porshnev et al. 1987).

In addition to the potential Fe-, Cl-, and P-containing compounds discussed above, non-acid sulfur-bearing aerosol constituents were suspected based on PVPL mass spectrometry (Hoffman et al. 1980; Mogul et al. 2021, 2025) and Vega 2 gas reaction gas chromatography (Gel'man et al. 1986; Porshnev et al. 1987). The release of $H_2S$ and OCS from captured aerosols (Hoffman et al. 1980; Porshnev et al. 1987; Zolotov et al. 2023) indirectly suggests the presence of inorganic and/or organic compounds that contain reduced sulfur. The apparent presence of abundant sulfur-bearing species ($SO_2$, S, $H_2S$, $SO_3$, etc.), suspected in the LNMS mass counts below ≈25 km, could be products of the thermal decomposition of non-$H_2SO_4$ aerosols (e.g., metal sulfates, cosmogenic organic matter) captured from clouds and sub-cloud hazes (Mogul et al. 2025; Zolotov et al. 2023). Mogul et al. (2025) assessed that ferric sulfates are significant components of aerosols in lower clouds (48–51 km altitude), with $Fe_2(SO_4)_3$ comprising 16 ± 3 wt% and a mass load of 1.0 ± 0.2 mg $m^{-3}$. Knollenberg and Hunten (1980) mentioned crystalline sulfates as candidates for Mode 3 grains.

### 16.2.2 Surface and Crustal Materials

The morphology of surface features seen in radar images indicates the dominance of basaltic materials on the plains and in other volcanic formations. Venera/Vega lander data

suggest typical sulfur abundance and mineralogy found in planetary basalts, with a notable contribution from exogenic sulfur. The abundance and mineralogy of sulfur in radar-bright, low-thermal-emissivity highlands and in enigmatic geological formations such as steep-sided domes, the Venera 8 site, the canali, and tessera terrain are much less specific. The critical question regarding constraints on sulfur abundances in non-basaltic crustal materials hinges on whether Venus experienced oxidation of crustal and mantle materials, an aqueous history, or plate tectonics (Sect. 16.4).

#### 16.2.2.1 Mafic Igneous Materials: Primary and Altered

Venera 13, Venera 14 (1982), and Vega 2 (1985) landers provided information on rock-forming elements via XRF analysis (Table 16.3). These measurements represent the only data available on the sulfur content in surface and crustal materials. The composition of rock-forming elements in these samples, along with the concentrations of U, and Th in the Vega 2 rocks (Table 16.4), suggests a mafic silicate composition. Aside from the abundance of bulk sulfur, the major element composition of rocks from Venera 14 and Vega 2 resembles that of olivine basalts found on oceanic islands, oceanic plateaus, and within greenstone belts, as well as in the N-MORB tholeiite series (e.g., Surkov et al. 1984; Barsukov 1986d; Barsukov 1992; Kargel et al. 1993; Basilevsky et al. 2007; Treiman 2007; Filiberto 2014). A factor analysis of Venus and oceanic igneous rocks by Ivanov (2016) indicates a similarity between the Venera 14 rock, MORB, and oceanic island basalts. Barsukov et al. (1986d) assigned the composition of the Vega 2 rocks as olivine-gabbro-norite, which is an igneous rock common in Precambrian layered intrusions and Mesozoic ophiolite complexes. Potassium-rich Venera 13 composition suggests an alkaline mafic rock (e.g., leucitic basalt, olivine leucitite) (Barsukov et al. 1982b; Barsukov 1992; Nikolaeva 1990; Kargel et al. 1993; Filiberto 2014; Treiman 2007). No terrestrial analogs for Venera 13 and Vega 2 rocks are inferred through factor analysis by Ivanov (2016). The large error bars for major elements limit the identification of better analogs among mafic igneous rocks. The lack of data on Na and trace elements also constrains a definitive determination of the petrological type of these three samples.

Layered and porous rocks at the landing sites of the Venera 13 (Fig. 16.1) and Venera 14 could be presented by physically weathered lava flows, pyroclastic deposits, or airborne deposits of impact-generated particles (Garvin et al. 1984; Basilevsky et al. 1985, 2004). The Magellan radar-based morphology of the landing site regions (Weitz and Basilevsky 1993; Basilevsky et al. 2007) is consistent with a basaltic composition. The bulk sulfur content in the Venera 13 and Vega 2 samples is significantly higher than that found in typical terrestrial, lunar, martian basalts, and eucrites (≈0.05–0.3 wt%), suggesting exogenic (Sect. 16.3.2) rather than cosmic (Sect. 16.4.3) sources of sulfur. Although high sulfur contents are expected in S(VI)-bearing magmas oxidized beyond the Ni-NiO (NNO) buffer (e.g., Baker and Moretti 2011), more reduced magmas with S(-II) are suggested for Venus' plain-forming basalts, as discussed below.

The overall low dielectric constant (relative dielectric permittivity) (4 to 6, Pettengill et al. 1988, 1992, 1997) of the surface materials below ≈6054 km radius and in the landing site regions does not indicate abundant phases such as Fe sulfides and/or oxides like magnetite. The low dielectric constant (<8) in older surface materials found in lowlands, including the landing sites of Venera 13, 14, and Vega 2, may indicate a scarcity of Fe sulfides or oxides

in mafic bedrocks and/or products of their chemical alteration. A low sulfide content in sulfur-rich Vega 2 materials is indirectly supported by in situ measured specific electrical resistance of $\approx 10^6$ Ωm, characteristic of terrestrial basalts at 770 K. In contrast, the electrical resistance of the Venera 13 and 14 materials (89 and 73 Ωm, respectively, Kermurdzhian et al. 1983) suggests highly conductive materials. However, these low electrical resistance values are inconsistent with ordinary radar emissivity and the overall low dielectric constant of these and other landing regions (Venera 8, 9, and 10, and Vega 1 and 2) reported by Garvin and Head (1986a, b) and Weitz and Basilevsky (1993). The abundance of Fe sulfides and/or oxides in the Venera 13 and 14 materials is inconsistent with their Fe content (Table 16.3), which is typical for mafic rocks. Therefore, the electrical resistance data for Venera 13 and 14 materials may be flawed.

The regional geology of the Venera 9, 10, 13, and 14 and Vega 1 and 2 landing ellipses (Weitz and Basilevsky 1993; Basilevsky et al. 2007) suggests a dominance of plains formed by high-yield eruptions of low-viscosity magmas (Head et al. 1992; Kargel et al. 1993; Crumpler et al. 1997; Ghail et al. 2024). The morphology, density, and hardness of rock fragments at the landing sites of Venera 9 and 10 are consistent with physically degraded lava fragments (Fig. 16.7) (Florensky et al. 1977, 1983; Garvin et al. 1984; Basilevsky et al. 1985). The low albedo in the visible and near-IR spectral ranges (Golovin et al. 1983) supports a mafic composition at both sites. In addition to the results of the XRF analyses (Table 16.3), the mafic composition of the surface rock is indicated by the concentrations of K, U, and Th at the landing sites of Venera 9 and 10 and Vega 1 and 2 (Table 16.4). Specifically, K, U, and Th concentrations resemble those found in sub-alkaline basalts and gabbro (Basilevsky et al. 1992; Kargel et al. 1993; Treiman 2007). However, no perfect match exists, particularly regarding U and/or Th. Nikolaeva (1995) demonstrated that all low-K Venus' rocks are significantly enriched in K, U, and Th compared to N-MORB. Nikolaeva (1997) noted that Venus' rocks (Table 16.4, except Venera 8) have higher U content than island-arc tholeiitic basalts, and Venera 9 material is also enriched in Th.

The morphology of geological features on the vast plains observed in orbital radar images (Fig. 16.8) is consistent with basaltic volcanism (Barsukov et al. 1986a; Head et al. 1992; Crumpler et al. 1997; Ghail et al. 2024). The common occurrence of small shield volcanoes ($<\approx 500$ m in height, 2–20 km in diameter) with gentle ($<5°$) slopes on the vast plains suggests mainly effusive eruptions of volatile-poor mafic melts (Aubele and Slyuta 1990; Head et al. 1992; Hahn and Byrne 2023). The morphology of 93 large volcanoes, with a mean diameter of about 280 km (and several sub-types with mean diameters of 140, 280, and 525 km and different aspect ratios), identified by Ivanov and Head (2025), indicates that more localized effusive basaltic eruptions occurred after the formation of volcanic plains. Vivid lava morphology at volcanic centers (Fig. 16.9) is also consistent with basalts, though $\approx 1$ μm surface daytime emission reported by Mueller et al. (2008) suggests the ultramafic composition of a volcanic rise.

Although sulfur has not been considered in Venus igneous petrology models and discussions, terrestrial analogs and experimental solubility data can be invoked to assess the fate of igneous sulfur. As on Earth, the abundance and mineralogy of magmatic sulfur should depend on the redox state of magma and the Fe(II) content (Wallace and Carmichael 1992; Baker and Moretti 2011; Boulliung and Wood 2022, 2023; Chap. 12: Simon and Wilke 2026). The cited works suggest that magmatic sulfur has a minimum solubility at 1.0–1.5

log $fO_2$ units above the quartz-fayalite-magnetite (QFM) buffer, which is near the Ni-NiO buffer. At this minimum solubility, the abundance of S(-II) equals that of S(VI). The solubility strongly increases at log $fO_2$ <IW-3 and can reach a few wt% sulfur (e.g., Kiseeva and Wood 2015; Namur et al. 2016) when sulfur forms complexes with Ca and Mg in the silicate melt. Such a high solubility may characterize molten enstatite chondrites and aubrites (McCoy et al. 1999; Berthet et al. 2009), as well as mafic magmas on Mercury (Chap. 15: Renggli et al. 2026), but it does not apply to mafic melts that formed Venus' plains. Wänke et al. (1973) noted that the FeO/MnO ratio in terrestrial, lunar, and achondrite samples correlates with $fO_2$ in the magmatic system. Schaefer and Fegley (2017) stated that the FeO/MnO ratio in surface materials (Table 16.3) suggests $fO_2$ of corresponding melts between terrestrial upper mantle (≈QFM buffer, Frost and McCammon 2008) and martian (≈IW to IW + 2 log $fO_2$ units, Wadhwa 2008) values. However, FeO vs. MnO plots from Kargel et al. (1993) with error bars of XRF data do not indicate a difference between Venus' and terrestrial mafic rocks. The Earth-like (at and slightly below QFM) moderately reducing conditions do not imply different sulfur solubility in magma than in terrestrial counterparts. The redox state of Venus' mafic igneous rocks implies Earth-like sulfide mineralogy (pyrrhotite, pentlandite, Fegley and Treiman 1992a) and ≈0.1 wt% S (likely within 0.05–0.2 wt%) in unaltered basalts (Moore and Fabri 1971; Wallace and Carmichael 1992; Chap. 13: Kiseeva et al. 2026). Neither magmatic sulfates (e.g., anhydrite) nor Mg/Ca sulfides (oldhamite, niningerite) that could contribute to elevated sulfur content are expected in low-viscosity lavas that formed plains and volcanic centers. Therefore, abundant sulfur in the Venera 13 and Vega 2 samples could be in secondary sulfates, consistent with the expected chemical alteration pathways of basaltic glasses and Ca-rich pyroxenes at Venus' conditions (Sect. 16.3.2).

Morphologically fresh crater outflows and corresponding airborne parabolic deposits of ejected materials (Basilevsky et al. 2004) are characterized by a slightly higher dielectric constant (≈8, Campbell et al. 1992), which could result from a higher abundance of Fe sulfides in recently excavated materials. Magellan radar data from Pettengill et al. (1992) regarding ejecta from large craters (e.g., Mead, Stanton, Boleyn) imply the delivery of materials with a high dielectric constant, suggesting elevated abundances of such minerals at depths of one kilometer or more. The decimeter-wave emission data indicate the presence of abundant Fe sulfides and/or oxides below 0.7 to 0.9 meters beneath the layer with a dielectric constant of ≈4.5 (Antony et al. 2022). However, interpreting radar data related to materials with varying dielectric properties is not straightforward. The physical structure—such as the shape, size, or distribution of rock fragments—can influence microwave emissivity (Ford and Pettengill 1983; Pettengill et al. 1992), particularly through multiple (volume) scattering (Tryka and Muhleman 1992; Pettengill et al. 1992, 1996; Campbell et al. 1999; Bondarenko and Kreslavsky 2018).

Several lines of observational evidence indicate gradual modification and alteration of pristine igneous mafic materials. For example, Arvidson et al. (1992) noted a general decrease in radar contrasts with stratigraphic age. Bondarenko et al. (2003) documented the correlation of the inferred dielectric permittivity of lava flows with stratigraphic age. These changes may reflect chemical weathering, mechanical degradation, or airborne sedimentation.

#### 16.2.2.2 Evolved Igneous Rocks and Venera 8 Site

The one-modal hypsometric curve on Venus (Ford and Pettengill 1992) does not indicate the presence of low-density felsic continental crust in large regions (Grimm and Hess 1997). The absence of stratovolcanoes and arc structures that signify subduction of lithospheric plates on Earth suggests a deficiency of evolved plutonic (diorites, granodiorites, granites) and volcanic rocks (andesites, dacites, rhyolites). However, evolved igneous rocks formed via differentiation of mafic melts in magma chambers should be present. MELTS magmatic code models (Shellnutt 2013) show that phonolites and rhyolites could result from anhydrous and hydrous fractional crystallization of mafic magmas with the composition of Venera 13 and Venera 14 rocks. Significant differentiation is expected in large, long-standing magma chambers, as it occurred on Earth in the Bushveld, Skaergaard, Stillwater, and other layered intrusions of bulk mafic or ultramafic composition within the continental crust. Geophysical models for Venus' crust (Head and Wilson 1992) suggest a more common occurrence of shallow and large magma chambers than on Earth, residing in neutral buoyancy zones. The slower solidification of such magma reservoirs in the hot crust favors igneous differentiation to felsic melts at the chamber's tops and ultramafic cumulates at the base.

**Steep-Sided Domes**

Although most $SiO_2$-enriched rocks could be preserved in crystallized plutons, some low-viscosity melts may have reached the surface from shallow magma batches. About 145 widely distributed pancake-shaped, steep-sided domes (Fig. 16.10) could form from such melts (Fink et al. 1993; McKenzie et al. 1992; Moore et al. 1992; Parvi et al. 1992). These features are morphologically similar to rhyolite and dacite domes. Still, they are 10 to 100 times wider and have volumes (25–3400 $km^3$) that are orders of magnitude larger than terrestrial analogs (Parvi et al. 1992). The volume of steep-sided domes implies large magmatic chambers, as Head and Wilson (1992) suggested. The rounded morphology of the domes indicates that their formation in a single eruption did not cause explosions, major degassing, or pyroclastic activity, which are suppressed at high ambient pressure (Sect. 16.4.3). The density of dome-forming lava (2400 to 2700 kg $m^{-3}$), estimated from local topographic bending, does not suggest that the lava is highly vesicular; however, it still does not provide precise information regarding its composition (Borelli et al. 2025). The domes are commonly associated with coronae, which are unique circular tectonic-volcanic structures up to 1000 km in width, supposedly formed above mantle plumes and large crustal magma reservoirs (Stofan et al. 1991, 1997; Head and Wilson 1992; O'Rourke and Smrekar 2018; Gülcher et al. 2025). If evolved melts form through closed-system differentiation in the magma chambers, abundant sulfides are not expected in the domes. Although sulfides of chalcophile elements (Ni, Cu) in association with pyrrhotite sometimes yield significant concentrations of Fe, Ni, and Cu in the lower sections of magma chambers (Barnes et al. 2017; Latupov et al. 2024; Maier et al. 2018), Fe sulfides typically remain dispersed in the inner regions of closed system crystallizing mafic magma chambers. The high viscosity of magma in Venus' domes should have prevented the accumulation of sulfides at the bases of the domes. The common association of steep-sided domes with plains adjusted to tessera terrain (Sect. 16.2.2.4) does not exclude the melting of tessera-composing country rocks surrounding mafic magma chambers that exist in a zone of neutral buoyancy (Parvi et al. 1992). This process could be significant if tesserae are composed of evolved magmatic, metamorphic, or sedimentary materials. Although such processes are not rare within the continental crust (Eichelberger 1978; Glazner and Usslet 1988), the scarcity of domes

within tesserae suggests a different pathway, and massive sulfide deposits are not expected in the domes.

On Earth, the accumulation of massive sulfides in magma chambers is linked to an involvement of crustal sulfur that formed through non-magmatic processes (Grinenko 1985). Massive sulfide deposits are rare in layered mafic intrusions (Latupov et al. 2024; Maier et al. 2001; Barnes et al. 2017), and exceptions (e.g., Uitkomst intrusion) likely involve the assimilation of country rocks (Maier et al. 2018). If Venus bypassed an early water ocean phase with an $O_2$-bearing atmosphere (Sect. 16.4.2), sedimentary sulfate reservoirs might not have formed or contributed to igneous processes. Conversely, magmatic systems with abundant sulfur-bearing minerals could have arisen. In one scenario, the assimilation of putative crustal sedimentary sulfur (e.g., in anhydrite) would lead to a reduction of sulfur and the formation of massive sulfide deposits in magma chambers, as suggested for Norilsk-type intrusions (Gorbachev and Grinenko 1973; Grinenko 1985). In another scenario, the assimilation of sulfate-rich rocks increases magma's $fO_2$ and sulfur solubility in the sulfate form. This process could result in the crystallization of magmatic sulfates (anhydrite) from oxidized silicic magmas, as observed in some subduction-zone magmatic systems (Masotta and Keppler 2015). In all cases, abundant sulfides are not anticipated in melts that form steep-sided domes.

### Venera 8 Site

The concentrations of U, Th, and K measured in situ by the Venera 8 lander (Table 16.4) suggest an evolved magmatic composition that falls between mafic and ordinary silicic rocks. There are no images or spectroscopic data from the landing site, and therefore, the petrological type and geodynamic settings of Venera 8 rocks remain speculative. There is no convergence among the proposed compositions of silicate rocks. Andesitic, high-K calc-alkaline, leucitic, shoshonitic, melasyenitic, syenitic, nepheline syenitic, and quartz monzodiorite rock types have been proposed since 1973 to match concentrations of K, U, and Th (Nikolaeva 1990; Kargel et al. 1993; Shellnutt 2019, and references therein). Nikolaeva (1990) found a good match between the Venera 8 composition and quartz monzodiorites and quartz syenites. The suggested compositions resemble felsic to intermediate igneous rocks (diorite/ granodiorite), which are typical of convergent margins of terrestrial lithospheric plates. All these terrestrial analogs were formed with significant involvement of crustal materials, which allowed Nikolaeva (1990) to suggest that Venera 8 magma formed within an Earth-like continental crust. Geochemical considerations of magma fractionation by Nikolaeva and Ariskin (1999) ruled out the formation of Venera 8 melts through fractional crystallization of mafic melts. They suggested that the composition of Venera 8 could result from melting eclogite. In contrast, the petrological modeling of Shellnutt (2019) with the MELTS code indicates that the Venera 8 rock compositions proposed by Nikolaeva (1990) could form through the fractional crystallization of hydrous Venera 14-type mafic melts at the Ni-NiO $fO_2$ (QFM $+\sim$0.7 log units) buffer. The silicate composition of Venera 8 materials, inferred by Shellnutt (2019), corresponds to magnesian, calc-alkalic trachydacite/granodiorite rocks that are common in Archean greenstone belts. A steep-sided dome, 23 km in diameter, located in a peripheral part of the Venera 8 landing region (Abdrakhimov 2001; Parvi et al. 1992), may consist of such rocks. Shellnutt (2019) suggested that the Venera 8 lander sampled a bimodal volcanic complex formed from a compositionally evolved magma chamber.

Magellan images and geological mapping indicate a high abundance of small volcanic shields within the landing ellipse (Abdrakhimov 2001; Basilevsky et al. 1992; Weitz and Basilevsky 1993; Basilevsky 1997; Basilevsky et al. 2007). The size (a few km across) and gentle morphology of volcanic shields suggest that low rates of magma emplacement could have occurred due to widespread, isolated, long-lived, and possibly shallow magma reservoirs (Head et al. 1992; Crumpler et al. 1997). These fields of small, shallow shields are inconsistent with silicic, viscous melts but may have formed from mafic lavas similar to alkaline oceanic island hotspots. Shield plains occupy about 10–15% of Venus' surface and are partially buried by plains with wrinkle ridges, a dominant surface type typical of lowlands (Ivanov and Head 2004). Therefore, Venera 8-type igneous rocks could represent a significant portion of the upper crust. Although the abundance and mineralogy of sulfur have not been discussed in the Venera 8 rock, its probable mafic alkaline composition suggests a sulfur fate like that of its terrestrial counterparts.

#### 16.2.2.3 Canali-Type Channels

Unusual lava channels (canali) and more complex channels and valleys are primarily observed on the volcanic plains (Baker et al. 1992, 1997; Komatsu et al. 1992, 1993; Bledsoe and Klimczak 2025) (Fig. 16.11), although rare canali are seen in tessera terrain (Fig. 16.12). Some features are outflow channels related to impact craters. Rare valley networks suggest supply from a shallow subsurface. Although most of the ≈200 channels are tens to hundreds of km long, one channel, Baltis Vallis, spans almost 7000 km. The channels are typically characterized by widths of 1 to 3 km and high apparent width-to-depth ratios that remain nearly constant along their length. The bottom of the Baltis Vallis is 20 to 100 m lower than the surrounding plains (Oshigami and Namiki 2007). Canali channels exhibit meanders, but cut-off meanders, tributaries, distributaries, and levee structures are rare. Their morphology indicates erosion by low-viscosity melts supplied from the subsurface through a single main conduit, maintaining steady effusion rates. Baker et al. (1997), Williams-Jones et al. (1998), and Oshigami and Namiki (2007) discussed both thermal and mechanical erosion by highly turbulent lava. The morphology of canali does not indicate significant crystallization of melts under ambient conditions at the time of formation, which constrains melt composition and past climate. In this context, silicate lava compositions are less favorable than non-silicate low-temperature melts, which do not solidify readily during emplacement.

Microwave or other remote sensing data provide minimal information on melt composition in the canali and valleys. Suggested low-viscosity silicate melts include ultramafic (komatiites) and alkaline mafic (picrites, olivine nephelinites) compositions (Baker et al. 1992; Komatsu et al. 1992; Gregg and Greeley 1993; Williams-Jones et al. 1998). The primary concern is their rapid cooling, which prevents the formation of prolonged morphologically uniform channels (Komatsu et al. 1992). Thermal erosion by hot ultramafic melts could be inconsistent with this rapid cooling (Gregg and Greeley 1993, 1994; Williams-Jones et al. 1998) and the topography profiles across Baltis Vallis (Oshigami and Namiki 2007). On Earth, ultramafic melts commonly assimilate sulfur from crustal and surface rocks, as evidenced by sulfur isotopes and other data (Arndt et al. 2008; Bekker et al. 2009; Kubota et al. 2022). The bottoms of some komatiite lava flows and sills reveal massive sulfide deposits that may have formed by incorporating sulfides from the underlying rocks affected by hot lava (Lesher 1989; Arndt et al. 2008; Bekker et al. 2009).

Similarly, ultramafic lava on Venus may have influenced sulfides and/or sulfates present on the volcanic plains through gas-solid reactions (Sects. 16.3.2 and 16.4.4). Whether ultramafic melts assimilated crustal sedimentary sulfur depends on an aqueous (Sect. 16.4.2.3) versus an anhydrous geological history. Regardless of the sulfur source in ultramafic melts, putative bottom sulfide deposits may remain unexposed due to the minimal erosion and surface modification rates since the lava emplacement (Arvidson et al. 1992; Kreslavsky and Bondarenko 2017; Carter et al. 2023). Indirect detection of subsurface sulfide deposits in canali and valleys revealed by penetrating radar from the EnVision Venus orbiter (Ghail et al. 2018, Sect. 16.5.2) would be consistent with ultramafic melts.

The non-silicate melts discussed include sulfur (Baker et al. 1992; Williams-Jones et al. 1998), carbonatite (Baker et al. 1992; Kargel et al. 1994), carbonatite-chloride (Williams-Jones et al. 1998), carbonate-sulfate (Kargel et al. 1994; Treiman 1995, 2009), and chloride (Zolotov and Mironenko 2009; Zolotov 2019) melts. The eutectic temperature of these salt mixtures is significantly lower than that of silicates, and certain complex salt melts could remain stable under current ambient conditions. These parameters are more consistent with mechanical erosion than with thermal erosion caused by hot silicate melts (Oshigami and Namiki 2007).

Native sulfur melts at 388 K at 1 bar. Sulfur lava flows are rare on Earth (Harris et al. 2000) and are suggested to occur on Io (Greeley et al. 1984; Kargel et al. 1999). The formation of native sulfur through the incongruent melting of pyrite and/or pyrrhotite was suggested for Io by Lewis (1982) and is considered possible on Venus (Baker et al. 1992). It remains unclear if Venus' crustal conditions allow for the production of abundant sulfur from sulfides, which is uncommon on Earth. Although liquid sulfur becomes more viscous upon cooling to Venus' surface temperature, it remains less viscous than ultramafic (komatiitic) melt. Under current surface conditions, liquid sulfur does not boil but undergoes rapid evaporation after emplacement due to high vapor pressure ($\approx$1.5 bar at 750 K), particularly if it erupts at a temperature above $\approx$750 K. Evaporation of liquid sulfur in the canali suggests decreasing discharge and flow widths with increasing distance from a source, which is not observed.

Carbonatites are considered suitable analogs for canali's lavas due to their lower viscosity compared to silicate melts and their association with alkali mafic and ultra-mafic rocks, which may form through a low-degree partial melting of carbonated mantle peridotite (Baker et al. 1992; Kargel et al. 1994; Treiman 1994). Mechanical erosion by low-viscosity carbonatite melts aligns with the channels' morphology and the melts' cooling models (Williams-Jones et al. 1998; Oshigami and Namiki 2007). Although typical carbonatites consist of Mg-Ca-Na-K carbonates, sulfates are minor components alongside chlorides, phosphates, and silicates. Alkali-rich carbonatites exhibit lower viscosities and solidus temperatures than Ca-Mg carbonatites (Williams-Jones et al. 1998). The ternary Ca-Na-K carbonate system with admixtures shows an eutectic between 850 and 938 K, and natrocarbonatite lavas from the Oldoinyo Lengai volcano (Tanzania) have an eruption temperature of approximately 770 to 780 K (Dawson et al. 1995), only slightly hotter than Venus' ambient temperature. These low melt temperatures could indicate a mixture of alkali chlorides, and alkali-carbonate-chloride melts were regarded as suitable valley-formed melts (Williams-Jones et al. 1998). Kargel et al. (1994) presented arguments that carbonatite-sulfate melts on Venus could form by melting crustal carbonate and sulfate deposits or from a carbonated mantle.

The common occurrence of carbonatites and carbonate-sulfate melts on Venus is questionable due to several factors. These factors include the high eutectic temperature (1250 K, Treiman 1995) and the high viscosity of the $CaCO_3$-$CaSO_4$ system (Kargel et al. 1994), the unlikely formation of crustal carbonate and sulfate deposits without surface liquid $H_2O$ (Sect. 16.4.2), doubts about a highly oxidized upper mantle with magmatic S(VI) (Sects. 16.4.1 and 16.4.2), and the thermal decomposition of crustal carbonates upon heating induced by a runaway greenhouse (evaporation of all surface $H_2O$ through greenhouse heating of an increasingly moist atmosphere; Ingersoll 1969, Kasting 1988; Sect. 16.4.2.5) or a greenhouse effect caused by global volcanic degassing (Sect.). Sulfate-carbonate melts have not been reported on Earth. The mass of atmospheric $CO_2$ exceeds that of terrestrial crustal carbonates (Sect. 16.4.3), and abundant carbonated materials are not expected in the interior of Venus. Finally, the instability of Ca, Na, and K carbonates regarding atmospheric $SO_2$, HCl, and HF (Sect. 16.3.2.2; Zolotov 2018) suggests chemical weathering of carbonatites. Therefore, no carbonates could be present in the surface materials.

Chemical models (Sect. 16.4.4.1) for a reduced early Venus water ocean suggest the subsequent formation of $CaCl_2$-NaCl evaporites and their involvement in shallow subsurface melting (Zolotov and Mironenko 2009; Zolotov 2019; Sect. 16.4.2). Low-viscosity Ca-Na chloride melts with a eutectic temperature of 786 K (Tian et al. 2021) could be responsible for the formation of canali, crater outflow valleys, valleys with chaotic terrain, and valley networks related to sapping from the subsurface. The greenhouse warming caused by degassing during the formation of volcanic plains (Solomon et al. 1999; Bullock and Grinspoon 2001; Sect. 16.4.4) may have led to the melting and mobilization of crustal chloride deposits. The superior ability of Ca-Na chloride melts to dissolve silicates may have contributed to the erosion and development of the unique morphological characteristics of canali. The presence of sulfates in chloride-rich bedrocks in canali and valleys would indicate an aqueous early history and oxidized surface water (Sect. 16.4.2). However, the instability of Ca, Mg, Na, and K chlorides regarding atmospheric $SO_2$ (Sect. 16.3.2.2) suggests the formation of a sulfate weathering crust on chloride-rich rocks if they are present in the canali and outflow valleys. Besides that, sublimation of alkali chlorides in lowlands could be faster than sulfatization (Zolotov 2025).

#### 16.2.2.4 Tessera Terrain and Highland Rocks

Tessera terrain (Figs. 16.12 and 16.13) is characterized by extensively tectonically deformed complexes that cover about 8% of Venus' surface area (Barsukov et al. 1986a; Bindschadler and Head 1991; Ivanov and Head 1996, 2011). Tesserae are elevated massifs of $\approx 10^2$ to $10^3$ km in size. Some tessera occupy parts of highland plateaus. Smaller, isolated tesserae are bordered by materials from the volcanic plains. Although the density of impact craters on tesserae is statistically indistinguishable from that of the surrounding ridged and densely lineated plains (Kreslavsky et al. 2015), stratigraphic relationships suggest that tesserae are the oldest terrains in the visible portion of geological history (Ivanov and Head 2011, 2015). However, an apparent improvement in some adjoining geological units in the tessera-forming tectonic deformations suggests overlapping formation times of tesserae and other heavily tectonized terrains (densely lineated and ridged plains, ridge belts) (Ivanov and Head 2015), consistent with the overlapping retention ages of the impact craters (Kreslavsky et al. 2015). Photogeological analysis of Magellan radar images by Ivanov

(2001) showed that tessera precursor material appears as plains and that some tesserae may have the same (basaltic) composition as the adjusted plains. A mafic composition of tessera, which consists of deformed, thick basaltic lava formations, has been hypothesized in some tectonic interpretations (e.g., Hansen 2006).

The eroded layered and folded rock structures in Magellan radar images of several tesserae (Figs. 16.12 and 16.13) suggest a sedimentary and/or volcanic origin. Linear structures may reflect deposition through the actions of impacts, wind, or liquid $H_2O$. Extended parallel linear features indicate the mafic/ultramafic composition of the lava flows and pyroclastic deposits in cases of volcanic origin. Large intrusive complexes (granitoids, anorthosites, layered mafic massifs) are inconsistent with these structures (Byrne et al. 2021).

The geomorphology of the plains within tesserae resembles that of the regional plains and suggests a basaltic composition (Ivanov 2001; Gilmore and Head 2018). However, some radar-dark areas in local lows may be the result of aeolian deposition of fine-grained material (Byrne et al. 2021). The basaltic composition of the intratessara plains aligns with only one steep-sided dome (Parvi et al. 1992) and with no signs of pyroclastic deposits, explosion-related calderas, or stratovolcanoes. Unlike prior morphology-based studies, the lava complex within the tessera highlands of the Ovda Region (Fig. 16.12) is suggested to be basaltic in composition based on the interpretation of Magellan radar images related to surface roughness (Wroblewski et al. 2019).

Some interpretations of tessera morphology, deformation patterns, high gravity-topography ratios, and geodynamic models do not exclude the silicic composition of the highland plateaus in supposedly exposed fragments of putative felsic continental crust (Nikolaeva et al. 1988, 1992; Romeo and Capote 2011; Resor et al. 2021). The interpretation of gravity-topography data by Maia and Wieczorek (2022) suggests an Airy isostatic regime for highlands, and a 15–40 km thick crust does not indicate crustal composition. However, the viscous relaxation models of Nimmo and Mackwell (2023) show that quartz-dominated compositions relax too quickly to serve as plateau-forming materials. Kreslavsky et al. (2000) noted that a lower dielectric permittivity in a tessera region compared to the plains suggests reduced density of the surface materials due to felsic and/or physically weathered rocks.

Interpretations of nighttime $\approx 1$ μm thermal emissivity of surface materials from Galileo (Hashimoto et al. 2008) and Venus Express (Mueller et al. 2008; Helbert et al. 2008; Basilevsky et al. 2012; Gilmore et al. 2015) do not rule out the presence of more felsic tessera compositions compared to materials found on the plains with basaltic lava morphology. Thermal surface emission in the visible range obtained during the Parker Solar Probe flyby is consistent with near-IR data and tentative compositional interpretations (Wood et al. 2021; Lustig-Yaeger et al. 2023). Experimental data on the thermal emission of terrestrial materials at surface temperatures of Venus (Dyar et al. 2020, 2021; Helbert et al. 2021; Treiman et al. 2021) suggest relatively low Fe(II) content in surface silicate materials on the tessera highlands. However, interpretations of the thermal emission from the highlands in terms of composition require confirmation because the apparent low thermal emission from high-altitude surfaces might be attributed to physical properties (e.g., grain size) and lower actual temperatures than those inferred from Magellan altimetry (Basilevsky et al. 2012; Gilmore et al. 2015, 2017). Another possible reason for low thermal emissivity is coating by light-toned products from chemical weathering that could have formed during the formation of the volcanic plains (Gilmore et al. 2015; Sect. 16.4.4) or in the current epoch as more stable

phases (e.g., Ca and Na sulfates) in the highlands (Zolotov 2018, Sect. 16.3.2.2). If low thermal emission from the tessera highlands reflects the composition of light-toned bedrock rather than a coating, the layered features (Fig. 16.13; Byrne et al. 2021) may indicate ancient sediments that are depleted in ferrous silicates and sulfides. Putative sedimentary rocks may include sandstones, clay deposits, or shales, while chemical sediments can consist of carbonates, phosphates, and salt-rich evaporitic deposits (chlorides, sulfates).

Although felsic silicate deposits could be aeolian (loess) or impact in origin, the formation of abundant felsic rocks likely required aqueous processes that favored granitization (Campbell and Taylor 1983; Taylor and McLennan 2008). The formation of the listed chemical sediments, which required liquid $H_2O$ on early Venus, is discussed in Sect. 16.4.2. Whether sulfur is present in sulfates depends on the fate of oxygen, which was strongly affected by hydrogen escape. Detection of sulfate-rich layered rocks in the tessera terrain with the following missions (Sect. 16.5) would indicate past aqueous and oxidizing conditions caused by hydrogen escape from a moist greenhouse atmosphere.

#### 16.2.2.5 Low Radar Emissivity Highlands

At Venus' highlands, roughly above a radius of 6054 km, low microwave emissivity (high reflectivity) suggests the presence of electrical semiconductor material with a high dielectric constant (Pettengill et al. 1982, 1988, 1992; Simpson et al. 2009) that might be associated with sulfides. The low-emission intensity pattern varies with elevation and location; however, local geology and geomorphology have no clear effect (Arvidson et al. 1994; Klose et al. 1992; Treiman et al. 2016), Fig. 16.14. The data suggest a dielectric constant of 50 or more in the surface materials (Pettengill et al. 1988); neither the composition nor the origin of low-emissivity materials is known. Other interpretations of these data include ferroelectric minerals (e.g., Shepard et al. 1994; Treiman et al. 2016) and volume scattering phenomena unrelated to the composition (Tryka and Muhleman 1992; Pettengill et al. 1992, 1996, 1997; Bondarenko and Kreslavsky 2018).

Both physical and chemical processes have been invoked to explain the low microwave emissivity phenomenon observed in highlands. Windblown low-density grains may have concentrated dense, high-dielectric phases (magnetite, Fe sulfides) in lag deposits (Greeley et al. 1991). Although Fe sulfides are often invoked to explain the low radar emissivity in highlands (Pettengil et al. 1982; 1988; Klose et al. 1992; Wood 1997), the thermodynamic stability domains for Fe and other sulfides may not be accurately constrained due to gas-gas chemical disequilibria at elevations (Zolotov 2018). Nevertheless, pyrite could be more stable than magnetite at elevations (Klose et al. 1992; Zolotov and Volkov 1992; Wood 1997; Zolotov 2018; Sect. 16.3.2.3), and the metastable existence of primary and/or secondary pyrrhotite cannot be entirely excluded (Zolotov 2018).

Considerations of gas-solid type chemical equilibria do not exclude high-altitude condensation of heavy metal sulfides, such as PbS and $Bi_2S_3$ (Schaefer and Fegley 2004) in highlands (Sect. 16.3.2.3). However, the spatially variable low-altitude boundary of low-emissivity highlands is inconsistent with atmospheric condensation, which should create a stricter transition altitude worldwide. The increased radar emissivity at the highest elevations (Klose et al. 1992; Arvidson et al. 1994; Treiman et al. 2016; Fig. 16.14) contradicts condensation above certain altitude-temperature levels.

In addition to compounds with an elevated dielectric constant, materials with

ferroelectric properties have been proposed to explain radar brightness in highlands and a decrease in brightness at some of the highest elevations (Arvidson et al. 1994; Shepard et al. 1994; Brackett et al. 1995). The microwave emissivity of the Ovda Region highland region (Fig. 16.12) does not contradict the presence of a ferroelectric phase at higher elevations, which becomes paraelectric at lower altitudes and higher temperatures. Although several ferroelectric phases have been proposed (e.g., chlorapatite, Treiman et al. 2016), the problem remains unsolved, and no sulfur-bearing minerals are included.

### 16.2.3 Bulk Venus, Its Core, and Mantle

Although the bulk composition and sulfur content of Venus are unknown, inferences can be drawn from Earth's composition, Venus' size, mean density, and the composition of its surface and atmosphere. Most estimations rely on models and reflect the physical-chemical processes that occurred on early Venus (Sect. 16.4.1). Physical models of planetary accretion suggest that the Earth and Venus formed from compositionally similar materials. The similarity of bulk planetary compositions is supported by comparable densities between Earth and Venus, the roughly Earth-like abundances of K, U, and Th in surface materials (Table 16.4; Sects. 16.2.2.1 and 16.2.2.2), the widespread volcanic features with basaltic lava morphology (Sect. 16.2.2.1), the matching of rock-forming elemental composition of sampled surface rocks (Table 16.3) with terrestrial igneous counterparts (Sect. 16.2.2.1), and the similarity in the masses of carbon and nitrogen in the upper envelopes of the planets (Sect. 16.4.3).

Strictly identical compositions of Venus and Earth would result in ≈1.9% greater density for Venus than observed (BVSP 1981). However, the pressure-corrected density of Venus is ≈3% less than that of Earth (Phillips and Malin 1983). This difference is often attributed to lower Fe and sulfur content, along with higher oxygen abundance on Venus. The higher volatility of sulfides in the hotter regions of the solar nebula may have contributed to a slightly lower sulfur content in Venus-forming bodies (BVSP 1981). However, this mechanism is inconsistent with sulfur-rich materials on Mercury's surface (Nittler et al. 2018; Chap. 15: Renggli et al. 2026) and with the likely formation of terrestrial planets from planetary embryos and differentiated planetesimals with metal-rich cores (Sect. 16.4.1). The stochastic accretion of compositionally diverse planetary embryos, including different core sizes and variable degrees of impact stripping, could have led to various planetary densities and decreased bulk Fe and sulfur contents on Venus. A higher basalt/eclogite ratio resulting from the warmer upper interior could also contribute to the lower density of Venus (BVSP 1981). Based on current data, it is impossible to make a definitive prediction about the bulk sulfur content, which primarily reflects the composition and size of the core.

In early models of the bulk composition of terrestrial planets, the bulk sulfur content was determined by both the abundance of sulfur in the core and the core's mass fraction. The latter was constrained by a planet's bulk density and by the Birch-Murnaghan equation of state for planetary materials to approximate their densities under high pressures (Table 16.5). Mantles and crusts were thought to be sulfur-free. The Ve2 model (BVSP 1981) is based on the results of equilibrium condensation calculations of solar nebular materials and is modified using feeding zones. This model features a sizable sulfur-rich core and a low-

density mantle with elevated pyroxene/olivine and Mg/Fe ratios. Anders (1980) used either the solar S/K ratio or the amount of Fe available after forming Fe-Ni metal and silicates in the solar nebula to constrain sulfur abundance in Venus (Ve3 model). The latter model has been commonly used for over 40 years (e.g., Xiao et al. 2021). Models Ve1, Ve2, and Ve3 contain no oxygen in the core. In contrast, the pyrolite Ve4 model features a small sulfur-poor and oxygen-rich core, along with an olivine-rich and FeO-rich mantle. The Ve5 model represents a Fe-deficient composition that accounts for the low density of Venus. This model features a low-density, sulfur-poor, and oxygen-rich core. Compared to similar models for the Earth (BVSP 1981), these assessments indicate slightly smaller mass fractions for the core and lower sulfur abundances in both the core and the entire planet. In agreement with these models, Aitta (2012) demonstrates that 4.8 wt% sulfur in the core of Venus is insufficient to explain core density; thus, other light elements are needed. Recent models propose that sulfur and Si are major light elements in the cores of Earth, Mercury, Mars, and the Moon (e.g., Huang et al. 2019, Terasaki et al. 2019; Steinbrügge et al. 2021; Chap. 14: Komabayashi et al. 2026), as well as Venus (Xiao et al. 2021; O'Neil 2021). However, the challenge is to distinguish between sulfur and Si in the core compositions of all terrestrial planets.

The range of possible core sulfur contents of 1–10 wt% corresponds to 0.3–3 wt% of sulfur in the bulk planet (Table 16.5). The latter values are lower than sulfur abundances relative to rock-forming elements (e.g., Si, Fe, Mg) found in the solar photosphere and in Ivuna-type carbonaceous chondrites (CI), which best represent solar system abundances within the meteorite collection (Lodders 2021), and in samples returned from the Ryugu carbonaceous asteroid (Yokoyama et al. 2022). The trapping of $H_2S$ gas into sulfides of Fe, Ni, and Mn in the solar nebula (e.g., Lauretta et al. 1997) may have been incomplete in the inner solar system. This deficiency of sulfides may also reflect their thermal vaporization in the inner solar system (Morgan and Anders 1980). These pathways align with the sub-solar S/(Si, Fe, Mg) ratios in ordinary and enstatite chondrites (Lodders and Fegley 1998), which could be significant building blocks of the planetary embryos that formed Venus. Further in-depth discussion of sulfur in the core is in Sect. 16.4.1.

There are no identified or suspected rock samples from Venus in the meteorite collection, and the elevated sulfur content in otherwise basaltic materials in surface samples (Table 16.3) could be due to chemical alteration at the surface (Sect. 16.3.2). The similarity in the Fe content in these samples with terrestrial basalts and only a moderately more reduced Venus' mantle, as suggested by the MnO/FeO ratio in surface samples (Sect. 16.2.2.1), implies Earth-like sulfide-saturated mafic melts with 0.05–0.2 wt% sulfur and their mantle source regions containing 100–300 ppmw (μg/ g) sulfur (Chap. 13: Kiseeva et al. 2026).

In the mantle transition zone on Earth, certain sulfides (e.g., Ni-poor) may exist in a liquid state (Chap. 13: Kiseeva et al. 2026). The hotter geotherm of Venus in the crust and upper mantle would promote the melting of sulfides in the upper mantle or lower crust. Consequently, the zone of potential sulfide melts could be closer to the surface and might be more extensive. Significant sulfide that accumulated and was trapped during the late stage of magma ocean crystallization (Sect. 16.4.1) may influence the physical properties of the upper interior (viscosity, density, etc.) and the concentration of chalcophile elements in the magma.

## 16.3 Sulfur in Current Planetary Processes

### *16.3.1* Atmospheric Processes

The current understanding of atmospheric chemical and physical-chemical processes is based on the photochemical and thermochemical modeling of selected reactions, chain reactions, and large multicomponent segments of the atmosphere. Major chemical processes involve sulfur-bearing gases and condensates. Sulfur plays a key role in cloud chemistry and the mesosphere. The sulfur cycle has received persistent attention since the discoveries of $H_2SO_4$ in cloud aerosol and $SO_2$ gas above and beneath clouds in the 1970s. In preparation for the Pioneer Venus mission, Prinn (1975, 1978) conducted influential assessments of sulfur chemistry throughout the atmosphere. In the 1980s, the models focused on interpreting compositional data obtained from the Pioneer Venus, Venera 11 to 15, and Vega spacecraft (Sect. 16.2.1) concerning photochemical and thermochemical processes as well as atmospheric dynamics (Winick and Stewart 1980; Krasnopolsky and Parshev 1981a, 1981b, 1981c, 1983; Yung and DeMore 1982; DeMore et al. 1985). Since the early 1990s, modeling efforts have aimed to understand the fates of sulfur-bearing gases observed through ground-based, orbital, and spacecraft flyby data. Significant understanding has been gained through comprehensive models that account for the kinetics of photochemical and thermochemical reactions in the hydrogen-carbon-nitrogen-oxygen-sulfur-Cl system, mass balances, and eddy transport in the middle and upper atmosphere (Mills 1998; Mills and Allen 2007; Zhang et al. 2010, 2012; Jessup et al. 2015; Krasnopolsky 2012). Krasnopolsky and Pollack (1994) evaluated the physical chemistry of the sulfur-oxygen-hydrogen system within the clouds alongside the lower atmosphere. Yung et al. (2009) and Krasnopolsky (2007, 2013) specifically quantified sulfur in the lower atmosphere. Bierson and Zhang (2020) first modeled the atmosphere from 112 km to the surface, a task later accomplished by Dai et al. (2024). Stolzenbach et al. (2023) conducted the first three-dimensional simulations of the coupled hydrogen-carbon-oxygen-sulfur-Cl system at altitudes ranging from 40 to 95 km. Atmospheric sulfur cycles have been discussed by von Zahn et al. (1983), Prinn (1985), Prinn and Fegley (1987), Esposito et al. (1997), Fegley et al. (1995, 1997a), Mills et al. (2007), Krasnopolsky (2007, 2013), and Bierson and Zhang (2020). Section 16.3.1.1 overviews the chemical and physical processes that govern atmospheric sulfur. Sections 16.3.1.2 to 16.3.1.10 provide details on the formation, transport, and consumption of sulfur-bearing compounds. Atmosphere-surface interactions are described in Sect. 16.3.2.

#### 16.3.1.1 Major Pathways and Sulfur Cycle

At and above the cloud top, photochemical processes govern the fate of chemical species. Thermochemical reactions dominate in the middle and lower clouds, extending from the cloud deck to the surface. Transport and mixing occur in the lower and middle atmosphere through eddy diffusion and a global Hadley-cell type circulation that includes low-latitude upwelling and high-latitude downwelling; polar and topographic upwellings are also possible. Eddy diffusion and wind transport both occur in the upper atmosphere. Photochemical dissociation of oxygen-bearing gases ($CO_2$, $SO_2$, $H_2O$, OCS) at and above cloud top creates oxidizing conditions that facilitate the oxidation of $SO_2$, OCS, and $H_2S$ into S(VI) species such as $SO_3$ and $H_2SO_4$. In the largely thermochemically controlled

lower atmosphere below ≈30 km, conditions are more reducing, S(VI)-bearing gases are absent, $SO_2$ may be less abundant than at mid-30 km, and OCS is much more abundant than at higher altitudes (Table 16.1). Within and around clouds, the behavior of chemical compounds is influenced by the competition between photochemical and thermochemical reactions, the mixing of species from photochemical and thermochemical origins via eddy diffusion and plumes, the condensation-evaporation processes of sulfuric acid and potential polysulfur aerosols, the dissolution-precipitation of mineral species such as metal sulfates, and the alteration of inorganic (Fe-metal, FeS, Mg silicates, etc.) and organic matter of cosmic origin. The short-term (e.g., hourly, daily) compositional variability of $SO_2$ and SO at cloud top and above is influenced by solar insolation, reaction rates, and competitive mixing. The causes of the long-term compositional variability of $SO_2$ at cloud top (Fig. 16.3) could be related to changes in atmospheric circulation patterns below and within clouds. The Hadley-cell circulation could account for latitudinal gradients of $SO_2$, OCS, and CO observed at 30 to 40 km. Observed and suspected vertical compositional gradients reflect changes in temperature, pressure, and density (Seiff et al. 1985), as well as the competitive rates of chemical reactions and gas mixing. In the deep lower atmosphere, rapid gas-phase thermochemical reactions are primarily balanced by reverse reactions, while differences in reaction rates lead to net production or loss of gases. In contrast to earlier chemical equilibrium models for the middle and lower atmosphere (Florensky et al. 1978; Oyama et al. 1980), the apparent lack of chemical equilibration between many gases above, within, and below clouds implies ongoing chemical reactions throughout the atmosphere. Thermochemical equilibration between major chemically active gases is assumed only at the surface (e.g., Krasnopolsky and Parshev 1979; Fegley et al. 1997b; Krasnopolsky 2013, Table 16.2), but data on the gas composition (Table 16.1) are insufficient to confirm this. Although there is no thermochemical equilibration between $SO_2$, OCS, CO, and $CO_2$ above a thin layer close to the modal Venus' radius (Krasnopolsky and Pollack 1994; Zolotov 1996), some compounds (e.g., $S_n$ species, $H_2SO_4 \cdot nH_2O$ (l), and $H_2SO_4$(g), Fig. 16.5) could equilibrate at certain altitudes and environments.

Surface rocks and minerals (Sect. 16.3.2), volcanic degassing, and space materials (Sect. 16.4.3) supply sulfur to the atmosphere. The net supply of reduced gases (OCS, $H_2S$, and $S_2$) through oxidation or thermal decomposition of pyrite ($FeS_2$), discussed by von Zahn et al. (1983), Prinn (1985), and Fegley et al. (1995, 1997a), remains vague due to the uncertain composition of the deep atmosphere and the possible stable existence of pyrite (Sect. 16.3.2.3). However, the oxidation of pyrrhotite ($Fe_{1-x}S$) by $CO_2$ and $H_2O$ is likely (e.g., Fegley et al. 1995, 1997a). Although the volcanic delivery of $SO_2$, OCS, $S_2$, $H_2S$, and sulfur-Cl gases could account for the sulfur atmospheric inventory, current fluxes remain ambiguous. A difference between the supposed volcanic and atmospheric gas compositions implies the chemical consumption of volcanic gases after degassing (Wilson et al. 2024). For example, abundant OCS, $S_2$, and sulfur-Cl compounds in volcanic gases can be oxidized to $SO_2$ (Sect. 16.4.3). Sulfur-bearing gases may be released through the decomposition of sulfides in crustal and surface materials (Sect. 16.3.2.3). Although the impact of space sources on the sulfur inventory in the atmosphere is negligible (Sect. 16.4.3), space materials could influence the composition of clouds (Mogul et al. 2025; Zolotov et al. 2023) and the mesosphere, as discussed below. In addition to net sources, sulfur-bearing gases could be released into the

atmosphere as products of gas-solid reactions at the surface (Sect. 16.3.2).

The net sink of atmospheric sulfur occurs through the formation of Ca, Na, and K sulfates in exposed surface materials and a potential formation of secondary sulfides (Sect. 16.3.2). A balance between volcanic degassing and atmosphere-surface interactions determines the mass of atmospheric sulfur over time (Sect. 16.4). However, much lower rates of gas-solid reactions than gas-phase reactions exclude the effects of atmosphere-surface reactions on the short-term behavior of atmospheric gases and their altitudinal and meridional gradients.

The atmospheric sulfur cycle is interconnected with the cycles of carbon, oxygen, hydrogen, nitrogen, chlorine, and iron, which involve various reactions between $SO_2$, SO, $SO_3$, $S_n$, OCS, CO, $CO_2$, $S_2O_2$, $S_2O$, CS, $CS_2$, O, $O_2$, $H_2O$, $H_2SO_4$, $H_2S$, HS, ClO, $S_xCl_yO_z$, $S_xCl_y$, $NO_2$, and other gases. The coupling of sulfur, carbon, and oxygen cycles occurs through the oxidation of sulfur-bearing species in and above upper clouds and the interactions between OCS, $SO_3$, CO, and $S_n$ in the lower atmosphere. The sulfur and chlorine cycles are linked because, similar to Earth's stratosphere, Cl-bearing species function as catalysts in upper clouds and the mesosphere, act as intermediate species in reaction chains, and contribute to the formation and decomposition of sulfur-Cl-bearing gases. HCl gas delivered to the upper clouds from the lower atmosphere is photochemically oxidized into oxygen-Cl compounds that inhibit the net recombination of atomic oxygen and CO into $CO_2$. HCl may participate in a series of reactions that produce $S_n$ species, $S_xCl_yO_z$, and $S_xCl_y$ gases. The net effect of photochemistry in the middle atmosphere is the consumption of $CO_2$, $SO_2$, $H_2O$, and HCl, along with the return of CO, $H_2SO_4$(g), and $SO_2Cl_2$ to the sub-cloud atmosphere (Krasnopolsky 2013). The net effect of thermochemical reactions in the lower atmosphere is the consumption of CO, $H_2SO_4$(g), and trace Cl-bearing gases, leading to OCS, $SO_2$, $S_n$, and HCl.

Although the key processes in the atmospheric sulfur cycle (Fig. 16.15) are established, the roles of chemical and physical processes and reaction pathways are debated and remain uncertain. The following description of major pathways mainly reflects the results of atmospheric modeling by Yung and DeMore (1982), Mills (1998), Mills and Allen (2007), Zhang et al. (2012), Krasnopolsky (2007, 2012, 2013), Bierson and Zhang (2020), Pinto et al. (2021), and Dai et al. (2024). The details and other references are provided in the following sections.

Reduced sulfur-bearing gases (OCS, $H_2S$, HS, CS, and $CS_2$) form in thermo-chemical processes in the deep sub-cloud atmosphere, where $SO_2$ is the most abundant sulfur-bearing gas. $SO_2$ and the reduced gases are transported upward toward clouds through eddy diffusion and upwelling. The Hadley cell circulation (Fig. 16.16) leads to low-latitude upwelling, that is supported by meridional gradients of $SO_2$ (Marcq et al. 2021; Oschlisniok et al. 2021), OCS (Marcq et al. 2008, 2023), and CO (Sect. 16.2.1). Both local topography-related (Marcq et al. 2020, 2023; Lefèvre et al. 2020) and polar upwellings (Oschlisniok et al. 2021; Marcq et al. 2023) are suspected. OCS is consumed through interactions with atomic S gas at upper 20s km and $SO_3$ at mid-30s km. $S_n$ gases supplied from below could condense in the lower clouds to form aerosols that are composed of or contain condensed native sulfur ($S_x$, polysulfur). A debated formation of $S_n$ gas species through a series of reactions in upper clouds may also contribute to $S_x$ aerosols. Photochemical dissociation of $CO_2$

$$CO_2 + h\nu \rightarrow CO + O \quad (16.1)$$

$$O + O + M \rightarrow O_2 + M \quad (16.2)$$

in the mesosphere (M stands for a third body, typically an inert molecule like $N_2$). A slower conversion from CO to $CO_2$ generates excess O and $O_2$, which are consumed through the oxidation of sulfur-, carbon-, nitrogen-, Cl-, and Fe-bearing species at the cloud top and above. $SO_2$, OCS, and $H_2S$ are consumed through photolysis and reactions with atomic oxygen and $O_2$. The net process characterizes the formation of $SO_3$ at the cloud top,

$$SO_2 + CO_2 \rightarrow SO_3 + CO \quad (16.3)$$

in which $SO_2$ is oxidized by O released in reaction 16.1. The interaction of $SO_3$ with $H_2O$ vapor produces sulfuric acid solution, $H_2SO_4 \cdot nH_2O$ (l), which is the main component of cloud aerosols. Gravitational subsidence of aerosol particles causes the evaporation of sulfuric acid solution and possibly native sulfur, leading to the formation of $H_2SO_4$, $H_2O$, and $S_n$ gases below the cloud deck.

Metal (Fe, Mg, Ca, Ni, Na, Al) sulfates could precipitate at the cloud deck and sink following the dissolution of space and possible basaltic dust in sulfuric acid aerosol and its evaporation (Zolotov et al. 2023; Mogul et al. 2025). In the lower atmosphere, thermal decomposition of cloud-sourced ferric sulfate particles releases $SO_3$ (and/or $SO_2$ + $O_2$) into the atmosphere (Mogul et al. 2025)

$$Fe_2(SO_4)_3 \rightarrow Fe_2O_3(\text{hematite}) + 3SO_3 \quad (16.4)$$

followed by $SO_3$ reduction to $SO_2$. The chemical instability of settled ferric sulfate particles at the surface (Zolotov 2021) also suggests their alteration to hematite and $SO_2$ via interaction with reduced gases such as CO or OCS. As the temperature increases below the cloud deck, $SO_3$ forms through the pyrolysis of $H_2SO_4$(g)

$$H_2SO_4(g) \rightarrow SO_3 + H_2O \quad (16.5)$$

At mid-30 km, thermochemical interactions of $SO_3$ with OCS significantly consume both gases. However, the reaction pathways remain uncertain (Sects. 16.3.1.2, 16.3.1.4, and 16.3.1.6). The consumption of $H_2SO_4$(g) and $SO_3$ is reflected in compositional profiles modeled in the sub-cloud atmosphere (Fig. 16.17). The photochemical production of CO above clouds via reaction 16.1 and the thermo-chemical consumption of OCS together with a poleward Hadley cell type circulation could cause decreasing *x*OCS and increasing *x*CO toward high latitudes (Fig. 16.16) observed in the sub-cloud atmosphere above 30 km (Sect. 16.2.1.3). Photodissociation of $CO_2$ (reaction 16.1), oxidation of $SO_2$ to $SO_3$ (reaction 16.3) and $H_2SO_4$ at cloud top, pyrolysis of $H_2SO_4$(g) to $SO_3$ (reaction 16.5), and reduction of $SO_3$ to $SO_2$ by OCS and possibly CO below clouds, together with the upward transport of $SO_2$ compete a short sulfur cycle (Fig. 16.15).

A long atmospheric sulfur cycle involves the high-latitude subsidence of OCS-depleted and CO-enriched gas, along with the partial thermochemical conversion of CO and sulfur gases (S, $S_2$) to OCS at the lower scale height (<16 km). OCS-rich near-surface gases are transported upward through eddy diffusion and equatorial rising, followed by the consumption of OCS in reactions with atomic S and $SO_3$, as noted above. Estimated durations of short and long sulfur atmospheric cycles are months and $3 \times 10^4$ a, respectively (Krasnopolsky 2013). Models for the current atmosphere require neither degassing from the

interior (Sect. 16.4.3) nor the supply of gases through slow gas-solid interactions (Sect. 16.3.2), which may not affect current atmospheric processes. Sources and sinks of sulfur-bearing gases are outlined in Tables 16.6 and 16.7, which depict the significant formation and consumption processes through gas-phase reactions.

#### 16.3.1.2 Sulfur Dioxide

Except for the upper mesosphere, $SO_2$ is the most abundant sulfur-bearing gas (Table 16.1). The low mixing ratio of $SO_2$ in the upper and middle atmosphere reflects the net $SO_2 \rightarrow H_2SO_4$ conversion in upper clouds. Atmospheric models suggest that $SO_2$ is consumed in the middle atmosphere and produced below clouds. Krasnopolsky (2013) inferred that the lifetime of $SO_2$ in the middle atmosphere is on the order of months. In the model of Bierson and Zhang (2020), the transport timescale of $SO_2$ is ~10 a near the surface and months at 90 km. The chemical loss timescale exceeds ~$10^2$ a in the lower and middle atmosphere, and it is significantly shorter in the upper middle atmosphere and above it.

#### Consumption and Production at the Cloud Top and Above

In the middle atmosphere, $SO_2$ is consumed through photodissociation, which mainly occurs below 70 km altitude and oxidation to $SO_3$ by net reaction 16.3, as inferred in many models (e.g., Yung and DeMore 1982; Krasnopolsky and Parshev 1983; Zhang et al. 2012; Krasnopolsky 2012, Fig. 16.18). Photodissociation is reversible, but the formation of $H_2SO_4$ results in a net loss of $SO_2$. An additional net loss may occur if $S_n$ gases and condensed $S_x$ form in upper clouds (e.g., Pinto et al. 2021; Fig. 16.19) and sink with aerosol particles (Sect. 16.3.1.7). Photolysis of $SO_2$ leads to the formation of SO, $S_2O_2$, and oxygen species, mainly O and $O_2$, which are involved in recombination reactions to $SO_2$ and in oxidation reactions to produce $SO_3$ that hydrates to $H_2SO_4$ (Yung and DeMore 1982; Mills and Allen 2007; Krasnopolsky 2012; Zhang et al. 2012),

$$SO_2 + h\nu \rightarrow SO + O \tag{16.6}$$

$$SO_2 + O + M \rightarrow SO_3 + M \tag{16.7}$$

$$SO_3 + H_2O + M \rightarrow H_2SO_4(g) + M \tag{16.8}$$

$$\text{Net process:}\quad 2SO_2 + H_2O \rightarrow SO + H_2SO_4(g) \tag{16.9}$$

In reaction 16.8, M stands for an inert molecule ($CO_2$, $N_2$) or $H_2O$. A high $H_2SO_4/SO_2$ concentration ratio at cloud top (Table 16.1) suggests a significant contribution of additional oxygen in the $SO_2 \rightarrow H_2SO_4$ conversion. The oxygen is supplied mainly through photolysis of $CO_2$ above clouds by reaction 16.1 and drives the oxidation of S(IV) to S(VI) by reaction 16.7. The primary net process of $SO_2$ consumption and $H_2SO_4$(g) production in the middle atmosphere is

$$SO_2 + CO_2 + H_2O \rightarrow H_2SO_4(g) + CO \quad (16.10)$$

Modeling and observations demonstrate that the formation of $H_2SO_4$(g) and liquid sulfuric acid (back reaction 16.5) in upper clouds reduces $x$$SO_2$ from ≈150 ppmv in the lower atmosphere to ≈0.1 ppmv at ≈70 km. Cl-bearing compounds (Cl, ClCO, ClO, etc.) play a crucial role in the oxidation of CO to $CO_2$, which impacts the net production of oxygen, and they may be involved in $SO_2$ oxidation (Yung and DeMore 1982; Mills and Allen 2007). DeMore et al. (1985) suggested an oxidation pathway that includes the interaction of $SO_2$ with Cl, leading to the formation of $ClSO_2$ (Sect. 16.3.1.10), its oxidation to $ClSO_4$ by $O_2$, and the formation of $SO_3$ through the interaction of $ClSO_4$ with Cl. This pathway of $SO_2$ consumption to $H_2SO_4$(g) corresponds to a net reaction

$$SO_2 + Cl + O_2 + H_2O \rightarrow H_2SO_4(g) + ClO \quad (16.11)$$

However, the rate constants for the corresponding reactions (such as $ClSO_2$ oxidation) remain uncertain, and subsequent models (Mills and Allen 2007; Krasnopolsky 2007, 2012, 2013; Bierson and Zhang 2020; Dai et al. 2024) do not support the critical roles of Cl-bearing species in $SO_2$ oxidation. Ubukata et al. (2025) experimentally demonstrated the possibility of $SO_2$ uptake into liquid sulfuric acid droplets through $SO_2$ oxidation by $NO_2$ dissolved in the droplets.

In addition to photodissociation and oxidation, Krasnopolsky (2007) considered the net reduction of $SO_2$ by CO,

$$SO_2 + 2CO \rightarrow 2CO_2 + S \quad (16.12)$$

that could be a source of $S_n$ in the clouds under oxygen-deficient conditions. His models suggested that the flux of $H_2SO_4$(g) (reaction 16.10) exceeds that of $S_n$ (reaction 16.12) by a factor of 4. An advanced model by Krasnopolsky (2013) demonstrates the overwhelming dominance of the net reaction 16.10 and suggests a supply of possible cloud sulfur species from the lower atmosphere rather than from the cloud top. The prevalence of reaction 16.10 aligns with the high $H_2SO_4/S_x$ mass ratio (≈7 to 10) in cloud aerosols, which is tentatively inferred from data obtained from the Vega reaction gas chromatography (Porshnev et al. 1987).

The modeled $SO_2$ consumption corresponds to its lifetime of several months in the middle atmosphere (Krasnopolsky 2007, 2013). A decrease in $x$$SO_2$ above clouds toward higher latitudes (Sect. 16.2.1.1) may reflect $SO_2$ loss as gases move toward the poles at cloud altitude (Marcq et al. 2013), with a typical $SO_2$ lifetime of a few Earth days. The observed decrease in $x$$SO_2$ above clouds at 70 to 80 km (Fig. 16.18) is attributed to photochemical oxidation to SO by reaction 16.6 (e.g., Zhang et al. 2012; Krasnopolsky 2012).

Despite the net $SO_2$ loss through photodissociation and oxidation below ≈80 km (reactions 16.6–16.12), $SO_2$ is regenerated by the recombination of SO and O as well as via the oxidation of SO, primarily in reactions with $NO_2$ and ClO

$$SO + O + M \rightarrow SO_2 + M \quad (16.13)$$

$$SO + NO_2 \rightarrow SO_2 + NO \quad (16.14)$$

$$SO + ClO \rightarrow SO_2 + Cl \quad (16.15)$$

(Mills and Allen 2007; Krasnopolsky 2006, 2012; Zhang et al. 2010, 2012; Dai et al. 2024). Another likely source of $SO_2$ is the oxidation of OCS (Sect. 16.3.1.6) by the net process

$$OCS + 2CO_2 \rightarrow SO_2 + 3CO \quad (16.16)$$

in the middle atmosphere (Krasnopolsky 2013) and above. Oxidation of $H_2S$ leads to $SO_2$ as well. Another source of $SO_2$ in the upper-middle atmosphere, and possibly higher, is the putative $SO_2$ exsolution from sulfuric acid aerosol addressed by Rimmer et al. (2021) and Dai et al. (2024).

**Production and Consumption in the Lower Atmosphere**

Thermochemical reduction of $SO_3$ by OCS and CO primarily contributes to the production of $SO_2$ in the lower atmosphere. It is unclear whether the reaction

$$SO_3 + CO \rightarrow CO_2 + SO_2 \quad (16.17)$$

is (Prinn 1975, 1985; von Zahn et al. 1983; Bierson and Zhang 2020) or not (Krasnopolsky 2007, 2013) important below the cloud deck because its rate has not been determined experimentally (Sect. 16.3.1.4). Krasnopolsky and Pollack (1994) and Krasnopolsky (2007, 2013) modeled $SO_2$ production through a coupled reduction of $SO_3$ and oxidation of OCS by a net process

$$SO_3 + 2OCS \rightarrow CO_2 + CO + SO_2 + S_2 \quad (16.18)$$

that may include reactions

$$OCS + SO_3 \rightarrow CO_2 + S_2O_2 \quad (16.19)$$

$$S_2O_2 + OCS \rightarrow CO + SO_2 + S_2 \quad (16.20)$$

Their thermochemical kinetic and eddy diffusion models suggest a substantial consumption of $SO_3$ at ≈36 km altitude (Fig. 16.17). The validity of this pathway remains to be tested through experimental assessments of reaction rates and detection of short-lived $S_2O_2$ at 30 to 40 km. Bierson and Zhang (2020) included the rapid rate of the $S_2O_2 \rightarrow 2SO$ conversion in their model and concluded that $S_2O_2$ may not accumulate significantly to make reaction 16.20 efficient. To avoid this problem, they modeled $SO_3$ loss via a three-body process

$$2SO_3 + OCS \rightarrow 3SO_2 + CO, \quad (16.21)$$

and reproduced the observed OCS gradients at 33 to 36 km (Sect. 16.2.1.3), inferring a positive CO gradient with altitude that roughly aligns with observations. Dai et al. (2024) adopted reaction 16.21 in their models and observed a match with *x*OCS data at 30s km. The interactions of OCS and $SO_3$, along with the formation of $SO_2$ and CO, are consistent with the observed anticorrelation between CO and OCS, as well as with the vertical and latitudinal gradients of OCS, CO, and $SO_2$ (Sect. 16.2.1). The accuracy of assessing contributions from reactions 16.19, 16.20, and 16.21 is limited due to the absence of experimental data on the reaction rates. However, regardless of $SO_3$ reduction reactions, all considered pathways suggest a source of $SO_2$ at ≈35 to 40 km. From the cloud deck to the lower 30s km, some $SO_2$ could also be produced by reducing $H_2SO_4$(g) by CO (Mills et al. 2007) and OCS. All considered interactions may hold greater significance in the high-latitude downwelling of Hadley cell circulation (Fig. 16.16), and the elevated $xSO_2$ at 33 km at higher latitudes (Marcq et al. 2021) might reflect $SO_2$ production through the reduction of $SO_3$ and $H_2SO_4$(g). If this is the case, then no $SO_2$ consumption in the deep lower atmosphere must be invoked to explain the observations.

Deep atmospheric $SO_2$ could be produced and consumed in thermochemical reactions considered in the kinetic models of Krasnopolsky and Pollack (1994), Krasnopolsky (2007, 2013), Bierson and Zhang (2020), and Dai et al. (2024). Although low $xSO_2$ values in the low latitude cloud (≈51–54 km, Oschlisniok et al. 2021) and sub-cloud atmosphere (≈33 km, Marcq et al. 2021) (Sect. 16.2.1.1) suggest equatorial upwelling of $SO_2$-depleted gas, no major $SO_2$ sink in the lower atmosphere has been modeled. Thermochemical reactions that consume $SO_2$ are likely compensated by back reactions (Krasnopolsky 2007, 2013) and could reach thermochemical equilibria at the surface (Krasnopolsky and Parshev 1979; Krasnopolsky and Pollack 1994; Zolotov 1996; Fegley et al. 1997b, Table 16.2). A slow geological sink of $SO_2$ via gas-solid type reactions (Fig. 16.15, Sect. 16.3.2) may not account for the apparent $SO_2$ deficiency in the deep low-latitude atmosphere. The net reduction of $SO_2$ by CO (reaction 16.12) at the lower scale height occurs through reactions

$$SO_2 + CO \rightarrow CO_2 + SO \quad (16.22)$$

$$SO + SO \rightarrow SO_2 + S \quad (16.23)$$

that mainly contribute to the loss of $SO_2$ in the models from Krasnopolsky (2013) and Dai et al. (2024). The interaction of two $S_2O$ molecules to form $SO_2$ and $S_3$ primarily compensates for $SO_2$ loss in the Dai et al. (2024) model.

Consumption of $SO_2$ through a net thermochemical process

$$SO_2 + 3CO \rightarrow OCS + 2CO_2 \quad (16.24)$$

may not be an essential pathway in the lower atmosphere, as indicated by kinetic modeling (Krasnopolsky 2007, 2013; Bierson and Zhang 2020; Dai et al. 2024) and experimental studies conducted under Venus' temperatures and 1 bar (Hong and Fegley 1997a). However, the stoichiometry of reaction 16.24 does not contradict a loss of 20 to 30 ppmv $SO_2$ (Marcq et al. 2021) and a gain of 20 to 30 ppmv OCS (Pollack et al. 1993; Krasnopolsky and Pollack 1993; Krasnopolsky 2007, 2013; Fegley et al. 1997b) in the deep low-latitude atmosphere. Possible chemical equilibration between $SO_2$, OCS, CO, and $CO_2$, only under the conditions of a modal planetary radius (Krasnopolsky and Pollack 1994; Zolotov 1996; Fegley et al. 1997b), may reflect the catalytic effects of surface materials (e.g., Fe oxides and sulfides, Sect. 16.3.2.3) that allow interaction of absorbed gas reactants and accelerate the reduction of $SO_2$. The uncertainty surrounding the potential depletion of $SO_2$ in the deep lower atmosphere will be clarified through the analysis of gas abundances conducted by the DAVINCI descent sphere, Zephyr (Garvin et al. 2022, Sect. 16.5.1).

#### Causes of Spatial and Short- and Long-term Variability

In the mesosphere and at the cloud top, both spatial and temporal variations in $x SO_2$, over a scale of a few days, show differences within two or more orders of magnitude (Fig. 16.2, Sandor et al. 2010; Belyaev et al. 2012, 2017; Encrenaz et al. 2012, 2016, 2019, 2020; Vandaele et al. 2017a, 2017b; Marcq et al. 2013, 2020; Mahieux et al. 2023). These variations likely reflect a short photochemical life of $SO_2$ rather than winds (e.g., Encrenaz et al. 2012). This aligns with a daily cloud top $SO_2$ cycle (Marcq et al. 2020) and higher $x SO_2$ and $SO_2$/SO ratios observed on the night side at 85 and 95 km (Sandor et al. 2010; Belyaev et al. 2017). The photochemical-dynamic model of Shao et al. (2022) demonstrates that a sizable day-night difference in $x SO_2$ above 85 km results from both photochemistry and the subsolar-to-antisolar circulation. Generally, the effects of wind and circulation on variability decrease with altitude. An irregular flux of cosmic materials may also impact sulfur-bearing gases in the mesosphere.

At the cloud top, $SO_2$ variability reflects the competition between photochemical destruction and supply via advection, which is currently higher at lower latitudes (Encrenaz et al. 2019; Marcq et al. 2020). A correlation of transient $SO_2$ enhancement at cloud top with local low-latitude upwellings (up to 3 m $s^{-1}$ measured by Vega balloons at 53 km, Linkin et al. 1986) agrees with the supply from the sub-cloud atmosphere through general meridional circulation (Encrenaz et al. 2019; Jessup et al. 2020; Marcq et al. 2013, 2020). Marcq et al. (2013) showed that even a tiny ($\approx$1%) variation in $SO_2$ cloud supply through low-latitude upwelling could cause latitudinal variations of tens of ppbv in mesospheric $SO_2$. Marcq et al. (2020) noted that the decrease in $x SO_2$ around local solar noon at low latitudes indicates that local $SO_2$-rich plumes cannot counterbalance the rapid midday photochemical depletion. The links between heat balance, momentum, convection, and turbulent mixing in clouds were first inferred in the three-dimensional models of Lefèvre et al. (2018, 2022). They reproduced a diurnal cycle in cloud convection and predicted a 7-km-thick convective layer at the cloud top caused by the absorption of solar UV light. $SO_2$, as a chemical tracer, was included by Leférve et al. (2022) to assess the effect of cloud dynamics on spatial and temporal variability. The estimated vertical eddy diffusion corresponded with assessments from in situ measurements, but it seemed to be orders of magnitude greater than the values used in 1D chemical modeling. The models of Kopparla et al. (2019) showed that the observed chemical variability patterns could be related to the 4-day Kelvin wave, 5-day Rossby waves, and the overturning circulation. The observations of Jessup et al. (2015) and Marcq et al. (2020) suggest that the circulation regime characterized by numerous low-latitude plumes and a declining $x SO_2$ toward higher latitudes (Fig. 16.2) could be replaced within a few Earth days by a regime with fewer plumes and a reversed latitudinal gradient.

An enrichment of cloud-top $SO_2$ and increased UV brightness above the downwind western slopes of Aphrodite Terra (not evident in the works of Jessup et al. 2015, 2020,

Lee et al. 2020, and Marcq et al. 2020) may indicate a vertical topography-related supply mechanism, possibly linked to the phase-shifted vertical winds (Bertaux et al. 2016). In turn, the typography-related winds are likely linked to stationary gravity atmospheric waves that correlate with topography (Peralta et al. 2017; Fukuhara et al. 2017; Koyama et al. 2019; Kitahara et al. 2019; Lefevre et al. 2020; Suzuki et al. 2023). Mountain gravity waves were suggested by the UV and mid-IR absorption patterns at the cloud top observed by Akatsuki (Fakuhara et al. 2017; Peralta et al. 2017). Kouyama et al. (2017) emphasized the impact of solar illumination, local time, and latitude on the appearance of waves in Akatsuki cloud top mid-IR data. The models of Leférve et al. (2018) revealed the effects of gravity waves on convection and mixing within clouds. Overall, the physical processes in the lower atmosphere could influence the dynamics (plumes and turbulent mixing) and short-term chemical variability throughout clouds, as shown in the models of Morellina et al. (2022), Leférve et al. (2018, 2022), and Kopparla et al. (2019). In lower clouds, the compositional variability of $SO_2$ and $H_2SO_4$(g) (Fig. 16.5) is primarily affected by turbulence in atmospheric circulation, while some $SO_2$ latitudinal dependence suggests $SO_2$-depleted plumes at low latitudes (Oschlisniok et al. 2021).

The decline of the disc-averaged cloud top $x$$SO_2$ from ≈500 ppbv in 1978 to ≈20 ppbv in 1995, and a drop from ≈200 ppbv in 2007 to ≈10 ppbv in 2014 (Fig. 16.3, Esposito et al. 1988; Encrenaz et al. 2013, 2016; Marcq et al. 2013, 2020; Vandaele et al. 2017b) has been discussed for 40 years. Esposito (1984) and Esposito et al. (1988) interpreted the decrease in $x$$SO_2$ observed with Pioneer Venus orbital UV spectroscopy in relation to volcanic degassing. They suggested that eruptions create buoyant plumes that lift $SO_2$ into the visible atmosphere and that there is an uneven supply of $SO_2$ due to volcanic degassing. However, observations of cloud-top $SO_2$ variability by the Venus Express (Marcq et al. 2013) and the Hubble Space Telescope (Jessup et al. 2015) during 2007–2012 did not show a correlation with the relatively uniform $SO_2$ mixing ratio of $(1 - 2) \times 10^{-4}$ measured via gas chromatography (Oyama et al. 1980; Gel'man et al. 1980) and remote observations in the lower atmosphere (Pollack et al. 1993; Bezard et al. 1993; Arney et al. 2014; Marcq et al. 2008, 2021, 2023) and lower clouds (Oschlisniok et al. 2021). The anti-correlation of $SO_2$ and $H_2O$ at cloud top observed by Encrenaz et al. (2020) suggests atmospheric rather than geological processes. Periods of high $SO_2$ abundance in the early 1980s and late 2000s could have been characterized by multiple low-latitude plumes, as suggested by Marcq et al. (2020). Indeed, Earth-based and HST observations of $SO_2$ in upper clouds over the last two decades have revealed both short- and long-term variability in plume intensity (Encrenaz et al. 2019; Jessup et al. 2015; Marcq et al. 2020). Other models explain the long-term $SO_2$ viability by changes in the effective eddy diffusion in clouds (Krasnopolsky 1986) and global circulation (Clancy and Muhleman 1991). Models by Kouyama et al. (2019) and Kitahara et al. (2019) demonstrate that the variability may reflect momentum deposition from propagating atmospheric gravity waves induced by topography. In other words, changes in the cloud circulation regime, such as in low-altitude plumes, could cause both temporal variability and long-term compositional trends observed at the cloud top.

Geological processes are unlikely to affect the long-term $SO_2$ variability. The flux of volcanic $SO_2$ on Earth ($\sim 2 \times 10^{10}$ kg $a^{-1}$, Carn et al. 2017; Schmidt and Carn 2022) is significantly lower than the $SO_2$ mass in the atmosphere of Venus ($10^{17}$ kg). Therefore, it is improbable that even major volcanic events over several years would significantly affect the atmospheric $SO_2$. The ambiguity surrounding current volcanism and geologically recent pyroclastic activity (Herrick et al. 2023; Filiberto et al. 2025, for reviews), along with suppressed magma degassing at Venus' ambient pressures (Sect. 16.4.3), casts doubt on the measurable volcanic gas contribution to atmospheric composition in recent decades. Volcanic plume modeling (Glaze et al. 1999, 2011; Airey et al. 2015; Ganesh et al. 2021) suggests that the dense lower atmosphere and limited volcanic degassing prevent volcanic plumes from reaching the cloud tops where $SO_2$ was observed. Possibly lower abundances of magmatic volatiles than on Earth (mainly $H_2O$ and $CO_2$, Sect. 16.4.3) also limit explosive eruptions and the formation of steady volcanic plumes. This is consistent with the lack of stratovolcanoes, cinder cone volcanoes, and the slim evidence for pyroclastic deposits on Venus (Head et al. 1992; Herrick et al. 2023).

The long-term anti-correlation of $SO_2$ and $H_2O$ vapor at cloud top observed from 2014 to 2019 by Encrenaz et al. (2020), but not in 2021 (Encrenaz et al. 2023), may indicate $H_2O$ involvement in processes that affect $SO_2$. The one-dimensional photochemical-diffusion cloud model of Shao et al. (2020) demonstrates close inter-relations of $SO_2$ and $H_2O$ in the middle and upper clouds (≈58–65 km) and provides results consistent with the anti-correlation. The modeling shows that eddy mixing transport alone cannot explain the observations and suggests a significant influence of sub-cloud processes. The models of Kopparla et al. (2020) demonstrate that long-term changes in cloud convection patterns could be caused by convective strength oscillations linked to the radiative effects of $H_2O$ vapor abundance at the cloud base. In their model, $x$H$_2$O at the convective cloud deck affects the clouds' thermal balance and convection, and vice versa. Kopparla et al. (2020) noted that the oscillation timescale of 3 to 9 years inferred in the model reflects the geometric mean of the radiative cooling time and the eddy mixing time at the deck of convective clouds. Shao et al. (2020) modeled the effects of periodic changes in chemical processes in upper clouds. They demonstrated that the supply of $H_2O$ vapor to the clouds through non-diffusive mechanisms (e.g., plumes) could influence $SO_2$ at cloud top, as well as a long-term anti-correlation between $SO_2$ and $H_2O$ observed at 64 km by Encrenaz et al. (2020). Rimmer et al. (2021) examined the effects of periodic long-term depletions in sub-cloud $SO_2$ concentrations and proposed exotic mechanisms that could account for them.

#### $SO_2$ Gradients

The cause of the observed increase in $x$SO$_2$ with altitude at ≈75 to 100 km (Sect. 16.2.1.1, Fig. 16.18) is not fully understood and conflicts with predictions from photochemical models of Yung and DeMore (1982) and Krasnopolsky (2012). Several researchers have suggested a contribution from an unknown sulfur reservoir that facilitates the production of $SO_2$ and SO through upper mesospheric photochemistry (Sandor et al. 2010, 2012; Belyaev et al. 2012, 2017; Vandaele et al. 2017a, b; Dai et al. 2024; Egan et al. 2025). Sandor et al. (2010) first discussed a supply of $SO_2$ from a mesospheric aerosol (Wilquet et al. 2009) that could consist of condensed sulfuric acid. The photochemical models by Zhang et al. (2010, 2012) demonstrated that the observed profiles of $SO_2$ and SO could be replicated by a significant increase in the concentration of $H_2SO_4$(g). However, the assessed observational upper limit for $x$H$_2$SO$_4$(g) above 85 km (Table 16.1) ruled out $H_2SO_4$(g) as a source of sulfur excess. Belyaev et al. (2012) and Mahieux et al. (2024) pointed out that a warmer upper mesosphere promotes the evaporation of sulfuric acid aerosol if it exists in that region. Belyaev et al. (2017) noted that $SO_2$ enrichment above 85 km correlates with the density of putative supersaturated sulfuric acid droplets in the upper haze at 70 to 90 km, as assessed by Luginin et al. (2016). Model sensitivity studies by Parkinson et al. (2015) demonstrated the possibility of increasing $xSO_2$ above 80 km at specific eddy diffusion profiles and $SO_2$ and $H_2O$ vapor contents in upper clouds. Sandor et al. (2012) proposed elemental sulfur as a probable reservoir in the upper mesosphere required by observed $SO_2$ and SO abundances to conserve sulfur atoms. Chemical and photochemical models by Zhang et al. (2012) and Bierson and Zhang (2020) reproduced the observed $SO_2$ and SO contents above 75 km by forcefully invoking a delivery of condensed polysulfur ($S_x$, $S_8$) to the mesosphere from above. However, no observations support the existence of an $S_x$ aerosol there. The nominal model of Dai et al. (2024) did not reveal elevated $xSO_2$ above 80 km, and the authors suggested a non-photochemical sulfur source. The photochemical model of Egan et al. (2025), both with and without meteoric sulfur sources, cannot reproduce the $SO_2$ gradient. Vandaele et al. (2017a) stated that no current photochemical model reproduces the $SO_2$ profile above ≈75 km without significant manipulation of the physical properties of the atmosphere. It remains to be estimated whether the daily delivery of 1 to 4 metric tons of cosmic sulfur (Sect. 16.4.3) could explain the mesospheric profile of $SO_2$ and the enhanced abundances of SO, $SO_3$, OCS, CS, and $CS_2$ (Table 16.1). Interestingly, an enhanced $xSO_2$ in the north polar mesosphere at 80 to 120 km (Vandaele et al. 2017b) is comparable to the elevated concentration of meteoric sulfur-bearing compounds in the polar upper stratosphere on Earth (Gómez Martín et al. 2017).

The latitudinal gradients of $SO_2$ at and above the cloud top (Marcq et al. 2011, 2013, 2020; Belyaev et al. 2012; Encrenaz et al. 2019; Jessup et al. 2015; Fig. 16.2), in the middle cloud (Oschlisniok et al. 2021), and in the lower atmosphere at 33 km (Marcq et al. 2021; Fig. 16.4) could reflect the fate of $SO_2$ in its transport through the Hadley cells (Fig. 16.16), which may involve the mesosphere. The often observed decrease in cloud-top $SO_2$ toward the poles could reflect consumption (reactions 16.6–16.9) and lesser plume activity at higher latitudes (Encrenaz et al. 2019; Marcq et al. 2020). In the mesosphere, the latitudinal gradient could manifest as $SO_2$ propagating to the upper mesosphere at low latitudes and circulating poleward, followed by downward transport to the lower mesosphere (Vandaele et al. 2017a).

The enhanced high-latitude $xSO_2$ at ≈33 km (Table 16.1) could reflect $SO_2$ formation through the reduction of $SO_3$ (reactions 16.17–16.21) and $H_2SO_4$(g) in a sub-cloud downwelling (Fig. 16.16). Alternatively, or additionally, lower $xSO_2$ at low latitudes at ≈33 km (Marcq et al. 2021) and at 51 to 54 km (Oschlisniok et al. 2021) does not exclude some consumption of $SO_2$ in the deep atmosphere followed by equatorial upwelling of an $SO_2$-depleted gas, as discussed above. Such an upwelling is roughly consistent with the interpretation of in situ UV spectra obtained from Vega probes at altitudes of 10 to 60 km, suggesting a decrease in $xSO_2$ toward the surface (Bertaux et al. 1996). However, lower atmosphere models (Krasnopolsky and Pollack 1994; Krasnopolsky 2007, 2013; Bierson and Zhang 2020; Dai et al. 2024) do not predict any $SO_2$ vertical gradient. Mogul et al. (2025) suggested that the reported Vega $SO_2$ profiles in the lower atmosphere were influenced by optical effects from captured and decomposed cloud aerosol. As mentioned earlier, the effectiveness of $SO_2$ reduction to OCS (reaction 16.24) in the near-surface atmosphere could be limited. The factors that may cause significant net consumption of $SO_2$ (in tens of ppmv) in the deep equatorial atmosphere, if it occurs, remain unclear.

#### 16.3.1.3 Sulfur Monoxide and Disulfur Dioxide

Although SO is observed only in the middle and upper atmosphere (Sect. 16.2.1.1; Table 16.1), its abundance and reaction pathways have been evaluated in photochemical and thermochemical kinetic models throughout the atmosphere. None of the three disulfur dioxide isomers ($S_2O_2$, cis, trigonal, and trans; often presented as SO dimers, $(SO)_2$, and OSSO) are detected, but kinetic modeling suggests that SO dimers represent significant intermediate species.

At cloud top and in the mesosphere, SO is the primary product of $SO_2$ photolysis (reaction 16.6). The formation of sulfuric acid is ineffective above ≈75 km, and SO is the main photochemical product of $SO_2$ in the mesosphere. An increase in $xSO_2$ and $x$SO above 80 km suggests a common source of sulfur (Sandor et al. 2010, 2012; Belyaev et al. 2012) (Sect. 16.3.1.2). A more efficient $SO_2$ photolysis in the upper mesosphere at 90 to 100 km likely causes elevated dayside $SO/SO_2$ mixing ratios there (Belyaev et al. 2012, 2017). On the night side, the suppressed photolysis accounts for lower $x$SO and $SO/SO_2$ ratios (Sandor et al. 2010). The only partial short-term, long-term, and spatial correlations of SO and $SO_2$ at cloud top and in the mesosphere (Sandor et al. 2012; Belyaev et al. 2012, 2017; Jessup et al. 2015; Encrenaz et al. 2015) suggest that the photolysis of sulfur-oxygen species is not the only source of sulfur in the upper atmosphere, as first noted by Jessup et al. (2015). The photochemical and dynamic models of Egan et al. (2025), which include meteoric sulfur sources, reproduce the observed increase in $x$SO above 80 km.

The high compositional and spatial variability of SO (Sandor et al. 2010, 2012; Belyaev et al. 2012; Encrenaz et al. 2015; Jessup et al. 2015) reflects its short chemical lifetime (hours or less, Bierson and Zhang 2020, Dai et al. 2024), along with other factors that influence $SO_2$ fluctuations within and above clouds (Sect. 16.3.1.2). These inferences have been assessed in photochemical models of Yung and DeMore (1982), Mills and Allen (2007), Yung et al. (2009), Zhang et al. (2012), Krasnopolsky (2012), Bierson and Zhang (2020), Dai et al. (2024), and Egan et al. (2025) in which SO and its dimers are involved in multiple reactions. Although SO is consumed via photolysis to atomic sulfur and oxygen, it primarily converts back to $SO_2$ (reactions 16.13–16.15). In the model of Mills and Allen (2007), reaction 16.15 provides a 10–20% SO loss at 66–80 km. The model of Krasnopolsky (2012) suggests the SO + $NO_2$ interaction 16.14 as the most likely SO loss below 75 km. That model also implies the conversion of SO to $SO_2$ via the intermediate $S_2O_2$ by reactions

$$SO + SO + M \rightarrow S_2O_2 + M \quad (16.25)$$

$$S_2O_2 + h\nu \rightarrow SO_2 + S \quad (16.26)$$

as the main pathway of atomic sulfur production in the middle atmosphere (Sect. 16.3.1.7), in which M is a neutral molecule.

The model of Dai et al. (2024) suggests a major SO sink via reaction 16.25 at 60 to 70 km, a significant consumption by reaction 16.14 at 75 to 80 km, and losses through reaction 16.13 and photolysis at 80 to 112 km

$$SO + h\nu \rightarrow S + O \quad (16.27)$$

Their nominal model revealed the observed increase in $x$SO in the upper mesosphere, suggesting increasing production with altitude above 80 km via reaction

$$S + O_2 \rightarrow SO + O \quad (16.28)$$

that provides the major SO source above ≈105 km. Dai et al. (2024) modeled that a slow release of dissolved $SO_2$ from sulfuric acid aerosols contributes to the supply of mesospheric $SO_2$ and the increase in $x$SO with altitude.

Significant attention has been given to SO dimers in upper clouds due to their decent spectral match with the spectra of clouds (Frandsen et al. 2016; Pérez-Hoyos et al. 2018) (Sect. 16.3.1.8). However, $S_2O_2$ species have too short photochemical life (2 to 5 s) to be abundant and contribute to the blue and UV parts of the spectrum (Frandsen et al. 2016, 2020; Egan et al. 2025). At altitudes of 60 to 70 km, the model of Dai et al. (2024) suggests that $x$SO is regulated by rapid conversions with $S_2O_2$ (reaction 16.25 and the back reaction). Neither the photochemical models (Krasnopolsky 2018; Bierson and Zhang 2020; Dai et al. 2024) nor observations (Marcq et al. 2020) indicate a sufficient SO to form an abundant $S_2O_2$ in the middle atmosphere. 3D photochemical and dynamical models by Egan et al. (2025) consider the behavior of major gases and several SO dimers at 50–170 km, both with and without sulfur sources from meteor ablation above 100 km. The modeled concentration of $S_2O_2$ species did not exceed ~5 ppbv in a thin layer at ~70 km, and it was several orders of magnitude lower above and below the cloud top. At altitudes of 40 to 70 km, $S_2O_2$ is consumed through photolysis and reactions with atomic oxygen. Below 40 km, $S_2O_2$ is destroyed via thermal decomposition following a collision with a third body ($CO_2$). In contrast to other models, Rimmer et al. (2021) assessed increasing $xS_2O_2$ from 80 km to the surface. The role of $S_2O_2$ in the formation of $S_n$ and $S_2O$ in upper clouds is emphasized in the photochemical models of Pinto et al. (2021) and Francés-Monerris et al. (2022) (Sects. 16.3.1.7 and 16.3.1.8).

In the lower atmosphere, SO could be an essential intermediate species in thermochemical reactions, as modeled by Krasnopolsky and Pollack (1994) and Krasnopolsky (2007, 2013). Although $S_2O_2$ has been considered an intermediate species in OCS + $SO_3$ interactions at ≈36 km (reactions 16.18–16.20; Krasnopolsky and Pollack 1994; Krasnopolsky 2007, 2013), the fast thermochemical consumption of $S_2O_2$ and SO (Mills 1998) could impede the role of $S_2O_2$ in OCS oxidation, as noted by Bierson and Zhang (2020). In the model of Krasnopolsky (2013), SO forms through the $SO_2$ + CO interaction 16.22 and reaction

$$SO_2 + S \rightarrow SO + SO \quad (16.29)$$

with maximal yields at 15 km and 5 km, respectively. SO is consumed via a slow SO+ SO interaction (reaction 16.23), which could be a major net source of atomic sulfur within the lowest scale height. Below 40 km, the model of Dai et al. (2024) suggests SO production via reactions 16.22 and 16.29, and the back reaction 16.25. Despite the significance of SO in thermochemical reactions, it remains an unmeasurable species with a modeled mixing ratio of 0.1 to 13 pptv below 40 km (Krasnopolsky et al. 2007). At surface conditions corresponding to the modal planetary radius, the chemical equilibrium $x$SO is between 17 and 41 pptv (Table 16.2).

#### 16.3.1.4 Sulfur Trioxide

Although $SO_3$ is only detected in the mesosphere above 75 km (Mahieux et al. 2023), it is an essential reactant above ≈35 km, as inferred from photochemical and thermochemical models. Surprisingly, $SO_3$ is the most abundant sulfur-bearing gas in the upper mesosphere (Table 16.1). The reported $xSO_3$ levels above 75 km are 1 to 3 orders of magnitude higher than those modeled by Zhang et al. (2012), Krasnopolsky (2012), Bierson and Zhang (2020), and Dai et al. (2024), which were based on measured $xSO_2$ and the upper limit of $H_2SO_4$(g). The observed increase in $xSO_3$ with altitude contradicts the modeling results from Krasnopolsky (2012), which were obtained without a forceful addition of sulfur-bearing species. Sensitivity modeling by Bierson and Zhang (2020) shows that $xSO_3$ is weakly affected by mesospheric $xH_2SO_4$(g) but strongly depends on an additional source of native sulfur ($S_x$, $S_8$) that may also cause gradients of $SO_2$ and SO above 80 km (Sects. 16.3.1.2 and 16.3.1.3). The inconsistency between the observed and modeled $SO_3$ remains unresolved. The supposed delivery of condensed sulfuric acid to the mesosphere (e.g., Zhang et al. 2010; Belyaev et al. 2012, 2017; Mahieux et al. 2024) could be a source of $SO_3$ via thermal evaporation and photochemical decay of $H_2SO_4$ gas and aerosol (Mahieux et al. 2023). In addition, photochemical oxidation of cosmic sulfur-bearing materials (mainly FeS and sulfur-bearing organic matter) may provide a reasonable explanation for positive gradients of mesospheric sulfur oxides.

In upper clouds, $SO_3$ is primarily produced through the oxidation of $SO_2$ by atomic oxygen (reaction 16.7), which is formed via the photolysis of $CO_2$ (reaction 16.1) and $SO_2$ (reaction 16.6). In the upper clouds, $SO_3$ is hydrated to $H_2SO_4$(g) by reaction 16.8 and converted to liquid sulfuric acid,

$$H_2SO_4(g) + nH_2O(g) \rightarrow H_2SO_4 \cdot nH_2O(l) \qquad (16.30)$$

Some $SO_3$ is consumed via photolysis to $SO_2$ and O and through reactions with SO and O (Zhang et al. 2012; Krasnopolsky 2012). The formation of sulfuric acid significantly reduces both $xSO_2$ and $xSO_3$, and $SO_3$ is modeled as a trace gas throughout clouds with a maximal mixing ratio of ≈1 ppbv at ≈67 km (Zhang et al. 2012; Krasnopolsky 2012). Below clouds, $SO_3$ forms via thermal dissociation of $H_2SO_4$(g) by reaction 16.5, as inferred in kinetic models (Krasnopolsky and Pollack 1994; Krasnopolsky 2007, 2013; Bierson and Zhang 2020). These models suggest that $xSO_3$ increases as altitude decreases from the cloud deck to ≈38 km (at low latitudes), where $xSO_3$ reaches ≈0.4 ppmv (Fig. 16.17). The decrease in $xH_2SO_4$(g) at lower altitudes observed by Kolodner and Steffes (1998) (Fig. 16.5) and Imamura et al. (2017) at 35 to 47 km likely reflects the thermal dissociation of this gas (Sect. 16.3.1.5) and $SO_3$ production.

Thermochemical kinetic models for the sub-cloud atmosphere imply $SO_3$ consumption through reactions with OCS and CO below the cloud deck down to ≈30 km. The models of Prinn (1985), Krasnopolsky and Pollack (1994), Krasnopolsky (2007, 2013), Bierson and Zhang (2020), and Dai et al. (2024) suggest significant $SO_3$ consumption in the mid-30 km through thermochemical interactions with CO and/or OCS (reactions 16.17–16.21). Although the validity of the modeled $SO_3$ consumption determined by reactions with OCS supports the observed OCS gradient at 30 to 48 km (Table 16.1), actual reaction pathways are uncertain because of insufficient experimental data on reaction rates. Krasnopolsky (2007, 2013) concluded that the $SO_3$-CO interaction (reaction 16.17), previously considered by Prinn (1985) and Krasnopolsky and Pollack (1994), provides only a minor (≈4 %) contribution to the loss of sub-cloud $SO_3$. A subordinate role of reaction 16.17 aligns with a consistent $x$CO at 30 to 40 km and a substantial decrease in $x$OCS with altitude, which could be caused by interactions of OCS with $SO_3$ (reactions 16.18–16.21). However, Bierson and Zhang (2020) showed that reaction 16.17 could be critical at 30 to 50 km, especially at altitudes where $SO_3$ is produced from $H_2SO_4$(g) via reaction 16.5. There is no observational data on CO at 42 to 50 km to verify the role of reaction 16.17, although Bierson and Zhang (2020) modeling shows that a decrease in $x$CO may not occur at those altitudes. The consumption of CO by reaction 16.17 could be offset by CO production via interactions of $SO_3$ with OCS (reactions 16.18–16.21). The modelers have relied on the rate of reaction 16.17, as estimated by Krasnopolsky and Pollack (1994). The precise contribution of reaction 16.17 to the consumption of $SO_3$ remains uncertain until its rate is experimentally determined. Despite the uncertain roles of CO and OCS, all models suggest an insignificant role for $SO_3$ in atmospheric chemistry below ≈30 km. Thermochemical equilibrium models for the near-surface atmosphere suggest only 0.3 to 0.9 pptv $SO_3$ (Table 16.2). Although the thermal decomposition of sinking cloud ferric sulfates produces $SO_3$ (Mogul et al. 2025), the current flux of space-sourced Fe to Venus (Carrillo-Sánchez et al. 2020) is insufficient to affect $SO_3$ content in the lower atmosphere.

#### 16.3.1.5 Sulfuric Acid Gas and Aerosol

Sulfuric acid vapor has been detected in clouds and below the cloud base at 30 to 55 km (Sect. 16.2.1.2, Table 16.1). Sulfuric acid is the second major sulfur-bearing gas within clouds, and down to ≈35–40 km. Its fate is linked to $SO_3$, $H_2O$, and sulfuric acid aerosol (Fig. 16.15), as evidenced by inferences drawn from photochemical and thermochemical models (Yung and DeMore 1982; Krasnopolsky and Parshev 1981a, 1981b, 1981c, 1983; Krasnopolsky and Pollack 1994; Mills 1998; Mills and Allen 2007; Yung et al. 2009; Zhang et al. 2012; Krasnopolsky 2007, 2012, 2013; Bierson and Zhang 2020). $H_2SO_4$(g) forms in upper clouds through the oxidation of $SO_2$ to $SO_3$ via reaction 16.7, followed by the hydrolysis of $SO_3$ to $H_2SO_4$(g) (reactions 16.8 and 16.30). This process leads to the formation of liquid sulfuric acid, $H_2SO_4 \cdot nH_2O$ (l), where $n$ ranges from 1.53 to 0.47 in 78 to 92% sulfuric acid by mass (Titov et al. 2018; Dai et al. 2025, for reviews). More concentrated acid is modeled in the lower cloud by Dai et al. (2022) and Shao et al. (2024). $H_2SO_4$ vapor coexists with sulfuric acid aerosol and forms at the cloud base via thermal evaporation of the aerosol (back reaction 16.30). It thermally decomposes to $SO_3$ and $H_2O$ below the cloud deck (e.g., by reaction 16.5) (Figs. 16.15 and 16.16). The temperature at the bottom of the primary cloud at ≈48 km (at low latitudes) agrees with the vaporization of liquid sulfuric acid aerosol particles (e.g., Krasnopolsky and Pollack 1994; Krasnopolsky 2015; Dai et al. 2022; Shao et al. 2024).

The suggested photodissociation of mesospheric $H_2SO_4$(g) (Sandor et al. 2010) to account for the positive $SO_2$ and SO gradients above 85 km (Sect. 16.2.1.1) is inconsistent with the $xH_2SO_4$(g) lower limit at 85 to 100 km (Sandor et al. 2012). However, photochemical modeling shows that the observed profiles could be reproduced by adding $H_2SO_4$(g) produced through evaporation of the sulfuric acid aerosol (Zhang et al. 2010, 2012; Bierson and Zhang 2020). Belyaev et al. (2012, 2017) and Mahieux et al. (2024) noted that elevated temperatures at 90 to 110 km favor the release of $H_2SO_4$(g) from aerosol particles that may exist in the mesosphere (Luginin et al. 2016). According to Mahieux et al. (2023), the frequency pattern observed for $SO_3$ in the mesosphere indicates a conversion between $SO_3$ and $H_2SO_4$(g). This conversion may be linked to condensed sulfuric acid, which is a potential source of $H_2SO_4$(g). The increased mixing ratios of $H_2O$ and HDO in the upper mesosphere (Mahieux et al. 2024) were interpreted in terms of the thermal dissociation of sulfuric acid aerosols coming from below.

Production of $H_2SO_4$(g) and sulfuric acid aerosol in upper clouds depends on the supply of $SO_2$ from the lower atmosphere via eddy diffusion and plumes. Models that considered eddy diffusion (Krasnopolsky and Pollack 1994; Mills 1998; Zhang et al. 2012; Krasnopolsky 2012, 2015; Bierson and Zhang 2020) suggest the formation of $H_2SO_4$ in a narrow layer within upper clouds. Consequently, Krasnopolsky's (2012) model for low latitudes indicates the formation of $H_2SO_4$ in a 3 km thick layer, with a peak at 66 km located just below the most productive $SO_2$ photolysis layer. The low latitude upwelling likely enhances $H_2SO_4$ production at higher altitudes than in the middle latitudes (Figs. 16.2 and 16.6). Photochemical modeling with circulation by Stolzenbach et al. (2023) and the 3D cloud model of Shao et al. (2024) both suggest the production of $H_2SO_4$ from $SO_2$ at higher altitudes in the equatorial upwelling region.

Within the main cloud layer, measured $xH_2SO_4$(g) generally corresponds to the vapor pressure in equilibrium with sulfuric acid solution (Kolodner and Steffes 1998; Imamura et al. 2017; Oschlisniok et al. 2021; Dai et al. 2022; Shao et at. 2024) (Fig. 16.15). The $H_2SO_4$(g) mixing ratio increases with temperature as it approaches the cloud base. According to the models of Hansen and Hovenier (1974) and Pollack et al. (1978), the concentration of $H_2SO_4$ in the sulfuric acid aerosol increases from 75 to 85% at ≈68 km toward the cloud base, where it reaches ≈98% by mass. In the models of Dai et al. (2022) and Shao et al. (2024), the aerosol composition changes from 78% $H_2SO_4$ at 62 km to 98% at the cloud base. Aerosol particles descend towards the cloud deck and evaporate, forming $H_2SO_4$(g) and $H_2O$ vapor. The fate of liquid sulfuric acid in clouds is consistent with the near-constant $xH_2O$ (30–50 ppmv) in clouds inferred from ground-based nightside near-IR observations (Pollack et al. 1993) and Venera 11, 13, and 14 in situ near-IR spectroscopy data (Ignatiev et al. 1997). In the model of Krasnopolsky and Pollack (1994), sulfuric acid aerosol evaporates at 48.4 km, consistent with in situ data from Pioneer Venus and Venera probes for the low-latitude atmosphere. In the improved cloud model for the $H_2SO_4 - H_2O$ system, Krasnopolsky (2015) determined cloud base altitudes at different latitudes: 47.5 km globally, 48.5 km at low latitudes, and 46 km at 60° latitude. He assessed the concentrations of $H_2O$ vapor, $H_2SO_4$(g), and sulfuric acid within clouds. Several other models (e.g., Zhang et al. 2012; Krasnopolsky 2012; Dai et al. 2022) predicted aerosol evaporation at ≈47–48 km.

Higher $H_2SO_4$(g) abundances and significant variability observed at equatorial and polar latitudes below 55 km (Sect. 16.2.1.2, Table 16.1) may reflect atmospheric dynamics that affect temperature at specific altitudes and conditions of aerosol evaporation (Sect. 16.3.1.2). The Hadley-type hot equatorial upwelling (Fig. 16.16) is suggested as a potential explanation for the presence of a sub-cloud $H_2SO_4$-rich layer at approximately 4 km higher altitude at equatorial latitudes compared to polar latitudes (Oschlisniok et al. 2021). The lower and less variable $xH_2SO_4$(g) observed below 55 km in the middle latitudes could be attributed to downwelling near the boundary of a Hadley circulation cell and a smaller polar circulation cell, as modeled by Oschlisniok et al. (2021). The observed accumulation of $H_2SO_4$(g) released from sulfuric acid aerosol at equatorial and polar latitudes has been reproduced by upward winds in their models. Dai et al. (2023) used Venus Express data on the $H_2SO_4$(g) abundance to estimate vertical eddy diffusion from the cloud base (43–48 km) to 55 km across various latitudes. Their approach included the condensation of $H_2SO_4$(g) and diffusion. The assessed eddy (turbulent) diffusion coefficient is an order of magnitude larger than other observation-based and model results. Significant latitudinal variations in eddy diffusion are consistent with global circulation. However, the elevated mixing in clouds has a minimal effect on atmospheric processes in both the upper and lower atmosphere (Dai et al. 2023, 2024). The 3D model of Shao et al. (2024) encompasses global circulation, transport within and around clouds, and chemical processes in the $H_2SO_4$-$H_2O$ system at altitudes of 40–85 km, demonstrating good agreement with measured $xH_2SO_4$(g), $xH_2O$, and aerosol properties at various latitudes and altitudes.

#### 16.3.1.6 Carbonyl Sulfide

OCS is the most abundant reduced sulfur-bearing gas observed at the altitude of ≈30 to 100 km (Sect. 16.2.1.3). The fate of OCS on Venus has been considered and modeled by Prinn (1975, 1978, 1985), von Zahn et al. (1983), Yung and DeMore (1982), Krasnopolsky and Parshev (1981a, 1981b, 1981c, 1983), Krasnopolsky and Pollack (1994), Mills (1998), Krasnopolsky (2007, 2012, 2013), Yung et al. (2009),Mills and Allen (2007), Zhang et al. (2012), Bierson and Zhang (2020), Sterzenbach et al. (2023), and Dai et al. (2024). OCS forms in the deep lower atmosphere, undergoes transport towards the clouds through eddy (turbulent) diffusion and a low-latitude upwelling, experiences thermochemical oxidation below the clouds, and is depleted by photolysis in and above the upper clouds. Some OCS could be transferred to the near-surface atmosphere through global circulation subsidence at higher latitudes (Fig. 16.16). This chemical and circulation pattern is supported by the increased OCS content at lower altitudes and the elevated $x$OCS at low latitudes above and below clouds.

#### Formation of OCS

The inverse relationship between the abundances of OCS and CO throughout the sub-cloud atmosphere, along with the increased $x$OCS/$x$CO ratio at low latitudes (Sect. 16.2.1.3), indicates OCS formation in a deep lower atmosphere, as discussed by Marcq et al. (2005, 2008, 2018, 2020) and modeled by Yung et al. (2009) and Marcq and Lebonnois (2013). Models by Krasnopolsky and Pollack (1994) and Krasnopolsky (2007, 2013) demonstrate that OCS production occurs at the lowest scale height (<16 km). They suggest that OCS forms through interactions of CO with $S_n$, where $S_n$ stands for atomic sulfur and $S_2$, the most abundant sulfur gas in the near-surface atmosphere (Sect. 16.3.1.7 and Fig. 16.17, Table 16.2). The models of Krasnopolsky (2007, 2013) suggest OCS formation primarily by reaction

$$S + CO + M \rightarrow OCS + M, \qquad (16.31)$$

which exhibits the highest yield at an altitude of 9 km. The results imply that OCS production by reaction

$$S_2 + CO \rightarrow OCS + S \quad (16.32)$$

could be offset by a reverse reaction

$$OCS + S \rightarrow CO + S_2 \quad (16.33)$$

which becomes increasingly significant with altitude and may account for net OCS loss at upper 20x km, as discussed below. In contrast, the modeling of Bierson and Zhang (2020) indicates OCS production by reaction 16.31 at 10 to 25 km and by reaction 16.32 below 10 km. The significance of reaction 16.32 was discussed by von Zahn et al. (1983) and Prinn (1985). Then, its feasibility was demonstrated experimentally at Venus' near-surface temperatures by Hong and Fegley (1997a). Fegley (2014) suggests reaction 16.32 as the most likely pathway for OCS formation. The reaction between $S_2$ and $CO_2$ is less efficient in OCS production (Krasnopolsky 2013). The formation of OCS through the reduction of $SO_2$ by CO (reaction 16.24) may not be significant due to its low rate assessed experimentally (Ferguson 1918; Hong and Fegley 1997a) and through kinetic evaluations (Fegley et al. 1997b). Besides gas-phase reactions, OCS could be supplied through interactions of $CO_2$ with sulfide minerals (Lewis 1970; von Zahn et al. 1983; Prinn 1985; Fegley and Treiman 1992a; Fegley et al. 1995) (Sect. 16.3.2.3).

The dust haze below 5 km (Grieger et al. 2004; Kulkarni et al. 2025) and the minerals in a permeable surface layer could enhance the formation of OCS through surface catalysis, which may promote chemical equilibration among OCS, CO, $SO_2$, and $CO_2$. To align with the observed concentrations of gases (Table 16.1), the equilibration is only feasible under the conditions of the modal planetary radius (Krasnopolsky and Pollack 1994; Zolotov 1996). Calculations of gas-phase chemical equilibria for the surface atmosphere, partially represented in Table 16.2, suggest $x$OCS levels ranging from 5 to 30 ppmv as follows: 20 ppmv (Krasnopolsky and Parshev 1981a), 3 ppmv (Zolotov 1985), 13 ppmv (Krasnopolsky and Pollack 1994), 28 ppmv (Zolotov 1996), 16–29 ppmv (Fegley et al. 1997b), 5 ppmv (Hong and Fegley 1997a), 20 ppmv (Krasnopolsky 2013), and 9 ppmv (Jacobson et al. 2017a). This discrepancy can be attributed to variations in the selected chemical systems, reactions, utilized $x$CO, $x$SO$_2$, and $x$S$_2$, as well as temperature and pressure among different studies. A lower $x$CO of ≈8 ppmv, obtained through extrapolation of Venera 12 data (Table 16.1, Gel'man et al. 1980) to the surface (Fegley et al. 1997a, b), corresponds to 4 to 5 ppmv OCS at equilibrium (Hong and Fegley 1997a; Model 9 of Table 16.2). Equilibrium calculations using measured $S_3$ (Zolotov 1985; Model 10 of Table 16.2) suggest only a few ppmv OCS in the near-surface atmosphere.

**Consumption of OCS**

Photochemical models do not indicate net production of OCS in the mesosphere and clouds due to photolysis and the overall oxidizing conditions it creates. According to Krasnopolsky (2013), net OCS consumption in the middle atmosphere occurs with the overall reaction

$$OCS + 2CO_2 \rightarrow SO_2 + 3CO, \quad (16.34)$$

while the overall decay to CO and $S_n$ is estimated to be insignificant. Actual pathways include photodissociation and interaction with atomic oxygen produced by the photolysis of $CO_2$, $SO_2$, $H_2O$, SO, and $SO_3$

$$OCS + h\nu \rightarrow CO + S \tag{16.35}$$

$$OCS + O \rightarrow SO + CO \tag{16.36}$$

(Krasnopolsky 2012; Zhang et al. 2012). Zhang et al. (2012) and Bierson and Zhang (2020) also considered OCS reactions with Cl, ClS, S, and $NO_3$ as possible loss routes, inferring reaction 16.35 as the main pathway for CO production at ~70 km.

Krasnopolsky (2007) estimated 14 ppbv OCS at 58 km, reflecting OCS consumption within clouds that follows eddy diffusion from the lower atmosphere. The modeled fate of OCS in upper clouds by Krasnopolsky (2012) agrees with the cloud top mixing ratio of ≈3 ppbv at 65 km and with a prominent decrease in *x*OCS with altitude from 64 to 72 km, as observed by Krasnopolsky (2008, 2010b). Significant consumption of OCS at and above the cloud top is modeled by Zhang et al. (2012), Bierson and Zhang (2020), Pinto et al. (2021), and Dai et al. (2024).

Most models do not predict an increase in *x*OCS with altitude above the cloud top (Mills 1998; Krasnopolsky 2012; Zhang et al. 2012; Bierson and Zhang 2020; Dai et al. 2024). Detecting a robust positive gradient at 65 to 100 km and ≈1 ppmv OCS at 100 km (Mahieux et al. 2023, Table 16.1) is inconsistent with modeling results for mesospheric OCS. Mahieux et al. (2023) noted that this inconsistency may indicate a source of OCS in the upper mesosphere. Here, I propose that a supply of FeS and sulfur-bearing organic matter from cometary dust (Carrillo-Sánchez et al. 2020) serves as a source of abundant mesospheric OCS, CS, and $CS_2$. This explanation aligns with the observed OCS gradient above cloud top, the sporadic detections of OCS at low latitudes, and its relatively stable detection pattern at higher latitudes (Mahieux et al. 2023). Considering the short photolysis lifetime of mesospheric OCS (≈6 hours), the observations of Mahieux et al. (2023) imply a steady supply of space materials. However, the observed variability in *x*OCS could indicate fluctuations in the supply of these materials.

Atmospheric models by Krasnopolsky and Pollack (1994), Krasnopolsky (2007, 2013), and Bierson and Zhang (2020) suggest OCS supply through eddy diffusion from the near-surface atmosphere (<20 km), with its net consumption occurring above ~25 km. This scheme aligns with strong OCS gradients inferred for the 30 to 36 km (Sect. 16.2.1.3, Table 16.1). Significant OCS consumption is modeled through oxidation by $SO_3$ at 30 to 40 km. There is no consensus on the specific reaction pathways involved in this process (reactions 16.18–16.21), and experimental data on reaction rates and mechanisms are needed to assess the thermochemical oxidation of OCS (Sects. 16.3.1.2 and 16.3.1.4).

Prinn (1975, 1978) proposed OCS interaction with atomic sulfur (reaction 16.33) as a significant pathway for OCS consumption and $S_2$ production in the middle and lower atmosphere. However, neither observation nor model suggests an elevated abundance of atomic sulfur needed to consume OCS at 30–36 km, where *x*OCS decreases sharply with altitude. The modeling by Krasnopolsky and Pollack (1994) and Krasnopolsky (2007) does not indicate that reaction 16.33 is efficient at a low $xS_n$ at 30–36 km. To avoid the problem of $S_n$ deficiency, Yung et al. (2009) suggested OCS interaction 16.33 with supposedly abundant atomic sulfur produced by soft-UV photolysis of $S_3$ and $S_4$ in the sub-cloud atmosphere

$$S_4 + h\nu \rightarrow S_3 + S \tag{16.37}$$

$$S_3 + h\nu \rightarrow S_2 + S \quad (16.38)$$

Whether this mechanism works can be tested by assessing the abundance of $S_n$ gases and by gaining a better understanding of UV irradiance below clouds. Krasnopolsky (2013) included photochemical reactions proposed by Yung et al. (2009) in his 2007 model and confirmed that the oxidation of OCS by $SO_3$ permits the principal OCS loss at 30 to 36 km, consistent with the anticorrelation of OCS and CO and their opposite vertical gradients (Sect. 16.2.1.3). However, the models of Krasnopolsky (2013) and Bierson and Zhang (2020) show that reaction 16.33 significantly contributes to OCS loss below 30 km. Krasnopolsky (2013) demonstrated that reaction 16.33 reaches its highest efficiency at 28 km and causes OCS loss in the upper 20s km, as suggested from observations by Pollack et al. (1993). Results of Bierson and Zhang (2020) show that reaction 16.33 is accountable for OCS loss at ≈5 to 30 km. Reaction

$$OCS + M \rightarrow CO + S + M \quad (16.39)$$

is modeled as accountable for OCS loss below 5 km (Krasnopolsky 2007; Bierson and Zhang 2020), while the reaction

$$OCS + S_2 \rightarrow CO + S_3 \quad (16.40)$$

provides a lesser contribution (Krasnopolsky 2007, 2013). The alteration of OCS to CO on the surface of hematite, likely a secondary mineral (Sect. 16.3.2.3), occurs too slowly to significantly influence OCS consumption, as Yung et al. (2009) estimate. Despite the assessed OCS loss being below ≈25 km through reactions 16.33, 16.39, and 16.40, net OCS production is commonly modeled in that region, as discussed above.

#### 16.3.1.7 Sulfur Gases and Condensates

Although sulfur gases ($S_n$) and condensates ($S_x$, $S_8$) could play significant roles in atmospheric chemistry, observational data are limited (Sects. 16.2.1.3 and 16.2.1.4), and there is no consensus regarding their occurrence, abundance, speciation, and chemical pathways. The current understanding of their behavior is based on models. The ambiguity of modeling results reflects the lack of rate constants for critical reactions.

$S_n$ gases can be produced above clouds (e.g., reaction 16.35) and through sub-cloud thermochemical reactions (e.g., reaction 16.18), influenced by photolysis from soft UV photons below 30 km (reactions 16.37 and 16.38), as well as consumed and formed in thermochemical reactions below 30 km (reactions 16.31, 16.32, 16.33, 16.39, and 16.40, Table 16.7). $S_2$ forms via $SO_2$-mineral reactions (Sect. 16.3.2.2) and is expected in volcanic gases (Sect. 16.4.3). Chemical equilibrium models for the sulfur system (San'ko 1981; Zolotov 1985; Krasnopolsky 2013) suggest increasing stability and relative abundances of heavier $S_n$ gases with altitude, with a dominance of $S_8$ at high altitudes. Barsukov et al. (1980a, 1982a), Dorofeeva et al. (1981), and Volkov et al. (1982) discussed the phase diagram of sulfur concerning condensation on Venus. They highlighted the potential for condensation at the cloud base at a total $S_n$ abundance of ≈0.2 ppmv suggested by gas-phase chemical equilibria at the surface (e.g., Models 4, 5 of Table 16.2). The phase diagram does not imply condensation at the cloud base at a total $S_n$ of ≈0.01 ppmv, as inferred from Venera spectrophotometer data below 25 km (Moroz et al. 1981; San'ko 1981; Sect. 16.2.1.3, Model 10 of Table 16.2).

The bulk $S_n$ content within and around clouds remains unknown, and it is unclear whether sulfur condenses. The major constraint is the diversity in model-based assessments of the bulk content and sources of sulfur in clouds. Both photochemical (from above) and thermochemical (from below) sources are considered in the literature. There is greater consistency in understanding $S_n$ gases below 30 km, where $S_3$ and $S_4$ are constrained by observations (Sect. 16.2.1.3, Table 16.1) and the dominance of $S_2$ in the near-surface atmosphere is suggested by chemical equilibrium assessments (Table 16.2). The bulk $S_n$ content and $xS_2$ of 0.1 to 0.4 ppmv correspond to chemical equilibria calculated from measured $CO_2$, CO, and $SO_2$ (Models 1–8 of Table 16.2). This bulk $S_n$ content allows $S_x$ condensation in clouds at $\approx$50 km (Krasnopolsky 2013). The consideration of extrapolated $x$CO or observational $xS_3$ data results in 0.01 to 0.02 ppmv $S_2$ and bulk $S_n$ in the near-surface atmosphere (Models 9 and 10), which could impede condensation in clouds.

#### Pathways of Gaseous and Condensed Sulfur in the Middle Atmosphere

The fate of sulfur species within and around clouds has been considered by Prinn (1975, 1985), Dorofeeva et al. (1981), Barsukov et al. (1982a), Young (1983), von Zahn et al. (1983), Krasnopolsky and Pollack (1994), Mills and Allen (2007), Zhang et al. (2012), Krasnopolsky (2012, 2016), Pinto et al. (2021), and Francés-Monerris et al. (2022). Figures 16.19 and 16.20 indicate that $S_n$ can be formed and consumed through multiple reactions. Currently, the prevailing chemical pathways and the amounts of $S_n$ produced in the middle atmosphere are insufficiently constrained. One reason for this ambiguity is the lack of data on reduced sulfur-bearing gases within clouds. Another reason is the inadequate information on the kinetics of various $S_n$ species reactions.

Before the Pioneer Venus mission, Prinn (1975, 1978) examined the formation of atomic sulfur and $S_2$ through the photolysis of OCS via reaction 16.35, the interaction of OCS with atomic sulfur (reaction 16.33), and the collision of two HS molecules formed through the photolysis of $H_2S$. His results show that the presence of oxidants, such as $O_2$, in upper clouds and above limits the net production of $S_n$ from supposedly abundant OCS and $H_2S$. In his models, atomic sulfur is consumed by reaction 16.28 with $O_2$ at and above the cloud top, and via reaction 16.33, which produces $S_2$ under deeper $O_2$-poor conditions and could result in $S_x$ upon polymerization and condensation. von Zahn et al. (1983) and Prinn (1985) further emphasized the overall formation of $S_n$ species through OCS photolysis and HS + HS interactions in upper clouds. Mills et al. (2007) mentioned the possible formation of free sulfur via the disproportionation of $SO_2$ to $S_n$ and $SO_3$ at the cloud top.

Mills and Allen (2007) considered $S_2$ formation in upper clouds through reaction cycles involving Cl-bearing species, such as SCl, $SCl_2$, $S_2Cl$, ClCO, Cl, and $Cl_2$. Their models suggest that atomic sulfur primarily forms through the photolysis of SO and SCl, followed by the formation of $S_2$ via the dissociation of $SCl_2$ and SCl. The consumption of atomic sulfur and $S_2$ is modeled through photochemical reactions in upper clouds involving SCl, $S_2Cl$, $SCl_2$, and oxygen-Cl species. At cloud top (63–70 km), atomic sulfur and $S_2$ gases are supposedly consumed by reactions 16.28 and 16.41, while reactions 16.42 and 16.43 are modeled as responsible for losses in more reduced upper clouds at 58 to 63 km

$$S_2 + O \rightarrow SO + S \quad (16.41)$$

$$S + Cl_2 \rightarrow SCl + Cl \quad (16.42)$$

$$2S_2 + M \rightarrow S_4 + M. \quad (16.43)$$

Mills and Allen (2007) noted a significant negative effect of $xO_2$ and the $O_2/Cl$ ratio on net $S_n$ production and modeled increasing losses of atomic sulfur and $S_2$ with altitude. Building on the work of Prinn (1975), Mills et al. (2007) posited that the fate of sulfur, whether it transforms into sulfuric acid or $S_x$ condensate, is governed by a competitive process: the oxidation of atomic sulfur by $O_2$ via reaction 16.28 versus its reaction 16.33 with OCS.

The production and fate of free sulfur in the middle atmosphere have been considered in the models of Krasnopolsky (2012) and Zhang et al. (2012). Their nominal models for clouds at 47–60 km suggest mixing ratios of $S_n$ (mainly $S_8$) of $10^{-12}$ to $10^{-9}$, which limit condensation. Zhang et al. (2012) demonstrated that the photolysis of OCS (reaction 16.35) and the interaction between OCS and S (reaction 16.33) could significantly contribute to the production of S and $S_2$ at 58–62 km if there is a substantial concentration of OCS. This abundance of OCS could be attributed to increased eddy diffusion from the lower atmosphere or other factors that are not yet clarified. Krasnopolsky (2012) demonstrated that the formation and photolysis of $S_2O_2$ (reactions 16.25 and 16.26) are more effective in producing atomic sulfur than the pathways listed above; however, $S_2O_2$ is not considered abundant in this and other photochemical models (Krasnopolsky 2018; Bierson and Zhang 2020; Pinto et al. 2021). In Krasnopolsky's (2012) model, atomic sulfur is easily lost by reaction 16.28 with $O_2$, which limits the buildup of sufficient $S_n$ gases to form $S_x$ aerosol in upper clouds. The formation of atomic sulfur in clouds through the thermochemical reduction of $SO_2$ by CO (reaction 16.12) is considered negligible under the overall oxidizing conditions of the upper clouds (Krasnopolsky 2013). In these scenarios, condensed sulfur is absent from upper clouds, zero net transport of free sulfur to the sub-cloud region is expected, and possible condensed $S_x$ in clouds is sourced from below.

Assuming sufficient formation of atomic sulfur in the initial reactions (e.g., reaction 16.26), several research groups addressed polymerization pathways to constrain condensation within clouds. Mills and Allen (2007) modeled the formation of polyatomic $S_n$ species via reactions 16.43 and 16.44,

$$S + S_2 + M \rightarrow S_3 + M \quad (16.44)$$

In contrast to the results of Zhang et al. (2012) and Krasnopolsky (2012, 2018), which do not predict much $S_2O_2$ and $S_n$ in upper clouds, photochemical models by Pinto et al. (2021) and Francés-Monerris et al. (2022) considered reactions of $S_2O_2$ that could produce $S_3$, $S_4$, and polysulfur species (Fig. 16.19). Their results for carbon-oxygen-nitrogen-sulfur-Cl species at 55 to 105 km suggest that $S_2$ is the most abundant $S_n$ gas at the cloud top (Fig. 16.21). Pinto et al. (2021) modeled the photodissociation of $S_2O_2$ through reactions such as reactions 16.26 and 16.45

$$\text{cis}-S_2O_2 + h\nu \rightarrow SO + SO \rightarrow S_2 + O_2 \tag{16.45}$$

and the polymerization of $S_2$ to $S_n$ as a source of condensed sulfur. Quantum-chemistry computations and photochemical models from Francés-Monerris et al. (2022) demonstrated that $S_2$ may not be a primary product of $S_2O_2$ photolysis and modeled $S_2$ formation via reactions

$$S_2O_2 + SO \rightarrow S_2O + SO_2 \tag{16.46}$$

$$ClS + SO \rightarrow S_2O + Cl \tag{16.47}$$

$$S_2O + SO \rightarrow S_2 + SO_2 \tag{16.48}$$

These reactions could facilitate the formation of polyatomic $S_n$ gases and condensed $S_x$ species from SO and cis-$S_2O_2$ via the $S_2O$ intermediate and maintain a significant fraction of condensed sulfur in cloud aerosols.

Models for the middle and lower atmosphere (Krasnopolsky 2007, 2013; Bierson and Zhang 2020) suggest a variable bulk sulfur content with altitude due to competitive consumption and production reactions. For the cloud base at 47 km, Krasnopolsky (2007) modeled the major $S_n$ gases as follows: $S_8$ (80%), $S_7$ (8.5%), and $S_6$ (10.5%). He estimated the $S_n$ number density to be near saturation at the cloud top, and that the complete evaporation of sulfur aerosol occurs at the cloud deck. The model of Krasnopolsky (2013) reveals the formation of $S_x$ aerosol through condensation in lower clouds (at $\approx$50 km). Krasnopolsky (2016) calculated the profile of condensed sulfur in the clouds based on an $xS_8$ of 2.5 ppm at the cloud deck (47 km), which was inferred from his 2013 model. It was shown that the lower cloud layer could contain 10 wt% sulfur (consistent with the Vega data, Porshnev et al. 1987) and that the mass loading of $S_x$ sharply decreases toward the cloud top. All models for the lower and middle atmosphere exclude $S_x$ from being an effective UV absorber in upper clouds (Sect. 16.3.1.8) if sufficient sulfur species do not form there.

### Sulfur Species in the Lower Atmosphere

If condensed sulfur is present in clouds (Sect. 16.2.1.4), the subsidence of aerosol particles due to gravity and circulation removes native sulfur from the middle atmosphere. The evaporation of $S_x$ within a few kilometers below the cloud deck would be followed by the sequential thermal decomposition of $S_8$ into lower-mass allotropes, with $S_2$ dominating at lower altitudes (Bierson and Zhang 2020; Krasnopolsky 2013; San'ko 1981; Zolotov 1985, Figs. 16.15 and 16.17) and in the near-surface atmosphere (Table 16.2).

Models suggest the thermochemical formation of atomic sulfur below 30 km through the photolysis of $S_3$ and $S_4$ (reactions 16.37 and 16.38; Krasnopolsky 1987, 2013), the interaction of SO molecules (reaction 16.23) formed via the interaction of $SO_2$ and CO (reaction 16.22) below 16 km, which together correspond to net sulfur production by reaction 16.12 (Krasnopolsky 2013), $S_2$ interaction with CO below 10 km (reaction 16.32; Bierson and Zhang 2020), and OCS dissociation by reaction 16.39 below 5 km (Krasnopolsky 2007, 2013; Bierson and Zhang 2020). $S_2$ is modeled as a byproduct of OCS oxidation by $SO_3$ at min-30s km (reactions 16.18–16.20) (Krasnopolsky and Pollack 1994; Krasnopolsky 2007, 2013); however, reaction 16.20 may be inefficient due to the short thermochemical life of $S_2O_2$ (Bierson and Zhang 2020). $S_2$ forms through OCS oxidation by atomic sulfur by reaction 16.33 at 5–30 km (Krasnopolsky 2013; Bierson and Zhang 2020), via $S_3$ photolysis via reaction 16.38 at 3–29 km, and through thermal decomposition of $S_4$ as assessed by Krasnopolsky (2013). $S_3$ forms through $S_4$ photolysis (reaction 16.37) at ≈18 km and via $S_2$ + OCS interaction 16.40 below 5 km (Krasnopolsky 2013).

The modeled and discussed consumption of $S_n$ gases included thermochemical oxidation by $SO_3$ (Mills and Allen 2007), photochemical dissociation by soft UV photons (reactions 16.37 and 16.38) (Yung et al. 2009; Krasnopolsky 2013; Bierson and Zhang 2020), and reactions 16.31 and 16.32 with CO. The loss of $S_n$ gases through the reduction of $SO_3$ to $SO_2$ is not supported by Krasnopolsky's (2013) and Bierson and Zhang's (2020) models. Photolysis of $S_3$ and $S_4$ to produce abundant $S_n$ in the sub-cloud atmosphere, proposed by Yung et al. (2009), has no observational support. Krasnopolsky (2013) and Bierson and Zhang (2020) included the kinetics of $S_4$ photolysis (reaction 16.37) in their models. However, they did not infer abundant $S_n$, which strongly affects OCS loss in the mid-30s km (Sect. 16.3.1.6). Models of Krasnopolsky (2007, 2013) and Bierson and Zhang (2020) suggest the consumption of atomic sulfur through the oxidation of OCS by reaction 16.33 below 30 km and via the reduction of CO to OCS by reaction 16.31 below 25 km. A lot more abundant $S_2$ is partially consumed through reaction 16.32 with CO in the near-surface atmosphere. At higher altitudes, reaction 16.33 could compensate for and overwhelm the loss of $S_2$. The net loss of $S_n$ gases from the lower atmosphere occurs through transport to the middle atmosphere. In Krasnopolsky's (2013) model, an upward eddy diffusion of $S_n$ is followed by condensation in lower clouds at 50 km.

#### 16.3.1.8 Unknown Blue-UV Absorber in Upper Clouds

In addition to gaseous $SO_2$, Venus' spectra suggest an unknown absorber or absorbers in the UV and blue (320 to 500 nm) spectral ranges (Esposito 1980). The featureless spectra of Venus in the near-UV range and in situ nephelometry data from the Pioneer Venus and Venera 14 probes suggest a condensed absorber (Ekonomov et al. 1984) occurring in the upper clouds above 57–60 km (Sect. 16.2.1.4). Among various potential compounds such as $FeCl_3$ (Kuiper 1969; Zasova et al. 1981; Krasnopolsky 2017), NO, and $Cl_2$, both gaseous and condensed sulfur-bearing species have been proposed to explain the absorption at short wavelengths. The spatial correlation of $SO_2$ with an unidentified UV-blue absorber, as deduced from Akatsuki UV observations, suggests the involvement of a sulfur-bearing absorber in atmospheric processes alongside $SO_2$ (Yamazaki et al. 2018). The candidate sulfur-bearing species considered are as follows: condensed elemental sulfur ($S_x$) (Hapke and Nelson 1975; Young 1983), $S_3$ and $S_4$ (Toon et al. 1982), $S_2O$ (Hapke and Graham 1989; Na and Esposito 1997), $CS_2$ (Barker 1978; Young 1978), $SCl_2$ (Krasnopolsky 1986), $NOHSO_4$ (Watson et al. 1979), irradiated $S_2O$ (Pérez-Hoyos et al. 2018), $S_2O_2$ gas (Frandsen et al. 2016; Pérez-Hoyos et al. 2018), and ferric sulfates (Jiang et al. 2024). However, many of these species are inconsistent with the entire UV-blue spectrum of Venus, observed cloud properties, and atmospheric and chemical models (Titov et al. 2018; Pérez-Hoyos et al. 2018; Limaye et al. 2018; Dai et al. 2025).

Condensed native sulfur may be present in lower clouds based on the Vega in situ data

(Sect. 16.2.1.4). However, the models of Zhang et al. (2012) and Krasnopolsky (2012) do not suggest the formation and accumulation of abundant $S_n$ species in upper clouds (Sect. 16.3.1.7). Krasnopolsky (2016) calculated the condensed $S_x$ profile through clouds and concluded that sulfur cannot be a significant UV absorber. The spectra of $S_4$, $SCl_2$, $NOHSO_4$, and $S_2O$ do not match the entire UV-blue spectrum of Venus, particularly between 0.4 and 0.5 μm (Pérez-Hoyos et al. 2018), although a mixture of absorbers may align with observations. The presence of $S_4$ in such a mixture would explain absorption at longer wavelengths (Pérez-Hoyos et al. 2018); however, $S_4$ is suggested from the blue absorption pattern only below ≈20 km (Krasnopolsky 2013). Another concern regarding $S_x$ serving as the UV absorber is its low solubility in sulfuric acid solutions, which suggests that sulfur exists as a separate phase that accumulates on the surfaces of aerosol droplets (Young 1983) and potentially forms separate aerosol particles. Consequently, the covering of liquid aerosol droplets by native sulfur should reduce the observed optical phenomenon of glory in upper clouds, as shown by Petrova et al. (2018).

Although all three SO dimers ($S_2O_2$, Sect. 16.3.1.3) are strong absorbers in the blue and UV wavelengths (Frandsen et al. 2016; Pérez-Hoyos et al. 2018), the photochemical models by Zhang et al. (2012), Krasnopolsky (2012, 2018), Zhang and Bierson (2020), and Egan et al. (2025) do not predict sufficiently abundant $S_2O_2$ isomers to be the main UV absorbers in upper clouds, which is consistent with the lack of $S_2O_2$ detection. Nevertheless, the upper cloud models of Pinto et al. (2021) and Frances-Monerris et al. (2022) suggest interactions involving $S_2O_2$ (reactions 16.45–16.48) that could produce abundant $S_2O$, polysulfur oxides, $S_3$, $S_4$, and condensed $S_x$ species (Fig. 16.19), all of which absorb in the blue-UV range.

Jiang et al. (2024) demonstrated that the interaction of Fe(III) sulfate with concentrated sulfuric acid leads to the formation of rhomboclase, $(H_5O_2)Fe(SO_4)_2 \cdot 3H_2O$, and acid ferric sulfate $(H_3O)Fe(SO_4)_2$. The authors showed that a mixture of these solid phases with liquid $Fe^{3+}$-bearing sulfuric acid sufficiently matches the spectrum of Venus. However, the possible presence of Fe in clouds suggested by in situ measurements (Petryanov et al. 1981b; Andreichikov et al. 1987; Mogul et al. 2025, Sect. 16.2.1.4) needs confirmation. It is unclear whether the supply of Fe from space (Fe-metal, FeS, ferrous silicates) and/or the surface compensates for the sinking of ferric sulfates toward the surface, where they irreversibly alter to stable hematite, $\alpha-Fe_2O_3$ (Mogul et al. 2025; Zolotov 2021; Zolotov et al. 2023). Today, there is no consensus on the composition of the UV-blue absorber in clouds.

#### 16.3.1.9 Polysulfur Oxides, Hydrogen Sulfide, and Carbon Sulfides

None of the reduced sulfur-oxygen and sulfur-hydrogen gases are detected in the middle and lower atmosphere. CS and $CS_2$ are only observed at or above cloud top (Table 16.1). Evaluating the fate of all these gases is model-dependent, and there is no consensus regarding the limited modeling results.

**Polysulfur Oxides ($S_xO$)**

Although no observational data on $S_xO$ gases are available, $S_2O$ is a potential condensable species in the middle atmosphere that absorbs in the UV range (Sect. 16.3.1.8). Yung and DeMore (1982) first modeled and discussed the formation of $S_2O$ in the middle and upper atmosphere via $S_2O_2$ + SO interaction (reaction 16.46) and its

consumption by reactions

$$S_2O + h\nu \rightarrow S + SO \quad (16.49)$$

$$S_2O + S \rightarrow S_2 + SO \quad (16.50)$$

$$S_2O + O \rightarrow 2SO \quad (16.51)$$

at 58 to 100 km. Na and Exposito (1997) modeled the formation of abundant $S_2O$ in upper clouds using the kinetics of reactions 16.46, 16.49, and 16.51 from Yung and DeMore (1982). The accumulation of $S_2O$ was attributed to its gradual depletion via reaction 16.51. However, the models of Mills (1998) suggest several orders of magnitude smaller $xS_2O$ than the evaluations of Na and Esposito (1997) due to the faster rate used for $S_2O_2$ loss via back reaction 16.25. The model results do not indicate that $S_2O$ is a major UV absorber in upper clouds.

The photochemical modeling of Pinto et al. (2021) and Frances-Monneris et al. (2022) suggests that the reactions of SO and $S_2O_2$ isomers are an essential source of $S_2O$ and other $S_xO$ species in the middle and upper atmosphere. Pinto et al. (2021) modeled the formation of $S_2O$ through the interaction of SO with $S_2O_2$ isomers (e.g., reaction 16.46) and ClS via reaction 16.47, as well as by reaction

$$ClO + S_2 \rightarrow S_2O + Cl \quad (16.52)$$

(Fig. 16.19). In their models, $S_2O$ reaches a ppbv level at the cloud top (≈ 65 km) and becomes less abundant in upper clouds (≈ 58 km) and the mesosphere. Calculated profiles of $S_3O$ and $S_4O$ show similar patterns at lower mixing ratios. At 100 km, the estimated mixing ratios of all $S_xO$ species are below $10^{-15}$. Frances-Monneris et al. (2022) modeled the formation of $S_2O$ through reactions 16.46 and 16.47, and the consumption of $S_2O$ through interaction with SO by reaction 16.48 at altitudes of 58 to 110 km. In their models, $S_2O$ reaches a maximum concentration at ≈ 63 km (Fig. 16.19). At 60 km, the modeled number densities of $S_2O$, cis-$S_2O_2$, and $S_2$ formed by reaction 16.48 are equal. The results suggest an essential role for $S_2O$ as an intermediate species in the formation of $S_n$ and other possible UV absorbers in upper clouds (Sects. 16.3.1.7 and 16.3.1.8).

#### Hydrogen Sulfide and Hydrogen Monosulfide

Although $H_2S$ has not been firmly detected (Sect. 16.2.1.3), it has been discussed as a key reduced gas by Prinn (1978, 1985) and von Zahn et al. (1983). They suggested $H_2S$ formation through the interaction between surface pyrite and $H_2O$ vapor (Sect. 16.3.2.3) and via thermochemical reactions in the deep atmosphere, as well as consumption by photochemical dissociation and oxidation in the middle atmosphere by the overall process

$$H_2S + 3CO_2 \rightarrow H_2O + 3CO + SO_2 \quad (16.53)$$

Yung and DeMore (1982) discussed $H_2S$ production and loss through photolysis and interaction with atomic hydrogen. However, they did not include the corresponding reactions in the model due to insufficient information on reaction rates. Krasnopolsky (2007, 2013) incorporated the rates of thermochemical reactions involving $H_2S$ in his models and estimated $xH_2S$ in the lower and middle atmosphere. The models assume that $H_2S$ equilibrates with other gases at the surface. Modeling results suggest the thermochemical formation of the HS through reactions

$$H + OCS \rightarrow CO + HS \quad (16.54)$$

$$H_2S + S \rightarrow HS + HS \quad (16.55)$$

$H_2S$ forms through reactions of HS with HCl, $H_2O$, HSCl, and $H_2$, along with the back reaction 16.55. The relatively fast reaction

$$HS + HCl \rightarrow H_2S + Cl \quad (16.56)$$

provides a major contribution. The formation of $H_2S$ is balanced by reaction 16.55 and by the flux toward clouds, where $H_2S$ undergoes oxidation. Maximum yields of reactions 16.55 and 16.56 are modeled at 18 km and 26 km, respectively. The modeled decrease in $xH_2S$ from an assumed 150 ppbv at 0 km altitude to a modeled 32 ppbv at the cloud deck at 47 km reflects the upward flux. The models of Bierson and Zhang (2020) and Dai et al. (2024) indicate a constant $xH_2S$ in the lower and middle atmosphere and a significant drop above 65 km. Their results are consistent with the spectroscopic upper limit for $xH_2S$ at the cloud top reported by Krasnopolsky (2008) (Table 16.1). Neither Pioneer Venus Large Probe mass spectrometry (Hoffman et al. 1980) nor Venera 13/14 entry probe gas chromatography data on $H_2S$ (Mukhin et al. 1983) are supported by these models. Most chemical equilibrium models for the near-surface atmosphere that assume 20 to 30 ppmv $H_2O$ vapor suggest an $xH_2S$ of 0.01 to 0.15 ppmv and 0.12 to 1.2 pptv HS (Table 16.2).

#### Carbon Sulfides

Although abundant CS and $CS_2$ are detected at and above cloud top (Table 16.1), they have yet to be included in atmospheric models, and the observed abundances of these reduced species in the overall oxidizing mesospheric environments remain unclear. The increasing mixing ratios of carbon sulfides with altitude correlate with $SO_3$, $SO_2$, SO, and OCS trends in the upper mesosphere (Sect. 16.2.1). As noted with respect to OCS (Sect. 16.3.1.6), one possible explanation for mesospheric CS and $CS_2$ is their formation through the photochemical alteration of extremely carbon-rich cometary dust (see Table 16.12 for carbon flux) that contains up to 50% organic matter by mass (Bardyn et al. 2017).

#### 16.3.1.10 Chlorine-Bearing Sulfur Gases

The formation and consumption of $ClSO_2$, $SO_2Cl_2$, SOCl, $ClSO_4$, SCl, $S_2Cl$, $SCl_2$, and HSCl have been considered since the early 1980s. Evaluations and models suggest that these species form in the middle atmosphere through HCl photolysis and are consumed in the sub-cloud atmosphere, reverting to HCl and sulfur-bearing gases. Strattan et al. (1979) and Yung and DeMore (1982) first discussed the role of the reaction

$$SO_2 + Cl + M \rightarrow ClSO_2 + M \quad (16.57)$$

in the chemistry of Cl and $ClO_n$ species that may affect the oxidation of $SO_2$ in upper clouds. DeMore et al. (1985) experimentally demonstrated the anoxic formation of $SO_2Cl_2$ through the photolysis of mixtures of $SO_2$ and $Cl_2$. Inferred reactions in the sulfur-oxygen-Cl system included the formation of the intermediate $ClSO_2$ via reaction 16.57 and its conversion to $SO_2Cl_2$ by reaction

$$ClSO_2 + ClSO_2 \rightarrow SO_2Cl_2 + SO_2 \quad (16.58)$$

Atmospheric models with corresponding reactions suggested that $SO_2Cl_2$ is a key Cl-bearing species in upper clouds with a mixing ratio of 4 ppmv. The oxidation of $ClSO_2$ by $O_2$ to $ClSO_4$ was identified as a critical pathway for $O_2$ consumption and $SO_2$ oxidation to $H_2SO_4$ (Sect. 16.3.1.2). Although subsequent modeling by Krasnopolsky (2012) and Bierson and Zhang (2020) did not support a significant role for Cl-bearing species in $SO_2$ oxidation, the role of $ClSO_2$ remains unclear due to uncertain reaction rates.

Mills (1998) considered reactions involving $SClO_x$ and $SCl_x$ occurring within and above upper clouds and provided the first quantitative estimation of their chemistry. $SCl_2$, SCl, $S_2Cl$, and $S_2Cl_2$ (the least abundant) are predicted to exist only within the upper cloud layer at ≈56 to 60 km. $ClSO_2$ is expected to be more abundant than $S_mCl_n$ gases. The model shows that $ClSO_2$ could serve as a buffer for $SO_2$, slowing its oxidation to $H_2SO_4$. At the same time, $S_mCl_n$ species could serve as intermediaries in the polymerization of atomic sulfur to $S_n$. Mills and Allen (2007) further discussed and evaluated the role of SCl, $S_2Cl$, and $SCl_2$ as intermediate species that affect $S_n$ gases. In their model, the formation of $S_2$ occurs via $S_2Cl$ and SCl (Sect. 16.3.1.7). As in the work of Mills (1998), the formation of sulfur-Cl gases was modeled to occur in the upper cloud layer that is depleted in $O_2$.

Zhang et al. (2012) modeled the fates of $ClSO_2$, SOCl, SCl, $ClS_2$, $SCl_2$, and $S_2Cl_2$ in the middle and upper atmosphere at 58 to 110 km. The formation of the most abundant $ClSO_2$ is modeled by reaction 16.57, in which M is $CO_2$. $ClSO_2$ is consumed via reactions with O, S, $S_2$, H, Cl, SO, SCl, and $ClSO_2$ (reaction 16.58), resulting in the production of $SO_2$. Estimated $x$$ClSO_2$ and $x$SOCl nearly reach ppbv levels at the cloud top at 65 km and are less abundant at both higher and lower altitudes. At 65 km, the concentrations of $SCl_2$, SCl, $ClS_2$, and $S_2Cl_2$ range from 1 to 100 pptv and decrease at higher altitudes due to oxidation. $SCl_2$ is more abundant than $ClSO_2$ in oxygen-depleted upper clouds at 58–60 km.

Krasnopolsky (2012, 2013) modeled $ClSO_2$, OSCl, SO2$Cl_2$, $SCl_2$, and SCl at altitudes of 47 to 112 km and from 0 to 47 km. $SO_2Cl_2$ is estimated to be HCl's most abundant photochemical product, with a mixing ratio of 8 ppbv at 47 km and 30 ppbv at 68 km. $SO_2Cl_2$ forms through reactions 16.57 and 16.58. Its net production in upper clouds occurs through the photolysis of $CO_2$ and HCl, along with reaction

$$2HCl + SO_2 + CO_2 \rightarrow SO_2Cl_2 + H_2O + CO \quad (16.59)$$

The modeling indicates a net supply of $SO_2Cl_2$ to the lower atmosphere and other photochemical products, such as CO and $H_2SO_4$(g). Below 30 km, $SO_2Cl_2$ is consumed by reactions

$$SO_2Cl_2 + S \rightarrow ClSO_2 + SCl \quad (16.60)$$

$$SO_2Cl_2 + Cl \rightarrow ClSO_2 + Cl_2 \quad (16.61)$$

The assessed $x$$SO_2Cl_2$ is 7 pptv at the surface, and a detectable constant concentration of 3.5 ppbv is estimated above ≈ 30 km because of the lack of gas sinks and sources (Fig. 16.17). Below 25 km, the primary sulfur-Cl-bearing gas is $ClSO_2$ at ≈ 20 pptv, while HCl is the most abundant Cl-bearing species at 0.5 ppmv in the lower atmosphere.

Bierson and Zhang (2020) modeled $SO_2Cl_2$ as the most abundant species among seven $S_xCl_yO_z$ and $SxCl_y$ gases at altitudes of 30 to 80 km, with a mixing ratio ranging from 0.1 to 10 ppbv at 70–80 km. Their model shows that $xSO_2Cl_2$ strongly decreases at lower altitudes. Below 30 km, $ClS_2$ is estimated to be the most abundant sulfur-Cl-bearing gas, with a mixing ratio of up to 1 ppbv at 0 km. Similarly, Dai et al. (2024) assessed $SO_2Cl_2$ as the major sulfur-Cl-bearing gas at 65 to 90 km at the pptv level. In the middle and lower atmosphere, they reported that $ClS_2$ is the most abundant gas at the ppbv level, followed by $SCl_2$, $S_2Cl_2$, and HSCl, which exist at the pptv level. The understanding of the fate of sulfur-Cl-bearing species is hindered by the lack of detection and the generally undetermined rates of gas-phase reactions.

## *16.3.2* Sulfur in the Atmosphere-Surface Interactions

### 16.3.2.1 Approaches, Constraints, and Major Pathways

Chemically active gases in the hot and dense near-surface atmosphere (Table 16.1 and 16.2) make gas-solid type chemical reactions unavoidable. $CO_2$ and $H_2O$ are likely oxidizing agents of Fe(II) in exposed materials such as silicates, sulfides, oxides, and silicate glasses (e.g., Barsukov et al.1982c; Fegley and Treiman 1992a, b; Fegley et al.1997a; Zolotov 2018). However, the fate of sulfur-bearing solids is more intricate and less understood. The prevailing view is that sulfur-bearing solids play a critical role in the atmosphere-surface system (von Zahn et al. 1983; Prinn 1985; Fegley and Treiman 1992a; Fegley et al. 1995, 1997a) and in the physical-chemical evolution of the system since the global volcanic resurfacing event (Bullock and Grinspoon 2001). It has been suggested that the chemical sequestration and release of gases at the atmosphere-surface interfaces affect both the abundance and speciation of $SO_2$, OCS, $H_2S$, $S_n$, and trace non-sulfur species such as CO throughout the atmosphere. The gas-solid interactions of sulfur-bearing compounds could be crucial for maintaining sulfuric acid clouds and, therefore, for the thermal and chemical structure of the entire gas envelope (Fegley and Prinn 1989; Bullock and Grinspoon 2001). Permeable geological materials (minerals and glasses in the bedrock, along with products of their physical and chemical weathering) may contain more sulfur than the atmosphere (Lewis and Kreimendahl 1980; Zolotov and Volkov 1992, Sect. 16.4.4.1) and may control the abundance and speciation of atmospheric sulfur via gas-solid reactions. Robert Mueller and John Lewis established the concept of chemical equilibria between chemically active atmospheric gases and crustal materials on Venus (Mueller 1963, 1964, 1965; Lewis 1968, 1970; Lewis and Kreimendahl 1980). Further investigations indicated the likelihood of chemical equilibria of certain trace atmospheric gases with crustal materials, which may have been established through gas-solid interactions over time. This section considers gas-solid reactions and equilibria at the current temperature, pressure, and assumed gas composition (Tables 16.1 and 16.2) below 10 km, the highest elevation on Venus above the 6052 km radius. Aspects of gas-solid interactions in the past and future are discussed in Sect. 16.4.4.

The elevated bulk sulfur abundance in the otherwise basaltic composition of the Venera and Vega samples (Tabl1e6.3) implies an exogenic (atmospheric and/or cosmic) source of sulfur. The sequestration of atmospheric sulfur through gas-solid weathering reactions is the standard explanation for these data. Sulfur dioxide likely plays significant role in such reactions, as inferred from evaluations of mineral stability and targeted experimental studies discussed below. These efforts led to lists of stable and unstable minerals in contact with $SO_2$, along with potential solid (sulfates, silica phases, silicates, and gaseous ($S_2$, CO, etc.) products of these interactions. Less conclusive and sometimes conflicting evaluations have emerged regarding the stability and fate of metal sulfides. Some sulfides are deemed unstable, releasing sulfur-bearing gases into the atmosphere. Stable sulfides may remain intact within the rocks and form at the surface through gas-solid reactions or condensation.

Currently, the study of both mineral stabilities and experimental evaluations of gas-solid interactions affecting sulfur-bearing compounds is hindered by various factors. A significant limitation is the absence of instrumental data on nearly all trace gases below ≈20 km (Table 16.1), which restricts understanding of near-surface gas chemistry and the assessment of key fugacities ($fO_2$, $fS_2$) that impact mineral stability. Existing models for gas composition (Table 16.2) establish boundary conditions for atmospheric models (e.g., Krasnopolsky 2007, 2013; Bierson and Zhang 2020) and assume gas-phase equilibration, which requires verification. The apparent absence of gas-phase equilibrium above the modal planetary radius, inferred from a comparison of equilibrium and supposed measurement-based concentrations (Krasnopolsky and Parchev 1979; Krasnopolsky and Pollack 1994; Zolotov 1996), complicates understanding of the directions and outcomes of gas-solid reactions in the highlands. Evaluations of gas-solid reaction pathways through assessments of mineral stability are limited by uncertainties regarding thermodynamic properties of pure solids and solid solutions, the typical formation of metastable reaction products, and the rates of solid-state diffusion of metals that may act as limiting factors for reactions at 650 to 750 K. Applications of gas-solid reaction laboratory experiments to Venus face challenges due to limited ability to measure gas composition and fugacity ($fO_2$, $fS_2$, $fSO_2$, etc.) in reaction vessels, the common use of steel vessels that alter gas composition, the short duration of experiments (less than a few months) relative to typical reaction rates, and the necessity to balance experimental pressure, temperature, and duration to investigate advanced alteration. As a challenge to both chemical equilibrium and experimental evaluations of weathering pathways, the phase and chemical composition of bedrocks and their weathering products remain poorly constrained beyond the apparent mafic lava composition of the extensive plains and volcanic centers (Sect. 16.2.2). Similarly, information on the weathering products from mafic materials at the Venera and Vega landing sites is limited to the increased sulfur content (Table 16.3) and the red slope in the near-IR spectral range (Golovin et al. 1983; Shkuratov et al. 1987). These observations suggest a coating by a fine-grained ferric phase, possibly hematite (Pieters et al. 1986). Current knowledge of atmospheric and surface conditions, thermodynamic properties of solids, and experimentally observed early stages of gas-solid reactions allows for only general inferences about the fate of sulfur-bearing compounds at the current atmosphere-surface interface.

Publications on sulfur in atmosphere-surface interactions on Venus have been reviewed by Volkov et al. (1986), Fegley and Treiman (1992a, b), Fegley et al. (1992, 1997a), Zolotov and Khodakovsky (1989), Zolotov and Volkov (1992), Wood (1997), Zolotov (2015, 2018, 2019), Gillmann et al. (2022), and Filiberto and McCanta (2024). Since the mid-1960s, evolving views on reaction pathways have reflected new data and models on atmospheric composition, advances in numerical and experimental modeling methods, and the growth of experimental efforts in the last decade. Somewhat divergent conclusions regarding gas-solid pathways have emerged over the past six decades due to variations in gas compositions, temperatures, and pressures used in theoretical analyses, as well as numerical and experimental models (Sects. 16.3.2.2 and 16.3.2.3). Although there is no consensus on the details, major pathways include gas-solid reactions of rock-forming minerals and glasses with $SO_2$, OCS, and $S_2$ (Fig. 16.15).

$SO_2$ reacts with basaltic glasses to form sulfates of Ca and Na, which are notable in the alteration products of alkaline samples (Sect. 16.3.2.2). $S_2$ gas forms via sulfatization reactions under strongly anoxic surface conditions (Table 16.2) through the disproportionation of sulfur

$$3S(IV)(\text{in } SO_2 \text{ gas}) \rightarrow 2S(VI)(\text{in metal sulfates}) + S^0(\text{in } S_2 \text{ gas}) \quad (16.62)$$

The disproportionation of sulfur in the sulfurization of solids by $SO_2$ has been suggested from experiments with calcite (Tarradellas and Bonnetain 1973), and Burnham (1979) proposed the disproportionation occurring in terrestrial volcanic environments. As an example, anhydrite forms through the alteration of Ca-rich pyroxenes,

$$CaSiO_3(\text{in pyroxene}) + 1.5SO_2(g) \rightarrow CaSO_4(\text{anhydrite}) + SiO_2(s) + 0.25S_2(g) \quad (16.63)$$

Plagioclases, particularly Na-rich compositions, exhibit greater stability concerning sulfatization based on evaluations of mineral stability and experimental results. The interaction of OCS with Fe(II)-bearing minerals (such as silicates, pyrrhotite, magnetite, etc.) and glasses may lead to pyrite formation. Gas-solid reactions can facilitate the equilibration of certain gases with mineral assemblages formed during weathering. In such a case, solids could buffer near-surface gas chemistry, and further alteration of geological materials could be impeded. The concentration of all sulfur-bearing gases could be buffered by magnetite-pyrite and/ or magnetite-pyrite-hematite mineral assemblages (Sect. 16.3.2.3).

#### 16.3.2.2 $SO_2$ Interactions with Minerals and Glasses

Hot $SO_2$ in the near-surface atmosphere is modeled as a significant gas that alters surface materials, leading to a net sink of atmospheric sulfur into secondary minerals and releasing gaseous by-products (Fig. 16.15). The formation of metal sulfates through overall reaction 16.62 contributes to the net oxidation of surface materials and reduces atmospheric gases by decreasing the $SO_2/S_2$ ratio. Considering the sulfur-rich compositions of surface materials (Table 16.3), the fate of $SO_2$ at the atmosphere-surface interfaces has been constrained through evaluations of mineral stability regarding gas-solid type reactions and experiments under simulated Venusian and other conditions. Volkov et al. (1986), Fegley and Treiman (1992a, b), Fegley et al. (1992, 1997a), and Zolotov (2015, 2018, 2019) reviewed early works on $SO_2$-mineral interactions. This section devotes more attention to recently published and insufficiently addressed works, as well as a variety of new experiments.

Long before measurements of trace chemically active gases in the lower atmosphere, Mueller (1963, 1964) linked the stability of crustal minerals to atmospheric composition. He demonstrated that gas-solid equilibria could impact atmospheric composition and that the measured composition of atmospheric gases could indicate the mineralogy of upper crustal materials interacting with the gas phase. Mueller (1965) estimated the abundances of sulfur-bearing atmospheric gases, assuming their equilibration with crustal Ca- and Fe-bearing minerals in the upper crust. He showed that chemical equilibration between wollastonite and anhydrite

$$CaSiO_3(\text{wollastonite}) + SO_3(g) = CaSO_4(\text{anhydrite}) + SiO_2(\text{quartz}) \quad (16.64)$$

at 700 K could only be achieved at $pSO_3$ of $10^{-13.8}$ bars. This implies the sulfatization of wollastonite at Venus' atmospheric $pSO_3$, which exceeds the equilibrium value. However, assessing $pSO_2$ at the $CaCO_3-CaSO_4$ equilibrium required data on atmospheric $fO_2$ to constrain the $SO_2/SO_3$ ratio. Mueller (1964, 1965) suggested that the redox state (e.g., $fO_2$, $CO_2/CO$, $SO_2/OCS$, $SO_3/SO_2$ ratios) of a hot, coupled atmosphere-upper crust system could be controlled by chemical equilibria with Fe-bearing minerals and that the redox state of the system may not exceed that controlled by the magnetite-hematite ($Fe_3O_4 - \alpha-Fe_2O_3$, Mag-Hem) mineral assemblage commonly used to maintain $fO_2$ in petrological experiments (Eugster 1957)

$$2Fe_3O_4(\text{magnetite, Mag}) + CO_2(g) = 3Fe_2O_3(\text{hematite, Hem}) + CO(g) \quad (16.65)$$

$$4Fe_3O_4 + O_2(g) = 6Fe_2O_3 \quad (16.66)$$

Mueller (1965) demonstrated that the magnetite-pyrite (Mag-Py) equilibrium

$$3FeS_2(\text{pyrite, Py}) + 2O_2(g) = Fe_3O_4 + 3S_2(g) \quad (16.67)$$

at $fO_2$ controlled by equilibrium 16.66 sets the $S_2/SO_3$ ratio that allows alteration of $CaSiO_3$ to $CaSO_4$ (reaction 16.64). Here, I note that $pSO_3$ at equilibrium 16.64 ($10^{-13.8}$ bars) at $fO_2$ from equilibrium 16.66 ($10^{-22.8}$ bars) corresponds to $pSO_2$ and $pS_2$ of $10^{-4.8}$ and $10^{-5.3}$ bars, respectively, assuming $SO_3$-$SO_2$-$S_2$-$O_2$ equilibria at 700 K. However, Mag-Py equilibrium 16.67 at $pO_2$ controlled by the Mag-Hem equilibrium 16.66 corresponds to partial pressures of $SO_3$, $SO_2$, and $S_2$ of $10^{-11.2}$, $10^{-2.3}$, and $10^{-10.3}$ bars, respectively, which are in the stability field of $CaSO_4$ at 700 K. Mueller (1965) stated that in a coupled atmosphere-upper crust system controlled by the Mag-Hem phase assemblage, sulfur would be stored in sulfates rather than in a $CO_2$-rich atmosphere.

Lewis (1968, 1970) considered gas-mineral equilibria of carbonates, silicates, and Fe sulfides and oxides to link the partial pressures of $SO_2$, $SO_3$, OCS, $H_2S$, $H_2O$, HCl, HF, CO, and $CO_2$, surface mineralogy, temperature, and pressure to interpret remote sensing data on the upper atmosphere of Venus. The $SO_2$ mixing ratio evaluated under various temperature-pressure conditions did not exceed 0.3 ppmv ($10^{-4.44}$ bars $SO_2$ at a total pressure of 120 bars) in the lower atmosphere. The latter $pSO_2$ value corresponded to the $CaCO_3$-$CaSO_4$ equilibrium at atmospheric $pCO_2$, presumably controlled by the calcite-quartz-wollastonite equilibrium

$$CaCO_3(\text{calcite}) + SiO_2(\text{quartz}) = CaSiO_3(\text{wollastonite}) + CO_2(g) \quad (16.68)$$

that was anticipated by Adamcik and Draper (1963) and Mueller (1963, 1964). Although the chemical consumption of atmospheric $SO_2$ has yet to be discussed, the works of Muller (1965) and Lewis (1970) imply the depletion of abundant atmospheric $SO_2$ (e.g., volcanic) to minute concentrations controlled by gas-mineral equilibria. Lewis and Kreimendahl (1980) noted that a mass of atmospheric sulfur corresponding to 100 ppmv of sulfur-bearing gases (similar to current estimates, Table 16.1) could be consumed in a ~50 cm layer of surface materials where all Ca-bearing phases convert to anhydrite. This mass balance estimate supports Mueller's (1965) idea that hot upper crustal materials control the atmospheric composition of sulfur-bearing gases.

Stability diagrams of sulfur-, Ca-, and Fe-bearing minerals developed by Lewis and Kreimendahl (1980) further demonstrated that the stability of calcite regarding sulfatization requires the dominance of reduced sulfur-bearing gases. This concurred with the estimations of Mueller (1965) and Lewis (1970) and with independent chemical models of the lower atmosphere developed by Prinn (1978). Lewis and Kreimendahl (1980) also considered a pyrite-calcite-anhydrite-wüstite $fO_2$ buffer ($10^{-22.6}$ bars at 750 K) consistent with ~90 bars $CO_2$ controlled by reaction 16.68. Although Lewis and Kreimendahl (1980) suggested stable coexistence of $CaCO_3$ and $CaSO_4$ based on preliminary Pioneer Venus data, refined gas chromatography data on $xSO_2$ (Oyama et al. 1980) implied instability of calcite.

Since 1980, the sulfatization of Ca-bearing minerals in contact with 130 to 185 ppmv $SO_2$, as reported by Oyama et al. (1980) and Gel'man et al. (1980), has been anticipated through a range of chemical equilibrium calculations of sulfatization reactions (e.g., reaction 16.63) and sulfur-bearing multicomponent systems. Barsukov et al. (1980a, b, 1982c) and Khodakovsky (1982) calculated the equilibrium phase compositions of rock-gas type systems, which included rocks (basalts, rhyolites, etc.) represented by their elemental compositions (hydrogen, carbon, oxygen, nitrogen, sulfur, Cl, F, Na, Mg, Al, Si, K, Ca, P, Mn, Ti, and Fe). They used fixed fugacities of $CO_2$, CO, $SO_2$, and $H_2O$ (open system with respect to these gases), along with surface temperatures and pressures obtained with the PVLP and Venera 12 probe. Equilibrium mineral assemblages calculated under Venus' lowlands typically contained 5–8 vol% anhydrite and, occasionally, pyrite, with sulfur content reaching 11 wt%. These models indicated that the sulfatization of freshly exposed rocks and glasses would not affect plagioclase, alkali feldspars, quartz, enstatite, fluorapatite, rutile, and titanite present in equilibrium assemblages. The presence of stable diopside in a $CaSO_4$-free assemblage, calculated with a minimal $xSO_2$ of 70 ppmv and a maximal $x$CO of 42 ppmv CO from Gel'man et al. (1980), suggested its relatively high resistance to sulfatization (Barsukov et al. 1980a, 1982c). Barsukov et al.'s (1980b) calculations demonstrated greater stability for felsic igneous rocks (rhyolites, granites) compared to mafic rocks. Following Mueller's (1964, 1965) concept, Barsukov et al. (1980b) proposed that their calculated solid phases represent the mineralogy of altered surface rocks and that the mineralogy of permeable surface materials buffers the partial pressures of trace, chemically active atmospheric gases ($SO_2$, OCS, $S_2$, CO, $H_2O$, $H_2$, etc.).

In 1982, the anticipated sulfatization of certain Ca-bearing minerals was supported by the detection of sulfur-rich solids at the landing sites of Venera 13 and 14 (Table 16.3). The sulfur/Ca atomic ratios below unity indicated an incomplete conversion of Ca-bearing phases to $CaSO_4$. Barsukov et al. (1983) employed the elemental composition of these samples in chemical equilibrium calculations for multicomponent gas-solid systems open to $SO_2$ and other atmospheric gases measured with PVLP and Venera 12 (Table 16.1). The resulting anhydrite-rich equilibrium assemblages contained more sulfur than was measured in surface materials, suggesting incomplete (ongoing or hindered) sulfatization of the surface materials. Equilibrium calculations that excluded sulfur-bearing atmospheric gases resulted in anhydrite-bearing assemblages consistent with the measured elemental compositions of surface materials. Other secondary phases in these assemblages included enstatite, anorthite, and minor diopside (they indicated unreacted Ca), oligoclase, nepheline (Venera 13), albite (Venera 14), magnetite, and quartz, as summarized by Volkov and Khodakovsky (1984) and Volkov et al. (1986).

Zolotov (1985) noted that sulfatization on Venus could release $S_2$ gas (reactions 16.62 and 16.63). This inference was based on experiments involving the anoxic sulfatization of calcite (Tarradellas and Bonnetain 1973), which reportedly occurred via the formation of a sulfite intermediate

$$CaCO_3 + SO_2(g) \rightarrow CaSO_3 + CO_2(g) \tag{16.69}$$

$$CaSO_3 + 0.5SO_2(g) \rightarrow CaSO_4 + 0.25S_2(g) \tag{16.70}$$

Although subsequent experiments do not fully support $CaSO_3$ as an intermediate phase (CaS is the likely intermediate at 1020 to 1170 K, Saadatfar et al. 2020), the overall sulfur disproportionation reaction (the sum of reactions 16.69 and 16.70) leading to anhydrite and $S_2$ has been confirmed. Phase equilibrium lines for silicates, carbonates, and sulfates of Ca, Mg, Mn, Na, and K, plotted in coordinates of $fSO_2$ and $fS_2$ (Fig. 16.22, Volkov et al. 1986; Zolotov 1985, 2018), demonstrated the instability of Ca-bearing pyroxenes (wollastonite, diopside, augite) and carbonates (calcite, dolomite) concerning sulfatization at all elevations on Venus at $xSO_2$ and $xS_2$ suggested by atmospheric measurements. Conditions in deep lowland trenches (e.g., at Dyana Chasma) and shallow, permeable subsurface appeared similar to those at the anorthite-anhydrite equilibrium

$$CaSi_2Al_2O_8(\text{anorthite}) + 1.5SO_2(g) = CaSO_4 + Al_2SiO_5(\text{andalusite}) + SiO_2(\text{quartz}) + 0.25S_2(g) \tag{16.71}$$

Phase diagrams developed in $fSO_2$ and $fS_2$ coordinates at 670 K and 52.1 bars (e.g., Zolotov 2018) indicate a more favorable sulfatization in highlands, where Ca-rich plagioclase is unstable regarding $SO_2$ and $S_2$. However, Mg-rich silicates (such as enstatite and forsterite) and magnesite ($MgCO_3$) should not react with $SO_2$ at any elevation.

Barsukov et al. (1986b, c, d) modeled equilibrium mineral assemblages at Venera 13, 14, and Vega 2 landing sites in a 16-component gas-solid system, considering thermodynamic data for binary solid solutions of Fe-Mg olivine, Fe-Mg orthopyroxene, Fe-Mg-Ca clinopyroxene, plagioclase, alkali feldspar, and scapolite. The measured elemental composition of solid materials, including sulfur, at the landing sites (Table 16.3) served as the input for the models. The mineral assemblages calculated at the contact with a $CO_2-CO-H_2O$ Venus' atmosphere contained 1.5 to 8 vol% $CaSO_4$, whereas solids equilibrated with a $CO_2-CO-H_2O-SO_2-Cl-HF$ atmosphere had 13 to 18 vol% $CaSO_4$ (Table 16.8). The latter values corresponded to 3.0 to 4.3 wt% sulfur in altered compositions and indicated a maximum possible amount of sequestered atmospheric sulfur. Besides anhydrite, mineral mixtures equilibrated with the atmospheric $SO_2$ contained K-dominated microcline (Na/K = 0.13/0.87), clinoenstatite (Fe/(Fe + Mg) < 0.1), magnetite, andalusite, kalsilite (Venera 13 and Vega 2), marialite, tephroite, quartz, and rutile. Diopside, plagioclase ($An_{20-30}$), and less abundant anhydrite in assemblages equilibrated with $CO_2-CO-H_2O$ atmospheres (Table 16.8) suggested incomplete sulfatization of Ca-bearing minerals at all three landing sites. Formed and/or primary alkali feldspars and Mg-rich silicates were identified as chemical weathering products at the landing sites. Varying sulfur contents in solid samples (Table 16.3) were interpreted as indicative of differing exposure of surface rocks, suggesting a shorter sulfatization time for sulfur-poor and less physically degraded rocks at the Venera 14 site (c.f., Zolotov and Khodakovsky 1989; Zolotov and Volkov 1992). Although calculations of multicomponent gas-solid systems have demonstrated a significant dependence of the evaluated equilibrium phase assemblages on gas composition and the thermodynamic data used, the continuous presence of anhydrite ensures the sulfatization of surface materials.

Using the same Gibbs free energy minimization code as in Barsukov's works, Klose and Zolotov (1992) reported the formation of Na-, K-, and Ca-sulfates in equilibrium assemblages formed through the alteration of evolved (alkali- and silica-rich) igneous rocks. Like Barsukov et al.'s (1980b, 1982c) work, the prevalence of abundant alkaline feldspars and quartz in equilibrium phases indicated the enhanced stability of evolved silicate rocks concerning sulfatization. Independent calculations of chemical equilibria in gas-basalt multisystem conditions, which also considered mineral solid solutions, by Klose et al. (1992) indicated the formation of anhydrite in all calculated assemblages and the stability of Na-rich plagioclase and Mg-rich silicates.

Semprich et al. (2020) utilized the Perple_X code (Connolly 2005) to calculate comprehensive phase equilibria, including solid solutions, in the $fS_2 - fO_2$ fields where rocks and gases were represented by rock-forming elements (Si, O, Ti, Al, Fe, Mg, Ca, Na, and K) and hydrogen-carbon-oxygen-sulfur gas, respectively. They accounted for diverse compositions of initial rocks and the physical-chemical conditions in the lowlands and highlands. They assessed the stability fields of silicates coexisting with anhydrite, pyrite, pyrrhotite, and Fe oxides across a wide range of $fO_2$ and $fS_2$ values, including the range proposed for the near-surface atmosphere (Fig. 16.23). Although Semprich et al. (2020) did not present results for silicate solid solutions, they used a pyrrhotite composition of $Fe_{0.88}S$ and only one sulfate ($CaSO_4$). Their calculations were roughly consistent with preceding models of multicomponent systems and considerations of individual gas-solid equilibria.

Fegley and Prinn (1989) have demonstrated the instability of calcite via the equilibrium

$$CaCO_3 + SO_2(g) = CaSO_4 + CO(g) \qquad (16.72)$$

at $xSO_2$ and $xCO$ obtained with the Pioneer Venus and Venera 12 probes (Table 16.1). They noted that the equilibrium $pSO_2$ in reaction 16.72 is two orders of magnitude lower than $pSO_2$ measured in the sub-cloud atmosphere. Subsequent comprehensive considerations of mineral stability by Fegley and Treiman (1992a, b) and Fegley et al. (1992) confirmed the instability of Ca-rich minerals (calcite, dolomite, diopside) and the stability of Mg-rich silicates in contact with the supposed near-surface abundances of $SO_2$ and CO. The discrepancy between atmospheric $fSO_2$ and low $fSO_2$ at mineral-anhydrite equilibria (e.g., reaction 16.72) was interpreted as ongoing sulfatization. Zolotov (2018) revised evaluations of phase diagrams reflecting mineral stability in terms of $fSO_2$, $fS_2$, and $fCO$. He demonstrated that plagioclase with an intermediate composition and an activity of anorthite between 0.2 and 0.6 could achieve equilibrium with $SO_2$, $S_2$, and CO in the lowlands. Although uncertainties in the thermodynamic data for minerals and atmospheric composition constrain the assessment of anorthite stability in the lowlands (Fig. 16.23), it was confirmed that the sulfatization of anorthite and Ca-rich plagioclase is thermodynamically more favorable at elevations where only Na-rich plagioclase can remain stable.

The consideration of chloride-sulfate equilibria indicates the instability of solid NaCl, KCl, $MgCl_2$, and $CaCl_2$ concerning sulfatization at most of Venus' surface (Zolotov 2025). Chlorides of Mg and Ca are more suitable for sulfatization at all hypsometric levels. Sulfurization of chlorides consumes $SO_2$ and $H_2O$, releasing HCl and $S_2$ into the atmosphere, as illustrated by the net reaction

$$2NaCl + H_2O(g) + 1.5SO_2(g) \rightarrow Na_2SO_4 + 2HCl(g) + 0.25S_2(g) \qquad (16.73)$$

All considered chlorides are unstable at low elevations (Fig. 16.24) but become slightly more stable at higher altitudes. NaCl and KCl could coexist with their corresponding sulfates at mountaintops. $KCl_2$, $MgCl_2$, and NaCl could sublime in the lowlands and condense in the highlands, and high-temperature sublimation might exceed sulfatization. Faster sublimation is evidenced by the disappearance of a halite sample in a 10-day gas-solid experiment under Venus' surface conditions (Longo 2024). NaCl, KCl, $MgCl_2$, and $CaCl_2$ are not expected to be products of chemical weathering in the lowlands, and chlorine may be present in other phases. The highlands could be relatively enriched in NaCl, KCl, $Na_2SO_4$, $K_2SO_4$, and $Mg_2SO_4$ through the condensation of chlorides and their subsequent sulfatization. Significant thermodynamic potential for the sulfatization of other minerals in the highlands (Zolotov 2018) also suggests a higher abundance of $CaSO_4$, $Na_2SO_4$, and $K_2SO_4$. The low near-IR emissivity of the highlands (Sect. 16.2.2.4) may reflect the presence of various low-albedo sulfates along with NaCl and KCl that have not been converted to sulfates due to thermodynamic or kinetic reasons.

In summary, multiple calculations of chemical equilibria under Venus' surface conditions suggest the stability of alkali feldspars, Mg- and Mn-rich silicates, and the sulfatization of Ca-rich pyroxenes and carbonates at ~100 to 200 ppmv $SO_2$ measured in the lower atmosphere. Under lowland conditions, the stability of rock-forming minerals concerning sulfatization decreases in the following sequence: microcline, albite, enstatite, forsterite, rhodonite, tephroite, magnesite, rhodochrosite, anorthite, sylvite, halite, diopside, dolomite, wollastonite, calcite, natrite, and K carbonate. If near-surface $fO_2$, $fSO_2$, and $fS_2$ are at the conditions determined by the Mag-Py equilibrium (Table 16.11), minerals less stable than anorthite should undergo sulfatization (Fig. 16.22). Similar conclusions can be drawn if near-surface $fSO_2$ corresponds to 100 to 200 ppmv $SO_2$ and $fO_2$ is in the range of $10^{-20}$ to $10^{-21.7}$, as estimated by Fegley et al. (1997b) for conditions of modal radius. Although diopside is unstable, the conditions of the diopside-anhydrite-enstatite-quartz equilibrium

$$CaMgSi_2O_6 + 1.5SO_2(g) = CaSO_4 + MgSiO_3 + SiO_2 + 0.25S_2(g) \qquad (16.74)$$

are close to the temperature-$f$ (gas) conditions suggested for the near-surface atmosphere (Tables 16.1 and 16.2). Therefore, the sulfatization of diopside could be hindered, which is consistent with some experiments discussed below.

**Inferences from Laboratory Experiments**

In addition to fundamental and technology-oriented studies of $SO_2$-solid reactions, the alteration of carbonates, silicates, and silicate glasses has been investigated in the context of volcanic and post-magmatic processes (Burnett et al. 1997; Li et al. 2010; Auris et al. 2013; Henley et al. 2015; Palm et al. 2018; Delmelle et al. 2018; Renggli and King 2018; Renggli et al. 2019a, b), and explicitly to constrain chemical weathering on Venus (Tables 16.9 and 16.10). The conditions on Venus' surface are challenging to replicate and monitor fully in the laboratory. Most Venus-focused experiments had insufficiently known partial pressures of $pSO_2$ and other gases ($S_2$, CO, OCS) that would affect the alteration of geological samples. Experiments conducted in pressure vessels often cannot be sustained for extended periods, limiting them to providing insights only on the initial stages of alteration. This contrasts with studies commonly found in the physical-chemical and chemical engineering literature, as experiments focused on Venus typically fall short in providing sufficient information to assess rates and mechanisms of advanced alteration.

Kinetic studies of $SO_2$-calcite interactions in $O_2$-bearing environments (sulfation) with applications to the sequestration of $SO_2$-rich industrial gases at ~700 to 1200 K (Tullin and Ljungstróm 1989; Iisa et al. 1992; Tullin et al. 1993; Qiu and Lindqvist 2000; Hu et al. 2007, 2008a, 2008b; Jeong et al. 2015) provide information on the reaction rate and mechanism. The rate-limiting step in the reaction mechanism may involve the formation of a $CaSO_3$ intermediate phase, which is subsequently oxidized to $CaSO_4$ by $O_2$. The rate and possible mechanism of the $SO_2 - CaCO_3$ reaction depend on $fCO_2$. Low $fCO_2$ and high temperature favor the calcination of $CaCO_3$ to CaO, which then reacts with the $SO_2$ gas (Tullin et al. 1993; Jeong et al. 2015). Meanwhile, elevated $fCO_2$ at lower temperatures enables a slower yet more efficient overall sulfatization of calcite without calcination (e.g., Tulin et al. 1993). Solid-state diffusion of $Ca^{2+}$ toward the calcite surface and inward diffusion of $SO_2$ through porous $CaSO_4$ have been utilized to model reaction kinetics. Although sulfation experiments and kinetic models may not directly apply to anoxic reactions 16.69 and 16.70, a potential rate-limiting step from reaction 16.69 for $CaSO_3$ formation under both $O_2$-rich and anoxic conditions suggests that the rates and activation energies of sulfation reactions can help assess sulfatization on Venus. Note that some experimental results on sulfatization (e.g., Hu et al. 2008b) do not indicate the formation of any intermediate phase. Regardless of the formation of intermediate sulfite, the $SO_2$-calcite reaction is constrained by a relatively slow initial step that is independent of $O_2$ content (Hu et al. 2008a; Jeong et al. 2015). Anoxic experiments conducted at 1020 to 1170 K indicate the formation of a CaS intermediate that reacts with $SO_2$ to produce the final reaction products, $CaSO_4$ and $S_2$ (Saadatfar et al. 2020). Their work suggests that the diffusion of $SO_2$ and $S_2$ may limit the reaction rate at an advanced stage of calcite alteration. It is unclear whether their results are applicable to Ca carbonates and silicates under Venus' temperature-$pCO_2-pSO_2-pS_2$ conditions. Although kinetic data on sulfation do not imply any long-term existence of calcite on the surface of Venus, the rapid anoxic sulfatization of calcite in simulated environments has been confirmed. Fegley and Prinn (1989) investigated calcite alteration using a 1 bar $CO_2-SO_2$ mixture characterized by Venus-like $pSO_2$ at temperatures ranging from 873 to 1123 K. A calcite crystal was covered by a ~10 μm-thick anhydrite layer at 1123 K within eight days. Although the sulfatization of calcite at 90 bars $CO_2$ may be slower than at 1 bar, the reaction rate (equivalent to 1 μm $a^{-1}$ of $CaSO_4$ formed), assessed through extrapolation of results to ~740 K, suggests geologically rapid alteration of calcite if it appears on the surface. Based on these values, Fegley and Treiman (1992a) noted that the depletion of current atmospheric $SO_2$ to $fSO_2$ at the $CaSO_4 - CaCO_3$ equilibrium 16.72 could occur in ~1.9 Ma. Aveline et al. (2011) observed the sulfatization of calcite exposed to 4.8 bars of pure $SO_2$(g) at 733 K. Sulfatization of calcite has been observed in several experiments at the NASA Glenn Extreme Environments Rig (GEER) facility, under conditions that simulate the temperature, pressure, and gas compositions of Venus' surface. Radoman-Shaw (2019) and Santos et al. (2023) reported the formation of 0.25 µm and 3 to 5 µm thick coatings of Ca sulfate in the GEER over 42 and 30 days, respectively. This discrepancy in the results may reflect varying degrees of $SO_2$ consumption in reactions with the stainless steel chamber walls, which is proposed to occur in the GEER experiments (e.g., Lukco et al. 2018). Slower sulfatization indicates greater $SO_2$ consumption and/or a more reduced gas phase, where reactions 16.69 and 16.70 are suppressed at elevated $S_2/SO_2$ ratios. Compositional mapping across a calcite alteration rind obtained by Santos et al. (2023) does not indicate sulfur diffusion into the calcite sample but suggests that $Ca^{2+}$ diffuses toward the surface. Sulfatization of calcite and natrite ($Na_2CO_3$) mineral surfaces under Venus' conditions was observed in a 60-day GEER experiment (Gilmore et al. 2025). Neither reaction rates nor the mechanism of calcite alteration has been inferred from experiments that closely simulate Venus' surface conditions.

Venus-relevant experiments (Tables 16.9 and 16.10) reveal that $SO_2$ reacts more slowly with Ca-rich silicates and silicate glasses compared to calcite, and the relative rates of sulfatization of silicates seem roughly consistent with evaluations of mineral stability. Fegley and Treiman (1992a) confirmed that diopside undergoes sulfatization more slowly than calcite, based on experiments by Fegley and Prinn (1989). The slower sulfatization of Ca silicates (wollastonite, diopside, augite, tremolite, etc.) than calcite has been reported by Aveline et al. (2011), Radoman-Shaw (2019), Santos et al. (2023), and Port et al. (2025). In a 42-day GEER experiment, Radoman-Shaw (2019) observed widespread sulfatization of Ca-bearing pyroxenes, resulting in the formation of anhydrite and the appearance of Na sulfates on aegirine and jadeite grains. Berger et al. (2019) did not report alterations in pyroxenes and plagioclase in basalt but noted some oxidation of olivine to Fe oxide(s). Santos et al. (2023) and Reid et al. (2024) did not indicate the sulfatization of plagioclases and pyroxenes at supposedly Venus-like $pSO_2$ levels. However, mineral phases were not explicitly analyzed in either set of experiments, so phase changes cannot be entirely dismissed. In experiments conducted by Aveline et al. (2011), Radoman-Shaw (2019), and Santos et al. (2023), plagioclase feldspar appeared to be the most $SO_2$-resistant Ca-bearing silicate, and no sulfatization was observed for undoubtedly thermodynamically stable phases such as Mg-rich silicates (enstatite, forsterite) and alkali feldspars.

Sulfatization of silicate glasses was observed in Venus-focused experiments conducted at various temperatures, pressures, gas compositions, and durations (Berger et al. 2019; Radoman-Shaw 2019; Radoman-Shaw et al. 2022; Santos et al. 2023; Reid et al. 2024). The sulfatization of Ca and Na in the glasses occurred significantly faster than the oxidation and/or sulfidation of ferrous minerals, which has not always been the case. More prominent sulfatization than oxidation was also observed in high-temperature experiments at elevated $pSO_2$, irrespective of Venus' environments (Renggli and King 2018). Different degrees of alteration were observed in studies involving compositionally similar MORB (Santos et al. 2023) and tholeiitic glasses (Radoman-Shaw et al. 2022; Reid et al. 2024). Reid et al. (2024) reported a greater degree of alteration in alkaline basaltic glass compared to tholeiitic glass after conducting a series of similar experiments with each material. The alteration of glasses by $CO_2-SO_2$ mixtures into sulfates and Fe oxides occurred significantly faster than with $CO_2$ alone (Teffeteller et al. 2022), which led to the oxidation of Fe(II) and the carbonation of glass surfaces.

The experiments involving the interaction of silicate glasses with $SO_2$-bearing gases resulted in sulfates of Ca and Na (anhydrite, thernadite, glauberite), along with less prominent Fe sulfides and/or oxides. Different proportions of anhydrite, Na sulfate, and Na-Ca sulfates were observed in the alteration products of compositionally variable basaltic glasses. Calcium sulfates were detected on the surface of basaltic glasses during a 42-day GEER run (Radoman-Shaw et al. 2022) and on a polished surface of MORB glass after a 30-day experiment in GEER (Santos et al. 2023). Sodium sulfate is a predominant alteration product of alkaline glasses (Berger et al. 2019; Radoman-Shaw et al. 2022; Reid et al. 2024). Reid et al. (2024) noted the formation of abundant Na sulfate through the alteration of alkaline basalt glass and the appearance of anhydrite on tholeiitic glass surfaces during 14- and 30-day runs. Radoman-Shaw et al. (2022) reported the formation of abundant thernadite on a synthetic Venera 13-like K-rich glass with added $Na_2O$. In contrast to high-temperature experiments conducted at high $pSO_2$ (Renggli and King 2018), the formation of $MgSO_4$ has not been observed in experiments relevant to Venus. This observation agrees with equilibrium thermodynamic assessments (Fig. 16.22).

The observed formation of sulfate crystals on glass surfaces affects their texture. It alters the composition of subsurface layers influenced by the migration of cations (e.g., Dyar et al. 2021). Compositional profiles through alteration rinds of alkaline glasses (Berger et al. 2019; McCanta et al. 2023; Reid et al. 2024) suggest migration of Ca and Na toward the surface, resulting in relative enrichments of Si and Al in the subsurface layer. Berger et al. (2019) and Reid et al. (2024) noted that the more prominent alteration of alkaline glasses likely reflects the faster diffusion of Na compared to Ca, as observed in silicate glasses (Behrens 1992). These inferences align with experiments on the interactions of silicate glasses and minerals with $SO_2$-rich high-temperature (>1000 K) gases, which suggest sulfatization through solid-state diffusion of cations (Ca, Na, and Mg), inward migration of electron holes and/or oxygen atoms, and modification of the Si-Al-O framework below the surface without forming secondary Al-Si rich phases during the initial stages of sulfatization (Auris et al. 2013; Henley et al. 2015; Delmelle et al. 2018; Palm et al. 2018; Renggli and King 2018; Renggli et al. 2019a, 2019b). Even in $SO_2$-rich gas experiments, the diffusion of cations through sulfate coatings surpasses the inward diffusion of $SO_2$, resulting in the formation of new sulfate on the surface of the coatings (Renggli et al. 2019a). Therefore, the diffusion of cations from the substrate to the surface is likely the rate-limiting factor for sulfatization, at least in glasses (e.g., Auris et al. 2013; Renggli et al. 2019a). Except for alkaline basaltic glasses, silicates and glasses generally remain partially uncoated by secondary phases in experiments conducted under Venusian conditions. Only relative sulfatization rates are suggested from the early stages of alteration, and additional experiments are needed (Sect. 16.5.2.1).

### Discussion and Summary on Sulfatization

The analyses of mineral stability and experimental results indicate the likelihood of sulfatization of Ca-, Na-, and K-bearing carbonates, silicates, and basaltic glasses at the present surface. These materials exhibit different thermodynamic affinities for sulfatization and alter at varying rates. Several Ca-bearing minerals (e.g., plagioclase, F- and Cl-apatite, titanite) are thermodynamically stable, and/or their alteration is kinetically impeded. The chemical compositions of solids at the landing sites (Table 16.3), compared to calculations of corresponding equilibrium mineral assemblages (Table 16.8), suggest incomplete and uneven sulfatization of the sampled materials. Given the exposure of sampled materials over several hundred Ma, the incomplete sulfatization is inconsistent with the experimental alteration of Ca-bearing phases and basaltic glasses observed within days to months (Table 16.9). This inconsistency could be attributed to slower alteration at advanced stages and the resistance of certain phases to sulfatization. Another reason is possible chemical equilibration between some minerals (e.g., labradorite, Barsukov et al. 1986c; Klose et al. 1992; Zolotov 2018), their sulfatization products, and gases on Venus. For example, the following chemical equilibrium could impede the alteration of plagioclase in basalts (Table 16.10),

$$Na_{0.4}Ca_{0.6}Al_{1.6}Si_{2.4}O_8 \textit{(labradorite)} + 0.9SO_2\,\textit{(g)} = 0.6CaSO_4 + 0.4NaAlSi_3O_8 + 0.6SiO_2 + 0.6Al_2SiO_5 + 0.15S_2\,\textit{(g)} \quad (16.75)$$

A complete gas-solid equilibration of exposed materials, as modeled by Barsukov et al. (1982c, 1986c), Klose et al. (1992), and Semprich et al. (2020), is not supported by the variable sulfur content and S/Ca ratio in the samples (Tables 16.3 and 16.8). The varying degrees of sulfur intake may reflect differences in exposure time, fugacities of reacting gases, surface area, grain size, and the surface/volume ratio in exposed materials, as well as the abundance of glass, phenocryst size, the plagioclase/pyroxene ratio, the amount of Na-rich phases (e.g., glasses) subjected to sulfatization, and the presence of secondary sulfides. Interestingly, the abundance of sulfur (Table 16.3) does not correlate with the Al/Ca ratio, which may reflect the plagioclase/pyroxene ratio, and with the Fe content that could affect the magmatic sulfide content and the degree of secondary sulfidation. However, the advanced physical weathering (formation of smaller-size particles) of the Venera 13 rocks (Fig. 16.1; Basilevsky et al. 1985; Garvin et al. 1984) compared to Venera 14 coincides with a higher sulfur content in the former sample (Zolotov and Volkov 1992). The faster alteration of alkaline basalt glasses compared to tholeiitic glasses (Reid et al. 2024) may have also led to a higher sulfur intake by the K-rich Venera 13 rock than by its Venera 14 counterpart. The increasing amount of secondary sulfates in exposed basaltic materials, associated with exposure age, affects physical properties (e.g., near-IR emissivity; Dyar et al. 2020, 2021; McCanta et al. 2024). These properties may be used in the future (Sect. 16.5.2) to map surface materials regarding the degree of sulfatization. Timing of sulfatization in the landing sites is discussed in Sect. 16.4.4.1.

The thermodynamic instability of Ca-rich plagioclase and the lesser stability of diopside, along with other Ca-bearing silicates and mafic glasses under highland conditions (Zolotov 2018), could lead to more pronounced sulfatization at higher elevations. Admixtures of anhydrite in crushed basalt samples (Dyar et al. 2021) and sulfate coatings of 0.1 mm have been shown to significantly reduce the near-IR emissivity of basalt (McCanta et al. 2024). Such admixtures and high-albedo coatings have the potential to create 'white' mountains. Putative sulfate coatings could have formed gradually since the inception of the highland rocks. Alternatively, or additionally, they could be a consequence of intense sulfatization under conditions of elevated volcanic $pSO_2$ and temperatures associated with the formation of the volcanic plains (Sect. 16.4.4.1). Regardless of the cause of possible high-altitude sulfatization, the low near-IR emissivity (high reflectivity) at high altitudes (Sect. 16.2.2.4) could be attributed to low-albedo alteration materials (Ca-, Na-, Mg-sulfates, alkali chlorides) of basalts and basaltic glasses.

#### 16.3.2.3 Stability and Reactivity of Sulfides

Despite analytical, numerical, and experimental studies, the fate of metal sulfides at the surface remains less clear than that of sulfates. This ambiguity reflects the uncertainty of the thermodynamic properties of minerals and their solid solutions, the unknown abundances of $SO_2$, OCS, $S_2$, and CO below 12 km (Table 16.1), model-dependent constraints on the composition of the near-surface atmosphere (Table 16.2), and the absence of proof of gas-phase equilibration in the lowlands. The lack of such equilibration above a thin atmospheric layer (Krasnopolsky and Pollack 1994; Zolotov 1996) restricts the evaluation of mineral stability in the highlands (Zolotov 2018). Consequently, $fS_2$ and $fO_2$ values, commonly used to assess the stability of coexisting sulfides and oxides, are ambiguous under conditions of modal planetary radius and may not be clearly definable at altitudes within the framework of equilibrium chemical thermodynamics. For the oxygen-sulfur-Fe system (Figs. 16.22, 16.24 and 16.25), even minor changes in atmospheric chemistry can affect the composition of stable solid phases. Since the 1970s, researchers have utilized various atmospheric compositions and assumptions regarding gas-phase and gas-solid equilibria, often resulting in inconsistent conclusions about sulfide stability.

##### Fe Sulfides and Oxides in the Surface Materials and Buffering of Atmospheric Gases

As mentioned in Sect. 16.3.2.2, Mueller (1965) first considered the stability of Fe sulfides and oxides under the presumed surface conditions and discussed their influence on the composition of trace sulfur- and carbon-bearing atmospheric gases. He demonstrated that sulfur-bearing gases should not be major atmospheric species in contact with Fe sulfides at ~700 K. In his opinion, the measured abundances of such gases and the $CO_2/CO$ ratio will inform the mineralogy of sulfides and oxides in upper crustal materials. Without developing mineral stability diagrams for the hydrogen-oxygen-sulfur-Fe and carbon-oxygen-sulfur-Ca-Si systems, he proposed that the highest possible atmospheric $CO_2/CO$ ratio could be controlled by coexisting hematite and magnetite (Mueller 1964) via equilibrium 16.65, which sets $fO_2$ of the near-surface atmosphere by equilibrium 16.66. Mueller (1965) noted that such oxidizing conditions do not exclude the coexistence of pyrite with Fe oxides; however, they do not allow for the stability of troilite and pyrrhotite. According to his viewpoint, the magnetite-pyrite-hematite (Mag-Py-Hem) equilibrium (e.g., a sum of reactions 16.66 and 16.67) could influence the partial pressures of trace sulfur-bearing atmospheric gases ($SO_2$, OCS, $SO_3$, $S_n$, etc.) and the oxidation state ($fO_2$, $CO/CO_2$) of the gas phase. He also considered the magnetite-pyrite (Mag-Py) equilibrium (e.g., reaction 16.67), which could control sulfur-bearing gases at lower $fO_2$ than the magnetite-hematite (Mag-Hem) equilibrium 16.66. The concept from Mueller (1965) remains valid in 2025. The atmospheric composition (Sect. 16.2.1, Table 16.1) aligns with Mueller (1965)'s inferences regarding the instability of reduced ferrous sulfides and the coexistence of Fe oxides with pyrite, which could constrain gas chemistry in the pore spaces of permeable crustal materials and the near-surface atmosphere in the lowlands. Because near-surface gas composition establishes boundary conditions for atmospheric models (Krasnopolsky and Parshev 1981a; Krasnopolsky and Polack 1994; Krasnopolsky 2007, 2013; Bierson and Zhang 2020), it likely affects the entire lower atmosphere. Table 16.11 (col. 3) depicts the assessed near-surface atmospheric composition corresponding to the measured abundances of $CO_2$, $SO_2$, and $H_2O$ at $fO_2$ and $fS_2$ controlled by the magnetite-pyrite equilibrium.

Lewis (1968) considered gas-solid equilibria involving troilite, pyrite, magnetite, ferrous sulfate, olivine, and enstatite to constrain the unknown atmospheric abundances of $H_2S$, OCS, and $SO_2$. The estimated mixing ratios of sulfur-bearing gases ($< 10^{-6}$) were below their observational upper limits, with reduced gases controlled by troilite-bearing mineral assemblages dominating over $SO_2$. As noted in Sect. 16.3.2.2, Lewis (1970) estimated possible mineral-buffered atmospheric compositions at the $fO_2$ set by either graphite-$CO_2$−CO or magnetite-silicate equilibria. The assessed mixing ratios of OCS and $H_2S$ (50 ppmv and 5 ppmv at 748 K and 120 bars) in equilibria with troilite appeared more abundant than $SO_2$ (0.3 ppmv) and $SO_3$ ($\sim 10^{-16}$), which were supposedly controlled by calcite-anhydrite equilibria in the surface/near-surface materials (equilibria 16.64 and 16.72). Following the work of Mueller (1965), these estimations demonstrated that stable troilite in the surface materials establishes low atmospheric $fSO_2$ and $fSO_3$ levels, which allowed for the stability of Ca-rich carbonates and silicates. While the mineral stability diagrams developed by Lewis and Kreimendahl (1980) reinforced this conclusion, the in situ data on the high $SO_2/(OCS + H_2S)$ ratio in the sub-cloud atmosphere published later in 1980 (Table 16.1) suggested a more oxidizing atmosphere. This atmosphere promotes the stability of pyrite over troilite or pyrrhotite and favors Fe oxides over ferrous silicates.

The consideration of the PVLP and Venera 12 data on $SO_2$, CO, and $CO_2$ in calculations of gas-solid multisystems by Barsukov et al. (1980a, b, 1982c) and Khodakovsky (1982) inferred the instability of troilite and pyrrhotite at the surface. Pyrite and/or magnetite appeared in their equilibrium mineral assemblages, depending on the gas composition, temperature, and pressure used in the models. Barsukov et al. (1980b, 1982c) concluded from these fluctuations that a pyrite-magnetite-anhydrite assemblage might be present in the surface materials, potentially buffering the composition of the near-surface atmosphere. von Zahn et al. (1983) and Prinn (1985) discussed the role of pyrite in linking the atmospheric and geological sulfur cycles. They anticipated the presence of pyrite in exposed igneous rocks, and its consumption through thermal decomposition and interactions with atmospheric $CO_2$ and $H_2O$, resulting in the formation of $S_2$, OCS, and $H_2S$, respectively. In their schemes, the photochemical oxidation of reduced gases to $SO_2$ was eventually followed by the sulfatization of exposed calcite (von Zahn et al. 1983) and Ca silicates (Prinn 1985) to anhydrite. The alleged reduction of anhydrite to pyrite in the crust to complete the geological sulfur cycle has not been detailed. Fegley et al. (1995, 1997a) slightly modified the geologic sulfur cycle of Prinn (1985) based on their experiments on pyrite → pyrrhotite → magnetite conversion. In their pathways, pyrite from igneous rocks thermally converts to metastable pyrrhotite by reaction

$$(1-x)FeS_2 \rightarrow Fe_{1-x}S\ (\text{pyrrhotite, Pyh}) + (0.5-x)S_2(g) \qquad (16.76)$$

Interactions of pyrrhotite with $CO_2$ and $H_2O$ produce magnetite and release OCS and $H_2S$ into the atmosphere, respectively. Here, plausible pyrrhotite oxidation is expressed by reactions

$$3Fe_{0.875}S + 3.5CO_2(g) \rightarrow 0.875Fe_3O_4 + 3OCS(g) + 0.5CO(g) \quad (16.77)$$

$$3Fe_{0.875}S + 3.5H_2O(g) \rightarrow 0.875Fe_3O_4 + 3H_2S(g) + 0.5H_2(g) \quad (16.78)$$

Anhydrite is not reduced to complete the cycle. In other words, Fegley et al. (1995) proposed net oxidation of sulfides to sulfates in upper crustal materials that occurs via photochemistry-driven oxidation of OCS, $H_2S$, and $S_n$ to $SO_2$. Although the stability and reactivity of pyrite remain questionable, this perspective aligns with the current understanding of the fate of sulfur in the atmosphere-surface system (Fig. 16.15).

Although the instability of pyrrhotite and troilite on the surface has been evident since 1980, calculations of chemical equilibria have led to varied conclusions regarding the stability of pyrite at the surface. The stability of pyrite in the lowlands has been inferred from calculations of chemical equilibria in gas-basalt multi-systems (Barsukov et al. 1980a, b, 1982c) as well as from calculations of selected gas-mineral equilibria (Zolotov 1991a, b, c, 1992b, 2018, 2019; Zolotov and Volkov 1992; Hashimoto and Abe 2005) such as equilibrium 16.67 and

$$Fe_3O_4 + 6SO_2(g) + 16CO(g) = 3FeS_2 + 16CO_2(g) \quad (16.79)$$

Calculations of basalt-atmosphere multi-systems by Klose et al. (1992) suggest a preferable stability of pyrite in the highlands and magnetite stability in the lowlands. Assuming the existence of stable pyrite in the low-temperature highlands, Hashimoto and Abe (2000) proposed a climate stabilization mechanism via reaction 16.79: increasing surface temperatures, facilitating the oxidation of pyrite, raising atmospheric $xSO_2$, and enhancing the albedo of clouds, leading to cooling. Secondary pyrite could form through the interaction of reduced sulfur-bearing gases with ferrous minerals and glasses, as well as via the oxidation of pyrrhotite by $CO_2$, $H_2O$, $SO_2$ (Zolotov 2018), and $S_2$,

$$FeO\ (\text{in minerals and glasses}) + 2OCS(g) \rightarrow FeS_2 + CO_2(g) + CO(g) \quad (16.80)$$

$$FeO\ (\text{in minerals and glasses}) + 2H_2S(g) \rightarrow FeS_2 + H_2O(g) + H_2(g) \quad (16.81)$$

$$2Fe_{0.875}S + CO_2(g) \rightarrow 0.25Fe_3O_4 + FeS_2 + CO(g) \quad (16.82)$$

$$Fe_{0.875}S + 0.375S_2(g) \rightarrow 0.875FeS_2 \quad (16.83)$$

However, concrete pathways need to be determined experimentally. Experimental data (e.g., Lv et al. 2015; Aracena and Jerez 2021; Liu et al. 2023) demonstrate that both thermal decomposition (reaction 16.76) and anoxic oxidation of pyrite occur through the formation of pyrrhotite. When applied to Venus, experiments by Treiman and Fegley (1991) and Fegley et al. (1995) revealed the alteration of pyrite in the presence of $CO_2$-rich gas mixtures at 1 bar (Table 16.9). Fegley et al. (1995) reported a more rapid pyrite-to-pyrrhotite conversion (reaction 16.76) compared to the subsequent oxidation of pyrrhotite to Fe oxides. Wood and Brett (1997) expressed concerns about the relevance of the $pS_2$, which was presumed but not measured in the experiments of Fegley et al. (1995), to the conditions on Venus. In his reply, Fegley (1997) argued for pyrite instability at supposedly low Venus' $fS_2$ and observation-based $fS_3$ (Krasnopolsky 1987), grounded on considerations within the sulfur-oxygen-Fe system, using the coordinates of $fS_3$, $fS_2$, and $fO_2$. Hong and Fegley (1997b) studied the kinetics of the pyrite-to-pyrrhotite conversion in various gas mixtures at 1 bar. They demonstrated the feasibility of the reaction at Venus' temperatures. Experimental data on $fS_2$ over pyrite, as measured and reviewed by Hong and Fegley (1998), further supported the feasibility of reaction 16.76 at 740 K and $fS_2 < 10^{-5.0\pm0.2}$ bars. It follows that pyrite in Venus' lowlands is unstable with regard to thermal decomposition into pyrrhotite at an $fS_2$ of $10^{-5.7}$ bars, which corresponds to ~20 ppbv $S_2$ at chemical equilibrium with 9–13 pptv $S_3$, assessed by Krasnopolsky (2013) for altitudes of 3 to 10 km (Table 16.1; Model 10 of Table 16.2), and at an $fS_2$ of $10^{-6.0}$at the chemical equilibrium with 8 ppmv CO (Model 9 of Table 16.2). However, the instability of pyrite at that low $fS_2$ does not necessarily indicate its weathering on Venus. This is because the mineral is not typically found in fresh basaltic materials that are more reduced than the QFM $fO_2$ buffer, as inferred for the three Venus' surface samples (Sect. 16.2.2.1).

Reid et al. (2024) used pyrite to introduce sulfur-bearing gases in Venus-focused basalt-weathering experiments, during which the magnetite-hematite buffer was employed to set the oxidation state ($fO_2$, $CO_2/CO$, $SO_2/S_2$, etc.) of the gas in the reaction vessel. Based on the experimental work of Lv et al. (2015), they anticipated that $SO_2$ would dominate among the sulfur-bearing gases produced through the pyrite-$CO_2$ interaction. However, pyrrhotite, hematite, and magnetite in post-run assemblages indicated incomplete pyrite oxidation. These four minerals cannot stably coexist (Figs. 16.22, 16.24 and 16.25), and their occurrence hampers a proper assessment of the gas phase during the runs. The presence of pyrrhotite may reflect its formation as an intermediate metastable phase via alteration of pyrite (Fegley et al. 1995; Lv et al. 2015; Aracena and Jerez 2021; Liu et al. 2023). Another possible explanation for the presence of pyrrhotite is a partial reduction of gases ($CO_2 \rightarrow CO$, OCS; $SO_2 \rightarrow$OCS, $S_2$) on the stainless steel walls of the reaction vessel. This reduction may also explain the limited oxidation of basalts observed in these and other experiments in steel vessels (Berger et al. 2019; Radoman-Shaw et al. 2022; Santos et al. 2023). Although the observed pyrite-to-pyrrhotite conversion aligns with previous works (Treiman and Fegley 1991; Fegley et al. 1995; Hong and Fegley 1997b; Lv et al. 2015), the results from Reid and colleagues may not determine the fate of pyrite in the lowlands.

If Venus' surface $fS_2$ is $10^{-4.4}$ to $10^{-4.7}$ bars (~0.2 to 0.4 ppmv $S_2$ at 95.6 bars total pressure), inferred from the supposed equilibration of major atmospheric gases (Models 5, 7, and 8 of Table 16.2), pyrite could coexist with magnetite (Figs. 16.22 and 16.25; Table 16.11). Radoman-Shaw (2019) noted the oxidation of pyrrhotite and pyrite to magnetite in the GEER runs. Santos et al. (2024) observed unaltered pyrite and noted the alteration of troilite, pyrrhotite, and Fe-metal to pyrite and magnetite during a 60-day GEER experiment conducted at 93 bars and 733 K. Again, we note that the use of stainless-steel chambers and vessels in the most Venus-relevant experiments to address the stability of sulfides (e.g., Berger et al. 2019; Kohler 2016; Port et al. 2016, 2018; Port and Chevrier 2017a, b, 2020; Radoman-Shaw 2019) could have altered the composition and redox state of reactive gases. The limited ability to monitor or regulate the gas composition in these experiments complicates the application of these findings to Venus. Regardless of the initial presence of pyrite and pyrrhotite in the exposed surface materials on Venus, their chemical pathways require accurate and meticulous experimental investigations under controlled gas composition.

The stability fields of solids in the sulfur-oxygen-Fe system indicate that environments in Venus' lowlands closely resemble the conditions under which pyrite, magnetite, and hematite coexist (Semprich et al. 2020; Zolotov 1991a, b, c, 1992b, 1996, 2018, 2019). These phase diagrams demonstrate that a definitive conclusion regarding mineral stability cannot be drawn without precise information on the atmospheric composition at the corresponding temperature, pressure, and altitude. Stability diagrams created by Fegley (1997) and Zolotov (1991b, c, 1992b, 2018, 2019) show that varying conclusions about sulfide stability were drawn based on the use of different instrumental data for atmospheric composition ($S_3$, $SO_2$, OCS, $H_2S$, and CO) and various assumptions regarding gas-phase equilibration. This is further illustrated in Fig. 16.25. Ambiguity also reflects the uncertainty of the thermodynamic data used for solid phases, leading to imprecise conditions for phase equilibria. This is evident in the Mag-Hem equilibrium (refer to equilibria 16.65 and 16.66, Figs. 16.22 and 16.25, and Fegley et al. 1997b) and the Py-Pyh boundary, which varies across the works of Hong and Fegley (1998), Semprich et al. (2020), and Zolotov (1992b, 2018, 2019).

The temperature, pressure, and suggested gas composition at volcanic plains closely align with the conditions established by the magnetite-pyrite assemblage discussed elsewhere (Mueller 1965; Barsukov et al. 1980b, 1982c; Klose et al. 1992; Hashimoto and Abe 2000, 2005; Wood 1997; Zolotov 1992b; Zolotov and Khodakovsky 1989; Zolotov and Volkov 1992). An assemblage like this in permeable surface materials could influence the ratios between oxidized and reduced gases (e.g., $SO_2/OCS/S_2$) in the near-surface atmosphere and shallow subsurface, provided that chemical equilibria 16.67 and/or 16.79 are achieved. The Mag-Py-Pyh and Mag-Py-Hem invariant equilibria

$$0.5Fe_3O_4 + 2FeS_2 = 4Fe_{0.875}S + O_2(g) \tag{16.84}$$

$$4Fe_2O_3 + FeS_2 = 3Fe_3O_4 + S_2(g) \tag{16.85}$$

$$8Fe_2O_3 + 0.5FeS_2 = 5.5Fe_3O_4 + SO_2(g) \tag{16.86}$$

set the range of fugacities of $O_2$, $S_2$, and other gases controlled by the Mag-Py equilibrium (Figs. 16.22, 16.24, and 16.25; Table 16.11).

The near-IR reflectance and color of the surface materials at the Venera landing sites (Pieters et al. 1986; Shkuratov et al. 1987, Sect. 16.2.2.1) suggest the presence of hematite in widespread rock coatings and fine-grained materials, aligning with the redox state of the surface atmosphere estimated from the atmospheric composition (Fegley et al. 1997b). The considerable error bar of $fO_2$ set by the Mag-Hem equilibrium 16.66, due to uncertainties in the thermodynamic properties of minerals (Fegley et al. 1997b), does not exclude the existence of stable hematite alongside magnetite and pyrite in the lowlands (e.g., Fig. 16.25). The impurities in Venus' minerals could also contribute to the $fO_2$ uncertainty at the natural Mag-Hem equilibrium. Thermo-dynamic calculations (Zolotov 1994, 1995b; Semprich et al. 2020) demonstrate that a hematite-ilmenite solid solution (Hem activity $<1$) in equilibrium with magnetite yields a lower $fO_2$ than the genuine Mag-Hem equilibrium. This outcome aligns more closely with near-surface atmospheric models (Table 16.2), constrained by measurements. If this is the case, the Mag-Py-Hem assemblage in the surface materials could control $fS_2$ ($< 10^{-4.17}$ bars, equilibrium 16.85), $fSO_2$ ($< 10^{-1.17}$ bars, equilibrium 16.86), $fO_2$ ($< 10^{-20.8}$ bars, equilibrium 16.66), and the corresponding fugacities of other gases. The concentrations of $SO_2$ and CO ($xSO_2 < 713$ ppmv, $xCO > 9$ ppmv) are more consistent with the measurements (Table 16.1) than with those calculated in Table 16.11 for the nominal Mag-Py-Hem buffer.

In addition to chemical weathering, non-silicate Fe-bearing phases could be delivered from clouds. Ferric sulfates and chlorides could serve as UV-blue absorbers in upper clouds (Sect. 16.3.1.8). Ferric sulfates were likely captured from lower cloud aerosols by the PVLP LNMS and decomposed thermally to hematite at altitudes below ~25 km (Mogul et al. 2025; Zolotov et al. 2023). Zolotov (2021) suggested that the accumulation of hematite on the surface occurs through alteration of unstable ferric sulfate (Fig. 16.25), which forms from cosmic Fe in the clouds (Fe-metal, FeS) and sinks to the surface as particles due to gravity. If this suggestion is accurate, the surface of Venus should appear red in the near-IR spectral range (~1 μm), irrespective of the oxidation of surface ferrous materials to hematite. Hematite sourced from space could help buffer the atmospheric composition. An increased Ni/Fe ratio in surface fines will suggest cosmic sources.

Although pyrite and Fe oxides are expected to be secondary minerals in the lowlands, the low dielectric constant of the lowland materials, estimated from the Magellan radar data and ground-based radiometric data (Sect. 16.2.2.1), does not indicate the presence of abundant Fe sulfides and magnetite formed through gas-solid weathering reactions in the current epoch, at least in the upper (<0.5 m) surface layer. The supposed presence of more abundant Fe sulfides and/or oxides in the deeper (>1 km, Pettengill et al. 1992) and shallower (>0.7 m, Anotony et al. 2022) subsurface, as suggested by the radar data, could indicate the formation of these minerals under the previous climatic conditions, followed by burial by lava and fine-grained materials (Sect. 16.4.4.2).

Suppose near-surface abundance of trace sulfur-oxygen-carbon-hydrogen gases is governed by the Mag-Py and/or Mag-Py-Hem assemblages (Figs. 16.22 and 16.25, Table 16.11). In that case, the role of anhydrite in the buffering process remains uncertain. Despite the limited sulfatization of surface materials at the landing sites (Table 16.8), the experimentally noted reactivity of basaltic glasses with $SO_2$ (Tables 16.9 and 16.10) and the thermodynamic favorability of sulfatization for Ca-bearing pyroxenes, there is no clear indication of equilibration for essential sulfatization reactions (Sect. 16.3.2.2). However, the apparent lack of complete equilibration does not necessarily exclude the participation of anhydrite in the buffering process. The kinetic inhibition of advanced sulfatization and/or the equilibration of plagioclase with its weathering products (reactions 16.71 and 16.75), as suggested by equilibrium models (Barsukov et al. 1986c; Klose et al. 1992; Zolotov 2018), could decelerate the depletion of atmospheric $SO_2$ to levels at which $CaSO_4$-bearing mineral assemblages are in equilibrium (e.g., equilibrium 16.75). The role of anhydrite in buffering by the Mag-Py assemblage, as suggested by Barsukov et al. (1982c), could be illustrated by the equilibrium expressed by Fegley and Treiman (1992a),

$$3FeS_2 + 6CaAl_2Si_2O_8 + 11O_2(g) = 6CaSO_4 + Fe_3O_4 + 6Al_2SiO_5 + 6SiO_2 \quad (16.87)$$

that sets $fO_2$ at $10^{-21.5\pm0.4}$ bars at 740 K. Because actual $O_2$ is absent from the near-surface temperature, here is another representation of equilibrium 16.87,

$$3FeS_2 + 6CaAl_2Si_2O_8 + 22CO_2(g) = 6CaSO_4 + Fe_3O_4 + 6Al_2SiO_5 + 6SiO_2 + 22CO(g) \quad (16.88)$$

The possibility of buffering by the assemblage in equilibria 16.87 and 16.88 is illustrated in Fig. 16.22, particularly if the activity of anorthite is below unity and the uncertainties of phase equilibrium lines due to thermodynamic data are considered. The current atmospheric $xSO_2$ of ~150 ppmv could indicate a slow ongoing sulfatization of Ca-bearing pyroxenes, while the corresponding $fSO_2$ of $10^{-1.84}$ bars might regulate concentrations and fugacities of trace sulfur-bearing gases at the Mag-Py equilibrium at the surface-atmosphere interface (Table 16.11; Fig. 16.22). The buffering of atmospheric gases by secondary minerals in lowlands may have resulted from a physical-chemical evolution of the atmosphere-surface system since the last global volcanic resurfacing event (Sect. 16.4.4.2).

#### Sulfur-Oxygen-Fe Minerals in the Highlands

The lack of equilibration among gases above a thin near-surface layer restricts evaluations of phase equilibria in the highlands, making such assessments heavily reliant on assumptions. If gas-phase equilibria are assumed in the highlands, decreasing $fO_2$ with altitude facilitates the stability of pyrite. Indeed, calculations of complete equilibria in multicomponent gas-solid systems indicate the stability of pyrite under highland conditions (Khodakovsky 1982; Klose et al. 1992; Semprich et al. 2020). If $fO_2$ is unaffected by gas equilibria and remains the same as in the lowlands, hematite is stable in the highlands (Fegley 1997; Fegley et al. 1997b; Semprich et al. 2020; Zolotov 1987, 1996; Zolotov and Khodakovsky 1989). However, when evaluating the stability of Fe sulfides and oxides, the results depend significantly on the chosen gas-solid chemical equilibria and the measured abundances of gases. Phase diagrams plotted in terms of $fS_2$, $fS_3$, $fSO_2$, $fOCS$, $fCO$, and $fH_2S$ in the studies mentioned above and in (Zolotov 2018, 2019) do not exclude the presence of hematite, magnetite, pyrite, or even pyrrhotite in the highlands. The confusing discrepancy arises from the deviation of extrapolated measured gas abundances (Table 16.1) from the conditions of gaseous equilibria as altitude increases above a narrow near-surface layer at the modal radius (Krasnopolsky and Pollack 1994; Zolotov 1996). Consequently, chemical equilibrium approaches may not be directly applicable for assessing the reactivity of solids exposed to the highland atmosphere. Regardless of further remote sensing and in situ data on surface materials, experimental studies on the kinetics of gas-mineral reactions involving selected gases or disequilibrium gaseous mixtures are necessary to understand the secondary mineralogy of the highlands, some of which could have formed during or shortly after the global volcanic resurfacing (Sect. 16.4.4).

#### Chalcogenides of Trace Elements in the Surface Materials

The stability of trace metal sulfides in the surface and upper crustal materials has been examined in the context of cloud composition, the low microwave emissivity of highland materials (Sect. 16.2.2.5), and the hazes in the lower atmosphere. Lewis (1968) was the first to consider the migration of metal-bearing compounds via sublimation from rocks and condensation in the presumed cold polar regions of Venus. Lewis (1968, 1969) investigated the possible condensation of sulfides and chalcogenides of Hg in the clouds. He concluded that the high surface temperature and limited abundance of atmospheric Hg prevent the condensation of HgS (cinnabar) anywhere on the surface. Similarly, Brackett et al. (1995) considered the vapor pressures of metal chalcogenides and halogenides as a function of temperature to determine the likelihood of their sublimation from lowland materials and condensation in the highlands and in the lower atmosphere below 15 km. For chalcogenides, they observed a decrease in vapor pressure at 740 K in the following sequence: $As_2S_2$, CuS, HgS, $NiS_2$, $Sb_2S_3$, $CoS_2$, SnS, PbS, $Ag_2S$, $Bi_2S_3$, and ZnS. Brackett et al. (1995) demonstrated the potential for sublimation and condensation of the compounds considered above and assessed the rates of sublimation, eddy transport, and condensation of solids. They concluded that a 0.1 to 1 cm thick layer of metal chalcogenides and chlorides could accumulate in the highlands over 1 to 10 Ma. Brackett et al. (1995) noted that the high dielectric constants (180 and 190) of potentially condensable $Sb_2S_3$ (stibnite) and PbS (galena) might contribute to the low microwave emissivity in the highlands. Their work suggests that condensation could influence the formation of near-surface (<6 km) atmospheric hazes suggested from in situ observations (Ragent and Blamont 1980; Gregier et al. 2004). Note that the interpretation of Venera 13 spectrophotometer data suggests basaltic composition of the near-surface haze (Kulkarni et al. 2025).

Schaefer and Fegley (2004) further investigated the potential condensation of volatile species as a function of altitude from the surface to ~51 km. They calculated solid-gas equilibria constrained by estimated partial pressures of volatile compounds of trace elements (Hg, Zn, Cd, Cu, Pb, Ag, Zn, Sb, Bi, As, etc.) that could be sublimated or volcanically outgassed from the crust. They used element/sulfur ratios in the terrestrial oceanic crust as a proxy for Venus' crust. They assumed that the atmospheric sulfur abundance (Table 16.12) represents the degassing degree of other elements. Schaefer and Fegley (2004) predicted condensation of $As_2S_3$, $Sb_2S_3$, and $Bi_2S_3$ (bismuthite) above altitudes of 27, 17, and 1.6 km, respectively, and condensation of PbS, $Cu_2S$, ZnS, $Ag_2S$, and CdS at all hypsometric levels. PbS and $Bi_2S_3$ could condense above ~2.6 km if the atmospheric abundances of Pb and Bi correspond to depleted crustal abundance and/or inefficient sublimation. In such a case, both sulfides could contribute to the low microwave emissivity of the highlands. Schaefer and Fegley (2004) also discussed the possible condensation of Pb-Bi sulfosalts (e.g., $PbBiS_4$, $Pb_3BiS_6$), known as fumarolic minerals. Port and Chevrier (2021) used the same approach to investigate the gas-solid equilibria for HgS and HgTe. They demonstrated that condensation in the highlands is improbable due to insufficient Hg and Te in the crustal rocks available for sublimation.

Inferences about the degassing and condensation of trace metal compounds are constrained by unknown atmospheric compositions, with assessed gas abundances likely representing upper limits. The apparent lack of hydrothermal processes, which can concentrate trace metals in upper crustal materials, also decreases the likelihood of an ample occurrence of trace element sulfide minerals that can sublimate. In other words, the occurrence of crustal trace elements in magmatic sulfides and rock-forming minerals (plagioclase, pyroxenes, olivine) and glasses, rather than in their mineral phases (e.g., PbS, HgS, ZnS), would have limited their degassing through sublimation. Lastly, releasing sulfur into the atmosphere from pyrrhotite oxidation (e.g., reactions 16.77 and 16.78; Fegley et al. 1995) may affect the ability to use atmospheric sulfur as an anchor in the estimations mentioned above.

Several qualitative exploration experiments have simulated the behavior of trace sulfur-bearing compounds in the atmosphere-surface system (Table 16.9). Kohler et al. (2013) and Kohler (2016) demonstrated the sublimation of HgS under simulated conditions resembling Venus' mountaintops (653 K, 55 bars $CO_2$) within two days. Port and Chevrier (2017a) and Port et al. (2018) noted the complete loss of HgS in the 1-bar $CO_2$ atmosphere at 653 to 733 K within one day, with lesser losses observed in $CO_2-SO_2$ and $CO_2-OCS$ gas mixtures. They observed the oxidation of galena to $PbSO_4$ (anglesite) in all the runs, and Port et al. (2019) reported stable anglesite in these gas mixtures at Venus' temperatures and pressures. Port et al. (2020) investigated the behavior of Te-Bi-sulfur mixtures and $Bi_2S_3$ under lowland and highland conditions in the presence of $CO_2$, $CO_2-SO_2$, and $CO_2-OCS$ gas combinations. They reported a preferential formation of $Bi_2Te_2S$ (tetradymite), with additional phases (e.g., $Bi_2S_3$, $Bi_2Te_3$) appearing depending on the starting sample composition, initial gas composition, temperature, and pressure. These experiments provide preliminary insight into the formation of exotic chalcogenides. However, further work with constrained $fS_2$ and $fO_2$ is required to confirm these reaction pathways.

## 16.4 Sulfur in the History of Venus

Sulfur on Venus may have played significant roles in differentiation and core formation, in the magma ocean, in magmatic degassing, in atmospheric processes and climate, in the trapping of degassed volatiles within minerals, and in the exchange of crustal and mantle materials through igneous, tectonic, and potential aqueous processes. The following questions remain unanswered, and this ambiguity limits the understanding of sulfur on Venus. How much $H_2O$ did Venus accrete during its formation? Did Venus lose $H_2O$ from its atmosphere, and if so, how and when? Did hydrogen escape lead to the oxidation of atmospheric, crustal, and mantle materials? Did $H_2O$ condense on the surface to form oceans? Did crustal materials subduct into the mantle? Did Venus support life? Understanding Venus' evolution is hampered by the lack of solid samples, and radar images provide insights into only about 1/10 to 1/5 of Venus' geological history. Although geological, climatic, and coupled atmospheric-interior evolution has been addressed in multiple works (Avice et al. 2022; Herrick et al. 2023; Ghail et al. 2024; Gillmann et al. 2022; O'Rourke et al. 2023; Rolf et al. 2022; and Salvador et al. 2023, for recent reviews), the behavior of sulfur has not been considered in detail. The behavior of sulfur remains a mystery, and its precise role in Venus' evolution requires further investigation. This section briefly describes current views on Venus' evolution, provides insights into the fate of sulfur, and discusses publications in which sulfur is considered.

### *16.4.1* Formation and Early History

Accretion models suggest that Earth and Venus formed from compositionally similar materials in the protoplanetary disk (e.g., Wetherill 1978; Hansen 2009; Morbidelli et al. 2012; Walsh and Levison 2016; Raymond and Izidoro 2017; Raymond et al. 2020). While accretion from planetesimals and planetary embryos is a commonly accepted pathway (Raymond et al. 2020; Raymond and Morbidelli 2022), accretion from pebble-sized materials is not ruled out (Johansen et al. 2021). The stochastic accretion of planetary embryos and planetesimals (e.g., Raymond et al. 2009; Sossi et al. 2022) suggests compositional differences between the planets; however, drastic variances are not expected, and pebble accretion could lead to more similar compositions. The compositional similarity of the planet-building materials of Earth and Venus aligns with their comparable planetary sizes, densities, moments of inertia, and the masses of carbon and nitrogen in the combined atmosphere-crustal materials (Table 16.12), as well as basaltic materials with similar compositions and oxidation states (Sect. 16.2.2).

Notable compositional differences between the planets, such as atmospheric and surface $H_2O$ content, could reflect divergent evolutionary patterns rather than the initial ones. The higher abundances of $^{20}$Ne and $^{36}$Ar in Venus' atmosphere may indicate the preservation of the primordial atmosphere due to fewer early giant impacts that may have stripped away an early terrestrial atmosphere (Genda and Abe 2005; Sakuraba et al. 2019). Venus' $H_2O$ deficiency is commonly attributed to the photodissociation of $H_2O$ vapor and the subsequent escape of hydrogen. This escape aligns with the anomalously high atmospheric D/H ratio of $(2.4 \pm 0.5) \times 10^{-2}$ (Donahue et al. 1997), although the amount of accreted $H_2O$ might not be directly inferred from these data (e.g., Grinspoon 1993; Grinspoon and Lewis 1988; Zahnle and Kasting 2023).

The potential accretion of Earth and Venus from Mars-mass differentiated planetary embryos (Morbidelli et al. 2012; Raymond et al. 2020; Raymond and Morbidelli 2022; Izidoro et al. 2021) suggests that sulfur was partitioned into the cores of those bodies as sulfides. While troilite is likely the sulfide present in the cores, its abundance and occurrence could vary depending on the bulk composition of the body and its differentiation pathways. Differentiated bodies with the composition of ordinary chondrites could have relatively troilite-poor metal cores, whereas differentiated carbonaceous chondritic bodies may even possess troilite cores (Bercovici et al. 2019), which have yet to be proven in meteorite samples. Although iron meteorites (many of which originate from planetesimal cores) commonly contain only ~0.5 to 1 wt% sulfur (Goldstein et al. 2009), concentrations of siderophile elements in iron meteorites suggest an initial ~0 to 19 wt% sulfur in the cores of asteroids and planetesimals (Chabot and Zhang 2022; Tornabene et al. 2023; Zhang et al. 2024). Some 'missing' sulfur might be lost through collision-induced volatilization of sulfides (Rubin et al. 2022), consistent with an apparent depletion in other moderately volatile elements (Cu, Ga, and Ge) in iron meteorite samples, or could be attributed to the lower mechanical resistance of sulfides to attrition in space, during atmospheric passage, or weathering on Earth (Kracher and Wasson 1982). Alternatively, some sulfur (in sulfides) could be in core parts that are underrepresented in the iron meteorite collection (Ni et al. 2020). Sulfides could be present in the putative inner cores, troilite core-sourced lavas, or mantle intrusions formed during the crystallization of the outer metal cores (Bercovici et al. 2019; Elkins-Tanton et al. 2020). Preserved core sulfides are a hypothesis, and the possible accretion of Venus and Earth from sulfur-depleted differentiated non-carbonaceous planetesimals does not imply the formation of sulfur-rich planetary cores. A low sulfur content in the solid inner core of the Earth, in the presence of Si in the outer core, is supported by the experiments of Sakai et al. (2023) on the partitioning of Si and sulfur between solid and liquid Fe metal under core pressure-temperature conditions. Their results show that the Earth's core includes oxygen and/or hydrogen, cannot consist of Fe-Si-sulfur alloys, could contain only 0.7 to 1.7 wt% sulfur, and that sulfur may not be an important light element in the solid inner core. Suer et al. (2017) conclude that Earth's core is sulfur-poor (< 1-2 wt% S), based on high-pressure–temperature metal–silicate partitioning experiments showing that mantle geochemical signatures cannot be reconciled with a sulfur-rich core. Similarly, low core sulfur contents have been estimated at 0.6 wt% and 1.9 wt% by Dreibus and Palme (1996) and McDonough (2014), respectively. Models of the Earth's composition, based on chondritic components by Alexander (2022), suggest sulfur core contents ranging from 0.70 to 0.82 wt%.

The Moon likely formed from a giant collision with differentiated Earth ~4.47 Ga ago, or about 100 Ma after the formation of calcium-aluminum-rich inclusions (CAIs) (Canup 2012). The potential absence of a Moon-forming collision on Venus may have contributed to the differences among the planets regarding core formation from embryonic cores, suggesting that Venus could have experienced less large-scale melting. However, Venus' slow retrograde rotation does not rule out a giant impact that altered its angular momentum (and spin direction) without resulting in the formation of a moon (Canup et al. 2001; Davies 2008). Jacobson et al. (2017b) do not dismiss the possibility of an early giant impact on Venus occurring about 11 Ma after CAIs, followed by less energetic collisions that did not disturb a stratified core, as discussed below. If a significant fraction of undifferentiated planetesimals and pebbles contributed to planet formation (Johansen et al. 2021; Brož et al. 2021), a delayed segregation of metal-sulfide cores involved the melting of silicates driven by the decay of long-lived radionuclides.

In the case of more likely accretion from planetary embryos, additional silicate-sulfide separation could have occurred in magma oceans on both planets (Rubie et al. 2016), which likely formed during the late stages of accretion (see Elkins-Tanton (2012), Solomatov (2015), Schaefer and Elkins-Tanton (2018), and Salvador et al. (2023), for reviews). The likelihood of magma ocean formation, their sizes, and evolution depend on the sizes and compositions of impactors and the timing. Impact-generated atmospheres impeded heat loss to space, allowing magma-atmosphere interfaces to exist, and extending the lifetime of magma oceans. Higher solar irradiation at lower heliocentric distances, as on Venus, resulted in elevated atmospheric and surface temperatures, thereby promoting the possible extended presence of a magma ocean (Hamano et al. 2013; Lebrun et al. 2013; Salvador et al. 2017). If Venus and Earth acquired equal amounts of $H_2O$, the more substantial thermal blanketing effect of Venus' steamy atmosphere could have played a significant role in the divergent evolution of the planets, affecting ocean solidification, degassing, mantle composition, and the fate of $H_2O$. However, the formation of the steam atmosphere over magma oceans on Earth and Venus can be hindered by high $H_2O$ solubility in the reduced ultramafic melts (Sect. 16.4.2.1).

The mass and fate of accreted $H_2O$ played a vital role in the physical-chemical evolution of magma oceans, affecting the behavior of elements during crystallization and degassing. Different amounts of $H_2O$ may have contributed to the divergent evolution of magma oceans on Earth and Venus (Salvador et al. 2023). $H_2O$ influences the composition of outgassed volatiles (Gaillard et al. 2022), the melt-solid partitioning of elements, the mineralogy of the forming mantle, magma ocean dynamics, and the planetary tectonic style established after crystallization (Schaefer and Elkins-Tanton 2018). Significantly, $H_2O$ could impact the oxidation state of magma and mantle.

$H_2O$ was present in hydrated and oxidized carbonaceous chondritic materials (e.g., CI/CM type chondrites), and impact-generated atmospheres of terrestrial planets are modeled to consist of $H_2$, $CO$, $H_2O$, $CH_4$, $NH_3$, $H_2S$, and $N_2$ (e.g., Schaefer and Fegley 2010; Zahnle et al. 2010, 2020). The modeled atmospheric compositions were dependent on the assemblage of impactors, magma-gas interactions, escape, and other factors (Schaefer and Fegley 2017). The late accretion of oxidized planetesimals may have elevated the mantle's $fO_2$, but it cannot account for the oxidation of Earth's entire mantle. Photo-dissociation of atmospheric $H_2O$, hydrogen escape, and the net accumulation of remaining oxygen may have caused the oxidation of atmospheric gases (CO to $CO_2$, $H_2S$ to $SO_2$, etc.) and the magma ocean tops, possibly affecting the entire magma oceans in cases of vigorous convection (Hamano et al. 2013; Wordsworth 2016; Wordsworth et al. 2018; Krissansen-Totton et al. 2021). However, an early quick solidification of the magma ocean top, even in the presence of a steamy atmosphere on Venus and Earth modeled by Selsis et al. (2023), suggests a different scenario for atmospheric $H_2O$ escape and consumption. Regardless of whether a magma ocean existed, hydrogen escape could have served as a significant oxidation pathway for the coupled atmosphere-crustal-mantle systems on Earth and Venus (Kasting et al. 1993; Sharp et al. 2013; Gillmann et al. 2009). Zahnle and Kasting (2023) found that oxidation caused by hydrogen escape on Venus could reach the upper mantle if the upper mantle existed as a stagnant lid or if the magma ocean was stably stratified (not convecting). Despite the common assertion regarding an effective atmosphere-magma ocean exchange, this process could be limited due to slower diffusion and higher viscosity in the lower temperature near-surface ocean, as well as the likely presence of scum or froth material acting as a stagnant lid. In addition to the mass of affected magma through magma-gas partition, the degree of oxidation depends on the amount of accreted $H_2O$, the oxidation of Fe species by $H_2O$, and $H_2$ production in the upper ocean through reactions

$$Fe^0\ (metal) + H_2O\ (gas\ or\ in\ melt) \rightarrow FeO\ (in\ silicates\ or\ melt) + H_2(g) \tag{16.89}$$

$$3FeO\ (in\ silicates,\ melt,\ etc.) + H_2O\ (gas\ or\ in\ melt) \rightarrow Fe_3O_4\ (magnetite) + H_2(g) \tag{16.90}$$

Another key factor is solar irradiation, which affects greenhouse temperature, the photodissociation of $H_2O$, and the escape of hydrogen and oxygen. The formation of magmatic sulfate was only possible through the oxidation by accumulated net atmospheric oxygen at the magma-atmosphere interface if the magma is not sufficiently mixed.

Regardless of what occurred to the $H_2O$, high-pressure disproportionation of Fe(II) to Fe-metal and Fe(III) (Hirschmann 2012; Schafer and Elkins-Tanton 2018; Armstong et al. 2019; Deng et al. 2020) in the lower magma oceans and sequestration of metal to the core could facilitate oxidation of silicate mantles on Earth-type planets. The disproportionation of Fe(II) in the melt seems to be a more suitable process than the melting and recrystallization of bridgmanite initially suggested by Frost et al. (2004) and Wade and Wood (2005). Oxidation via Fe disproportionation does not exceed the QFM buffer (Deng et al. 2020) and may have resulted in similar oxidation states in the mantles of Earth and Venus. The near-constant $fO_2$ of mantle rocks throughout Earth's history aligns with this idea (Schafer and Elkins-Tanton 2018). The MnO/FeO ratio of Venus' surface samples (Table 16.3) does not indicate oxidation beyond the $fO_2$ conditions of the QFM buffer (Schaefer and Fegley 2017; Sect. 16.2.2), suggesting S(-II) speciation in the magma ocean and sulfide mineralogy (pyrrhotite, pentlandite, chalcopyrite, troilite) in the mantle formed upon crystallization.

On Venus and Earth, cores likely formed and evolved compositionally during planetary accretion. Jacobson et al. (2017b) modeled a multi-stage core formation on Earth-like planets, predicting the accumulation of oxygen and Si in the outer cores, and suggesting that Venus could have better preserved such stratification due to the absence of a late giant impact that supposedly homogenized Earth's core ~100 Ma after CAIs (~4.47 Ga). They hypothesized that the absence of a detectable internally driven magnetic field on Venus reflects the chemical stratification of the core, which prevented convection.

The supposed planet-forming materials contain at least 0.5 wt% sulfur as follows: ~2% in ordinary chondrites, ~2 to 6 % in enstatite and carbonaceous chondrites, and ~0.5 to 1% in iron meteorites (Jarosewich 1990; Goldstein et al. 2009; Lodders and Fegley 1998). The bulk sulfur content of the Earth is estimated to be 0.64 wt% (McDonough 2014), 0.9 wt% (Kargel and Lewis 1993), 2.9 wt% (Morgan and Anders 1980), and $0.61 \pm 0.15$ wt% (Wang et al. 2018, concordance estimate). However, the assessed sulfur content in the bulk silicate Earth (BSE) is 200 to 270 ppmw (e.g., O'Neill 1991; Kargel and Lewis 1993; McDonough and Sun 1995; Palme and O'Neill 2014; Wang et al. 2018; Chap. 13: Kiseeva et al. 2026). Sampled mantle and mantle-sourced igneous rocks suggest 120 to 250 ppmw sulfur in various mantle reservoirs (Chap. 13: Kiseeva et al. 2026). The discrepancy between sulfur content in the Earth-building materials, bulk silicate Earth, and the mantle implies the segregation of sulfide melt to the core during a global silicate-Fe(metal) differentiation that likely occurred at the magma ocean stage. The depletion of chalcophile and highly siderophile elements (HSE) in mantle-derived rocks compared to abundances in chondrites is generally consistent with sulfide-silicate and metal-silicate partitioning during core formation (e.g., Wood et al. 2014; Kiseeva and Wood 2015).

A thorough segregation to the core remains to be reconciled with the chondritic-like sulfur/Se/Te ratios and elevated sulfur, Se, Te, and HSE abundances in the mantle rocks. The standard explanation of mantle sample sulfur, Se, Te, and HSE abundances is that they result from the accretion of a veneer following core-mantle differentiation (e.g., Rose-Weston et al. 2009; Wang and Becker 2013). Varas-Reus et al. (2019) demonstrate that the Se isotopic composition of mantle peridotites is consistent with a thin veneer of CI-type chondritic material delivered from the outer solar system. The tellurium stable isotope composition of mantle-derived rocks does not contradict the presence of a veneer of CM and/or enstatite chondritic material (Hellmann et al. 2021). Alexander (2022) modeled that a thin (~0.2 wt% BSE) veneer of CM- or EL-like chondritic material could explain the abundance of elements but not the isotopic composition of sulfur and Se. Wang et al. (2023) modeled that the chalcogen isotopic ratios in the BSE could be influenced by evaporation from planetesimals. They concluded that the late veneer cannot exceed 0.2 wt% of the mantle's mass. Even a tiny veneer could be a significant nitrogen source in the BSE. An analogous late veneer is likely present on Venus, where the absence of mantle homogenization due to a late giant impact does not exclude a more substantial contribution of veneer materials to the upper mantle composition and overall nitrogen content. The abundant nitrogen in Venus' atmosphere (Table 16.12) supports this idea. To determine whether a later veneer of chalcogen-, HSE-, and nitrogen-rich materials accreted, chemical and isotopic analysis of Venus' mantle-derived rocks and atmospheric chalcogens is required (Sect. 16.5).

Although the segregation at the magma ocean stage likely involved separating immiscible FeS liquid from the silicate melt, various pathways regarding silicate-sulfide partitioning in the magma ocean were modeled (e.g., O'Neill 1991; Rubie et al. 2016;

Laurenz et al. 2016; Suer et al. 2021; Steenstra et al. 2022). The quantity of precipitated sulfides depended on the bulk magma ocean's sulfur abundance and sulfur content at sulfide saturation (SCSS), which is influenced by temperature, pressure, melt composition (especially FeO content), and oxidation state (e.g., Baker and Moretti 2011; Smythe et al. 2017; Chap. 12: Simon and Wilke 2026). Experiments below 24 GPa indicate a decrease in SCSS with increasing pressure and decreasing temperature (e.g., Blanchard et al. 2021). The low sulfur solubility at the depth of a magma ocean supposedly allowed for the exsolution of droplets of liquid FeS (so-called "the late Hadean matte") and their segregation to the core (O'Neill 1991). The Earth's magma ocean crystallization model by Zhang Zhou et al. (2024), based on generalized SCSS experimental data at <24 GPA ($<$ 746 − 970 ppmw), also suggests late-stage sulfide precipitation from evolved ($FeO-SiO_2-Al_2O_3$ enriched) magma at depths of 120 to 220 km that does not suggest sulfide percolation through the crystallized mantle. Overall, the percolation of liquid sulfides to the core remains the standard explanation for the abundance of chalcophile and siderophile elements in mantle-derived rocks (Chap. 13: Kiseeva et al. 2026).

Data on sulfur solubility at deep magma ocean pressures (43 to 53 GPa) from Steenstra et al. (2022) revealed a high SCSS (0.3 to 0.4 wt% sulfur). Their model suggests a consistent SCSS throughout the ocean, allowing sulfide segregation only during the final stages of magma crystallization and implying the formation of a relatively sulfur-rich mantle. In this case, the abundance of siderophile elements in mantle rocks may be due to Fe disproportionation rather than sulfide segregation to the core. If the contentious model of Steenstra et al. (2022) is correct, sulfide-rich regions in the upper parts of the mantle, with FeS trapped in the silicate matrix, could have formed during the late stages of ocean crystallization. The slow solidification of Venus' magma ocean (~100 Ma compared to Earth's ~5 Ma, Hamano et al. 2013) may have influenced the formation of a sulfide-enriched mantle and possibly enhanced the development of exceptionally sulfur-rich upper mantle regions.

On Earth and Venus, the magma ocean stage and subsequent occurrence of mantle sulfides were influenced by the accretion of various sulfur-rich impactors, differentiated (core-bearing) and chondritic. These impactors enhanced sulfide segregation and established sulfur, chalcophile, and siderophile elements abundances in the mantle (Rubie et al. 2016; Suer et al. 2021; Li et al. 2016). Diverse pathways are modeled for multi-stage (Rubie et al. 2016) and homogeneous (Suer et al. 2021) accretion. The potential absence of a giant impact homogenization of Venus' mantle may have preserved a sulfur-rich mantle and sulfide-rich upper mantle regions formed via magma ocean crystallization. This hypothesis could be tested through in situ measurements of sulfur, chalcophile, and siderophile elements in unaltered basalts (Sect. 16.5.2). Potential sulfide-rich upper mantle regions could be inferred from gravity data and by investigating large impact crater ejecta (Sect. 16.5.2).

The different mantle convection patterns on Venus (Rolf et al. 2022) and Earth may have further influenced the segregation of sulfides to the cores. On Venus, sulfides formed through magma ocean crystallization could remain in the lithospheric lid. If a basal magma ocean existed throughout Venus' history (O'Rourke et al. 2020), it might be enriched in dissolved sulfur at the SCSS level and contain accumulated FeS liquid at the bottom, similar to some mafic magma chambers on Earth (Sect. 16.2.2.2). If Venus did not possess a large-scale magma ocean, the melting of silicates in its interior due to radioactive heating would have enabled the separation of sulfide-metal liquids and their gravitational accumulation into the cores. Such a pathway, with $^{26}$Al decay as an energy source, likely formed metal-sulfide cores on many differentiated planetesimals without the formation of magma oceans (e.g., McCoy and Bullock 2016; Elkins-Tanton 2016).

The composition and oxidation states of atmospheres formed through magma ocean degassing have been a subject of interest. If convection was restricted in the upper magma ocean, the outgassing of oxidized volatiles ($CO_2$, $H_2O$) may have resulted from the oxidation of the ocean's surface due to hydrogen escape. If the deep ocean had been oxidized through the disproportionation of Fe(II), reduced magmatic gases (CO, $H_2$, OCS, etc.) could have been released into the atmosphere. However, vigorous mixing of an ocean equilibrated with Fe-metal at depth might have established a positive $fO_2$ gradient below approximately 50 to 200 km, allowing for the release of oxidized magmatic gases ($CO_2$, $H_2O$) at $fO_2$ around the QFM buffer (Armstrong et al. 2019; Deng et al. 2020), and $SO_2$ could also be among the gases released.

The chemical co-evolution of the atmospheric-ocean system may have led to more oxidized atmospheric compositions in the later stages of accretion. Hirschmann (2012) suggested a gradual depletion of atmospheric $CO_2$ through its dissolution in more oxidized near-surface magma, followed by the precipitation of graphite or diamond in the interior of the convecting magma ocean. The higher abundance of degassed $CO_2$ on Venus (Table 16.12) could indicate less trapping than on Earth (Armstrong et al. 2019), possibly due to the absence of a large magma ocean created by violent impact(s). By analogy with $CO_2$, an excessive amount of atmospheric $SO_2$ could dissolve at the atmosphere-magma interface and be reduced to sulfides at depth.

Gaillard et al. (2022) included $SO_2$ and $H_2S$ in outgassing models for shallow magma oceans. Under oxidizing conditions ($>$ IW + 2, i.e., $>$ 2 log $fO_2$ units above the IW buffer), they predicted the exsolution of $CO_2$, $SO_2$, $N_2$, and $H_2O$ into a dense ~100-bar atmosphere. Sulfur-poor and $H_2$-CO dominant gases were modeled for reduced magmas ($<$ IW $-$ 2). Even moderately reduced melts ($<$ IW + 1) release only trace $H_2S$ (<0.05 vol%) due to the high solubility of sulfur as an FeS complex in the melt. Gaillard et al. (2022) illustrated that the Earth's crustal inventory of sulfur ($1.2 \times 10^{22}$ kg, Canfield 2004) could be explained by degassing from an oxidized ($\sim$ IW + $2.5$) magma ocean. Applying this model to Venus suggests that the apparent accumulation of sulfur in atmospheric and crustal materials (Sect. 16.2) required relatively oxidizing ($>$ IW + 2) conditions in the upper magma ocean and/ or in the history of volcanic degassing (Sect. 16.4.3). The occurrence of supposedly volcanic and abundant sulfur species in the present atmosphere-surface system aligns with relatively oxidized magma and parent mantle rocks.

### *16.4.2 Aqueous and Anhydrous Pathways*

The potential for $H_2O$ condensation and the fate of a possible hydrosphere on early Venus have been evaluated (Ingersoll 1969; Kasting and Pollack 1983; Kasting 1988; Abe and Matsui 1988; Way et al. 2016; Salvador et al. 2017; Way and Del Genio 2020; Krissansen-Totton et al.

2021; Turbet et al. (2021). Sulfur has not been specifically discussed in Venus' aqueous conditions, and the following inquiries remain to be addressed. Were volcanic-sourced $SO_2$, OCS, and $H_2S$ oxidized to $H_2SO_4$ in a moist atmosphere, leading to the production of native sulfur followed by the rainout of sulfate and elemental sulfur, as modeled for early anoxic Earth (e.g., Kasting et al. 1989) and Mars (Sholes et al. 2017)? Was $SO_2$ disproportionated to sulfate and native sulfur in aqueous environments at the planetary surface? What were the sulfur isotopic signatures of such processes? Was the trapping of volcanic-sourced $SO_2$ into minerals sufficient to prevent the formation of sulfuric acid clouds? Did the atmosphere become oxidized sufficiently to allow for sulfate accumulation in oceanic water and the formation of sulfate-bearing sediments? Did pyrite and/or native sulfur form via aqueous oxidation of magmatic sulfides or sulfate reduction? Was native sulfur reduced, oxidized, or disproportionated to sulfide and sulfate? If $H_2O$ avoided condensation, the inquiries relate to sulfur in a hot, evolving atmosphere-crustal system with varying $H_2O$ content. These inquiries include the fate of sulfuric acid clouds; the oxidation of atmospheric and crustal materials driven by photochemistry and hydrogen escape; and the partitioning of volcanically degassed sulfur between the atmosphere and solid materials, along with the corresponding isotopic fractionation. This section outlines the likelihood of liquid $H_2O$ being present on Venus and discusses the implications for sulfur on a presumably abiotic planet.

#### 16.4.2.1 Did $H_2O$ Condense?

The possibility of similar $H_2O$ initial abundances on Earth and Venus (Sect. 16.4.1), the presence of moderately oxidized igneous rocks at the landing sites of the Venera and Vega spacecraft (Sect. 16.2.2.1), along with the high atmospheric D/H ratio, indirectly point to a period in Venus' history when $H_2O$ was abundant in the outer shells of the planet. However, no data indicates the presence of past liquid $H_2O$, and there is no consensus on this topic.

Venus might have accreted less $H_2O$ than Earth, which would have limited $H_2O$ condensation if $H_2O$ was released into the atmosphere (e.g., Kasting, 1988). The planet could have been desiccated through $Fe^0$–steam reactions during a collision of large embryos that also altered Venus' spin, as hypothesized by Davies (2008). Such a pathway implies $H_2$ formation via reactions 16.89 and 16.90 and loss before the solidification of the planet and oxidation of its interior. Gillmann et al. (2016) illustrated that post-accretion impacts from 400 to 800 km bodies could have triggered volcanism, heating, and atmospheric erosion, placing the planet in a runaway greenhouse. An early atmospheric $H_2O$ could be lost via photodissociation due to intense solar UV flux from the young Sun, allowing for the escape of hydrogen and, under certain conditions, oxygen (Zahnle and Kasting 1986; Chassefiére 1997; Kulikov et al. 2006; Lammer et al. 2008, 2011; Gillmann et al. 2009; Lichtenegger et al. 2016; Zahnle and Kasting 2023). A fraction of atmospheric steam could be lost through the oxidation of Fe-metal and silicate's Fe(II) in impact-generated gases, steam-magma, and steam-silicate interactions via reactions 16.89 and 16.90, followed by hydrogen escape.

In one modeled scenario, Hamano et al. (2013) demonstrated that thermal blanketing of steam atmosphere and solar irradiation at ~0.7 AU facilitated the slow solidification of a magma ocean within ~100 Ma. They modeled that the corresponding gradual $H_2O$ outgassing could have desiccated the planet through hydrodynamic hydrogen escape. Their models indicated that the $H_2O$ reservoir in the mantle becomes more extensive upon ocean solidification than the $H_2O$ reservoir in the atmosphere, limiting the chances for condensation. Models from Hier-Majumder and Hirschmann (2017), Miyazaki and Korenaga (2022), Bowert al. (2022), Sossi et al. (202,3) and Maurice et al. (2024) also suggested $H_2O$ depletion due to inefficient magma ocean outgassing during crystallization. Gaillard et al. (2022) noted the suppressed degassing of high-solubility magmatic $H_2O$ in a 100-bar $CO_2$/CO-rich atmosphere. As described in Holloway's (2004) study of submarine lava, trapped magmatic $H_2O$ could be consumed via oxidation of the melt (reaction 16.90) at the uppermost layer of the magma ocean, contributing to the oxidation of the mantle. This presents an unaddressed mechanism of $H_2O$ consumption in the upper magma ocean on Venus and elsewhere.

Contrary to Earth, possibly fewer large-scale collisions contributed neither to the formation of an extensive magma ocean nor aided in $H_2O$ degassing (Sect. 16.4.1). After the crystallization of Venus' magma ocean, the potential establishment of a stagnant lid mantle convection pattern (Rolf et al. 2022) may have limited the magmatic degassing of $H_2O$. The restricted degassing could be balanced by hydrogen escape without $H_2O$ condensation (Miyazaki and Korenaga 2022). Retention of $H_2O$ in the mantle is consistent with the petrological interpretation of the Vega 2 rock composition (Table 6.3) (Barsukov et al. 1986d; Barsukov 1992). A few oceanic masses could be stored within Earth's nominally anhydrous mantle phases (e.g., Kaminsky 2018), and Venus' mantle might be similar (Zolotov al. al. 1997).

The high atmospheric D/H ratio can be explained without invoking liquid $H_2O$ at the surface (Grinspoon and Lewis 1988; Donahue et al. 1997; Zahnle and Kasting 2023). Oxidation of Venus' upper interior could have occurred without hydrogen escaping from the steam atmosphere (Sect. 16.4.1), and moderately oxidized igneous rocks at landing sites (Sect. 16.2.2.1) may not indicate past $H_2O$-rich environments. Venus' one-modal hypsometric curve does not signify the existence of large cratons (Sect. 16.2.2.2) that could have formed in the presence of an ocean of water (Campbell and Taylor 1983). The supposedly layered rock formations in the tessera terrain (Sect. 16.2.2.4, Fig. 16.12 and 16.13) might be of volcanic origin rather than sedimentary rocks deposited by water. Climate models remain ambiguous regarding $H_2O$ condensation. For example, the 3D modeling of Turbet et al. (2021) shows that a day-night cloud asymmetry could have prevented condensation on early Venus.

#### 16.4.2.2 Atmospheric and Crustal Sulfur Without Water Condensation

The fate of sulfur in the runaway greenhouse atmosphere and upper crustal rocks depended on the amount of atmospheric $H_2O$ vapor, solar UV irradiation, volcanic degassing, and resurfacing through volcanism, tectonics, and aeolian processes. The amount of atmospheric water was determined by accretion, early evolution (Sect. 16.4.1), ongoing volcanic degassing, and consumption via hydrogen escape and steam-rock reactions, which caused oxidation and hydration (e.g., the formation of phyllosilicates). The levels of atmospheric $H_2O$, cloud composition ($H_2O$ vs. $H_2SO_4-H_2O$) and coverage primarily dictated the atmospheric and surface temperature. The concentration of atmospheric $H_2O$ and solar UV flux influenced the photodissociation of $H_2O$ vapor, the fluxes of hydrogen and oxygen to space, and the net mass of oxygen available for the oxidation of atmospheric and crustal materials (e.g., Zahnle and Kasting 2023). Regardless of the redox state of the lower atmosphere and surface, oxidation conditions in the upper atmosphere, due to $CO_2$ and $H_2O$ photolysis, favored the oxidation of volcanism-sourced sulfur-bearing gases ($SO_2$, OCS, $H_2S$, $S_2$) to S(VI) species ($SO_3$, $H_2SO_4$(g), and sulfuric acid aerosol). As at present (Table 16.1), $SO_2$ likely was the main sulfur-bearing atmospheric gas, and a higher abundance of $H_2O$ enabled more oxidizing conditions via hydrogen escape. Consequently, sulfatization rather than sulfidation of surface materials provided a sink for atmospheric sulfur from a moist atmosphere.

Because the amounts of Ca and Na in exposed volcanic rocks exceed the mass of sulfur-bearing gases released from their corresponding magmas, most of the degassed sulfur must have been trapped in secondary minerals. This imbalance likely made complete sulfurization of permeable volcanic rocks improbable. Any increase in atmospheric sulfur mass due to volcanic degassing was followed by consumption through gas-solid reactions. Prolonged periods without volcanism led to lower $fSO_2$ values that approached those controlled by equilibria with sulfates (e.g., reactions 16.74, 16.75, and 16.88). As in the present epoch, even trace amounts of volcanic $SO_2$ in the atmosphere likely hindered the formation of carbonates through weathering reactions. If low $fSO_2$ allowed carbonate stability, their formation would be impeded by slow $CO_2$-silicate reactions (Tanner et al. 1985). In weathering crusts, carbonates could not withstand episodes of large-scale volcanic degassing of sulfur-bearing compounds (Sect. 16.4.3) or the oxidation of sulfides in the lava (reactions 16.77 and 16.78), which enhanced bulk sulfur and $SO_2$ abundances in the atmosphere.

In the case of an $H_2O$-poor atmosphere, the greenhouse temperature and the fate of sulfur species in the atmosphere-crust system might not have differed significantly from those in the current epoch that began after the last global volcanic resurfacing (Sect. 16.4.4). The disproportionation of sulfur in $SO_2$-solid interactions led to the formation of metal sulfates and $S_2$ gas (e.g., reaction 16.62). Pyrite could have formed via pyrrhotite oxidation and the sulfidation of Fe(II) in exposed glasses and silicates (Sect. 16.3.2.3). Photochemical oxidation of $SO_2$, OCS, and $H_2S$ led to both S(VI) species and elemental sulfur. Even if sulfur-MIF occurred in an $O_2$-less upper atmosphere, the corresponding signature might not have been recorded in the rocks due to the evaporation of condensed $S_x$ in the lower atmosphere. In the latter case, atmospheric haze of condensed $S_x$ acted as an effective absorber of UV radiation (Kasting et al. 1989).

Both elevated $H_2O$(g) content in the atmosphere and moderate solar UV flux enhanced hydrogen escape while limiting oxygen escape (Zahnle and Kasting 2023). These factors contributed to increased S(VI)/(S(IV) + S(0) + S(-II)) and $CO_2$/CO ratios in the atmospheric gases, ultimately favoring the accumulation of atmospheric $O_2$. $SO_2$ was more abundant than $SO_3$, except under extreme $O_2$-rich (> 1 bar) and hot (> 1000 K) conditions (Fig. 16.26). If oxygen escape was limited, the evolution of a steam atmosphere could have led to the oxidation of a significant mass of Fe(II) in the rocks (Zahnle and Kasting 2023), and the supposed atmospheric $O_2$ should have caused geologically instantaneous oxidation of magmatic sulfides (pyrrhotite, pentlandite, etc.) in the exposed materials,

$$2Fe_{0.875}S + 3.3125O_2 \rightarrow 0.875Fe_2O_3 + 2SO_2 \qquad (16.91)$$

According to experimental data (e.g., Alksnis et al. 2018; Aracena and Jerez 2021; Liu et al. 2023), the oxidation of ferrous sulfides to hematite is favored by elevated temperatures and occurs through the formation of magnetite as an intermediate phase. Although the released $SO_2$ contributed to the atmospheric inventory of sulfur and promoted the formation of sulfuric acid clouds, this inventory was regulated by a balance between sources and sinks related to minerals. The sinks were enhanced by elevated surface temperatures and occurred through the exemplary net reaction

$$SO_2 + 0.5O_2 + CaSiO_3(\text{in pyroxenes and glasses}) \rightarrow CaSO_4 + SiO_2 \qquad (16.92)$$

in which $SO_2$ and $O_2$ could be replaced by $SO_3$. In contrast to current anoxic surface environments (Sect. 16.3.2), sulfatization occurred without sulfur disproportionation, and sulfur did not return to the atmosphere. The crustal thickness affected by anhydrous sulfatization and the degree of sulfatization depended on volcanic and tectonic resurfacing; however, possible aqueous environments (Sect. 16.4.2.3) allowed for more efficient resurfacing through erosion and sedimentation.

Hydrogen escape and limited oxygen escape from the atmosphere, along with one Earth's ocean mass, likely caused the large-scale oxidation of crustal materials (e.g., Kasting 1988; Zahnle and Kasting 2023). A significant greenhouse effect in the steam-rich atmosphere favored crustal melting (Kasting 1988), while the increased geothermal gradient facilitated upper mantle melting and volcanic activity. Although widespread surface magma allowed for the trapping of net atmospheric oxygen through oxidation by $CO_2$, $H_2O$, and $O_2$, the atmospheric $fO_2$ could not have strongly exceeded the $fO_2$ values controlled by the Mag-Hem buffer (reaction 16.66), which ranges from $10^{-5}$ to $10^{-3}$ bars. The plausible buffering of the atmospheric redox state by the Mag-Hem equilibrium in surface magmas established the atmospheric $SO_3/SO_2$ ratio at ~$10^{-3}$ (Fig. 16.26), and the $CO_2/CO$ ratio, allowing sulfur oxidation in the affected silicate melts. Under more oxidizing conditions than the IW $fO_2$ buffer, magmatic sulfur is more soluble in the S(VI) form (Chap. 12: Simon and Wilke 2026), and the oxidation of surface melts in contact with a hot steam atmosphere enhances sulfur solubility. Magmatic S(-II) was oxidized to S(VI), and atmospheric sulfur ($SO_2$, $SO_3$) was trapped in the sulfate melt complexes. The enhanced solubility also limited the degassing of magmatic sulfur supplied to magma surface ponds. Together with the high-temperature sulfurization of the surface, this trapping of volcanogenic sulfur oxides restricted the formation of sulfuric acid clouds. Solidification of surface melts led to anhydrite-bearing igneous rocks. In addition to abundant ferric oxides (Zahnle and Kasting 2023), anhydrite is expected to be present in rocks that consume the oxygen mass, as in the Earth's ocean, and no sulfur-MIF signature is anticipated. The detection of such rocks would indicate large-scale oxidation facilitated by steam greenhouse environments.

#### 16.4.2.3 Water Condensation and Aqueous Environments

A plausible Earth-like initial $H_2O$ inventory, an accumulation of $H_2O$ in a primordial atmosphere through impact and magma degassing (Sect. 16.4.1), and the young Sun's low luminosity favored the condensation of atmospheric $H_2O$ after the accretion and cooling of the surface. While $H_2O$ vapor and planetary albedo were crucial in determining surface temperature and condensation, limited $CO_2$ degassing or its sequestration as graphite (Hirschmann 2012) slightly mitigated greenhouse heating. Radiative-convective atmospheric modeling (Kasting and Pollack 1983; Kasting et al. 1984; Kasting 1988; Abe and Matsui 1988) demonstrated that the early solar flux at Venus' orbit was near the critical value needed to trigger the runaway greenhouse effect (Ingersoll 1969; Rasool and de Bergh 1970). Abe et al. (2011) and Salvador et al. (2017) indicated that an albedo slightly higher than that of today's Earth was necessary to support an ocean on early Venus. In the models of Kasting (1988), Way et al. (2016), Way and Del Genio (2020), and Salvador et al. (2017), high-albedo clouds cooled the surface and promoted condensation. Models indicate that clouds and the faint early Sun could have permitted a long (~0.5–3.8 Ga) presence of liquid $H_2O$ (Pollack 1971; Kasting 1988; Abe and Matsui 1988). However, these were 1-D models that had to make unconstrained assumptions about albedo. Yang et al. (2014), Way et al. (2016), and Way and Del Genio (2020) emphasized the importance of Venus' slow rotation, rather than solar luminosity or $H_2O$ mass, to enable surface liquid $H_2O$ over extended periods of its history. The possible chloride melts that formed canali (Sect. 16.2.2.3) and the speculated felsic composition and layered rocks of the tessera terrain (Sect. 16.2.2.4) indirectly support aqueous deposition of sediments and evaporites.

If atmospheric $H_2O$ condensed (Pollack 1971; Kasting 1988; Abe and Matsui 1988; Yang et al. 2014; Way et al. 2016; Way and Del Genio 2020; Salvador et al. 2017; Krissansen-Totton et al. 2021), Venus' surface temperature would be below the critical temperature of $H_2O$ (647 K or higher for salty solutions). The cloud-free moist greenhouse models suggest temperatures below ~500 K, while models that consider the elevated albedo of the clouds result in temperatures as low as ~300 K (Pollack 1971; Kasting 1988). Global 3-D simulations incorporating the clouds, slow rotation, and topography suggest temperatures ranging from ~276 to 420 K, depending on location, elevation, and surface water coverage (Way et al. 2016; Way and Del Genio 2020). Therefore, ambient conditions can range from sub-critical hydrothermal (~600 K) to environments typical for most of the Earth's history (~280–350 K).

In the moist greenhouse, atmospheric warming is constrained by the increasing planetary albedo at higher solar luminosity. Along with diffusion-limited hydrogen escape, this limited warming prolongs the existence of liquid $H_2O$, according to Kasting (1988). In his model, hydrogen escape and net oxygen accumulation become efficient above a surface temperature of ~350 K. Kasting (1988) and Abe et al. (2011) noted that a significant fraction of the oceanic $H_2O$ could be lost through hydrogen escape before the surface temperature runs away (true runaway occurs). Oxygen remaining after hydrogen escape was supposedly consumed through the oxidation of carbon-, Fe-, and sulfur-bearing species in the atmosphere, oceanic water, and the crust.

As on Earth, aqueous processes occurring on land, in oceans, and hydrothermal settings should encompass the erosion, transport, and deposition of solids, as well as the chemical alteration (dissolution, hydration, oxidation, ion exchange) of minerals and glasses, transport of solutes, and precipitation of secondary minerals in weathering rinds and crusts from solutions carried by surface streams and groundwater. The compositions of surface and oceanic waters were formed by the dissolution of minerals and glasses that supplied major cations ($Na^+$, $Ca^{2+}$, $Fe^{2+}$, etc.) and through volcanic or impact degassing that provided anions ($Cl^-$, $HS^-$, $HCO^-$, etc.). In the case of an initial anoxic atmosphere, NaCl-type ocean water with dissolved inorganic carbon (DIC, $CO_2$ + $HCO^-$ + $CO_3^{2-}$) was likely. The concentration and speciation of DIC depended on atmospheric $fCO_2$, temperature, and *pH*, which reflected acid-base balances in the $H_2O$-rock systems. The oceanic *pH* was likely neutral to alkaline, as carbonic acid formed via $CO_2$ dissolution would be neutralized by cations, leading to carbonate precipitation at neutral or alkaline *pH*. A $CO_2$-rich atmosphere could not be sustained in the long run ($\sim 10^7 - 10^8$ a) when in contact with mafic/ultramafic rocks, specifically if new rocks were introduced and carbonates were removed from the surface through subduction or burial (e.g., Walker 1977; Sleep and Zahnle 2001; Zahnle et al. 2007; Zolotov 2020).

#### 16.4.2.4 Sulfur in Aqueous Environments

Sulfur behavior on early Venus may differ from that on early Earth due to the absence of life (e.g., sulfate-reducing, sulfide-oxidizing, $Fe^{2+}$-oxidizing, and $O_2$-producing organisms) and biological organic matter. However, abiotic and organic-poor high-temperature settings may not significantly differ from their Earth's counterparts. As on Earth, the solubility of Fe sulfides likely controlled the concentration of major sulfur-bearing solutes ($H_2S$ and $HS^-$, depending on the *pH* and temperature) in surface and oceanic water. The solubility increases with temperature (see Chap. 4: Pokrovski et al. 2026), and sulfides in igneous rocks (pyrrhotite, pentlandite) could be strongly affected by dissolution in ambient high-temperature ($>\approx$ 500 K) environments and hydrothermal systems. Higher greenhouse temperatures and lower *pH* levels at elevated atmospheric $fCO_2$ enhanced the concentrations of dissolved $H_2S$ in surface and ocean-top water. Secondary Fe sulfides (pyrite, marcasite, pyrrhotite) and chalcophile metals (Cu, Zn, Pb, etc.) could have precipitated under high-temperature settings on the land surface and in Earth-like basalt-hosted hydrothermal systems throughout the globe. As in Earth's hydrothermal systems (Edmond et al. 1979; Kump and Seyfried 2005), the supply of $Fe^{2+}$ via local high-temperature sulfide dissolution favored the precipitation of secondary sulfides under lower temperatures and more alkaline settings. Pyrite could be the most significant secondary sulfide precipitated at elevated bulk sulfur content and higher *pH* (e.g., Garrels and Christ, 1965; Rickard 2012a, b, 2014).

The aqueous oxidation of sulfides on Venus has not been evaluated, and only qualitative inferences can be made based on analogs and insights from equilibrium chemical thermodynamics and the kinetics of abiotic redox reactions. Generally, the sulfate concentration in surface and oceanic water may have been influenced by a dynamic balance involving volcanic/impact degassing of sulfur-bearing species, their photochemical and chemical oxidation, sulfide oxidation, and trapping through precipitation and high-temperature abiotic reduction.

On Earth, sulfate content in Archaean seawater did not exceed ~2.5 micromoles (Crowe et al. 2014), despite some initial oxidation of sulfides on land. During the Great Oxidation

Event (GOE, Holland 2006) at ~2.4 Ga, the atmospheric $O_2$ mixing ratio abruptly reached ~$10^{-2}$ after nearly 0.5 Ga of oxygenic photosynthesis in the late Archaean (Lyons et al. 2014). The large-scale oxidation of sulfides on land occurred during this event and enhanced oceanic sulfate content. Whether free $O_2$ accumulated on Venus depends on the relative rates of hydrogen escape and net oxygen sinks through the oxidation of atmospheric and volcanic (Sect. 16.4.3) gases (CO, $SO_2$, OCS, $H_2S$, $S_n$, and $H_2$), aqueous ($Fe^{2+}$, $H_2S$, $HS^-$), and crustal ferrous materials. The relative masses of escaped hydrogen, Fe(II), and sulfides available for oxidation played crucial roles in determining the fates of oxygen, Fe, and sulfur in the coupled atmosphere-ocean-crust systems.

Some sulfates could have formed regardless of $O_2$ accumulation in the lower atmosphere. As on the early Earth (Walker and Brimblecombe 1985; Kasting et al. 1989; Farquhar et al. 2001; Ono 2017; Catling and Zahnle 2020) and Mars (Franz et al. 2014; Sholes et al. 2017; Chap. 17), photolysis of $SO_2$, OCS, or $H_2S$ in an anoxic atmosphere could have caused the formation of native sulfur and S(VI) species ($SO_3$, $SO_4^{2-}$), leading to a sulfur-MIF isotopic signature in surface waters and sediments. The photochemical oxidation of $SO_2$ (reactions 16.6 and 16.7) and other sulfur-bearing gases (Sect. 16.3) likely produced sulfuric acid aerosols and acid rain, which could have contained native sulfur (Kasting et al. 1989). However, this process could be limited if the sink of atmospheric sulfur exceeds the supply of reduced compounds ($SO_2$, OCS, $S_n$, $H_2S$). An efficient supply of volcanic $SO_2$ (Sect. 16.4.3) and $SO_2$-bearing impact plumes generated by hydrated carbonaceous chondrite impactors (Schaefer and Fegley 2010) and/or in $H_2O$-bearing targets, followed by acid rains, could have caused acid sulfate weathering on land, somewhat like that on early Mars (Chap. 17; Zolotov and Mironenko 2016). The dissolution of $SO_2$ in water caused the formation of unstable sulfurous acid ($H_2SO_3$) and bisulfite ion ($HSO_3^-$) in acid rain and surface reservoirs. Abiotic disproportionation of S(IV) in these compounds led to $SO_4^{2-}$ and $H_2S$ and/or native sulfur, depending on temperature, pressure, *pH*, and catalytic conditions

$$4SO_2(aq) + 4H_2O(l) \rightarrow 3SO_4^{2-} + 6H^+ + H_2S(aq) \tag{16.93}$$

$$4SO_2(aq) + 4H_2O(l) \rightarrow 3SO_4^{2-} + 6H^+ + S^0(\text{s or l}) + H_2(g) \tag{16.94}$$

As on Earth, S(IV) disproportionation could have occurred in ambient (surface waters), diagenetic, and hydrothermal (<~700 K) environments. Although abiotic low-temperature disproportionation was less efficient than diverse bio-mediated processes on Earth (Finster 2008), the overall reaction 16.93 consumed atmospheric and dissolved $SO_2$ to form coexisting aqueous sulfate and sulfide species. Native sulfur formed via overall reaction 16.94 occurred in crystalline ($S_x$, $S_8$) or liquid forms at temperatures <718 K (at 1 bar total pressure). The kinetic estimates of Ranjan et al. (2023) indicate a slow disproportionation of S(IV) ($SO_3^{2-}$, $HSO_3^-$) to $SO_4^{2-}$ and $S^0$ under low-temperature aqueous conditions on early Earth. Their work suggests a dominant role of aqueous photolysis in the consumption of S(IV) species in surface water. It indicates the potential occurrence of S(IV) species in surface aqueous environments on early anoxic planets, only if photolysis is not efficient. Their work also suggests that photolysis of sulfites limits the buildup of $S_x$ haze in the atmosphere, unless there is rapid volcanic degassing of sulfur-bearing species. Their work does not rule out the presence of sulfites in deep surface reservoirs and groundwater. Concerning early Venus, the results of Ranjan et al. (2023) emphasize the significance of integrating atmospheric and geochemical modeling to address volcanism, photolysis, and aqueous reactions at the surface.

Redox reactions involving sulfates and sulfides were limited at lower temperatures for kinetic reasons. However, acid-base reactions influenced both aqueous chemistry and secondary mineralogy. Sulfates of Ca, Mg, Na, and K precipitated in the weathering crusts alongside clay minerals, migrated in surface and ground waters, deposited via evaporation (c.f., Zolotov and Mironenko 2016), and delivered $SO_4^{2-}$ ions to the oceans. The precipitation of Fe sulfides (e.g., pyrite) consumed aqueous sulfide species across a wide range of temperatures and geological settings (e.g., surface, diagenetic, oceanic, and hydrothermal). Kinetically inhibited aqueous sulfate reduction below ~500 K (Ohmoto and Lasaga 1982) allowed for the metastable coexistence of $SO_4^{2-}$ with $H_2S$ and $HS^-$, which were supplied through the dissolution of sulfides and volcanic $H_2S$, as well as $SO_2$ disproportionation. Although sulfates could be more abundant at the ocean surface, the vertical gradients of sulfate and sulfides may not be as pronounced as in the Black Sea, which is rich in biological organic compounds found in subsurface waters and bottom sediments.

Except under alkaline conditions, the oxidation of aqueous sulfides to $SO_4^{2-}$ requires lower $fO_2$ than $Fe^{2+}$ oxidation to poorly soluble ferric oxyhydroxides (e.g., Garrels and Christ 1965). It follows that sulfide oxidation in Venus' weathering crusts was possible before a significant increase in atmospheric $fO_2$ and thorough oxidation of oceanic $Fe^{2+}$. Based on mass balance estimations, the stable coexistence of trace sulfate, which is formed through the photooxidation of volcanic sulfur-bearing gases, with abundant Fe(II) species was envisioned for a prebiotic ocean on early Earth by Walker and Brimblecombe (1985) and applicable for early Venus. An increase in Venus' atmospheric $fO_2$ via hydrogen escape boosted the oxidation of sulfides to $SO_4^{2-}$ at the ocean top and on land, following the transfer of sulfate- and $Fe^{2+}$-bearing surface waters to the oceans, where $SO_4^{2-}$ and $Fe^{2+}$ coexisted until further $fO_2$ increase. Even low-temperature oxidation of aqueous sulfides by $O_2$ occurs rapidly (Millero et al. 1987), and there was no inhibition of these processes in the geological timescale.

During the GOE, the influx of sulfates from continents formed via the oxidation of sulfides facilitated the microbiological reduction of oceanic sulfate and increased the sulfide content in deeper waters. It boosted pyrite formation in organic-rich bottom environments (Reinhard et al. 2009). The biological reduction of sulfate may have limited the accumulation of oceanic sulfate, which remained scarce during the Archaean (Habicht et al. 2002; Rickard 2014), consistent with the sulfur-MIF rather than mass-dependent fractionation of sulfur in Archaean rocks (Crowe et al. 2014). Despite the supply of sulfates to the ocean, deep oceans partially remained reduced ($Fe^{2+}$-rich, sulfidic, and sulfate-poor) until after ~1.8 Ga (Isley 1995; Kump and Seyfried 2005). Unlike Earth's marine sediments (Rickard 2012b, 2014; Emmings et al. 2022), the apparent lack of biological sulfate reduction and abundant organic matter on Venus favored the accumulation of oceanic sulfates and may have extended the lifetimes of $SO_4^{2-}$ and $Fe^{2+}$ in sulfate-poor oceanic water before a significant $O_2$ buildup.

On both planets, oceanic sulfate could be withdrawn through reduction at temperatures above ~470 to 520 K in hydrothermal systems hosted in mafic and ultramafic rocks (Alt et al. 1989) and the precipitation of pyrite

$$SO_4^{2-} + 2H+ + 4H_2(aq) \rightarrow H_2S(aq) + 4H_2O(l) \quad (16.95)$$

$$Fe^{2+} + 2H_2S(aq) \rightarrow FeS_2 + H_2 + 2H^+ \quad (16.96)$$

More reduced Venus' rocks (Sect. 16.2.2.1) promoted sulfate reduction by providing additional $H_2$ (reaction 16.90) and $Fe^{2+}$ for reactions 16.95 and 16.96, respectively. If Venus had higher ambient and hydrothermal system temperatures, this would favor abiotic sulfate reduction (reaction 16.95), followed by the precipitation of metal sulfides, such as pyrite.

Another aqueous sulfate sink is a high-temperature anhydrite formation in basalt-hosted oceanic hydrothermal systems, where $Ca^{2+}$ is sourced from basalts (Edmond et al. 1979; Alt et al. 1985, 1989; Chen et al. 2013). This is also inferred in water-rock interaction models for early Venus (Zolotov and Mironenko 2009), in which precipitation of $CaSO_4$ prevents the formation of hot (320 to 720 K) sulfate-rich oceanic water even in oxidizing conditions. The high temperature promoted anhydrite deposition in surface and hydrothermal environments, as well as the precipitation of highly soluble Na and Mg sulfates through evaporation in lagoons. The latter process provided the major sink of sulfates upon exhaustion of surface water.

As on Earth (Walters et al. 2020; Chap. 9: Schwarzenbach and Evans 2026), the subduction of oceanic lithospheric plates could have delivered sulfur to the mantle, affecting its redox state and sulfur content. However, the uncertainty surrounding the likelihood of plate tectonics on Venus (Rolf et al. 2022; Ghail et al. 2024) hinders implications. Despite sulfate formation in photochemical and low-$f$ $O_2$ aqueous settings, the concentration of oceanic sulfates could be comparable to that in Archaean oceans (Habicht et al. 2002; Crowe et al. 2014; Rickard 2014). Similar to Earth's history, the formation of sulfate-rich seawater necessitated substantial oxidation of land sulfides alongside oceanic $HS^-$ and $H_2S$. Such oxidation is questionable without a significant accumulation of atmospheric $O_2$.

The GOE occurred due to an imbalance between the net $O_2$ sink from oxidation and the production resulting from the burial of organic carbon and the escape of hydrogen (Catling 2014; Lyons et al. 2014; Catling and Kasting 2017). Venus likely lacked oxygenic photosynthesis, and various factors could have limited the $O_2$ supply from the upper atmosphere. The slow and prolonged net oxygen production due to limited hydrogen escape in the moist greenhouse atmosphere (Kasting 1988) could have allowed for the rapid abiotic oxidation of aqueous $Fe^{2+}$ (Davison and Seed 1983) and sulfide species (Millero et al. 1987), both on land and on the ocean's surface. Without felsic continents (Sect. 16.2.2), the prevalence of Fe(II)-rich mafic and ultramafic rocks across the planet favored $O_2$ consumption. Dominant mafic and ultramafic rocks on the land and ocean floors provided more Fe(II) for oxidation than those in the Archaean. More reduced mafic rocks on Venus (Sect. 16.2.2.1) offered a greater capacity to consume $O_2$ than their counterparts on Earth. On early Earth (Isley 1995; Kump and Seyfried 2005), the seafloor hydrothermal supply of $Fe^{2+}$ strongly contributed to the oceanic inventory of Fe(II). Higher temperatures and the rock's Fe(II)/Fe(III) ratios on Venus suggest a more advanced hydrothermal supply of $Fe^{2+}$, $H_2S$, and $H_2$ (reaction 16.90) than on Earth. As on early Earth (Walker and Brimblecombe 1985), the elevated Fe/sulfur ratio in rocks affected by aqueous processes limited the sequestration of $Fe^{2+}$ to sulfide minerals, and $Fe^{2+}$ likely remained abundant in the ocean, more so than dissolved sulfides, until after a significant supply of $O_2$.

Whether Venus' atmospheric $O_2$ reached a percentage mixing ratio depended on the balance between the $O_2$ flux from the upper atmosphere and the supply of reducing agents. Unless there was a scarcity of $H_2O$, a significant accumulation of $O_2$ could have occurred with a limited supply of volcanic rocks and hydrothermal $Fe^{2+}$, $H_2S$, and $H_2$. Limited resurfacing on land and the absence of plate tectonics that remove oxidized materials favored this accumulation. Other factors that could limit oxidation include oceanic stratification and the absence of $Fe^{2+}$ oxidizing microorganisms. Enhanced $fO_2$ facilitated the precipitation of ferric oxyhydroxides in the ocean and the oxidation of sulfides, both on land and in the ocean. On Earth, episodic precipitation of ferric species through biological and abiotic oxidation of oceanic $Fe^{2+}$ led to banded iron formations (BIF) that formed before, during, and after the GOE (Isley 1995; Yin et al. 2023). Similarly, Venus' BIF-like formations could have developed under a range of atmospheric $fO_2$ conditions.

Mass balance assessments indicate that trapping the oxygen remaining after hydrogen escape from the Earth's ocean mass requires the oxidation of Fe(II) to hematite in a global layer several tens of km thick (Kasting and Pollack 1983; Lécuyer et al. 2000; Zahnle and Kasting 2023). Aqueous environments provided much more favorable conditions for consuming the vast mass of oxygen than gas-solid interactions. If Venus' atmosphere lost such an amount of $H_2O$ vapor during the moist greenhouse, one would expect the development of abundant BIF-like formations via the oxidation of $Fe^{2+}$-rich oceanic water and their significant burial in the crust and/ or the mantle. The formation of sulfate-rich oceanic water would likely occur if a water mass comparable to Earth's ocean had been lost. Fe(II) oxidation over several tens of km implies a substantial oxidation of sulfides in those rocks.

#### 16.4.2.5 The Cessation of an Aqueous Period

If $H_2O$ condensation ever occurred, the subsequent consumption of surface water would have influenced the fates of sulfur and other elements (Cl, Na, Fe, Si, etc.) involved in aqueous processes. Net consumption of liquid $H_2O$ likely occurred via hydrogen escape from the moist greenhouse atmosphere (Kasting 1988) and the hydration of crustal materials. Oxidation by $H_2O$ contributed to the drying, though the produced $H_2$ (e.g., reaction 16.90) must escape to affect the redox state of the whole system. Some $H_2O$ might have been removed by the subduction of hydrated slabs into the mantle, thereby preserving hydrogen in high-pressure phases (Albarède 2009). These processes desiccated increasingly warmer environments, facilitating evaporation, brine formation, and precipitation. Alteration of rocks by sub-critical oceanic water strongly affected the compositions of water and rock. The precipitation of low-solubility anhydrite further depleted oceanic sulfates, and dissolved silica became more abundant as water temperature increased. The evaporation of the remaining $H_2O$ placed the planet into a runaway greenhouse state (e.g., Ingersoll 1969; Kasting 1988). The consumption of liquid $H_2O$ led to silica-rich rock and chloride evaporites forming on dry ocean floors (Zolotov and Mironenko 2009) that could be distinct from sulfate-rich facies deposited earlier at lower temperatures.

How the aqueous period ended is still being determined. In one commonly considered pathway, the surface temperature approached the critical point for $H_2O$ as solar luminosity increased. In Way and Del Genio's (2020) model, increasing solar luminosity does not significantly affect the global mean surface temperature of ocean-bearing Venus. They suggested that volcanic degassing episodes increased the greenhouse temperature and caused ocean evaporation. One could propose that a temperature spike caused by an impact might trigger a runaway greenhouse. Extreme runaway greenhouse heating (e.g., Kasting 1988) favored decarbonization (e.g., Höning et al. 2021) and dehydration of crystal materials that contributed to warming.

Silicate rocks could have melted at the surface due to the extreme greenhouse effect of an $H_2O$-rich atmosphere (Kasting 1988) and in the upper mantle, triggering volcanism. Partial or complete melting of chlorides, silica phases, sulfides, silicates, sulfates, and Fe oxides (e.g., from BIF-like rocks) favored the assimilation of compounds formed during the aqueous period. Formed silicate melts could exhibit higher Fe(III)/Fe(II) and S(VI)/S(−II) ratios, along with greater sulfur, Cl, and $SiO_2$ contents compared to the original basalts. The Fe(II) in magma could be oxidized by oxygen formed via hydrogen escaping from the atmospheric $H_2O$ (Warren and Kite 2023). Additional magmatic sulfates could have formed if near-surface magma was oxidized beyond ∼ QFM + 1. If these processes occurred, the solidification of the second magma ocean could have led to compositionally unique igneous rocks. If formed, sulfide- and chloride-rich melts can be separated from silicate magmas and crystallized independently. During lesser runaway greenhouse heating in the post-aqueous period (e.g., Kasting 1988), only some crustal materials could have been metamorphosed via dehydration and decarbonatization. Some materials (e.g., chloride evaporites, felsic rocks) could have been melted, mobilized, and redeposited. Sulfates and sedimentary sulfate-rich rocks could survive melting and assimilation. Despite global volcanic and tectonic resurfacing (Sects. 16.2.2 and 16.4.4.1), some morphological, mineralogical, chemical, and isotopic indicators of post-aqueous metamorphic and igneous processes may still be preserved on Venus' surface. Ejecta from several large impact craters (e.g., Mead, Stanton, Boleyn), characterized by a high dielectric constant (Pettengill et al. 1992), could provide insights into the composition of ancient rocks at depths of several kilometers. The composition of layered rock formations in the tessera plateaus, canali, and outflow channels (Sects. 16.2.2) may provide insights into past environments relevant to sulfates, sulfides, and other phases (chlorides, ferric oxides, silica minerals) and rock formations such as BIF's analogs formed in the presence of liquid $H_2O$. Sulfates, phosphates, silica phases, and ferric oxides in sediments (or as secondary phases) supposedly deposited from aqueous solutions are stable on the current surface. Common chlorides and carbonates (except $MgCO_3$) undergo sulfatization by $SO_2$ (Sect. 16.3.2) and may not be remotely detected. Oxygen isotopic composition in sedimentary sulfates, especially when contrasted against silicates and $CO_2$, can constrain when hydrogen escape occurred and the extent and nature of crustal weathering and/or mantle convection at the time of this escape. The oxygen isotope signature is likely to be significant if escape occurred before 3.5 Ga and if the mantle had become isolated by a stagnant lid from the atmosphere, hydrosphere (if any), and crust (Zahnle and Kasting 2023, K. Zahnle, private communication).

### *16.4.3* Volcanic Degassing, Space Sources, and Mass Balances

Roughly comparable masses of carbon and nitrogen in the outer shells of Earth and Venus (Table 16.12) suggest similar degassing pathways. However, higher relative abundances of $^{20}Ne$, $^{36}Ar$, and $^{84}Kr$, along with a lower mass of radiogenic $^{40}Ar$ in the atmosphere of Venus, imply differences in degassing between the planets (Donahue and Pollack 1983; Avice et al. 2022; Gillmann et al. 2022, for reviews). In particular, the lower $^{40}Ar$ content (25–30% of Earth's) suggests reduced and/or earlier degassing of carbon and nitrogen. The early degassing aligns with the atmospheric abundances of carbon, nitrogen, and $^{40}Ar$ that have accumulated through the decay of $^{40}K$ by the time of degassing. A limited late degassing of these elements could reflect the establishment of the present stagnant lithosphere early in history (O'Rourke and Korenaga 2015; Rolf et al. 2022).

The sulfur inventory within Earth's atmosphere-hydrosphere-crust system suggests that volcanic degassing is the primary source of sulfur, rather than sulfide weathering, and sulfur is considered an 'excess volatile' in this system (Rubey 1951) alongside nitrogen, carbon, Cl, and $H_2O$. Rubey (1951) noted the similarity between the relative bulk compositions of 'excess' and volcanic volatiles. Although terrestrial volcanic gas sulfur is primarily sequestered in seawater and sulfate minerals, biological sulfate reduction and the subduction of oceanic lithospheric plates (Bekaert et al. 2021) restrain the evaluation of the mass of degassed sulfur throughout Earth's history. Nevertheless, subducted slab-delivered fluids are depleted in $^{34}S$ ($\delta^{34}S = -2.5 \pm 3‰$), suggesting a negligible amount of subducted sulfates (Li et al. 2020) and the current abundance of non-magmatic sulfur (mainly in sedimentary and oceanic sulfates) may roughly constrain the amount of degassed sulfur.

The comparable abundances of nitrogen and carbon in Earth's atmosphere, ocean, and crust and Venus' atmosphere may suggest the existence of an 'excess' mass of sulfur in Venus' interior delivered from volcanic degassing. Volcanic activity is the widely accepted source of sulfur in the current atmosphere (Fegley 2014), although a portion of volcanic sulfur is reportedly supplied through the oxidation of lava's sulfides by atmospheric $CO_2$ and $H_2O$ (reactions 16.77 and 16.78; von Zahn et al. 1983; Prinn 1985; Prinn and Fegley 1987; Fegley and Treiman 1992a; Fegley et al. 1995). However, the delivery of sulfur from sulfides in permeable surface rocks is limited by the sulfur abundance in fresh basalt (~0.1 wt%), and alteration of a 40-meter-thick global layer of basalt is required to account for 150 ppmv $SO_2$ in the current atmosphere. A thicker layer is necessary to match the sulfur mass in the atmosphere and the abundant secondary sulfates (Table 16.8) in surface materials. In Rubey's (1951) terminology, sulfide-derived sulfur does not contribute to the 'excess' sulfur. Therefore, true volcanic gases ($SO_2$, OCS, $H_2S$, and $S_2$) likely represent a valuable net source of sulfur inventory in the current atmosphere and chemically altered surface materials. In contrast to $CO_2$ and $N_2$, which remained in Venus' atmosphere, a large majority of degassed sulfur was sequestered in crustal materials, as indicated by the low sulfur/carbon ratio in the atmosphere. As an illustration, Venus' atmosphere growth model by Weller and Kieffer (2025) via several tectonically induced volcanic degassing events resulted in much higher sulfur/carbon ratios than those in the current atmosphere.

Table 16.12 depicts estimates of the amount of 'excess' sulfur on Venus derived from the masses of carbon, nitrogen, and sulfur in terrestrial materials and Venus' atmospheric abundances. These assessments indicate that the possible molar amount of outgassed sulfur on Venus is an order of magnitude less than nitrogen and two orders of magnitude less than carbon. However, Venus' atmosphere has much higher nitrogen/sulfur and carbon/sulfur ratios than assessed. Compared to the estimated amount of degassed sulfur, the negligible mass of Venus' atmospheric sulfur implies a substantial reservoir of sequestered sulfur within the crust and/or in subducted materials. The prevalence of igneous rocks on the surface of Venus (Sect. 16.2.2) implies a significant capture of degassed sulfur before the last global volcanic resurfacing event. Trapped sulfur could be stored in igneous rocks containing assimilated crustal materials, sedimentary rocks formed via aqueous (Sect. 16.4.2) or anhydrous (e.g., aeolian) deposition, and metamorphic (Semprich et al. 2025; Chap. 10: Harlov et al. 2026) formations. Investigating the abundance, mineralogy, and isotopic composition of sulfur in these rocks could provide insights into past atmospheric conditions (such as via sulfur-MIF) and into the environmental context (aqueous vs. anhydrous, aqueous oxidizing vs. aqueous reducing) during the sequestration of atmospheric sulfur of volcanic origin, magmatic assimilation of crustal materials, and the transport of sulfur-bearing materials to the mantle via subduction and/or other pathways (Rolf et al. 2022; Ghail et al. 2024; Semprich et al. 2025).

The amount of sulfur-bearing species in volcanic gas depends on the solubility and initial abundance of sulfur in magma. Temperature and pressure enhance solubility, while the magma composition has complex effects, primarily reflecting the redox state and abundance of Fe(II) (Wallace and Campbell 1992; Carroll and Webster 1994; Oppenheimer et al. 2011; Boulliung and Wood 2022, 2023; Chap. 11: Casas et al. 2026; Chap. 12: Simon and Wilke 2026). The mafic composition of Venus' volcanic rocks and the oxidation state at the landing sites (Sect. 16.2.2.1) suggest sulfur solubility comparable to tholeiitic and alkaline basalt melts. Pressure is the main factor affecting the exsolution of sulfur-bearing gases and determining the amount of degassed sulfur. Although Venus' atmospheric pressure (~47 to 110 bars at different elevations, Seiff et al. 1985) suppresses the degassing of silicate magmas (Head and Wilson 1986), low-solubility Venus' species ($CO_2$, CO, $H_2$, $N_2$) degas preferentially compared to high-solubility $H_2O$ (Garvin et al. 1982; Holloway 1992). In contrast to other volatiles, assessments of sulfur degassing yield controversial results due to the factors affecting gas-melt partitioning and the experimental challenges involved. Part of this controversy arises from the decreasing solubility of sulfides (decreasing SCSS) as pressure drops during magma ascent, which reduces the sulfur content in the melt available for degassing. Conversely, sulfur degassing may lead to melts being sulfur-undersaturated regarding sulfides, providing sulfur to both the melt and gas phase.

Gaillard and Scaillet (2014) modeled the gas-melt separation of supposed Earth-like ($CO_2$, $H_2O$, sulfur) volatiles in basaltic magma as a function of vent pressure. They inferred very low $(SO_2 + H_2S)/CO_2$ and $H_2O/CO_2$ gas ratios at $10^2$ bars that could characterize volcanic vents currently on Venus. The S/C atomic ratio of $\sim 3 \times 10^{-3}$ was assessed for basalt-sourced gases at the QFM $fO_2$ buffer and ~100 bars, while the initial magmatic ratio was ~5. However, newer experiments on $SO_2$ solubility and models by Boulliung and Wood (2023) for basaltic melts suggest that at least 80% sulfur degasses before the ascending magma reaches a pressure of $10^3$ bars. Comparable amounts of $CO_2$ and $SO_2$ in basaltic gases at ~100 bars are also inferred using the Sulfur_X model of Ding et al. (2023) for degassing in $CO_2$−$H_2O$-sulfur-bearing volcanic systems. Models of Boulliung and Wood (2023) and Ding et al. (2023) imply thorough degassing of $SO_2$ and $CO_2$ from Venus' basalts. The lack of substantial suppression of $SO_2$ degassing at ~100 bars is supported by the often sulfur-depleted glasses in submarine basalts (Peterson et al. 2017; Wallace and Edmonds 2011; Moore and Lewis 1976) and the interpretation of Kilauea 2018 eruption samples (Lerner et al. 2021). Note that no significant sulfur degassing is inferred from oceanic plateau basalt at ocean floor pressures of 200 to 300 bars, likely reflecting sulfide formation (Reekie et al. 2019). Regardless of the results of degassing models, abundant atmospheric $SO_2$, surface sulfates, along with trace sulfides in erupted basalts, are inconsistent with a significant suppression of sulfur degassing under Venus' surface conditions. This indicates similar $CO_2$ and $SO_2$ levels (within an order of magnitude) in volcanic gases. Furthermore, if Venus' mafic magma is depleted in carbon due to early profound $CO_2$ degassing (indicated by the low $^{40}Ar/CO_2$ ratio in the atmosphere) and/or limited crustal recycling to the mantle, magmatic gases could have a higher S/C ratio than their terrestrial counterparts.

Zolotov and Matsui (2002) calculated gas-phase thermochemical equilibria to model the speciation of potential Venus' volcanic gases as a function of the carbon/hydrogen/sulfur/Cl atomic ratios and $fO_2$. The nominal bulk gas composition was based on the Kilauea 1918–1919 lava lake gas composition (Gerlach 1980), incorporating a comparable amount of sulfur and carbon (sulfur/carbon ratio of ~0.5). Their models, which exhibit low hydrogen/carbon ratios, can refer to the initially $H_2O$-poor planet (Prinn and Fegley 1987), echo the current atmospheric composition ($xH_2O/xCO_2 = 3 \times 10^{-5}/0.965$), and reflect the greater solubility of $H_2O$ compared to $CO_2$ at $10^2$ bars (Fig. 16.27). At the QFM $fO_2$ buffer, modeled hydrogen-poor gases are rich in $CO_2$, while sulfur-bearing gases include $SO_2$, $S_2$, $S_2O$, OCS, $S_2Cl$, $SSF_2$, $S_3$, SO, $SCl_2$, and $S_2Cl_2$. At the IW buffer, hydrogen-deficient gases include CO, OCS, $S_2$, $CS_2$, and $S_3$, and various sulfur-halogen species. Sulfur-bearing species in volcanic gases, where the hydrogen/carbon ratio exceeds $10^{-2}$, are primarily denoted by $SO_2$, $S_2$, $H_2S$, $S_3$, and SO at the QFM buffer, and by OCS, $S_2$, $H_2S$, and $S_3$ at the IW buffer. Hydrogen-poor volcanic gases, containing similar amounts of sulfur and carbon, primarily consist of $CO_2$, $SO_2$, CO, OCS, and $S_2O$ (Fig. 16.28). Gases with a low sulfur/carbon ratio, as proposed by Gaillard and Scaillet (2014), consist of $CO_2$, CO, and $SO_2$, with $SO_2$ strongly dominating other sulfur-bearing gases.

The bulk and chemical composition of the near-surface atmosphere (Tables 16.1 and 16.2) differ from all modeled volcanic gases. Assuming net volcanic sources of atmospheric carbon, sulfur, Cl, and F, higher atmospheric carbon/(sulfur, Cl, F) ratios compared to those in volcanic gases suggest significant sequestration of degassed sulfur, Cl, and F in minerals (Sect. 16.3.2), along with an absence of exogenic carbonates in the crust (c.f., Zolotov 2018). Most sulfur-bearing volcanic gases modeled for Venus (Fig. 16.27) are more abundant than the gases in the near-surface atmosphere (Fig. 16.29). Exceptions include $SO_2$ and $SO_3$ in reduced (IW $fO_2$ buffer) and hydrogen-rich volcanic gases. The notable compositional difference between volcanic and near-surface gases would allow for distinguishing volcanic plumes using remote or in situ methods to constrain current volcanic activity (Wilson et al. 2024; Filiberto et al. 2025) and the physical-chemical conditions in magma and volcanic vents. This difference also suggests post-eruption gas-phase thermochemical reactions that diminish the abundances of sulfur-bearing volcanic gases (except $SO_2$ in reduced plumes) through oxidation to $SO_2$ by atmospheric $CO_2$ and $H_2O$.

The vast volume of mafic lava assessed from the mapping of Magellan radar images (140–250 Mkm$^3$, Ivanov and Head (2013), $(4.1 - 7.3) \times 10^{20}$ kg at 2900 kg m$^{-3}$ density) implies the release of sulfur-bearing gases during a geologically short period of widespread volcanic activity. Head et al. (2021) estimated that ~1/3 of current atmospheric sulfur could have been degassed from corresponding melts. Their estimations were based on the model of Gaillard and Scaillet (2014), in which 16 ppmw sulfur degassed from magma with an initial 1000 ppmw sulfur content. More sulfur could be degassed using the model of Boulliung and Wood (2023) mentioned above. If 10–100% sulfur is degassed from an initial magma containing 800 to 1700 ppmw sulfur, $3 \times 10^{16}$ to $10^{18}$ kg sulfur could have been released from 140 to 250 Mkm$^3$ of basaltic melt. This amount is comparable to the atmosphere's sulfur mass and the potential mass of 'excess' sulfur in the combined atmosphere and crust (Table 16.12), supposedly trapped in minerals after volcanic degassing. In contrast to sulfur, similar abundances of sulfur and carbon in volcanic gases do not suggest a significant contribution from the most recent global volcanic activity to the atmospheric $CO_2$ inventory, aligning with the estimations of Head et al. (2021) and López et al. (1998). It is plausible that current atmospheric sulfur-bearing species are products of global volcanic degassing, which were partially sequestered in sulfur-bearing minerals in the surface materials (Sect. 16.3.2.2, Table 16.8). Even if volcanic activity occurs in the current epoch (Herrick et al. 2023; Filiberto et al. 2025), these estimates suggest that such activity may not significantly impact the abundance of sulfur in the atmosphere. Considering the annual flux of volcanic $SO_2$ on Earth ($\sim 2 \times 10^{10}$ kg, Carn et al. 2017; Schmidt and Carn 2022), the $SO_2$ mass in Venus' atmosphere ($10^{17}$ kg) may not be affected by even major volcanic events for years (see also Sect. 16.3.1.2). The supply of reduced sulfur-bearing gases through the oxidation of exposed sulfides (e.g., reactions 16.77 and 16.78) may be more critical than current volcanic degassing. Complete oxidation of basalt's pyrrhotite to magnetite in a global 1 m thick surface basalt layer with 0.1 wt% sulfur yields approximately $1.3 \times 10^{15}$ kg sulfur (in OCS and $H_2S$) or about 2.5% of atmospheric abundance. Note that oxidation of pyrrhotite to pyrite (e.g., reaction 16.82) does not impact atmospheric sulfur, and the pyrrhotite-to-pyrite conversion (reaction 16.83) consumes atmospheric sulfur.

Fegley and Treiman (1991) estimated that the current atmospheric $SO_2$ mass could deplete to $f SO_2$ at the $CaSO_4 - CaCO_3$ equilibrium 16.72 in ~1.9 Ma, if $SO_2$ is consumed in reactions with calcite at the rate experimentally inferred by Fegley and Prinn (1989) (Sect. 16.3.2.2). To match this sequestration rate and maintain the $x SO_2$ level, Fegley and Prinn (1989) estimated the current volcanism rate of 0.4 to 11 km$^3$ a$^{-1}$, with a nominal value of 1 km$^3$ a$^{-1}$. Climate evolution modeling by Bullock and Grinspoon (2001), which included $SO_2$ consumption through reactions with calcite (Sect. 16.4.4.2), suggests that the volcanic supply of $SO_2$ over the past 20 to 50 million years has been essential to maintaining the current $H_2SO_4 - H_2O$ clouds. However, using calcite-based models to constrain the degassing rate is questionable because unstable and reactive calcite is unlikely to form under surface conditions (Zolotov 2018, for review), along with the prevalence of widespread basaltic materials (Sect. 16.2.2.1) instead of calcite-bearing sedimentary or metamorphic rocks.

Fegley and Treiman (1991) noted that estimates based on a slow sulfatization of diopside imply significantly lower rates of volcanic $SO_2$ degassing. Further use of atmospheric $SO_2$ abundance to constrain the volcanic degassing requires data on the kinetics of advanced stages of gas-solid reactions in basalt, particularly Ca-bearing pyroxenes and glasses (Sect. 16.3.2.2).

#### Space Sources

Space sources to Venus, primarily in cometary dust (Carrillo-Sánchez et al. 2020), provide a flux of $(1.1 - 4.2) \times 10^3$ kg of sulfur per day (Table 16.12). This estimate aligns with an assessment based on data regarding sulfur-bearing species in the Earth's upper atmosphere (~$10^3$ kg of sulfur per day, Gómez Martín et al. 2017). This cosmic flux on Venus accumulated over the last 0.5 Ga (Table 16.12) may represent only a tiny fraction of the atmospheric sulfur, suggesting predominantly geological sources of sulfur in the atmosphere and surface materials. The residence time of space-delivered sulfur, derived by dividing atmospheric sulfur mass by today's flux, is estimated to be $70 \pm 41$ Ga. In contrast, the residence time of all cosmic materials in the clouds is $3 \pm 2$ Ma (Zolotov et al. 2023). These residence times imply that sulfur-bearing and other space-delivered species had ample time to undergo alteration in the clouds and throughout the atmosphere. This suggests that the clouds are saturated with chemically altered space-sourced compounds and that these compounds have reached the surface during the current geological epoch, contributing to the composition (e.g., hematite) of surface materials (Mogul et al. 2025; Zolotov 2021; Zolotov et al. 2023).

## 16.4.4 Atmosphere-Surface Interactions During and After the Global Volcanic Resurfacing

### 16.4.4.1 The Global Volcanic Resurfacing

The Magellan radar data (Figs. 16.8, 16.9, 16.10, 16.11, 16.12 and 16.13) suggest that any evidence of early geological history before 0.3–0.9 Ga (McKinnon et al. 1997; Korycansky and Zahnle 2005; Herrick et al. 2023) has been obliterated by global tectonic and volcanic events over a relatively short geological timescale. These processes led to the formation of tessera plateaus, volcanic plains, and volcanic-tectonic structures, such as ridges, volcanic centers, and both small and large volcanoes, as well as coronas (Sect. 16.2.2, Strom et al. 1994; Basilevsky et al. 1997; Ivanov and Head 2011, 2013, 2025; Ghail et al. 2024). While the possibility of current volcanic activity is not ruled out (see Herrick et al. 2023 and Filiberto et al. 2025 for reviews), most volcanic rocks observed were emplaced during global resurfacing, and surface modification has been minimal since then (Carter et al. 2023).

Greenhouse heating is expected from volcanic degassing of $H_2O$ in the models of Phillips et al. (2001) and Gillmann and Tackley (2014). However, incorporating $SO_2$ and its effects on cloud properties into thermal balance models (Solomon et al. 1999; Bullock and Grinspoon 1996, 2001) reveals a more complex picture. Bullock and Grinspoon (2001) modeled that having $xH_2O$ at 100 times the present atmosphere, given the current $xSO_2$ level, prevented the formation of high-albedo low-altitude $H_2SO_4-H_2O$ clouds, resulting in greenhouse heating of the surface to ~900 K. Hashimoto and Abe (2000) demonstrated that increasing $xSO_2$ in the lower atmosphere at low $xH_2O$ results in cooling due to the increasing albedo of growing $H_2SO_4-H_2O$ clouds. The models from Bullock and Grinspoon (2001) for degassing of both $H_2O$ and $SO_2$ during the formation of volcanic plains suggested the formation of massive high-albedo $H_2SO_4-H_2O$ clouds, causing surface cooling down to ~650 K, followed by gradual warming as $SO_2$ was sequestered to minerals faster than hydrogen escaped (see Sect. 16.4.4.2).

The models propose $SO_2$ as the primary sulfur-bearing volcanic gas on Venus, at least during the last global volcanic activity (Sect. 16.4.3) at $fO_2$ between the QFM and IW buffers (Sect. 16.2.2.1). The degassing of sulfur-bearing gases during that brief period significantly contributed to the current atmospheric sulfur (Sect. 16.4.3). During the period of global volcanic resurfacing, the increase in atmospheric $fSO_2$ facilitated reactions between $SO_2$ and solids (Sect. 16.3.2.2). Although $SO_2$-solid reactions could have formed an array of minerals, the prompt formation of Ca and Na sulfates was unavoidable at elevated $fSO_2$. Following the event, a considerable amount (potentially 50 ± 40%) of the degassed sulfur was sequestrated in surface materials. This trapping is substantiated by the abundant sulfur found in surface samples (Table 16.3), the favorable reactions of $SO_2$ with basaltic glasses and Ca-rich pyroxenes (Sect. 16.3.2.2), and the equal masses of sulfur in the atmosphere and within a global 2 to 6 m thick layer of partially altered surface materials containing 1 to 3 wt% sulfur (c.f., Lewis and Kreimendahl 1980; Zolotov and Volkov 1992). These estimates indicate a moderate reduction in the mass of atmospheric sulfur, the mixing ratios of sulfur-bearing species, and changes in cloud properties since the global volcanic resurfacing.

The varying degrees of sulfur sequestration observed in surface materials (Tables 16.3 and 16.8) suggest efficient $SO_2$-solid reactions during global volcanic activity. Volcanic materials placed earlier, possibly exemplified by Vega 2 rocks, may have trapped more sulfur than those at the Venera 13 site (Fig. 16.1) and particularly at the Venera 14 site, which exhibits fewer physically altered rocks (Zolotov and Volkov 1992). Morphological analysis of Magellan images and geological mapping (Abdrakhimov 2005; Basilevsky et al. 2007; Ivanov and Head 2011; Weitz and Basilevsky 1993) has shown that the landing regions of Venera 13, 14, and Vega 2 are characterized by a widespread type of volcanic plains (e.g., plains with wrinkle ridges) formed during global resurfacing. The position of these plains in the middle of the stratigraphic sequence (Ivanov and Head 2011) implies a higher degree of sulfur sequestration in units placed before them. This suggests that these preceding units could be more altered than the sulfur-rich Vega 2 material. The varying sulfur content in three landing sites, which are believed to represent lava formations from a geologically brief global volcanic resurfacing event, indicates a significant capture of atmospheric sulfur during this resurfacing rather than in later periods.

In addition to the interaction of $SO_2$ with surface materials, such reactions should have affected ash and mineral dust particles in the lower atmosphere, which originate from volcanic, aeolian, and impact processes. The delivery of volcanic ash and mineral dust via volcanic-induced plumes or other means to sulfuric acid clouds resulted in the formation of metal sulfates (Mg, Fe, Ca, Al, etc.) and their saturation in the sulfuric acid solution. The gravitational settling of sulfate-rich particles from the lower atmosphere and clouds contributed to the preferential sulfatization of early-formed surfaces.

Gilmore et al. (2015) proposed that tessera plateaus, which formed shortly before a notable phase of plain-forming volcanism (Ivanov and Head 2011; Kreslavsky et al. 2015), may have undergone considerable chemical weathering in atmospheric environments due to widespread volcanic activity. They suggested that the alteration products formed contribute to the low near-IR emissivity of the tesserae surfaces (Sect. 16.2.2.4). In addition to elevated $xSO_2$ levels, lower temperatures in the highlands increased the thermodynamic affinity for sulfatization, similar to the current epoch (Sect. 16.3.2.2). Sulfatization of the highlands during the formation of the volcanic plains and structures aligns with an elevated near-IR emissivity of western Alpha Regio (Gilmore et al. 2015). This region may have developed after the surrounding volcanic plain (Gilmore and Head 2000) and been exposed to a compositionally and thermally distinct atmosphere following the end of significant volcanic activity and a partial consumption of volcanic gases (e.g., $SO_2$ trapping via gas-solid reactions, Bullock and Grinspoon 2001). Consequently, this region experienced less chemical weathering than the older areas of the Alpha tessera. Likewise, the surfaces of lava on large volcanoes formed after the emplacement of the volcanic plains (Ivanov and Head 2025) could be less altered than those on tesserae and plains.

In addition to the greater sulfatization of minerals and glasses on mountains formed before volcanic plains and large volcanoes, the condensation of volcanically degassed and/or sublimated chlorides, followed by their partial or complete sulfatization (Zolotov 2025), could have contributed to the formation of 'white mountains'. The presence of salts (sulfates, alkali chlorides) with high albedo and low thermal emissivity in the highlands that formed before emplacement of volcanic plains is the hypothesis that future missions can test (Sect. 16.5.2). The degree of sulfatization of geological formations (tesserae, stratigraphically distinguished volcanic plains, and younger large volcanoes), estimated with upcoming remote and in situ data, would help constrain the relative timing of the placement of the surface materials and sulfur-trapping reactions.

#### 16.4.4.2 Atmospheric Evolution and Gas-Solid Interactions

Bullock and Grinspoon (2001) developed climate evolution models spanning a period of 0.7 Ga, following global volcanic degassing that supposedly occurred within 100 Ma. They modeled changes in atmospheric $xSO_2$ and $xH_2O$ due to volcanic degassing and hydrogen escape, along with the sequestration of $SO_2$ through reactions with calcite, considering a rate-limited $SO_2$ diffusion in permeable crustal materials. The modeled cases included only hydrogen escape and the combined effects of hydrogen escape and $SO_2$ sequestration. In all instances, the cessation of volcanic degassing was followed by initial uneven cooling and the desiccation of the atmosphere, which led to the dissipation of $H_2SO_4-H_2O$ clouds and reduced planetary albedo. In the scenario where $xSO_2$ was set to the current value, surface temperatures between 710 and 750 K were established relatively early because the reducing planetary albedo partially offsets the greenhouse effect caused by decreasing atmospheric $xH_2O$. With both hydrogen escape and $SO_2$ trapping, the removal of $SO_2$ occurred much faster than that of hydrogen. Consequently, dense $H_2SO_4-H_2O$ clouds vanished within 100 to 600 million years, depending on the volume of degassed lava. Lower-opacity, $H_2O$-dominated clouds formed at higher altitudes. The reduction in cloud albedo at remaining elevated $xH_2O$ led to increased surface temperatures up to 870 K, although moderate cooling occurred once $xSO_2$ dropped below the $10^{-5}$ level. The modeling indicated that maintaining $H_2SO_4-H_2O$ clouds over the modeled period requires low-level volcanic degassing of $H_2O$ and $SO_2$. However, differing conclusions might arise if calcite is absent from basalt-dominated surface materials. Much slower $SO_2$ sequestration in reactions with silicates (Sect. 16.3.2.2) could have prolonged the lifespan of clouds without volcanic degassing, as Fegley and Treiman (1991) noted. Thus, volcanic degassing of $SO_2$ after global volcanic resurfacing may not be necessary to sustain the thermal structure of the atmosphere with $H_2SO_4-H_2O$ clouds.

Despite the work of Bullock and Grinspoon (2001), the fates of Fe sulfides and oxides in surface materials require attention. Pyrite, magnetite, and hematite are believed to be present in altered surface material at present, possibly buffering the composition of trace chemically active gases in the near-surface atmosphere (Sect. 16.3.2.3). Current conditions of the modal radius are comparable to those determined by the Mag-Hem, Mag-Py, or Mag-Py-Hem mineral buffers (Sect. 16.3.2.3, Figs. 16.22 and 16.25, Table 16.11). However, the ambiguity of the atmospheric composition, a substantial error bar (~0.8 log $fO_2$ units) at the Mag-Hem equilibrium 16.66 (Fegley et al. 1997b), and the presence of impure minerals in the buffering assemblages (e.g., Ti-bearing specimens, Zolotov 1994, 1995a, b) complicate distinguishing potential buffering assemblages at present. At a specific temperature, the Mag-Py-Hem equilibrium determines $fO_2$, $fS_2$, and the fugacities of sulfur-oxygen-carbon gases, except $CO_2$. The Mag-Py equilibrium (e.g., reactions 16.67 and 16.79) sets the $fO_2/fS_2$, $SO_2/OCS$, $SO_2/S_2$ fugacity ratios within a range between the Py-Pyh and Mag-Hem buffers, while the Mag-Hem equilibrium assemblage regulates $CO/CO_2$ fugacity ratio and $fO_2$ (equilibria 16.65 and 16.66; Fig. 16.25, Table 16.11). Swift gas interactions with Mag-Py-Hem mixtures and individual phases, as suggested by Venus-focused experiments (Reid et al. 2024; Santos et al. 2024; Gilmore et al. 2025; Sect. 16.3.2.3), ensure gas-solid interactions and equilibration in the oxygen-sulfur-Fe system over geological timescales. Despite some aeolian activity (Greeley et al. 1997; Kreslavsky and Bondarenko 2017; Carter et al. 2023), the surface modification rate since the global resurfacing ($\sim 2 \times 10^{-3}$ µm per year, Arvidson et al. 1992) has been orders of magnitude lower than that on Earth. This has allowed atmospheric gases to interact with the same permeable materials over several hundred million years, promoting gas-solid type equilibration. The larger masses of Fe- and sulfur-bearing minerals in permeable materials (at least ten meters) compared to those of trace gases ($SO_2$, OCS, CO, $S_2$, etc.) support the feasibility of buffering initially suggested by R. Mueller (1963, 1964, 1965). The decent match of atmospheric conditions with those controlled by major envisioned products from chemical weathering (pyrite, magnetite, hematite, anhydrite, Sect. 16.3.2.3) is unlikely to be accidental. It may result from the physical-chemical evolution of the coupled atmosphere-surface layer system following global resurfacing (Zolotov 1992a, b, 1995a, b, 2015). Below is the probable pathway of the atmosphere-surface system, following global resurfacing, that addresses the changing mineralogy of the oxygen-sulfur-Fe system.

A fraction of ferric oxides formed via high-temperature oxidation of volcanic materials during global volcanism (Warren and Kite 2023) has remained in the surface material. The atmosphere was cooled through partial consumption of degassed $H_2O$ and $SO_2$ via oxidation and sulfatization of basalts (reactions 16.62, 16.63, 16.71, 16.74, 16.75, and 16.90). Hydrogen escape (e.g., Chaffin et al. 2024; Gillmann et al. 2022), following photodissociation and thermochemical reactions (e.g., reaction 16.90) of $H_2O$ vapor, provided net oxygen for further Fe(II) oxidation in exposed silicates, glasses, and magmatic sulfides. Once hematite formed in surface materials, the Mag-Hem oxygen fugacity was maintained in the near-surface gas phase and pore spaces. Given that $O_2$ was virtually absent in the near-surface atmosphere, the Mag-Hem assemblage established the $CO_2/CO$ ratio (equilibrium 16.65). Also, it affected the mixing ratios of $SO_2$ to reduced sulfur gases (COS, $H_2S$, and $S_n$). Like the laboratory buffer, $fO_2$ remained set by the Mag-Hem buffer but varied with surface temperature. The gas phase cannot become more oxidized than the magnetite-hematite equilibrium until after all Fe(II) in permeable crustal materials has been oxidized. The latter is unlikely due to the abundance of Fe(II) in widespread and thick basaltic formations and kinetic inhibition of the magnetite-to-hematite transition at near-equilibrium conditions. The rate of hydrogen escape limited the supply of net oxygen (in $CO_2$, $H_2O$, and $SO_2$) that was mainly consumed in the formation of magnetite rather than hematite. However, the continuous accumulation of hematite from cosmic sources (Mogul et al. 2025; Zolotov 2021; Zolotov et al. 2023) may have contributed to the buffering assemblage(s) and aligns with the red spectral slope observed at Venera landing sites (Pieties et al. 1986; Shkuratov et al. 1987).

#### Fate of Fe Sulfides and Formation of Pyrite

Compared to sulfates (Sect. 16.4.4.1), the alteration and formation of sulfides through gas-solid reactions are less apparent. This ambiguity partly reflects the unclear trajectory of surface temperature following global degassing (Bullock and Grinspoon 2001). The elevated temperature facilitated the oxidation of ferrous and sulfide phases by $CO_2$. In addition to contributing to greenhouse heating, degassed $H_2O$ vapor oxidized Fe(II) in silicates and glasses (reaction 16.90; Warren and Kite 2023) as well as sulfides in volcanic products. During the period of volcanic degassing, pyrrhotite in exposed basalt was likely oxidized to magnetite by $CO_2$ and $H_2O$ (reactions 16.77 and 16.78). This oxidation may have occurred along the Mag-Pyh phase boundary (Figs. 16.22 and 16.25), potentially influencing the abundance of sulfur-bearing gases in the lower atmosphere.

Pyrite is more stable in the cooler highlands, while ferric oxides are more stable in the hot lowlands (Sect. 16.3.2.3, Klose et al. 1992; Zolotov 2018, 2019). At a constant temperature, a higher $f\,SO_2$ favors the stability of pyrite, magnetite, or hematite, depending on the fugacities of other gases ($S_2$, CO, $O_2$) (Fig. 16.22, Zolotov 2018, 2019). Although the temperature and fugacities remain only roughly constrained during the volcanic resurfacing period (Solomon et al. 1999; Bullock and Grinspoon 2001; Phillips et al. 2001), a severe greenhouse heating of $H_2O$-enriched atmosphere favored the decomposition of any pre-existing pyrite into $S_2$ and pyrrhotite (reaction 16.76), which could be oxidized to Fe(III) oxides (Sect. 16.3.2.3). The formation of ferric oxides was more plausible than that of pyrite, particularly at significantly elevated temperatures that allowed crustal structures to relax (Solomon et al. 1999). Secondary pyrite may have formed in low-temperature highlands exposed to an $SO_2$-rich atmosphere that cooled as $xH_2O$ decreased due to $H_2O$(g)-solid reactions and, to a lesser extent, hydrogen escape. In addition to the interaction of OCS and $H_2S$ with ferrous compounds (reactions 16.80 and 16.81), pyrite might have formed through the oxidation of pyrrhotite by $CO_2$ (reaction 16.82) and $H_2O$, as well as the sulfurization of pyrrhotite (reaction 16.83). Pyrite formed in this manner could contribute to the low microwave emissivity observed in the highlands (Sect. 16.2.2.5). As proposed by Hashimoto and Abe (2000), the presence of pyrite may have helped control the surface temperature through $SO_2$-albedo feedback. Increasing greenhouse warming led to higher $xSO_2$ over pyrite-bearing assemblages (e.g., Py-Mag), which increased cloud reflectance and stabilized the temperature.

Given the current conditions of $fSO_2$ and temperature, along with their proximity to the pyrite stability field (Sect. 16.2.2.3), it is conceivable that pyrite may have formed on the volcanic plains at elevated $fSO_2$ levels, particularly as the temperature significantly dropped toward current values (~730–745 K) or below them. The possible presence of high-dielectric-constant material in the shallow subsurface (Antony et al. 2022) on these plains may suggest the formation of secondary pyrite on lava during or shortly after global volcanic activity. The low dielectric constant in the upper decimeters of the plains (Pettengill et al. 1992) suggests fine-grained sedimentary deposits atop altered lava containing pyrite. This layered and porous material, observed at the landing sites of Venera 13 (Fig. 16.1) and Venera 14, could represent lithified fine-grained deposits (Florensky et al. 1983) resulting from impacts (Basilevsky et al. 2004), explosive eruptions (Garvin et al. 1984; Basilevsky et al. 1985), wind action (Greeley et al. 1997; Kreslavsky and Bondarenko 2017; Bondarenko and Kreslavsky 2018), and other sedimentary processes reviewed by Carter et al. (2023) (Sect. 16.2.2.1). Note that supposed high-dielectric-constant material in the shallow subsurface could be magnetite formed through Fe(II) oxidation by trapped magmatic $H_2O$ upon crystallization of basalts (Holloway 2004) and the low microwave emissivity could reflect volume scattering (Campbell et al. 1999; Tryka and Muhleman 1992) of rough subsurface materials rather than composition. The low emissivity of several large impact craters (Mead, Boleyn, etc.), volcanic edifices, and some tectonically formed features (e.g., ridges) reported by Pettengill et al. (1992) aligns with the latter explanation.

Wood (1994, 1997) suggested that pyrite could have formed through $SO_2$-solid reactions on the plains at 680 to 720 K (as opposed to the current 740 K) and in the low-temperature highlands before global volcanism and associated $SO_2$ degassing and greenhouse warming. According to Wood, weathering reactions following the global volcanic event and the partial consumption of $SO_2$ did not lead to pyrite formation on the plains. Wood's hypothesis aligns with materials that have high dielectric constants in the ejecta of large impact craters, the high microwave emissivity of volcanic plains, and the low emissivity reported in the highlands by Pettengill et al. (1992). However, the survival of pyrite is unlikely during periods of potentially severe greenhouse heating (e.g., Solomon et al. 1999; Bullock and Grinspoon 2001) associated with global volcanism. If pyrite did not survive, the high-dielectric-constant material in the deep subsurface could be represented by Fe(III) oxides formed via oxidation at gas-magma (Sect. 16.4.1), gas-solid, or aqueous (Sect. 16.4.2) settings.

Additional data (Sect. 16.5) are required to understand the potential formation of Fe sulfides and oxides before, during, and after the global volcanic resurfacing event.

Following the suggested formation of pyrite on the cooled plains, atmospheric and surface evolution may have occurred along the Mag-Py-Hem buffer (equilibria 16.85 and 16.86) (c.f., Zolotov 1992a, 1995a), which controlled the fugacities of trace sulfur-oxygen-carbon gases. Establishing conditions for the Mag-Py-Hem equilibrium, which depends solely on surface temperature, logically follows from the evolution of the atmosphere-surface system, resulting in the current environments. This evolution, which includes a later addition of pyrite to the Mag-Hem assemblage, is more suitable than developing the Mag-Py equilibrium assemblage alongside the atmospheric evolution.

Both $SO_2$ degassing and elevated greenhouse temperatures during global volcanic activity favored the sulfatization of exposed Ca- and Na-carbonates and the thermal decomposition of crustal carbonates, if they existed. If Fe sulfide-oxide equilibria began buffering trace gases after that activity, lower temperatures corresponded to lower $fO_2$ and $fSO_2$ values (Zolotov 1992a, 1995a, Fig. 16.30), which were less favorable for sulfatization. Zolotov (1995c) explored the effects of temperature, $fCO_2$, and $fSO_2$ buffered by the Mag-Py-Hem assemblage on carbonate stability. The models do not exclude stable Ca-bearing carbonates and their sulfatization if temperature and $fSO_2$ increase toward the present environments. Whether carbonates became stable remains unclear, and this inquiry requires comprehensive modeling.

#### 16.4.4.3 Future of the Atmosphere and Crust

Future Venus will likely be characterized by the physical-chemical co-evolution of the atmosphere, surface materials, and upper interior. The future of the atmosphere and permeable surface materials was modeled by Bullock and Grinspoon (2001) in the framework of a climate model covering the period after the global volcanic resurfacings at a constant solar luminosity (Sect. 16.4.4.2), briefly addressed by Gillmann et al. (2022), and is further discussed here.

It is unclear whether the decreasing interior radiogenic heating (Smrekar et al. 2022) will cause another pulse of volcanic resurfacing, degassing, and a related greenhouse response. If a global volcanic event occurs, the processes will roughly resemble those during and after the last global resurfacing (Sect. 16.4.4.2). Gas-solid type reactions will again proceed toward gas-solid equilibria. As in the present epoch (Sect. 16.3.2), Fe oxides and sulfides, along with Ca- and Na-sulfates, will participate in buffering reactions where gas concentrations depend on the surface temperature. Ultimately, the exhaustion of radionuclides will terminate the volcanic supply of rocks and gases. This cessation will further promote the establishment and maintenance of gas-solid equilibria in permeable surface materials that regulate atmospheric composition (except for $CO_2$, $N_2$, and noble gases). Regardless of any additional global volcanic events, further evolution will be influenced by increasing solar luminosity, gas-solid reactions in permeable surface materials, hydrogen escape, and planetary albedo changes resulting from clouds' altered composition and altitude. The oxidation of Fe(II) and S(-II) in exposed solids (primarily by hot $CO_2$) suggests the establishment of gas-solid equilibria involving relatively oxidized sulfur-bearing minerals (sulfates of Ca, Na, and K, and possibly pyrite).

On the one hand, the continued formation of sulfates (mainly Ca-, Na, Ca-Na-, and K-sulfates) via the sulfur disproportionation reaction 16.62 will decrease $xSO_2$ and bulk sulfur content toward the conditions (e.g., $SO_2/S_2$ ratio) corresponding to solid-gas equilibration. On the other hand, in the case of hematite-magnetite and/or hematite-magnetite-pyrite equilibration, one would expect changes in the atmospheric $SO_2/COS$, $SO_2/H_2S$, $CO_2/CO$, and $H_2O/H_2$ ratios. These changes reflect the increasing $fO_2$ (but higher $CO/CO_2$ ratio) and $fSO_2$ with temperature (Fig. 16.30). The suppressed gas-solid reaction rates under near-equilibrium conditions would not lead to thorough oxidation (including sulfatization) of exposed rocks, particularly given the limited oxygen enrichment due to hydrogen escape from the $H_2O$-poor atmosphere. As in the present epoch, $fSO_2$ controlled by equilibria with pyrite and/or sulfates would not allow the formation of carbonates.

Further heating would promote the decomposition of pyrite to pyrrhotite via reaction 16.76 and result in the complete oxidation to magnetite through the overall reaction

$$9FeS_2(\text{pyrite}) + 16CO_2(g) \rightarrow 3Fe_3O_4(\text{magnetite}) + 2SO_2(g) + 16OCS(g) \quad (16.97)$$

not permitting the buffering of atmospheric sulfur-bearing gases by the pyrite-bearing assemblages. Elevated temperatures favored pyrrhotite oxidation to magnetite (reactions 16.77 and 16.78). As temperatures rise, the net sulfur released into the atmosphere from the oxidation of sulfides may surpass the loss of $SO_2$ through sulfatization. This would enhance the bulk atmospheric sulfur abundance, increase the (OCS, $S_2$, CO)/$SO_2$ ratios in the lower atmosphere, and influence the properties of sulfuric acid clouds and the radiation balance. As in the present epoch, atmospheric $SO_2$ will not allow for the formation of carbonates. $CO_2$ and $N_2$ will remain the primary atmospheric gases. Without a volcanic supply of $H_2O$, the atmosphere will become desiccated through hydrogen escape and the oxidation of Fe(II) in the exposed rocks (reactions 16.78 and 16.90). On the one hand, the drying of the atmosphere will reduce the atmospheric greenhouse effect caused by $H_2O$ vapor. On the other hand, the loss of low-altitude $H_2SO_4-H_2O$ clouds and the corresponding decrease in planetary albedo will contribute to warming, as modeled by Bullock and Grinspoon (2001). These initial suggestions require evaluation by considering atmosphere-crust interactions, including mineral stability, element and compound mass balances, and the effects of greenhouse heating as solar luminosity increases.

In a few Ga, the Sun will become a red giant star. Before Venus immerses in the outer layer of the red giant Sun (Schröder et al. 2008), greenhouse heating will increase the geothermal gradient in the upper interior, enhancing sub-solidus mantle convection, causing convection in the crust (Solomatov and Jain 2025), melting in the upper mantle, and global volcanic degassing, which contributes to greenhouse heating. Crustal heating will result in interior and then surface melting of possible 'oceanic' chlorides, sulfides (e.g., remaining pyrrhotite, pentlandite), silicate glasses, Na-rich silicates, sulfates ($CaSO_4$, $Na_2SO_4$), and finally Mg-Fe silicates. In contrast to an early magma ocean, crustal and surface melts will have $fO_2$ between QFM and IW buffers (as in sampled Venus' basalts, Sect. 16.2.2.1). Despite the melting of sulfates, the prevalence of re-melted and newly erupted basalts suggests the dominance of S(-II) over S(VI) in the magma ocean. At corresponding $fO_2$ levels in the magma, $SO_2$ continues to prevail among sulfur-bearing magmatic gases (Sect. 16.4.3) that influence the composition of the lower atmosphere. These events and related surface melt-atmosphere interactions will alter the atmospheric composition due to a reduced magma rich in Fe(II)- and S(-II)-bearing melt complexes. The atmospheric ratios of $CO_2$/CO and $SO_2$/($H_2S$, $S_2$, COS) are expected to decrease from high values (e.g., from values controlled by the Hem-Mag equilibrium) to lower $fO_2$ values corresponding to the magma ocean and the related magmatic gases. The gas-magma equilibria at the surface of the magma ocean will control the composition of its atmosphere, where $CO_2$, CO, and $N_2$ could be the major gases (Gillmann et al. 2022). Additionally, melt-gas interactions will influence atmospheric $fS_2$ and bulk sulfur content. Determining whether these inferences are reasonable is unclear without quantitative considerations that involve thermal balances, the melting of geological materials, the solubility of volatiles in magma, magma degassing, and interactions between the atmosphere and a second or third magma ocean in the planet's history. Some insight could be gained through telescopic observations of rocky exoplanets near red giants.

## 16.5 Outstanding Questions and Future Exploration

Understanding questions about Venus' sulfur involves examining the abundances of sulfur-bearing compounds and their roles in current atmospheric and geological processes throughout history. The limited data on sulfur-bearing species, particularly in the lower atmosphere and geological materials, hinders our understanding of the fate of sulfur. This information scarcity restricts the use of sulfur-bearing species as indicators of atmospheric processes, atmosphere-surface interactions, and the coupled geological and climatic evolution. To address these gaps, new measurements of sulfur-bearing gases and solids are crucial for enhancing our comprehension. Looking ahead, significant progress in Venus' exploration is anticipated before 2035, thanks to four selected space missions (DAVINCI, Garvin et al. 2022; VERITAS, Smrekar et al. 2022; EnVision, de Oliviera et al. 2018, European Space Agency 2021; and Venus Orbiter Mission, VOM, I.S.R.O. 2024), along with several proposed projects such as Venera-D (Zasova et al. 2017), aimed at exploring the planet's atmosphere, surface, and interior. Leveraging a new generation of ground-based and space telescopes holds promises for refining our understanding of atmospheric composition. Details on selected and prospective missions, as well as telescopic observations and modeling efforts, can be found in publications by Glaze et al. (2018), Limaye and Garvin (2023), and Wiedemann et al. (2023). Table 16.13 outlines the plans and approaches to investigate the abundance and speciation of sulfur in atmospheric and crustal materials. In addition to upcoming observations, numerical and experimental modeling of atmospheric processes, gas-solids interactions, and processes within the deep and shallow interior, such as magmatism, volcanism, and metamorphism, can contribute to further progress. Advancements can also be achieved through new laboratory data on the physical-chemical properties of atmospheric, surface, and interior materials.

### *16.5.1* Abundances and Pathways of Atmospheric Sulfur

In the upper mesosphere, one challenge lies in understanding the sources of sulfur that seemingly lead to increasing mixing ratios of $SO_2$, SO, $SO_3$, OCS, CS, and $CS_2$ above 80 kilometers. There are two potential sources of sulfur: one from beneath (such as sulfuric acid aerosol) and the other from space, possibly as a component of interplanetary dust (Sects. 16.3.1.2 to 16.3.1.4). Further measurements of sulfur-bearing compounds can provide insights into their contribution to the mesospheric composition. Investigating whether there is a correlation between space fluxes, plumes, and mesospheric composition could provide clues about the origin of sulfur. Observers can look for compounds becoming abundant above 80 km and an abundance of elements (carbon, Fe, Si, Mg, Ni, etc.), indicative of space delivery. Telescopic observations of Venus' optical flashes, likely caused by meteors (Blaske et al. 2023), could serve as a proxy for assessing the intensity of space fluxes. Considering that fluxes of space materials are evident on Earth and Venus (Carrillo-Sánchez et al. 2020), it is essential to account for meteor ablation and photochemical alterations of cosmic materials in the mesospheric models. In the coming years, data on the upper mesosphere could be obtained via telescopic observations in the mm and sub-mm wavelengths and solar occultation data in the near-IR range (VIRAL spectrometer on the Venus Orbiter Mission, Patrakeev et al. 2022). Beyond observations above ~75 km, the underlying factors driving both short- and long-term variability of sulfur-bearing species within the mesosphere could be elucidated through flyby and orbital investigations of the cloud tops, as discussed below.

The primary challenges in cloud studies involve characterizing and understanding sulfur-bearing compounds, including their speciation and abundance. Constraining species distribution within the clouds and sub-cloud hazes, both vertically and latitudinally, while also understanding sulfur fluxes from the lower and upper atmosphere, is crucial. Key topics of interest include (1) the causes of long-term variability of $SO_2$ at the cloud tops; (2) the composition and origins of the UV-blue absorber in the upper clouds; (3) the composition of aerosol particle size modes; (4) the presence of non-$H_2SO_4$ aerosol inorganic species, such as $S_x$, and metal sulfates; (5) the composition of solutes in sulfuric acid, for example, metal sulfate/bisulfate complexes; (6) the potential for $S_n$ gas species formation above the clouds and the feasibility of $S_n$ supply from the lower atmosphere; (7) the origins, stability, fate, and sinks of possible metal sulfates in the clouds and hazes; and (8) the possibility of sulfur-bearing and other organic compounds, such as acid-insoluble organic matter (IOM) and poly-cyclic aromatic compounds of cosmic origin. Potential approaches to tackling these topics comprise long-term global measurements of cloud top composition (sulfur-, carbon-, nitrogen-, Cl-bearing gases, and $H_2O$ vapor) from telescopes and orbital platforms, remote and in situ investigations of cloud dynamics (such as convective plumes and meridional circulation), obtaining high spectral and spatial resolution spectra of clouds, and direct measurements of gas and aerosol composition within the cloud layers and hazes.

Several planned cloud and haze observations are relevant to these tasks (Table 16.13). In addition to telescopic observations, the mission plans include gathering compositional data on $SO_2$ and SO at the cloud tops and upper cloud aerosol in the UV spectral range using the DAVINCI (flybys), EnVision, VOM, and Venera-D orbiters, as well as the Venera-D Lander Module. The EnVision VenSpec-H spectrometer will provide data to constrain the abundances of $SO_2$ and OCS through spectral windows within 1 to 2.5 μm, both below and above the clouds. As observed previously (Table 16.1), $SO_2$ and OCS will be probed at 30 to 40 km altitudes in the night side spectral windows near 2.4 μm. Data from the Venera-D VIKA orbital near-IR spectrometer will provide constraints on sulfur-bearing compounds in the clouds, while the thermal IR Fourier spectrometer (5–40 μm) will deliver information on $SO_2$ at altitudes of 55 to 75 km. Abundances of sulfur-bearing gases and sulfuric acid aerosol at 45–55 km will be retrieved from radio occultations with EnVision and VERITAS (Akins et al. 2023). The Venera-D MM-radiometer will be used to measure $H_2SO_4$(g) and $SO_2$ both within and below the clouds. Sampling and analysis of aerosol particles are planned with a chromato-mass spectrometer onboard the Venera-D Lander Module. Baines et al. (2021) proposed analyzing the cloud aerosols at 52–62 km with a mass spectrometer via a balloon-borne gondola instrument package. In addition to observations, enhancing our understanding of sulfur in the middle atmosphere will involve developing cloud models that incorporate new measurements, chemical kinetics, interactions of photochemical and thermochemical processes, the physical chemistry of interactions among gas, liquid, and solid phases, material delivery from above and below the clouds, sinks of particulates, vertical and lateral dynamics, and global circulation.

In the lower atmosphere, understanding sulfur-bearing compounds relies on data regarding the abundance of $SO_2$, $SO_3$, $H_2SO_4$, OCS, CO, $S_n$, and $H_2S$ gases, which vary with altitude and latitude. Essential measurements yet to be made include the mixing ratios of $SO_2$, OCS, CO, and $S_n$ from the cloud deck to the surface, along with data on $H_2SO_4$ and $SO_3$ gases at ~30 to 48 km altitude. This information will help constrain the consumption mechanisms of $SO_3$ and OCS at 30 to 40 km, evaluate the origins of any existing $SO_2$ altitudinal and latitudinal gradients, enhance understanding of such gradients for OCS and CO, reveal the role of CO in reactions affecting sulfur chemistry at various altitudes and latitudes, and contribute to global circulation knowledge. In the near-surface atmosphere, it is crucial to measure the concentrations of chemically active gases ($CO_2$, $SO_2$, OCS, CO, $H_2O$, $H_2S$, $S_n$) to assess the degree of gas-phase chemical equilibration, examine the hypothesis of such equilibration occurring in the lowlands (Table 16.2), constrain the atmospheric redox state from $CO_2$/CO and $SO_2$/(COS, $H_2S$, $S_2$) ratios, and understand the pathways of gas-solid reactions, as well as the probability of atmospheric gas buffering by mineral assemblages. In the 2030s, corresponding measurements will be conducted using a mass spectrometer (VMS) and a tunable laser spectrometer (VTLS) onboard the DAVINCI descent sphere, Zephyr (Garvin et al. 2022). The VMS will analyze $SO_2$, OCS, $H_2SO_4$(g), $H_2S$, $S_n$, and $H_2O$ from the cloud deck to the surface. The VTLS measures $SO_2$, OCS, CO, $CO_2$, and $H_2O$, as well as stable isotopes of hydrogen, carbon, oxygen, and sulfur in these gases. The 2% accuracy of VTLS measurements for $SO_2$ and OCS, and VMS sampling every 0.1 to 1 km (depending on the altitude) will provide profiles of these gases in the lower atmosphere. Data from the Venus Oxygen Fugacity (V*f*Ox) sensor aboard the Zephyr sphere could enhance information on gases and help constrain the chemistry and degree of chemical equilibration in the lower atmosphere. Data from the VenSpec-H near-IR spectrometer onboard the EnVision orbiter will help assess concentrations of $SO_2$, OCS, CO, and $H_2O$ in the night-side lower atmosphere. The Venera-D Lander Module will be equipped with a gas chromato-mass spectrometer (VCS), a tunable laser absorption spectrometer (ISKRA-V), and a UV spectrometer (DAVUS) to measure the composition of $SO_2$, SO, OCS, CO, and $H_2O$ during the descent from 70 km to the surface. Beyond measurements, efforts should be made to advance coupled thermochemical-photochemical models, including global and local eddy diffusion and atmospheric circulation influenced by topography. The next logical steps involve developing 2D (e.g., meridional) and 3D chemical-transport models before obtaining data from upcoming missions. To ensure the modeling is fruitful, it is also necessary to experimentally determine rates of gas-phase reactions that are important for key scientific questions, such as OCS consumption at 30s km or $S_n$ production.

Sulfur isotopes ($^{32}S$, $^{33}S$, $^{34}S$, and minor $^{36}S$) in terrestrial atmospheric aerosols and gases ($SO_2$, OCS) provide valuable insights into the origins and processes that influence sulfur (Lin et al. 2018; Angert et al. 2019). Evidence of mass-independent fractionation of sulfur in tropospheric sulfate aerosols provides insights into stratospheric photochemistry. The $^{32}S/^{34}S$ ratio in atmospheric $SO_2$ and OCS implies mass-dependent fractionation on the Earth and sheds light on past and current biological processes, fossil fuel combustion, and volcanic degassing. Similarly, sulfur isotopes in Venus' atmospheric samples will provide insight into photochemical processes, circulation, and mixing, and the contribution of isotopically distinct sulfur sources from the planet's interior and space. The $^{32}S/^{34}S$ ratio deduced from the Pioneer Venus Large Probe data (Sect. 16.2.1.1) might be inaccurate, and spectroscopic measurements of $^{32}S$, $^{33}S$, and $^{34}S$ in $SO_2$ and OCS by the VTLS instrument are planned for the DAVINCI mission. Independent data could be collected using the ISKRA-V and VCS spectrometers on the Venera-D Lander Module, as well as from a mass spectrometer designed for the Aerosol-Sampling Instrument Package (ASIP), which is intended for a balloon mission (Baines et al. 2021).

Another question pertains to the impact of current volcanic activity on the composition of sulfur-bearing and other atmospheric gases and aerosols. Wilson et al. (2024) and Filiberto et al. (2025) discussed remote and in situ data on atmospheric composition that can indicate ongoing or recent volcanic activity. Fig. 16.30 illustrates that sampling a volcanic plume can provide valuable insights, allowing us to constrain the composition of volcanic gases and their interactions with atmospheric gases. In addition to compositional data, further observations in the near-IR (e.g., Arney et al. 2014) and microwave ranges can be used to detect ash-rich volcanic plumes in the lower atmosphere.

### *16.5.2* ***Abundances and Pathways of Sulfur in Surface and Interior Materials***

Like on Earth (Chap. 5: Eldridge 2026; Chap. 6: Turchyn et al. 2026; Chap. 10: Harlov et al. 2026; Chap. 11: Casas et al. 2026; Chap. 13: Kiseeva et al. 2026), Mars (Chap. 17: Franz and King 2026), the Moon, Mercury (Chap. 15: Renggli et al. 2026), and the moons in the outer solar system (Chap. 19: Lodders and Fegley 2026), the abundance, speciation, and isotopic composition of sulfur from Venus' surface and interior materials likely reflect the interplay between endogenic and exogenic processes throughout history. Given the limited data on sulfur in surface materials (Table 16.3), further insights into the composition of surface and interior materials could significantly enhance our understanding of sulfur abundance, speciation, and its role in current and past geological and atmospheric processes.

#### 16.5.2.1 Atmosphere-Surface Interactions

Essential inquiries concerning gas-solid reactions at the surface include the extent of sulfur intake, as well as the occurrence and composition of secondary sulfates and sulfides in the alteration products of rocks, minerals, glasses, and fine-grained materials, both at the surface and in buried rocks. This includes specific geological features and locations across various latitudes and altitudes. One aspect under consideration is the potential presence of abundant sulfides and/or sulfates in the highlands, and whether these minerals formed during the current epoch or alongside magma degassing during past global volcanic resurfacing (Sects. 16.4.4). Information on the abundance and mineralogy of sulfur in the highlands, such as the tessera terrain, will provide insights into the reasons for the low microwave and near-IR emissivity of the surface materials (Sects. 16.2.2.4 and 16.2.2.5). The preferential sulfatization of alkaline basaltic glasses (Sect. 16.3.2.2) offers an opportunity to differentiate alkaline basaltic lava flows from near-IR emissivity data.

The secondary mineralogy of sulfur will be tentatively constrained through remote sensing from several spacecraft (Table 16.13). Sulfides of Fe and trace metals from surface materials could be inferred from high-resolution radar studies conducted by the EnVision and VOM orbiters, which will constrain dielectric properties. Near-IR surface emissivity data will be collected globally, regionally, and locally from the VERITAS, EnVision, VOM, and Venera-D orbiters, as well as the DAVINCI flyby bus (CHRIS) and Zephyr sphere. These overlapping observations will be used to identify the types of surface rocks and determine whether these rocks are coated with low-albedo salts (sulfates, chlorides) or opaque substances like Fe sulfides and oxides (Dyar et al. 2020, 2021; Treiman et al. 2021; Helbert et al. 2021; McCanta et al. 2024). While major types of uncoated rocks (mafic vs. felsic) can be identified, the specific mineralogy may not be determined from remote observations. However, examining thermal emission in the near-infrared range over several years will help detect surface changes and identify recent and ongoing volcanic activity, as well as potential chemical alteration of fresh volcanic materials. Repeated radar imaging will also inform such activity (Filiberto et al. 2025).

In situ studies of surface materials conducted by landers and rovers can serve as ground truth for remote sensing investigations. Measurements from surface missions include assessments of sulfur abundance using XRF, Alpha Particle X-ray Spectrometry (APXS), Laser-Induced Breakdown Spectroscopy (LIBS), and gamma-ray spectrometry with neutron activation, along with evaluations of phase composition through XRD (X-ray diffraction), as well as Mössbauer, Raman, UV, and near-IR spectroscopy in $CO_2$ spectral windows. Microscopic optical and Raman imaging would yield insights into the altered texture of rock fragments, encompassing altered and unaltered mineral grains. Investigations of dust-coated, brushed, and drilled samples would help establish the degree of alteration and whether there has been sulfur uptake or loss. Analyzing the $^{32}S/^{33}S/^{34}S$ isotopic ratios in secondary phases will clarify the involvement of atmospheric sulfur. These measurements have been proposed for various NASA Venus Flagship Mission concepts. They could be carried out with the proposed Venera-D Lander Module equipped with the XRF/XRD spectrometer, an APXS spectrometer, a Mössbauer spectrometer, and a laser ablation mass spectrometer (Table 16.13, Widemann et al. 2023).

Another set of inquiries pertains to the direction of gas-solid reactions driven by chemical disequilibria, which depends on the composition of exposed materials, atmospheric composition, temperature, and pressure. DAVINCI data on the composition of the near-surface atmosphere will enhance evaluations of mineral stability and potential directions of weathering reactions (Garvin et al. 2022). Even without new data on near-surface environments, we can still progress in evaluating the stability and reactivity of solid phases on the current surface (refer to Fegley and Treiman 1992a; Zolotov 2019, 2018 for details). Promising research areas include investigating the stability of unexplored sulfides, sulfates, and minerals that could undergo sulfurization. We should also consider studying silicate solid solutions of rock-forming minerals (pyroxenes, feldspars, etc.) and the environmental conditions suggested by climate and integrated atmosphere-interior models.

Many initial reconnaissance and exploratory experiments have been conducted (Table 16.9; Sec. 16.3.2), laying the foundation for focused investigations of solid phases and glasses. These focused investigations can advance the field by deriving data on reaction rates, rate-limiting steps, and mechanisms throughout different stages of alteration, which can then be incorporated into numerical models to assess the alteration. Further experimental studies of gas-solid reactions could focus on thermodynamically plausible interactions. Investigating the advanced stages of alterations and the effects of proximity to equilibrium conditions on reaction rates is essential. In addition to longer runs, further evaluation of the rates and mechanisms of sulfatization may necessitate gas-solid experiments at higher temperatures and/ or elevated $p$SO$_2$, along with careful extrapolation of results to align with Venus' values. One possible method is to conduct 1 bar runs at SO$_2$ molecular number density (molecules per unit volume) corresponding to Venus, as Fegley and Prinn (1989) did. The preliminary experiments have also highlighted two key improvements for the design of weathering experiments: the use of chemically inert reaction vessels to maintain gas composition (for instance, by regulating gas flows or using mineral buffers), and the analysis of gaseous products from the gas-phase and gas-solid interactions. Alongside experiments, insights could be gained from terrestrial analog studies in volcanic (McCanta et al. 2014; D'Incecco et al. 2024) and metamorphic settings.

#### 16.5.2.2 Sulfur in Geological and Climatic Evolution

Outstanding, big-picture questions about sulfur relate to the formation and differentiation of the planet and its early evolution, which may have involved a magma ocean, the oxidation of atmospheric, crustal, and mantle materials, plate tectonics, and the existence of a hydrosphere (Sect. 16.4). More Venus-specific inquiries include the history of volcanic resurfacing and degassing and their impacts on physical-chemical processes in the atmosphere, as well as sulfur sequestration in crustal materials. These topics could be explored through investigations of bedrock using remote sensing, in situ sampling, and geophysical studies of shallow and deep interior via radar sounding, gravity measurements, topography (Table 16.13), and seismic measurements.

Understanding the fraction of accreted sulfur sequestered in the core versus that remaining in the mantle remains an outstanding question in Venus' exploration. In situ measurements of sulfur content in unaltered mafic rocks could provide insights into the mantle-core partitioning of sulfur, and elevated sulfur content in these rocks may indicate a sulfur-enriched silicate mantle. Concentrations of chalcophile elements in basalts also indicate sulfur partitioning to the core. Another promising approach involves analyzing copper isotopes in a basalt to provide insights into the segregation of sulfide liquids during core formation (Savage et al. 2015). The gravity data from the VERITAS (Smrekar et al. 2022; Cascioli et al. 2023) and EnVision (Rosenblatt et al. 2021) missions will address questions about the interior structure, including whether the core is liquid or solid, the presence of a basal silicate magma ocean (O'Rourke et al. 2020), and may offer constraints on the distribution of sulfur within the interior. However, direct inference of core sulfur content may prove elusive, akin to challenges faced with other terrestrial planets (Chap. 14: Komabayashi et al. 2026). First-order questions regarding sulfur in magmatic systems include major rock types (Sect. 16.2.2), such as tholeiitic basalts, alkaline mafic rocks, evolved silicate rocks, and potential non-silicate rocks in the canali and crater outflows. In addition to examining the abundance and mineralogy of sulfur in these formations, these inquiries cover the occurrence of cumulative magmatic and/or associated (e.g., hydrothermal) sulfide deposits. Near-IR and microwave emissivity measurements will tentatively constrain the composition of surface igneous rocks, and sulfide-rich secondary or primary formations may be inferred. Although not directly assessing composition, high-spatial-resolution radar images could be used to evaluate lava viscosity or to reveal patterns of physical weathering that depend on composition. Radar sounding conducted by the EnVision and VOM orbiters could infer the presence of massive sulfide formations at depth. New radar data are necessary to verify the occurrence of abundant Fe sulfides and/or oxides below ~0.7 m, as suggested by Antony et al. (2022).

Questions surround whether sulfide sulfur has undergone oxidation due to residual oxygen after hydrogen escape. We seek to understand the scale and context of such oxidation. The specific mechanisms remain unclear regarding whether sulfur, Fe(II), and other elements have experienced oxidation. Did this process occur at the top of a magma ocean, through later gas-solid interactions, or via aqueous processes near the planetary surface (Sect. 16.4)? These pathways could intertwine, creating a fascinating puzzle. Further inquiries arise regarding the fate of oxidized rocks. Did they accumulate in the upper crust or descend through the subduction of lithospheric plates? Perhaps they were submerged (e.g., Elkins-Tanton et al. 2007) or buried by basaltic lava throughout history. We also ponder whether aqueous processes led to the formation of sulfate-bearing rocks and if these rocks are exposed. Our inquiries extend to understanding when, where, and how sulfates underwent reduction and whether putative biotic sulfate reduction led to the fractionation of $^{32}S$ and $^{34}S$. Redox- and paleoclimate-related inquiries can be addressed through investigations of bedrocks within diverse geological formations. Attention could focus on ejecta from Mead, Stanton, and Boleyn impact craters that contain rock fragments from several kilometers deep. The abundance, mineralogy, and oxidation state of sulfur (S-II), S(-I), S(VI), and other elements (Fe, Mn, Ni, V, Eu, Ce, U, Mo, etc.) in igneous rocks are crucial for constraining the interconnected magmatic and climatic history. As observed in the surface rocks on Mercury (Nittler et al. 2018; Chap. 15: Renggli et al. 2026), a few wt% of the sulfide's sulfur in unaltered igneous rocks would suggest significantly reduced mantle rocks ($fO_2 < 3$ log units below the IW buffer, e.g., Namur et al. 2016) and a scarcity of $H_2O$ in Venus-forming materials. Conversely, detecting sulfates in silicate and/or non-silicate rocks would indicate oxidation of Venus' upper interior due to the accretion of abundant $H_2O$ and subsequent hydrogen escape. The detection of igneous rocks with sulfate sulfur and abundant Fe(III)—rather than sedimentary or metamorphosed chemical sediments like banded iron formations—hints at an anhydrous oxidation during a magma ocean stage.

The occurrence of putative sedimentary, metamorphic, or hydrothermally altered rocks within tessera terrain (Sect. 16.2.2.4) will yield additional constraints for our understanding of these processes. Layered deposits rich in sulfates would indicate the presence of $O_2$ in the atmosphere-hydrosphere system that preceded or accompanied aqueous (possibly evaporitic) deposition. The composition of metal (e.g., Ca, Mg, Na, K) sulfates will constrain the aqueous deposition conditions, such as salinity, temperature, and *pH*. Deposits rich in chlorides and/or sulfides (e.g., pyrite) but lacking sulfates would suggest aqueous deposition in anoxic, possibly hydrothermal environments. The presence of sedimentary sulfides in Venus' rock formations, the mineralogy of sulfides and trace metal content, along with the isotopic composition of sulfur and trace metals (Mo, Re, etc., Lyons et al. 2014), will help determine the characteristics of aqueous environments, redox conditions, and the fate of sulfur on land and in the ocean. Sedimentary sulfide would also assess the likelihood of biological sulfate reduction. The sulfur-MIF signature would suggest an anoxic atmosphere. Investigations into the elemental, isotopic, and phase composition of solid samples (Table 16.13) from at least two sites (plains and tessera terrain), along with remote sensing data, will be essential in clarifying the fate of sulfur during the interior-atmospheric evolution.

### *16.5.3 Conclusion*

Our understanding of sulfur on Venus is constrained by the minimal data available below 20 kilometers of altitude in the atmosphere. Only one gas species ($S_3$) has been tentatively detected in the near-surface atmosphere, and the total sulfur content in the surface materials has been quantified with considerable error bars at three locations. However, the abundance of measurements of sulfur-bearing gases at altitudes of 30 to 120 km, the presence of thick sulfuric acid clouds, and the sulfur-rich composition of otherwise basaltic rocks all indicate a significant role for sulfur in the atmosphere and in the interactions between the atmosphere and surface materials. Due to sulfur's role in photochemical, thermochemical, volcanic, magmatic, mass transfer, and isotope fractionation processes, the abundance and speciation of sulfur in potentially detectable materials (gases, aerosols, surface solids) signify global processes both currently and throughout history. For instance, elevated sulfur abundances in the atmosphere and surface samples imply past volcanic degassing of sulfur-bearing species, which affected cloud formation and the sequestration of atmospheric gases in minerals. If detected, sulfate and/or sulfide coatings in the highlands would suggest sulfurization in the current epoch and/or during global volcanic degassing several hundred million years ago. The mineralogy and abundance of sulfur and chalcophile elements in pristine igneous rocks will provide insights into the fate of sulfur during planetary differentiation and igneous processes. Sulfate in igneous rocks would indirectly indicate the presence of abundant early $H_2O$ that affected the oxidation of the mantle via hydrogen escape and/or the formation and assimilation of sulfate-rich crustal rocks, such as evaporites. Similarly, detecting layered sulfate-rich deposits in the tessera terrain would suggest deposition in oxygen-rich aqueous environments, potentially indicating hydrogen escape from a moist atmosphere. The discovery of pyrite associated with Fe oxides in surface materials would imply the buffering of sulfur-bearing gases and the redox state of the near-surface atmosphere in the current epoch. Such buffering indicates the past and future evolution of the coupled atmosphere-surface system, where gas composition is influenced by volcanic degassing, resurfacing by lava, hydrogen escape, and related climate changes. Consequently, prioritizing the study of sulfur-bearing compounds is essential in exploring Venus. Continued progress in our understanding of Venus' sulfur hinges on observations from terrestrial and space telescopes as well as flyby and orbital platforms. In situ data from atmospheric probes and landers are vital for deciphering deep atmospheric chemistry, gas-solid interactions, and the fate of sulfur in the planet's interior. Significant limitations can arise from measuring the fundamental properties of sulfur-bearing compounds and physicochemical systems, along with laboratory modeling of physicochemical processes in the atmosphere, on the surface, and below it. Lastly, developing and applying updated and innovative numerical models is indispensable for linking these efforts with observations of sulfur species and aligning with other astronomical, geophysical, petrological, and geological information about the Earth's sister planet.

**Acknowledgments** This chapter benefits from comments provided by Gleb Pokrovski, Daniel Harlov, Kevin Zahnle, Emmanuel Marcq, Alison Santos, Larry Esposito, Alexander Basilevsky, Mikhail Kreslavsky, Michael Way, Laura Schaefer, Lindy Elkins-Tanton, Ekaterina Kisseva, Molly McCanta, Joseph O'Rourke, Nancy Chabot, Rakesh Mogul, Sara Port, and Julia Semprich. Over the years, my work on Venus has been motivated and supported by Igor Khodakovsky, Vladislav Volkov, Bruce Fegley, Jr, and Alexander Basilevsky. The comprehensive perspective on planetary geochemistry was inspired by Alexey Yaroshevsky at Lomonosov Moscow State University in the early 1980s. The NASA Discovery (DAVINCI mission) and Solar System Workings programs supported the work on this chapter.

**Table 16.1** Volume mixing ratios of sulfur-bearing gases and CO in the atmosphere of Venus derived from remote sensing and in situ data

| Mixing ratio | Altitude, km | Methods and references |
|---|---|---|
| $SO_2$ | | |
| 0–76 ppbv | 80–100 | Ground sub-mm. Sandor et al. (2010, 2012) |
| 50–175 ppbv | 60–80 | Ground IR. Encrenaz et al. (2012) |
| 12.0 ± 3.5 ppbv | >88 | Ground mm-wave. Encrenaz et al. (2015) |
| 40–103 ppbv<br>200–500 ppbv (equator)<br>50–100 ppbv (North polar region) | 85–105<br>68–70<br>68–70 | Venus Express near-IR and UV solar occultation. Belyaev et al. (2012) |
| 100–300 ppbv (night)<br>50 ppbv (night)<br>10–30 ppbv (night)<br>150–200 ppbv (terminator) | 100<br>95<br>85<br>95 | Venus Express UV stellar occultation. Belyaev et al. (2017) |
| $10^2$–$10^3$ ppbv | 70–78 | Venus Express near-IR solar occultation. Belyaev et al. (2008) |
| 1 ppmv<br>0.02 ppbv | 100<br>70–90 | Venus Express near-IR solar occultation. Mahieux et al. (2023) |
| 50–430 ppbv<br>(~400 ppbv global mean) | 65–70 | Pioneer Venus UV. Stewart et al. (1979), Esposito et al. (1979, 1988), Esposito (1980, 1984) |
| 50–380 ppbv<br>(50 ± 20 ppbv global mean) | 70 | International UV Explorer. Na et al. (1990) |
| 120 ± 70 ppbv (average)<br>20 ± 10 ppbv (<30°)<br>400 ± 100 ppbv (>45°)<br>3–20 ppmv | 69<br>62 | Venera 15 IR. Zasova et al. (1993) |
| 8 ± 40 ppbv (aver. 1988)<br>60 ± 30 ppbv (equator, 1988)<br>300 ± 150 ppbv (50 °, 1988)<br>20 ± 60 ppbv (aver. 1991) | 69 | Rocket-borne UV. Na et al. (1994) |
| 100–550 ppbv | 65 | Venus Express near-IR solar occultation. Belyaev et al. (2008) |
| <10–1000 ppbv; median 20 ppbv | 70 | Venus Express UV. Marcq et al. (2011, 2013, 2020) |
| 30–700 ppbv | ~64 | Ground IR. Encrenaz et al. (2012, 2016, 2019, 2020, 2023) |
| 300–400 ppbv | 72 | Ground near-IR. Krasnopolsky (2010b) |
| 20 ± 10 ppbv | 70 | HST UV. Na and Esposito (1995) |
| 9–250 ppbv; aver. 197 ppbv | 74–81 | HST UV. Jessup et al. (2015) |
| 0.6–5.2 ppmv | 51–58 | PVPL LNMS. Mogul et al. (2025) |

| Mixing ratio | Altitude, km | Methods and references |
|---|---|---|
| 90 ± 60 ppmv (equator lat.)<br>150 ± 50 ppmv (polar lat.) | 51–54 | Venus Express radio occultation. Oschlisniok et al. (2021) |
| 180 ± 70 ppmv | 37–52 | Ground near-IR. Pollack et al. (1993) |
| 130 ± 40 ppmv | 35–45 | Ground near-IR. Bezard et al. (1993) |
| 130 ± 35 ppmv | ≤42 | Venera 12 GC. Gel'man et al. (1980) |
| 185 ± 43 ppmv | 22 | PVLP GC. Oyama et al. (1980) |
| 150 ppmv<br>125 ppmv<br>38 ppmv<br>25 ± 2 ppmv | 52<br>43<br>22<br>12 | Vega 1 UV. Bertaux et al. (1996) |
| 130 ± 50 ppmv | 30–40 | Venus Express near-IR. Marcq et al. (2008) |
| 140 ± 37 ppmv (2009)<br>126 ± 32 ppmv (2010) | 30–40 | Ground near-IR. Arney et al. (2014) |
| 130 ppmv (15 °S)<br>210 ppmv (>35 °N)<br>180 ppmv (latit. average) | 30–40<br>33 | Ground near-IR. Marcq et al. (2021) |
| 190 ± 40 ppmv | ~35 | Venus Express near-IR. Marcq et al. (2023) |
| SO | | |
| 10–30 ppbv | 70 | International UV Explorer. Na et al. (1990) |
| 12 ± 5 ppbv | 64–96 km | Rocket-borne UV. Na et al. (1994) |
| 0–31 ppbv | 70–100 | Ground sub-mm. Sandor et al. (2010, 2012) |
| 35–80 ppbv | 85–95 | Venus Express UV solar occultation. Belyaev et al. (2012) |
| 6–10 ppbv | >88 | Ground mm-wave. Encrenaz et al. (2015) |
| 1–15 ppbv | 74–81 | Ground sub-mm. Jessup et al. (2015) |
| 23 ppbv | 74–81 | HST UV. Jessup et al. (2015) |
| $H_2SO_4$(g) | | |
| <3 ppbv | 85–100 | Ground sub-mm. Sandor et al. (2012) |
| 1–2.5 ppmv | Cloud and sub-cloud, disk-average | Ground microwave. Butler et al. (2001) |

| Mixing ratio | Altitude, km | Methods and references |
|---|---|---|
| 0–30 ppmv | 42–64 | Pioneer Venus radio occultation. Jenkins and Steffer (1991) |
| 0–2 ppmv<br>18–24 ppmv<br>3–5 ppmv | 50<br>39<br>36 | Magellan radio occultation. Jenkins et al. (1994) |
| ~0.1–14 ppmv | 34–55 | Mariner 10 and Magellan radio occultation. Kolodner and Steffes (1998) |
| 0–9 ppmv | 30–55 | Ground microwave. Jenkins et al. (2002) |
| 1–5 ppmv<br><1 ppmv | 50–55, 0°–70° S<br>50–55, >70° S | Venus Express radio occultation. Oschlisniok et al. (2012) |
| 0–12 ppmv | 56–38 | Akatsuki radio occultation. Imamura et al. (2017) |
| 2–9 ppmv | 46–50 | Akatsuki radio occultation. Ando et al. (2024) |
| <12 ppmv (equator lat.)<br>5–7 ppmv (middle lat.)<br><9–12 ppmv (polar lat.) | 47<br>43–47<br>43 | Venus Express radio occultation. Oschlisniok et al. (2021) |
| $SO_3$ | | |
| 10 ppmv<br>1 ppmv<br>0.5 ppmv<br>0.1 ppmv | 95<br>85<br>80<br>75 | Venus Express near-IR solar occultation. Mahieux et al. (2023) |
| OCS | | |
| 2 ppbv<br>14 ppbv | 70<br>65 | Ground near-IR. Krasnopolsky (2008) |
| 0.3–9 ppbv (av. 3 ppbv) | 65 | Ground near-IR. Krasnopolsky (2010b) |
| <1.6 ± 2 ppbv | 70–90 | Venus Express near-IR solar occultation. Vandaele et al. (2008) |
| 1 ppbv<br>1 ppmv | 65<br>100 | Venus Express near-IR solar occultation. Mahieux et al. (2023) |
| 0.25 ppmv<br>10 ppbv | <50<br>>65 | Ground near-IR. Bezard et al. (1990) |
| 4.4 ± 1 ppmv | 33 | Ground near-IR. Pollack et al. (1993) |
| 0.1 ppmv<br>10 ppmv | 48<br>30 | Ground near-IR. Marcq et al. (2005) |
| 0.55 ± 0.15 ppmv<br>5–20 ppmv | 36<br>30 | Ground near-IR. Marcq et al. (2006) |

| Mixing ratio | Altitude, km | Methods and references |
|---|---|---|
| 2.5 ± 1 ppmv<br>4 ± 1 ppmv | 33 (60° S)<br>33 (10° S) | Venus Express near-IR. Marcq et al. (2008) |
| 0.52 ± 0.17 ppmv | 36 | Ground near-IR. Arney et al. (2014) |
| <40 ppmv<br><10 ppmv<br><2 ppmv | 52<br>42<br>22 | PVLP GC. Oyama et al. (1980) |
| 40 ± 20 ppmv | 29–37 | Venera 13 and 14 GC. Mukhin et al. (1982, 1983) |
| CO[1] | | |
| 30 ± 18 ppmv<br>20 ± 3 ppmv | 42<br>22 | PVLP GC. Oyama et al. (1980) |
| 30 ± 7 ppmv<br>17 ± 5 ppmv | 36<br>12 | Venera 12 GC. Gel'man et al. (1980), Krasnopolsky (2007) |
| 23 ± 5 ppmv | 36 | Ground near-IR. Pollack et al. (1993) |
| 24 ± 2 ppmv | 36 | Ground near-IR. Marcq et al. (2006) |
| 24 ± 3 ppmv<br>31 ± 2 ppmv | 33 (60° S)<br>33 (10° S) | Ground near-IR. Marcq et al. (2008) |
| 25 ± 3 ppmv (2009)<br>22 ± 2 ppmv (2010) | 35 | Ground near-IR. Arney et al. (2014) |
| 37 ppmv (poles)<br>22 ppmv (equator) | 35 | Venus Express near-IR. Tsang and McGouldrick (2017) |
| 35 ± 5 ppmv (global) | 36 | Venus Express near-IR. Marcq et al. (2023) |
| $S_3$ | | |
| 40 ± 10 pptv<br>15 ± 5 pptv | 15<br>3 | Venera 11 and 12 UV. Moroz et al. (1981) |
| 27 ± 30 pptv | 5–25 | Venera 11 and 13 UV. Krasnopolsky (1987) |
| 80 ± 30 pptv | 5–25 | Venera 14 UV. Krasnopolsky (1987) |
| 100 pptv<br>30 pptv | 19<br>3 | Venera 11 UV. Maiorov et al. (2005) |
| 18 ± 3 pptv<br>11 ± 2 pptv | 10–19<br>3–10 | Venera 11–14 UV. Krasnopolsky (2013) |
| $S_4$ | | |
| 6 ± 2 pptv<br>4 ± 4 pptv | 10–19<br>3–10 | Venera 11–14 UV. Krasnopolsky (2013) |
| $H_2S$ | | |
| <23 ppbv | 64–70 | Ground near-IR. Krasnopolsky (2008) |

(continued)

| Mixing ratio | Altitude, km | Methods and references |
|---|---|---|
| <40 ppmv<br><10 ppmv<br><2 ppmv | 52<br>42<br>22 | PVLP GC. Oyama et al. (1980) |
| 3 ± 1 ppmv<br>~1 ppmv | 0–24<br>47–64 | PVPL LNMS. Hoffman et al. (1980), von Zahn et al. (1983) |
| 80 ± 40 ppmv | 29–37 | Venera 13 and 14 GC. Mukhin et al. (1982, 1983) |
| Suspected | 50–55 | PVPL LNMS. Mogul et al. (2021) |
| CS | | |
| 40 ppmv<br>0.1 ppmv | 100<br>65 | Venus Express near-IR solar occultation. Mahieux et al. (2023) |
| $CS_2$ | | |
| 5 ppmv<br>0.03 ppmv | 90<br>70 | Venus Express near-IR solar occultation. Mahieux et al. (2023) |

[1]CO data does not list all measurements
GC: Gas Chromatography
HST: Hubble Space Telescope
IR: Infrared
LNMS: Large Neutral Mass Spectrometer
PVPL: Pioneer Venus Large Probe
UV: Ultraviolet
'ground': Telescopic observations from Earth
ppmv: Parts per million by volume
ppbv: Parts per billion by volume
pptv: Parts per trillion by volume

**Table 16.2** Mixing ratios of chemically active gases in the near-surface atmosphere of Venus, according to gas-phase chemical equilibrium models[1]

| Gas | | 1 | 2 | 3 | 4 | 5 | 6 | 7 | 8 | 9 | 10 |
|---|---|---|---|---|---|---|---|---|---|---|---|
| *T*, K | | 740 | 740 | 750 | 750 | 740 | 735 | 740 | 740 | 740 | 740 |
| *P*, bars | | ~96 | ~96 | 96.1 | 96.1 | 95.6 | 90 | 95.6 | 95.6 | 95.6 | 95.6 |
| $CO_2$ | | (0.96) | (0.96) | (0.97) | (0.97) | (0.965) | 0.97 | (0.965) | (0.965) | (0.965) | (0.965) |
| $N_2$ | | (0.034) | (0.34) | (0.03) | (0.025) | (0.035) | 0.03 | (0.035) | (0.035) | (0.035) | (0.035) |
| $SO_2$ | ppmv | (150) | (130) | (130) | (130) | 120 | (150) | (130) | (150) | (150) | (150) |
| $H_2O$ | ppmv | (30) | (200) | (20) | (20) | 33 | (30) | (30) | (30) | (30) | (30) |
| OCS | ppmv | 20 | 20 | 10 | 23 | 16 | 9 | 36 | 41 | 4.3 | 2.8 |
| CO | ppmv | (15) | (15) | (14) | (17) | 17 | 12 | (17) | (17) | (8) | 6.9 |
| $S_2$ | ppmv | 0.1 | 0.1 | 0.1 | 0.18 | 0.21 | 0.11 | 0.31 | 0.41 | 0.020 | 0.011 |
| $H_2S$ | ppmv | 0.5 | 0.3 | | 0.05 | 0.1 | 0.05 | 0.13 | 0.15 | 0.016 | 0.010 |
| $S_2O$ | ppbv | | | | | 31 | 17 | 32 | 42 | 4.4 | 2.8 |
| $H_2$ | ppbv | 3.0 | 20 | 3.0 | 2.4 | 3.6 | 3.0 | 3.7 | 3.7 | 1.8 | 1.5 |
| $S_3$ | ppbv | | | | | 0.63 | | 1.6 | 2.4 | 0.026 | (0.011) |
| $CS_2$ | pptv | | | | | 61 | | 106 | 141 | 1.5 | 0.64 |
| SO | pptv | | | | | 29 | | 36 | 41 | 19 | 17 |
| $S_4$ | pptv | | | | | 9.0 | | 13 | 23 | $5 \times 10^{-2}$ | $1.7 \times 10^{-2}$ |
| $S_5$ | pptv | | | | | 0.56 | | 7.8 | 16 | $9 \times 10^{-3}$ | $2 \times 10^{-3}$ |
| HS | pptv | | | | | 1.1 | | 1.0 | 1.2 | 0.18 | 0.12 |
| $SO_3$ | pptv | | | | | 0.30 | | 0.33 | 0.38 | 0.80 | 0.92 |
| $S_6$ | pptv | | | | | 0.09 | | 0.56 | 1.3 | $2 \times 10^{-4}$ | $3 \times 10^{-5}$ |
| $\log_{10} fS_2$ | bars | −5.02 | −5.02 | −5.02 | −4.76 | −4.70 | −5.00 | −4.53 | −4.40 | −5.71 | −5.97 |
| $\log_{10} fO_2$ | bars | | | −20.8 | −21.0 | −21.3 | | −21.36 | −21.36 | −20.71 | −20.58 |

[1]Chemical equilibrium among near-surface atmospheric gases remains hypothetical. Concentrations in parentheses are based on instrumental data (Table 16.1) and are used as anchors to assess the mixing ratios of other gases

1: Krasnopolsky and Parshev (1979)

2: Krasnopolsky and Parshev (1981a)

3: Barsukov et al. (1980a), their $H_2S$ value appears to be incorrect and is omitted

4: Barsukov et al. (1982c), Dorofeeva et al. (1981), Khodakovsky (1982)

5: Fegley et al. (1997b), the Gibbs free energy minimization method based on JANAF data; the results align with those of Zolotov (1996)

6: Rimmer et al. (2021) employed the Gibbs free energy minimization method with initial mixing ratios of 0.96 $CO_2$, 20 ppmv CO, 150 ppmv $SO_2$, 5 ppmv OCS, 10 ppbv $H_2S$, and 3 ppbv $H_2$

7–10: this work. In 7, $xSO_2$ is from Gel'man et al. (1980) and Marqc et al. (2021). In 9, the CO concentration of 8 ppm is obtained by extrapolating the CO gradient at 12 to 36 km (Table 16.1) to the surface. In 10, $xS_3$ corresponds to the data for 3 to 10 km from Krasnopolsky (2013). In models 7 and 8, the gas composition resembles that of the magnetite-pyrite buffer (equilibria 16.67 and 16.79; Table 16.11). In models 9 and 10, the gas composition resembles that determined by the magnetite-hematite buffer (equilibria 16.65 and 16.66; Table 16.11). Before acquiring new data on the near-surface atmospheric composition, models 8 and 9 can be regarded as nominal reduced and oxidized cases, respectively

**Table 16.3** Chemical composition of the surface materials on Venus at the Venera and Vega landing sites (wt%)[1]

| | Venera 13 | Venera 14 | Vega 2 |
|---|---|---|---|
| $SiO_2$ | 45.1 ± 3.0 | 48.7 ± 3.6 | 45.6 ± 3.2 |
| $Al_2O_3$ | 15.8 ± 3.0 | 17.9 ± 2.6 | 16.0 ± 1.8 |
| FeO | 9.3 ± 2.2 | 8.8 ± 1.8 | 7.7 ± 1.1 |
| MnO | 0.2 ± 0.1 | 0.16 ± 0.08 | 0.14 ± 0.12 |
| MgO | 11.4 ± 6.2 | 8.1 ± 3.3 | 11.5 ± 3.7 |
| CaO | 7.1 ± 0.96 | 10.3 ± 1.2 | 7.5 ± 0.7 |
| $K_2O$ | 4.0 ± 0.63 | 0.2 ± 0.07 | 0.1 ± 0.08 |
| $TiO_2$ | 1.59 ± 0.45 | 1.25 ± 0.41 | 0.2 ± 0.1 |
| $SO_3$ | 1.62 ± 1.0 | 0.88 ± 0.77 | 4.7 ± 1.5 |
| Sum | 96.1 | 96.3 | 93.4 |
| Sulfur (bulk) | 0.65 ± 0.4 | 0.35 ± 0.3 | 1.9 ± 0.6 |

[1]Surkov et al. (1984, 1986b). The data were obtained through X-ray fluorescent analysis, which does not provide speciation. Uncertainties are 1σ. Sulfur concentration is presented in both oxide and elemental forms. All Fe is reported as FeO. Na has not been measured

**Table 16.4** Potassium, uranium, and thorium contents in surface materials at the landing sites of the Venera and Vega landers[1]

| Lander | K, wt% | U, ppmw | Th, ppmw |
|---|---|---|---|
| Venera 8 | 4 ± 1.2 | 2.2 ± 0.7 | 6.5 ± 0.2 |
| Venera 9 | 0.47 ± 0.08 | 0.6 ± 0.16 | 3.65 ± 0.42 |
| Venera 10 | 0.30 ± 0.16 | 0.46 ± 0.26 | 0.70 ± 0.34 |
| Vega 1 | 0.45 ± 0.22 | 0.64 ± 0.47 | 1.5 ± 1.2 |
| Vega 2 | 0.40 ± 0.20 | 0.68 ± 0.38 | 2.0 ± 1.0 |

[1]Surkov et al. (1987b), passive gamma-ray spectroscopy method
[2]ppmw, part per million by weight

**Table 16.5** Sulfur contents in models of Venus' interior composition (BVTP 1981)

| | Ve1 | Ve2 | Ve3 | Ve4 | Ve5 |
|---|---|---|---|---|---|
| Core mass, wt% | 30.2 | 30.9 | 32 | 23.6 | 28.2 |
| Sulfur in core, wt% | 0 | 10 | 5.1 | 4.9 | 1.0 |
| Oxygen in core, wt% | 0 | 0 | 0 | 9.8 | 8.0 |
| Sulfur in bulk planet, wt% | 0 | 3.09 | 1.62 | 1.16 | 0.28 |

Ve1: Equilibrium condensation models of the solar nebula. The absence of sulfur in Ve1 is accounted for by an elevated nebula temperature at the radial distance where Venus-forming materials condensed
Ve2: Equilibrium condensation models of the solar nebula that include feeding zones
Ve3: Morgan and Ander's (1980) model is based on chemically modified nebular condensates that resemble stony and iron meteorites. The K/U ratio from Mars 5 data serves as a K/U proxy for volatile/refractory elemental ratios
Ve4: Pyrolite models developed by Ringwood were based on analogs of Earth's model
Ve5: a Fe-deficient model

**Table 16.6** Major sources and sinks of sulfur-bearing species in the atmosphere of Venus[1]

| Compound | Sources | | | Sinks | | | Notes |
|---|---|---|---|---|---|---|---|
| $SO_2$(g) | Volcanic degassing (net source), oxidation of OCS, reduction of $SO_3$, S-O-Cl, and S-Cl gases | | | Sulfatization of minerals and glasses (net sink), photolysis to SO, oxidation to $SO_3$, dissolution in sulfuric acid aerosol | | | Consumption in the upper and middle atmosphere and production in the lower atmosphere |
| | S, LA | | | | MA | UA | |
| SO(g) | Photolysis of $SO_2$ | | | Oxidation to $SO_2$, formation of $(SO)_2$ | | | Formation and loss mainly in the upper and middle atmosphere |
| | | MA | UA | | MA | UA | |
| $SO_3$(g) | Photochemical oxidation of $SO_2$, pyrolysis of $H_2SO_4$(g) | | | Hydrolysis to $H_2SO_4$, reduction to $SO_2$ and $S_2$ | | | $SO_3$ is reduced by OCS and/or CO in the sub-cloud atmosphere |
| | LA | MA | UA | LA | MA | | |
| $H_2SO_4$(g) | Hydrolysis of $SO_3$, vaporization of sulfuric acid solution | | | Hydration and condensation to sulfuric acid, pyrolysis to $SO_3$ | | | Production in upper clouds, an equilibration with sulfuric acid in clouds, and pyrolysis below clouds |
| | LA | MA | | LA | MA | | |
| $H_2SO_4 \cdot nH_2O$ (l), sulfuric acid | Hydrolysis of $H_2SO_4$(g) and condensation | | | Vaporization to $H_2SO_4$(g) | | | Condensation in upper clouds, complete vaporization at the cloud deck |
| | | MA | | LA | | | |
| $S_x$(condensed) | Polymerization of $S_n$ and condensation of $S_8$ | | | Sublimation to $S_n$ gases | | | Polymerization in clouds, sublimation below the cloud deck |
| | | MA | | LA | | | |
| OCS(g) | Volcanic degassing and oxidation of pyrrhotite (net sources), reduction of S and $S_2$ by CO, alteration of cometary dust in the mesosphere | | | Photolysis, oxidation by $SO_3$, $(SO)_2$, O, S, and $S_2$. Decomposition to CO and S. Formation of metal sulfides (net sink) | | | Thermochemical formation below ~25 km, thermochemical consumption below and within clouds, photolysis at and above cloud top |
| | S | | UA | S, LA | MA | | |
| $S_{1-8}$ (g) | Volcanic degassing, metal sulfide decomposition, sulfatization of minerals (net sources), sublimation of $S_x$, photochemical and thermochemical reactions | | | Reactions with CO, OCS, and $H_2S$, condensation to $S_x$ | | | Competitive production and loss in upper clouds, production in the sub-cloud atmosphere, soft UV photolysis of $S_3$ and $S_4$, thermochemical consumption below 30 km |
| | S, LA | MA | | S, LA | MA | | |
| $H_2S$(g) | Volcanic degassing and oxidation of pyrrhotite by $H_2O$(g) (net sources), thermochemical reactions in the lower atmosphere | | | Formation of metal sulfides (net sink), photolysis, oxidation | | | Formation in the lower atmosphere, oxidation in the lower and middle atmosphere |
| | S, LA | | | S, LA | MA | | |

[1]Additional cells in 'sources' and 'sinks' designate atmospheric regions
S: surface and near-surface (<10 km)
LA: lower atmosphere (below clouds)
MA: middle atmosphere (clouds)
UA: upper atmosphere (mesosphere)

**Table 16.7** Major chemical reactions involving sulfur-bearing gases in the atmosphere of Venus[1]

| Gas | Sources | Sinks |
|---|---|---|
| $SO_2$ | 16.13: $SO + O + M \rightarrow SO_2 + M$ (73 km)<br>16.14: $SO + NO_2 \rightarrow SO_2 + NO$ (70 km)<br>16.26: $S_2O_2 + h\nu \rightarrow SO_2 + S$ (70 km)<br>16.15: $SO + ClO \rightarrow SO_2 + Cl$ (74 km)<br>16.17: $SO_3 + CO \rightarrow CO_2 + SO_2$ (37 km)<br>**16.20**: $S_2O_2 + OCS \rightarrow CO + SO_2 + S_2$ (36 km)<br>**16.21**: $2SO_3 + OCS \rightarrow 3SO_2 + CO$ (35–41 km)[3]<br>16.23: $SO + SO \rightarrow SO_2 + S$ (14 km, 74 km) | 16.6: $SO_2 + h\nu \rightarrow SO + O$ (72 km)<br>16.7: $SO_2 + O + M \rightarrow SO_3 + M$ (66 km)<br>16.22: $SO_2 + CO \rightarrow CO_2 + SO$ (15 km) |
| SO | 16.28: $S + O_2 \rightarrow SO + O$ (100 km)[2]<br>**16.6:** $SO_2 + h\nu \rightarrow SO + O$ (72 km)<br>16.36: $OCS + O \rightarrow SO + CO$ (65 km)<br>16.22: $SO_2 + CO \rightarrow CO_2 + SO$ (15 km) | 16.27: $SO + h\nu \rightarrow S + O$ (105 km)[2]<br>16.13 : $SO + O + M \rightarrow SO_2 + M$ (73 km)<br>**16.14** : $SO + NO_2 \rightarrow SO_2 + NO$ (70 km)<br>16.15 : $SO + ClO \rightarrow SO_2 + Cl$ (74 km)<br>**16.25**: $SO + SO + M \rightarrow S_2O_2 + M$ (70 km)<br>**16.23**: $SO + SO \rightarrow SO_2 + S$ (14 km, 74 km) |
| $SO_3$ | 16.7: $SO_2 + O + M \rightarrow SO_3 + M$ (66 km)<br>**16.5**:<br>$H_2SO_4 + H_2O \rightarrow SO_3 + H_2O + H_2O$ (36 km) | **16.8**:<br>$SO_3 + H_2O + H_2O \rightarrow H_2SO_4 + H_2O$ (66 km)<br>16.17: $SO_3 + CO \rightarrow CO_2 + SO_2$ (37 km)<br>**16.19** : $OCS + SO_3 \rightarrow CO_2 + S_2O_2$ (36 km)<br>**16.20** : $S_2O_2 + OCS \rightarrow CO + SO_2 + S_2$ (36 km)<br>**16.21** : $2SO_3 + OCS \rightarrow 3SO_2 + CO$ (35–41 km)[3] |
| $H_2SO_4$ | **16.8**:<br>$SO_3 + H_2O + H_2O \rightarrow H_2SO_4 + H_2O$ (66 km) | **16.5**:<br>$H_2SO_4 + H_2O \rightarrow SO_3 + H_2O + H_2O$ (36 km) |
| OCS | **16.31**: $S + CO + M \rightarrow OCS + M$ (9 km)<br>**16.32**: $S_2 + CO \rightarrow OCS + S$ (5 km) | **16.35**: $OCS + h\nu \rightarrow CO + S$ (59 km)<br>16.36: $OCS + O \rightarrow SO + CO$ (65 km)<br>**16.19** : $OCS + SO_3 \rightarrow CO_2 + S_2O_2$ (36 km)<br>**16.20** : $S_2O_2 + OCS \rightarrow CO + SO_2 + S_2$ (36 km)<br>**16.21** : $2SO_3 + OCS \rightarrow 3SO_2 + CO$ (35–41 km)[3]<br>**16.33**: $OCS + S \rightarrow CO + S_2$ (28 km)<br>16.40: $OCS + S_2 \rightarrow CO + S_3$ (3 km)<br>16.39: $OCS + M \rightarrow CO + S + M$ (2 km) |
| S | 16.27: $SO + h\nu \rightarrow S + O$ (105 km)[2]<br>**16.26**: $S_2O_2 + h\nu \rightarrow SO_2 + S$ (70 km)<br>16.35: $OCS + h\nu \rightarrow CO + S$ (59 km)<br>16.38: $S_3 + h\nu \rightarrow S_2 + S$ (14 km)<br>16.37: $S_4 + h\nu \rightarrow S_3 + S$ (18 km)<br>**16.23**: $SO + SO \rightarrow SO_2 + S$ (14 km)<br>**16.32**: $S_2 + CO \rightarrow OCS + S$ (5 km)<br>16.39: $OCS + M \rightarrow CO + S + M$ (2 km) | 16.28: $S + O_2 \rightarrow SO + O$ (100 km)[2]<br>**16.33**: $OCS + S \rightarrow CO + S_2$ (28 km)<br>**16.31**: $S + CO + M \rightarrow OCS + M$ (9 km) |

(continued)

| $S_2$ | **16.20**: $S_2O_2 + OCS \rightarrow CO + SO_2 + S_2$ (36 km)<br>**16.33**: $OCS + S \rightarrow CO + S_2$ (28 km)<br>16.38: $S_3 + hv \rightarrow S_2 + S$ (14 km) | 16.40: $OCS + S_2 \rightarrow CO + S_3$ (3 km)<br>**16.32**: $S_2 + CO \rightarrow OCS + S$ (5 km) |
|---|---|---|
| $S_3$ | 16.37: $S_4 + hv \rightarrow S_3 + S$ (18 km)<br>16.40: $OCS + S_2 \rightarrow CO + S_3$ (3 km) | 16.38: $S_3 + hv \rightarrow S_2 + S$ (14 km) |
| $H_2S$ | **16.56**: $HS + HCl \rightarrow H_2S + Cl$ (26 km) | **16.51**: $H_2S + S \rightarrow SH + SH$ (18 km) |
| $S_2O$ | 16.46: cis $-$ $S_2O_2 + SO \rightarrow S_2O + SO_2$ (65 km)<br>16.47: $SCl + SO \rightarrow S_2O + Cl$ (64 km)[3] | 16.48: $S_2O + SO \rightarrow S_2 + SO_2$ (~65 km)[4] |
| $SO_2Cl_2$ | 16.58: $ClSO_2 + ClSO_2 \rightarrow SO_2Cl_2 + SO_2$ (24 km) | 16.60: $SO_2Cl_2 + S \rightarrow SO_2 + SCl$ (9 km)<br>16.61: $SO_2Cl_2 + Cl \rightarrow SO_2 + Cl_2$ (4 km) |

[1]The most significant reactions are in bold. Values in parentheses represent the mean altitudes of reaction yield from Krasnopolsky (2012, 2013). M denotes a neutral molecule ($CO_2$, $N_2$)
[2]The reactions and altitudes are sourced from Dai et al. (2024)
[3]The reaction and altitude are derived from Bierson and Zhang (2020) and Dai et al. (2024)
[4]Reactions are cited from Francés-Monerris et al. (2022)

**Table 16.8** Estimated degree of sulfatization of solid materials sampled at the landing sites of Venera 13, Venera 14, and Vega 2[1]

| | Venera 13 | | Venera 14 | | Vega 2 | |
|---|---|---|---|---|---|---|
| | 1 | 2 | 1 | 2 | 1 | 2 |
| Magnetite, vol% | 4–5 | | 4–5 | | 3–4 | |
| Anhydrite, vol% | 13 | 2.8 | 18 | 1.5 | 14 | 8.1 |
| Sulfur, wt% | 3.0 | 0.65 ± 0.4 | 4.3 | 0.35 ± 0.3 | 3.4 | 1.9 ± 0.6 |
| Sulfur, wt% (in sulfate)[2] | | 0.57 ± 0.4 | 4.3 | 0.27 ± 0.3 | 3.4 | 1.82 ± 0.6 |
| Ca, wt% | 5.1 ± 0.68 | | 7.4 ± 0.86 | | 5.4 ± 0.50 | |
| Degree of sulfatization[3] | | | | | | |
| Sulfur (factual in sulfate)/Ca | 0.14 ± 0.12 | | 0.046 ± 0.056 | | 0.42 ± 0.18 | |
| Sulfur (factual in sulfate)/Sulfur (at equilibrium) | 0.19 ± 0.13 | | 0.063 ± 0.070 | | 0.54 ± 0.18 | |

[1]The estimates are based on measurements from Table 16.3 and calculations of chemical equilibria in multicomponent gas-solid systems informed by the composition of solids at landing sites and the atmospheric composition of Venus (Barsukov et al. 1986c). Case (1) corresponds to the mineralogy of surface materials in equilibrium with the atmosphere open (fixed $f$ gases) with respect to $CO_2$, CO, $H_2O$, $SO_2$, HCl, and HF. The calculated sulfur content indicates the sequestration of atmospheric sulfur. Case (2) is for mineralogy equilibrated with the atmosphere open to $CO_2$, CO, and $H_2O$. In this case, the sulfur content corresponds to the measured values
[2]The estimated amount of sulfate sulfur in solid samples reflects 0.08 wt% sulfur in magmatic sulfides, assumed by analogy with the glass parts of submarine erupted basalts (Moore and Fabbi 1971)
[3]The degree of sulfatization corresponds to atomic ratios of supposedly sulfate sulfur in analyzed samples to Ca and sulfur in mineral assemblages equilibrated with atmospheric $SO_2$. The higher degree of sulfatization in the last row reflects the presence of plagioclase ($An_{20-30}$) in equilibrium assemblages and relates to the alteration of phases that are unstable with respect to sulfatization (e.g., Ca-bearing pyroxenes)

**Table 16.9** Experimental studies relevant to interactions of sulfur-bearing gases with geological materials on the surface of Venus[1]

| Temperature, pressure, and duration | Reacting gases and mineral buffers | Reacting solids | Alteration products | Notes | References |
|---|---|---|---|---|---|
| 873–1123 K, 1 bar, 8 days | $SO_2$, $CO_2$, trace $O_2$ | Calcite, diopside | Anhydrite | Faster sulfatization of calcite | Fegley and Prinn (1989), Fegley and Treiman (1992a) |
| 733 K, 4.8 bars, 4–5 days | $SO_2$ | Calcite, pyroxenes, basaltic glass, dunite, anorthosite | Anhydrite | Faster sulfatization of calcite, unaltered dunite | Aveline et al. (2011) |
| 733 K, 92 bars, 42 and 80 days | $CO_2$, $SO_2$, OCS, $H_2O$, CO, $H_2S$, HCl, HF | Calcite, wollastonite, diopside, enstatite, aegirine, jadeite, labradorite, silicate glasses, Py, Pyh | Ca, Na, Cu, and ferrous sulfates, halite, sylvite, Fe oxides | Na sulfates formed from Na-rich pyroxenes and glasses. Unreactive enstatite, labradorite, olivine, and Mag | Radoman-Shaw (2019), Radoman-Shaw et al. (2022) |
| 748 K, ~90 bars, 2–30 days | $CO_2$, CO, $SO_2$, $H_2S$, ± $H_2O$ | Basalts, basaltic pumice and glasses, and obsidian | Anhydrite, thenardite, Ca-Na sulfate, Fe oxides | No/minor alteration of plagioclase and pyroxene. | Berger et al. (2019) |
| 973 K, 90 bars, 14 days; 743 K, 90 bars, 22 and 30 days | $CO_2$, Mag-Py-Hem assemblage | Tholeiitic and alkaline basaltic glasses | Anhydrite, thenardite, minor Fe oxides, Pyh | All glasses were altered, alkaline glasses altered more | Reid et al. (2024) |
| 733 K, 93 bars, 11 and 30 days | $CO_2$, $SO_2$, OCS, $H_2O$, CO, $H_2S$, HCl, HF | Calcite, Mag, Hem, Py, chalcopyrite, MORB basaltic glass, basalt, granite | Anhydrite, Fe sulfides | Anhydrite from calcite, sulfidation of Fe oxides, minor glass change, unaltered pyrite, olivine, pyroxene, feldspars | Santos et al. (2023) |
| 733 K, 1 bar, 6 days | $CO_2$, $SO_2$ | Calcite, wollastonite, anorthite, tremolite | Anhydrite | Calcite is the most reactive, anorthite is less reactive | Port et al. (2025) |

(continued)

| Temperature, pressure, and duration | Reacting gases and mineral buffers | Reacting solids | Alteration products | Notes | References |
|---|---|---|---|---|---|
| 733 K, 93 bars, 60 days | $CO_2$, $SO_2$, OCS, $H_2O$, CO, $H_2S$, HCl, HF | Troilite, Pyh, Py, Mag, Hem, Fe-metal | Py, Mag | Secondary solids approached the Mag-Py buffer | Santos et al. (2024) |
| 733 K, 93 bars, 10 days | $CO_2$, $SO_2$, OCS, $H_2O$, CO, $H_2S$, HCl, HF | Rock-forming minerals and igneous rocks | Fluorite | Aragonite conversion to fluorite, halite disappeared | Longo (2024) |
| 664–804 K, 1 bar, 0–20 days | $CO_2$, CO-$CO_2$, $H_2$-$CO_2$, CO-$CO_2$-$SO_2$ | Py | Pyh, Mag, Hem, maghemite | Faster Py-to-Pyh transition than oxidation of Pyh | Fegley et al. (1995) |
| 665–864 K, 1 bar, 0.04–7 days | He, $N_2$, CO, $CO_2$-CO, $CO_2$-$SO_2$, $CO_2$-CO-$SO_2$, $H_2S$-$H_2$ | Py | Pyh | Assessed reaction kinetics | Hong and Fegley (1997b) |
| 709–864 K, 1 bar | He, $N_2$, $CO_2$ | Py | Pyh | Measured $pS_n$ over Py | Hong and Fegley (1998) |
| 797–1060 K, 1 bar | $CO_2$ | Pyh | Mag | Assessed reaction kinetics | Treiman and Fegley (1991) |
| 653 K, 42 bars; 773 K, 92 bars, 1–4 days | $CO_2$, $CO_2$-$SO_2$, $CO_2$-OCS | Pyh | Pyh | $fO_2$ may have been too low to destabilize Pyh | Port and Chevrier (2020) |
| 653–733 K, 1 bar, 1 day | $CO_2$, $CO_2$-$SO_2$, $CO_2$-OCS | Pyh | Hem, Mag, troilite, $Fe_2(SO_4)_3$ | Ferric sulfate and troilite are unstable at Venus' conditions | Port and Chevrier (2017a, 2017b) |
| 653–733 K, 1 bar, 1 day | $CO_2$, $CO_2$-$SO_2$, $CO_2$-OCS | Galena (PbS), metacinnabar (HgS) | Anglesite ($PbSO_4$) | Loss of HgS in pure $CO_2$ atmosphere | Port and Chevrier (2017a), Port et al. (2018) |
| 653 K 95 bars<br>698 K, 75 bars<br>733 K, 95 bars, 1 day | $CO_2$, $CO_2$-$SO_2$, $CO_2$-OCS | $Bi_2S_3$ + $Bi_2Te_3$<br>$Bi_2S_3$ + Te<br>Bi + Te + S | $Bi_2Te_2S$ | $Bi_2Te_2S$ and $Bi_2S_3$ from Bi-Te-S mixtures | Port et al. (2020) |

[1]Reactions in $O_2$-bearing conditions are not included. Mag, magnetite; Hem, hematite; Py, pyrite; Pyh, pyrrhotite

**Table 16.10** Alteration products of minerals and glasses resulting from chemical weathering experiments carried out under temperatures, pressures, and gas compositions of the Venus' surface[1]

| Mineral or glass | Major alteration products | References |
|---|---|---|
| Calcite | Anhydrite | Radoman-Shaw (2019), Radoman-Shaw et al. (2022), Santos et al. (2023), Gilmore et al. (2025) |
| Wollastonite | Anhydrite | Radoman-Shaw (2019), Radoman-Shaw et al. (2022) |
| Diopside | Anhydrite | Radoman-Shaw (2019), Radoman-Shaw et al. (2022) |
| Enstatite | No change | Radoman-Shaw (2019), Radoman-Shaw et al. (2022) |
| Ca-Mg-Fe pyroxene | No/minor alteration | Berger et al. (2019), Reid (2021), Santos et al. (2023) |
| Aegirine, jadeite | Na-bearing sulfates | Radoman-Shaw (2019), Radoman-Shaw et al. (2022) |
| Olivine | Fe oxides | Berger et al. (2019) |
| | No alteration | Radoman-Shaw (2019), Radoman-Shaw et al. (2022), Reid (2021) |
| Anorthite | No alteration | Santos et al. (2023) |
| Labradorite | No/minor alteration | Radoman-Shaw (2019), Radoman-Shaw et al. (2022), Berger et al. (2019), Reid (2021), Santos et al. (2023) |
| Alkali feldspar | No alteration | Santos et al. (2023) |
| Basaltic glass | Anhydrite, minor Fe oxides and sulfides | Reid (2021), Reid et al. (2024), Radoman-Shaw et al. (2022), Santos et al. (2023) |
| Alkaline basaltic glass | Thenardite, Na-Ca sulfates, Fe oxides, Fe sulfides | Radoman-Shaw (2019), Radoman-Shaw et al. (2022), Berger et al. (2019), Reid et al. (2024) |
| Obsidian | Thenardite, Ca-Na sulfate | Berger et al. (2019) |
| Pyrrhotite | Fe oxides | Radoman-Shaw (2019) |
| | Pyrite, magnetite | Santos et al. (2024) |
| Pyrite | Fe oxides | Radoman-Shaw (2019) |
| | No alteration | Santos et al. (2023, 2024) |
| | Pyrrhotite | Reid et al. (2024) |
| Chalcopyrite | Cu, Fe, S and Ag, Cl phases | Santos et al. (2023) |
| Magnetite | No alteration | Radoman-Shaw (2019), Radoman-Shaw et al. (2022) |
| | Fe sulfides | Santos et al. (2023) |
| Hematite | Magnetite | Santos et al. (2024) |
| Fe-Ti oxide | Fe, S phase, Fe, S, O phase | Santos et al. (2023) |
| Biotite | Fe sulfide, Fe oxide | Santos et al. (2023) |
| Apatites | No/minor change | Santos et al. (2023), Kohler et al. (2023) |
| Actinolite | No change | Santos et al. (2023) |
| Quartz | No change | Santos et al. (2023) |

[1]Experiments performed at 1 bar and with the participation of one gas are not listed

**Table 16.11** Mixing ratios and fugacity of chemically active gases under the conditions of the magnetite-pyrite equilibrium at 740 K and 95.6 bars[1]

| Gas | Magnetite-pyrite-pyrrhotite | Magnetite-pyrite[2] | Magnetite-pyrite-hematite |
|---|---|---|---|
| $CO_2$ | (0.965) | (0.965) | (0.965) |
| $SO_2$, ppmv | 15 | (150) | 713 |
| $H_2O$, ppmv | (30) | (30) | (30) |
| CO, ppmv | 38 | 16 | 8.9 |
| OCS, ppmv | 46 | 34 | 28 |
| $S_2$, ppmv | 0.10 | 0.32 | 0.70 |
| $H_2S$, ppmv | 0.17 | 0.12 | 0.10 |
| $S_2O$, ppbv | 4.6 | 35 | 137 |
| $H_2$, ppbv | 8.4 | 3.5 | 2.0 |
| $S_3$, ppbv | 0.3 | 1.7 | 5.4 |
| $CS_2$, pptv | 173 | 97 | 66 |
| SO, pptv | 9.2 | 39 | 103 |
| $S_4$, pptv | 1.4 | 13.7 | 65 |
| $S_5$, pptv | 0.48 | 8.5 | 60 |
| HS, pptv | 0.87 | 1.0 | 1.1 |
| $SO_3$, pptv | 0.017 | 0.40 | 3.4 |
| $S_6$, pptv | 0.020 | 0.62 | 6.5 |
| $\log_{10} fS_2$ | −5.01[3] | −4.51 | −4.17 |
| $\log_{10} fSO_2$ | −2.85 | −1.84 | −1.17 |
| $\log_{10} fO_2$ | −22.06 | −21.31 | −20.80[4] |

[1]The gas content, as determined by the Mag-Py buffer (equilibria 16.67, 16.79, and 16.97), is bracketed by the fugacities and concentrations established by the Mag-Py-Pyh (equilibria 16.82 and 16.84) and Mag-Py-Hem (equilibria 16.85 and 16.86) assemblages (Figs. 16.22, 16.23, 16.24, 16.25, and 16.30). Concentrations in parentheses are based on instrumental data (Table 16.1) and serve as anchors for assessing the concentrations and fugacities of other gases. Note that significant uncertainties in gas abundances at the buffers are shown in the listed figures. The assumed surface conditions of Venus fall within the uncertainty range of the Mag-Py-Hem buffer

[2]The gas composition corresponding to the Mag-Py equilibrium at 150 ppmv $SO_2$ aligns with the in situ measurement of $x$CO at 16 ppmv (Table 16.1). This composition is a nominal gas model for Venus' modal radius, consistent with other models (Fegley et al. 1997b; Table 16.2)

[3]$fS_2$ is determined by the Py-Pyh equilibrium 16.83 after Hong and Fegley (1998)

[4]$fO_2$ is set by the Mag-Hem equilibrium 16.66. The error bar for $fO_2$ at equilibrium 16.66 is 0.8 log units, due to thermodynamic data (Fegley et al. 1997b)

**Table 16.12** Abundance and sources of nitrogen, carbon, and sulfur in the atmosphere-crust system of Venus

| | Nitrogen | Carbon | Sulfur |
|---|---|---|---|
| Earth's atmosphere, crust, and hydrosphere (kg, g/g, moles normalized to carbon)[1] | $5.3 \times 10^{18}$<br>$8.9 \times 10^{-7}$<br>$4.0 \times 10^{-2}$ | $1.1 \times 10^{20}$<br>$1.9 \times 10^{-5}$<br>1 | $2.2 \times 10^{18}$<br>$3.7 \times 10^{-7}$<br>$7.3 \times 10^{-3}$ |
| Venus' atmosphere and crust (kg, g/g, moles normalized to carbon)[2] | $1.1 \times 10^{19}$<br>$2.3 \times 10^{-6}$<br>$7.7 \times 10^{-2}$ | $1.3 \times 10^{20}$<br>$2.6 \times 10^{-5}$<br>1 | $(2.4–4.6) \times 10^{18}$<br>$(5.0–9.5) \times 10^{-7}$<br>$(0.73–1.4) \times 10^{-2}$ |
| Venus' atmosphere (kg, moles normalized to carbon)[3] | $1.1 \times 10^{19}$<br>$7.7 \times 10^{-2}$ | $1.3 \times 10^{20}$<br>1 | $5.4 \times 10^{16}$<br>$1.1 \times 10^{-8}$ |
| Venus' space sources (kg per day, kg per 0.5 Ga)[4] | $(1.4–5.4) \times 10^{2}$<br>$(2.4–9.6) \times 10^{13}$ | $(3.3–13) \times 10^{3}$<br>$(0.6–2.4) \times 10^{15}$ | $(1.1–4.2) \times 10^{3}$<br>$(2.0–7.8) \times 10^{14}$ |

[1,2,3]Data for nitrogen and carbon for Earth and Venus are sourced from Volkov (1992). For Earth, sulfur values reflect the 'excess' mass after Rubey (1951). The 'g/g' values are normalized to planetary masses

[2]Venus' crust is assumed to be free of nitrogen and carbon. The values for sulfur correspond to the 'excess' amount calculated by the equations $S_{V-Ex} = (N_V/N_E) \times S_{E-ex}$ and $S_{V-Ex} = (C_V/C_E) \times S_{E-ex}$ represents the mass of 'excess' sulfur on Venus; $N_E$ and $C_E$ denote the masses of nitrogen and carbon in the Earth's atmosphere and crust, while $N_V$ and $C_V$ indicate the masses of nitrogen and carbon in Venus' atmosphere. $S_{E\text{-}ex}$ refers to the mass of 'excess' sulfur in the upper envelopes of Earth. The $S_{V\text{-}Ex}$ values correspond to the sulfates and sulfides that have formed from the sequestration of degassed sulfur over time

[3]The mass of sulfur in the atmosphere of Venus corresponds to 150 ppmv $SO_2$ (Volkov 1992)

[4]The fluxes correspond to a flux of space material of $(3.1 \pm 1.8) \times 10^4$ kg/day (Carrillo-Sánchez et al. 2020). The nitrogen, carbon, and sulfur values are evaluated by assuming that the flux of cometary dust (~94% of the infall) consists of two-thirds carbon- and nitrogen-free inorganic mass fraction with 7.9 wt% sulfur and one-third organic ($C_{100}H_{80}O_{30}N_{3.5}S_3$) mass fraction containing 5.1 wt% sulfur

**Table 16.13** The planned and proposed investigations that could enhance our knowledge of sulfur and sulfur-bearing compounds in the atmosphere, on the surface, and in interior materials of Venus[1]

| Methods | Missions, instruments, and details on methods | Science targets relevant to sulfur |
|---|---|---|
| *Telescopic studies* | | |
| UV spectroscopy | Earth-orbiting CLOVE cubsat, 0.32 μm to near-IR | $SO_2$ and aerosols at cloud top, UV absorber |
| Near-IR Earth-based spectroscopy | E.g., the high-resolution iSHELL spectrometer at the NASA IRTF facility, 2.32 and 2.46 μm spectral windows | $SO_2$, OCS, and CO at ~33–48 km |
| Mid-infrared spectroscopy | TEXES (7.4 and 19 μm) and CSHELL (4.535 μm) spectrometers at NASA ITIR facility | $SO_2$ in the upper clouds, sulfuric acid aerosol |
| Millimeter wave and sub-mm spectroscopy | James Clerk Maxwell Telescope (JCMT), Atacama Large Millimeter Array (ALMA) facility | $SO_2$ and SO at 70–120 km |
| *Orbital and flyby studies* | | |
| UV imaging | DAVINCI CHRIS flyby, VISOR, 0.355–0.375 μm, 20–30 km/pixel | Global dayside coverage, UV absorber, cloud structure and composition |
| UV spectroscopy | EnVision, VenSpec-U, 0.19–0.38 μm 2 nm spectral resolution; 0.205–0.235 μm 0.3 nm spectral resolution | $SO_2$ and SO in the mesosphere, upper clouds UV absorber, cloud dynamics |
| UV-visible spectroscopy | DAVINCI CHRIS flyby, CUVIS, 0.19–0.56 μm, 0.2 nm spectral resolution | $SO_2$ and SO in the mesosphere and upper clouds, UV cloud absorber, cloud features, dynamics, and composition |
| | Venera-D Orbital Module, VOLNA, 0.19–0.59 μm, 0.5 nm spectral resolution | |
| UV-visible imaging | VOM, VCMS | Cloud dynamics, UV absorber |
| UV-visible photometry | VOM, SPAV for solar occultation | Upper cloud structure |
| UV, visible, and near-IR imaging | Venera-D Orbital Module, VMS, 0.365 μm, 0.513 μm, 0.965 μm, 1 μm | Cloud aerosol and UV absorber, surface thermal emissivity and composition |
| Near-IR imaging | DAVINCI CHRIS flyby, VISOR, 0.93–0.938 μm, 0.947–0.964 μm, 0.990–1.030 μm | Regional (Alpha, Ovda) surface thermal emissivity and composition |
| | DAVINCI Zephyr sphere, VenDi, 0.74–1.04 μm, 0.98–1.04 μm; ~1–200 m/pixel | Local surface thermal emissivity and composition |

(continued)

**Table 16.13** (continued)

| Methods | Missions, instruments, and details on methods | Science targets relevant to sulfur |
|---|---|---|
| Near-IR spectroscopy | VERITAS, VEM, six surface[2] bands within 0.86–1.18 μm | Global surface thermal emissivity, morphology, and composition |
| | EnVision, VenSpec-M, six surface bands within 0.86–1.18 μm | Global surface thermal emissivity and composition, atmospheric $H_2O$ vapor |
| | VOM, VSEAM, ~1 μm | Surface thermal emissivity and composition |
| | EnVision, VenSpec-H, 1–2.5 μm, R~8000 | $SO_2$, OCS, CO, and $H_2O$ below the clouds (night side) and above the clouds (dayside) at 0–75 km |
| | VOM, VIRAL for solar occultations, 1.6 μm, 2.3–4.4 μm | $SO_2$, SO, OCS, and $H_2S$ above clouds |
| | Venera-D Orbital Module, VIKA, 1.05–1.65 μm, 2.3–4.3 μm | Gas and aerosol composition |
| | VOM, VASP polarimeter | Cloud microphysics, cloud top altitude |
| IR spectroscopy | Venera-D Orbital Module, SVET Fourier spectrometer, 5–40 μm | $SO_2$, $H_2O$, and cloud composition at 55–75 km |
| Radio occultation sounding | VERITAS, two band (X/X+KaKa) | $SO_2$, $H_2SO_4$(l), and $H_2SO_4$(g) at 45–55 km |
| | EnVision, one-way X-Ka band | |
| Millimeter-wave radiometry | Venera-D Orbital Module, MM-radiometer, 3–30 mm | $H_2SO_4$ (g) and $SO_2$ below clouds |
| Radar imaging | VERITAS, VISAR, 3.8 cm, 15–30 m/pixel | Global topography, morphology, and surface roughness |
| | EnVision, VenSAR, 9.4 cm, 10 m/pixel | Topography, morphology, and roughness |
| | EnVision, VenSAR, polarimetry (30 m/pixel) and microwave radiometry | Microwave emissivity and dielectric properties |
| | EnVision Subsurface Radar Sounder (SRS) | Subsurface structure down to ~1 km, 20 m vertical resolution |
| | VOM, VARSIS, 10–33 m, capable of polarimetric measurements | Microwave emissivity, dielectric properties, subsurface structure down to ~1 km, 10–25 m vertical resolution |
| Radio science | VERITAS radio science/gravity experiment | Gravity field, *k2* tidal Love number, and moment of inertia factor |
| | EnVision radio science/gravity experiment | |
| *In situ atmospheric studies* | | |
| Mass spectrometry | DAVINCI Zephyr sphere, VMS | $SO_2$, OCS, $H_2SO_4$, $H_2S$, and $S_n$, $^{34}S/^{32}S$ in $SO_2$ |

**Table 16.13** (continued)

| Methods | Missions, instruments, and details on methods | Science targets relevant to sulfur |
|---|---|---|
| | Aerosol-Sampling Instrument Package (ASIP) for a balloon mission | Chemical and isotopic composition of cloud aerosols and gases |
| Tunable laser spectrometry | DAVINCI Zephyr sphere, VTLS | $SO_2$, OCS, CO, $^{34}S/^{33}S/^{32}S$ in $SO_2$ and OCS |
| | Venera-D Lander Module, ISKRA-V | |
| $O_2$ sensor | DAVINCI Zephyr sphere, V*f*Ox | $O_2$ fugacity |
| Chromato-mass spectrometry | Venera-D Lander Module, VCS | Chemical and isotopic composition of gases and aerosols at 0–70 km |
| UV spectroscopy | Venera-D Lander Module, DAVUS, 0.25–0.4 μm | $SO_2$, SO, and UV absorbers at 0–70 km |
| *In situ studies of near-surface materials* | | |
| Raman-laser induced breakdown spectroscopy | NASA R-LIBS instrument | Elemental and phase composition of solid materials |
| Alpha-particle X-ray spectrometry | Venera-D Lander Module, APXS-V | Elemental composition of solids and gases |
| X-ray fluorescence spectrometry | Venera-D Lander Module, XRD/XRF | Elemental composition of solids |
| X-ray diffraction | Venera-D Lander Module, XRD/XRF | Solid phase composition |
| Laser ablation mass spectrometry | Venera-D Lander Module, LMS | Elemental and isotopic composition |
| Mössbauer spectroscopy | Venera-D Lander Module, MIMOS II | Mineralogy of Fe oxides and sulfides |

[1]Details can be found in Widemann et al. (2023), Smrekar et al. (2022), Garvin et al. (2022), Zasova et al. (2017), European Space Agency (2021), and I.S.R.O. (2024)

[2]Other bands will be used for cloud correction/investigations and $H_2O$ surface mapping

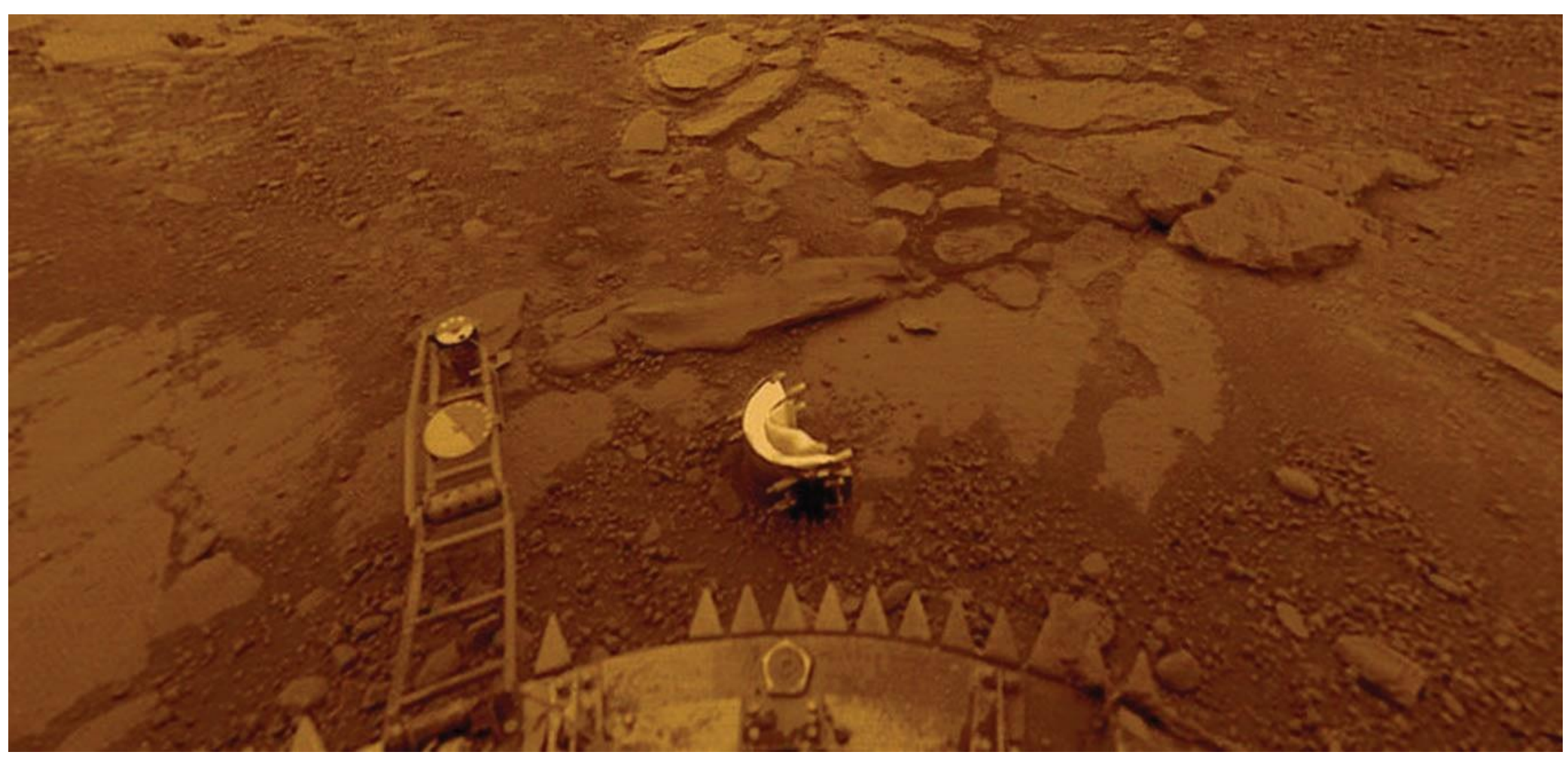

**Fig. 16.1** Layered rocks and rock fragments at the landing site of Venera 13. The original color images (Selivanov et al. 1983; Bokshteyn et al. 1983) were digitally mastered by Don Mitchel. The surface appears black under Venus' conditions, and the color reflects the surface at room temperature (cf., Pieters et al. 1986). The composition of the material is consistent with potassium-rich mafic silicate rock with elevated sulfur content (Table 16.3)

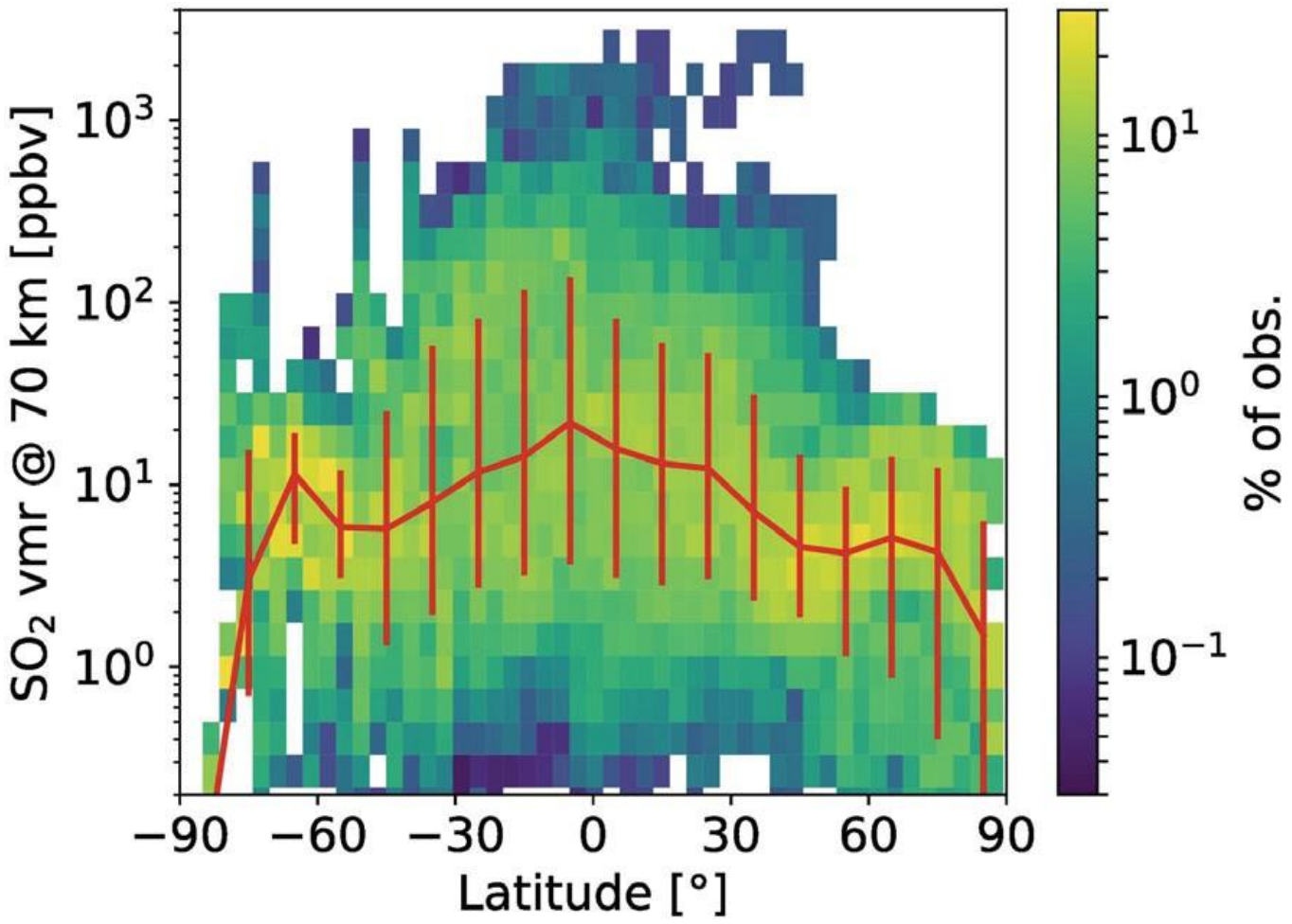

**Fig. 16.2** $SO_2$ mixing ratio at 70 km in the atmosphere of Venus as a function of latitude (Marcq et al. 2020). The red line represents the median value; the red bars indicate one standard deviation. The color pattern depicts the percentage of observations. The data were obtained from the Venus Express SPICAV-UV observations (Table 16.1)

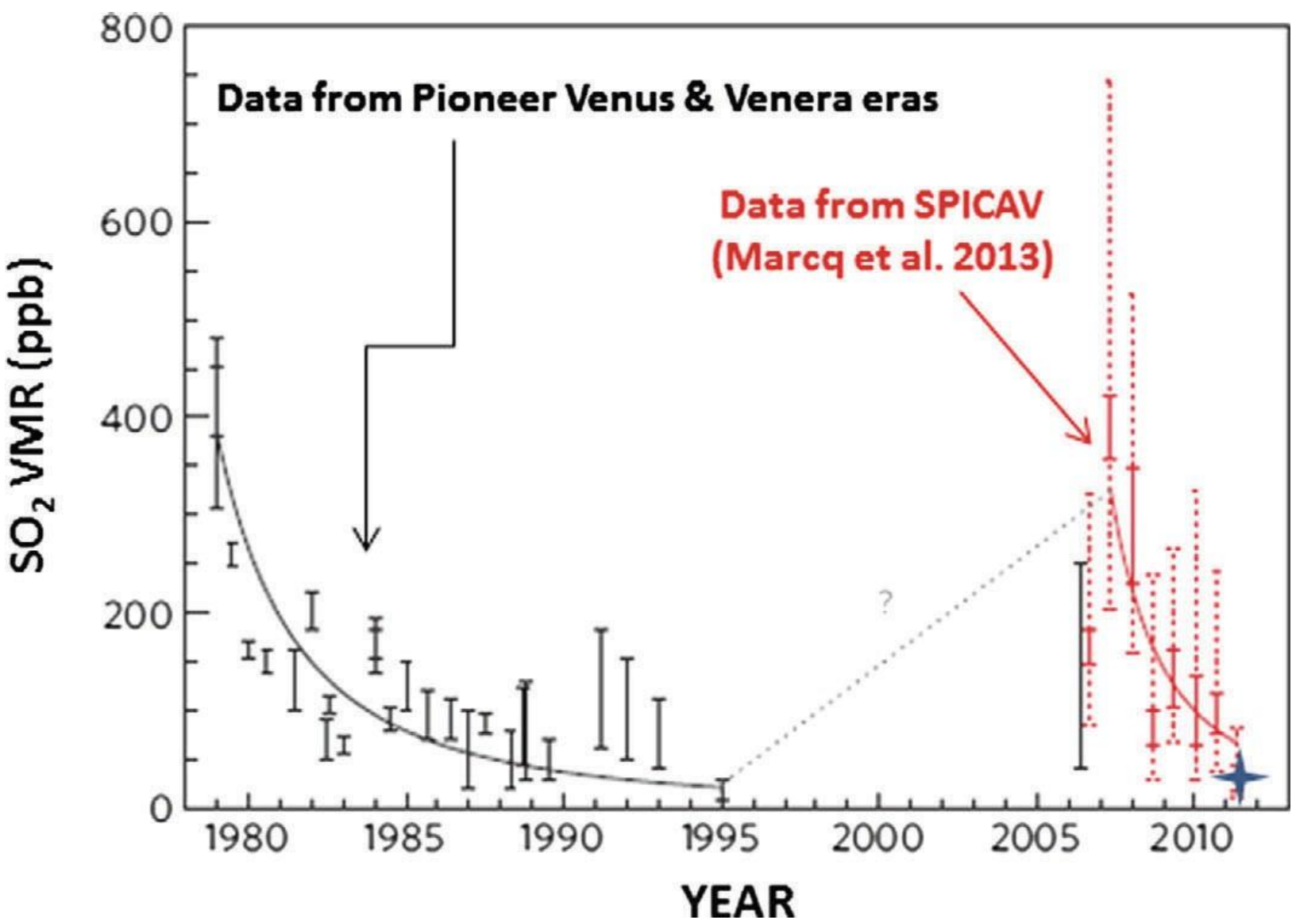

**Fig. 16.3** Long-term changes in the $SO_2$ volume mixing ratio at the cloud top of Venus' atmosphere (Marcq et al. 2013; Jessup et al. 2015). The blue cross represents the average value derived from STIS-HST observations

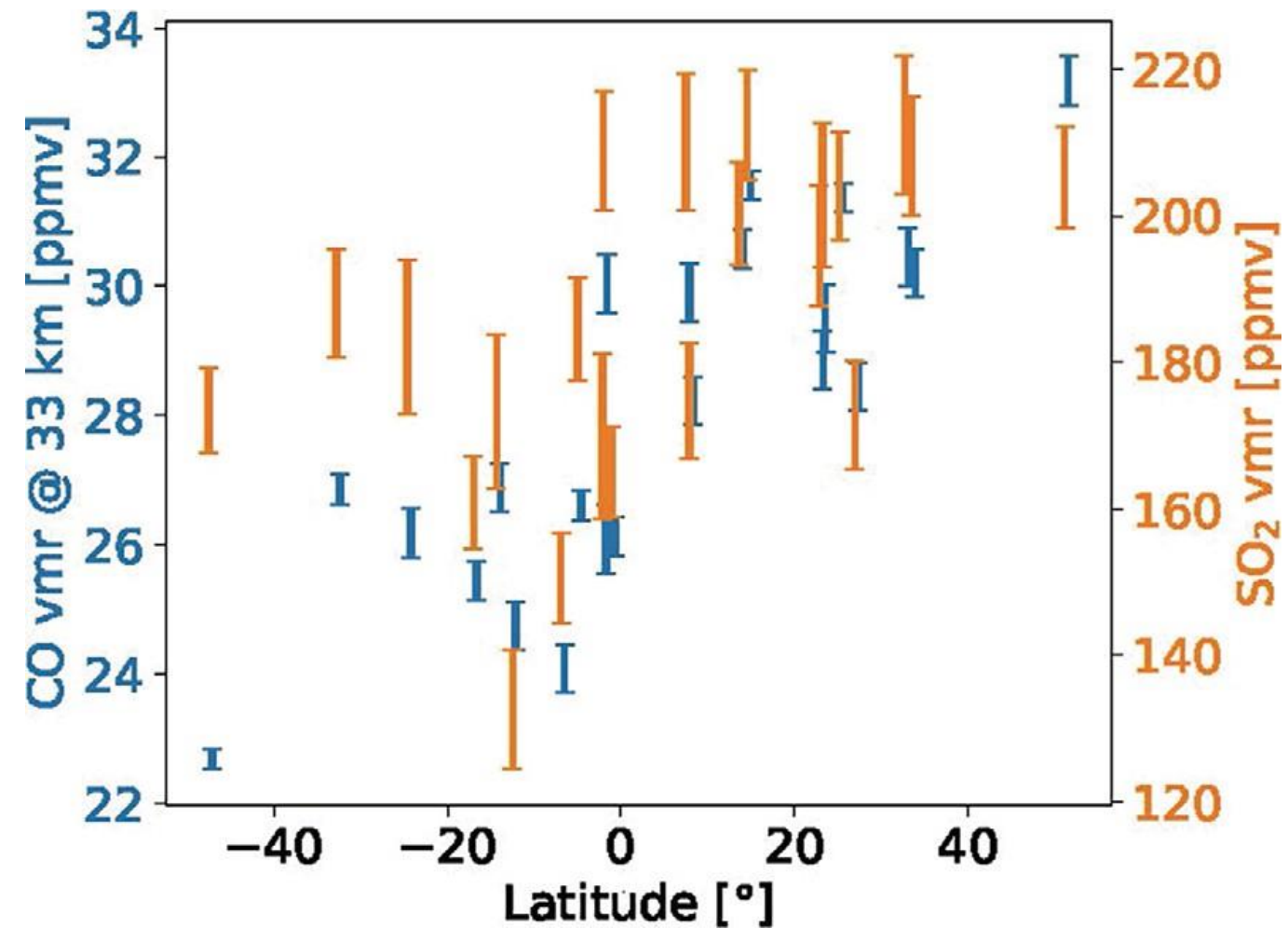

**Fig. 16.4** Latitudinal dependence of $SO_2$ and CO volume mixing ratios at 33 km altitude (Marcq et al. 2021). The data were sourced from ground-based near-IR nightside observations

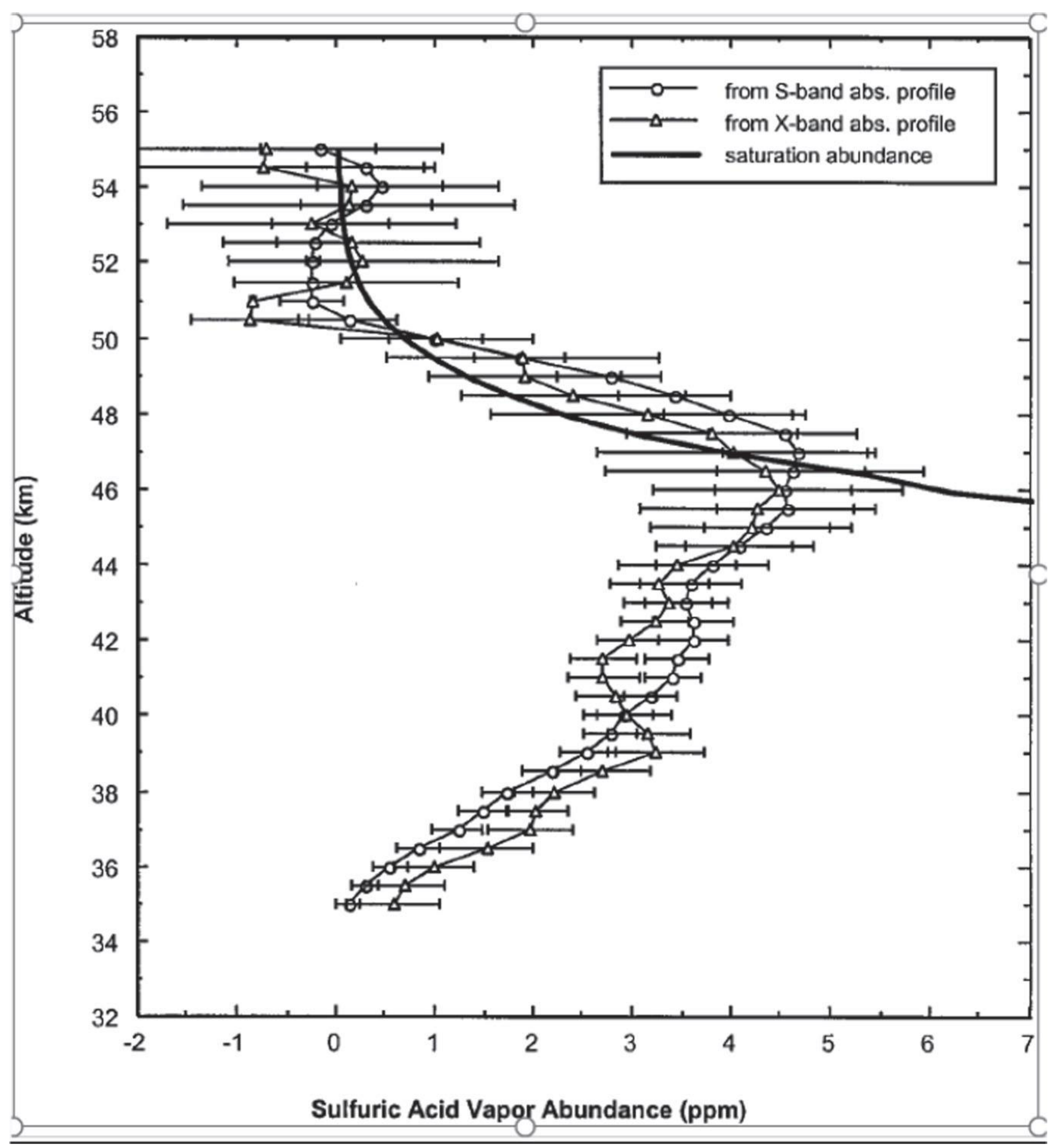

**Fig. 16.5** Abundance profile of $H_2SO_4$ vapor in the lower and middle cloud regions (at 48–55 km) and below clouds derived from the Magellan radio occultation data (Kolodner and Steffes 1998). Within clouds, the $H_2SO_4(g)$ mixing ratio aligns with the $H_2SO_4(g)$ saturation curve (black curve). The decrease in $xH_2SO_4(g)$ below clouds indicates its thermal decomposition into $SO^3$ and $H_2O$ vapor

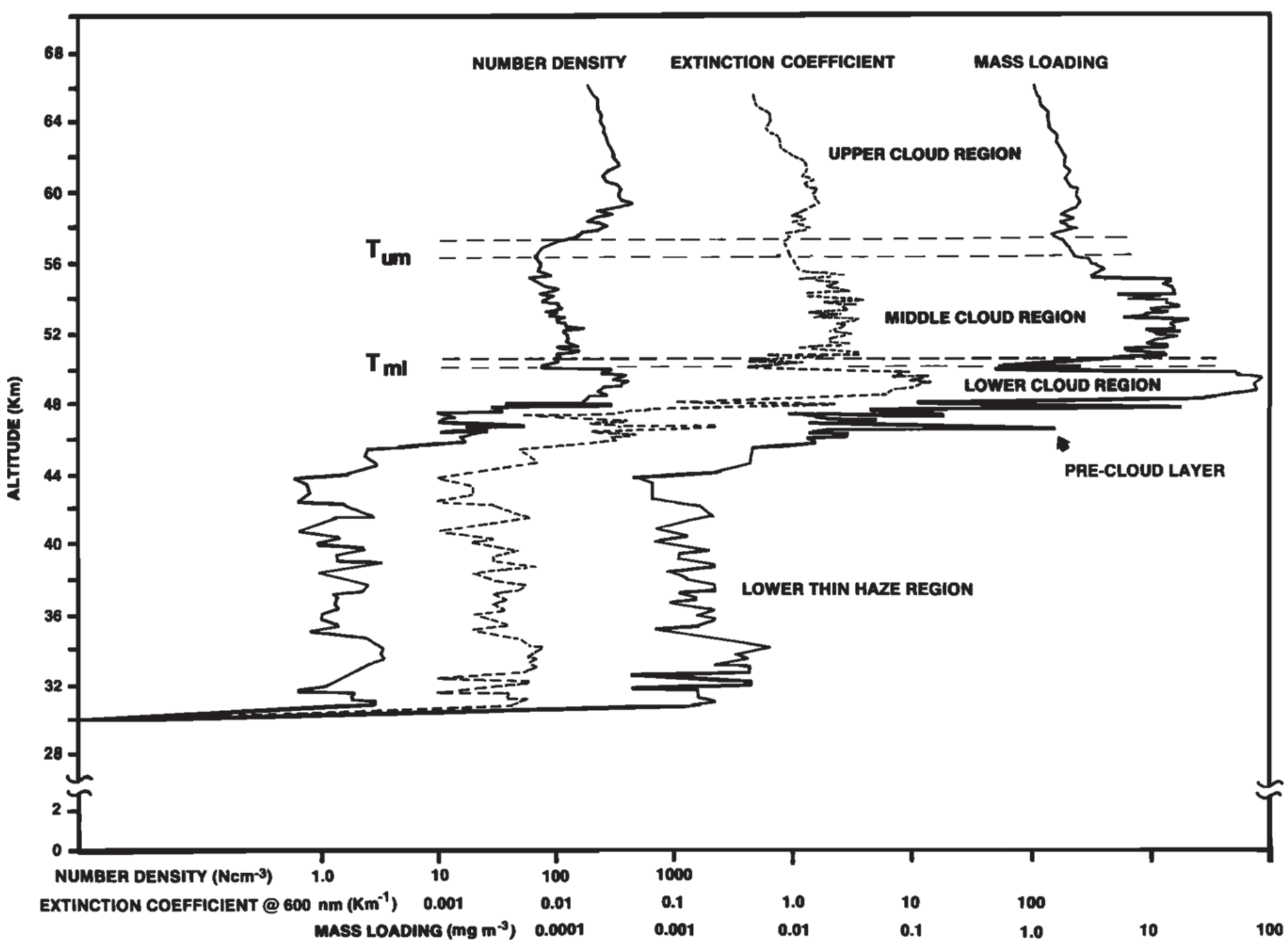

**Fig. 16.6** Vertical structure of Venus' clouds obtained with the Pioneer Venus probes (Knollenberg and Hunten 1980)

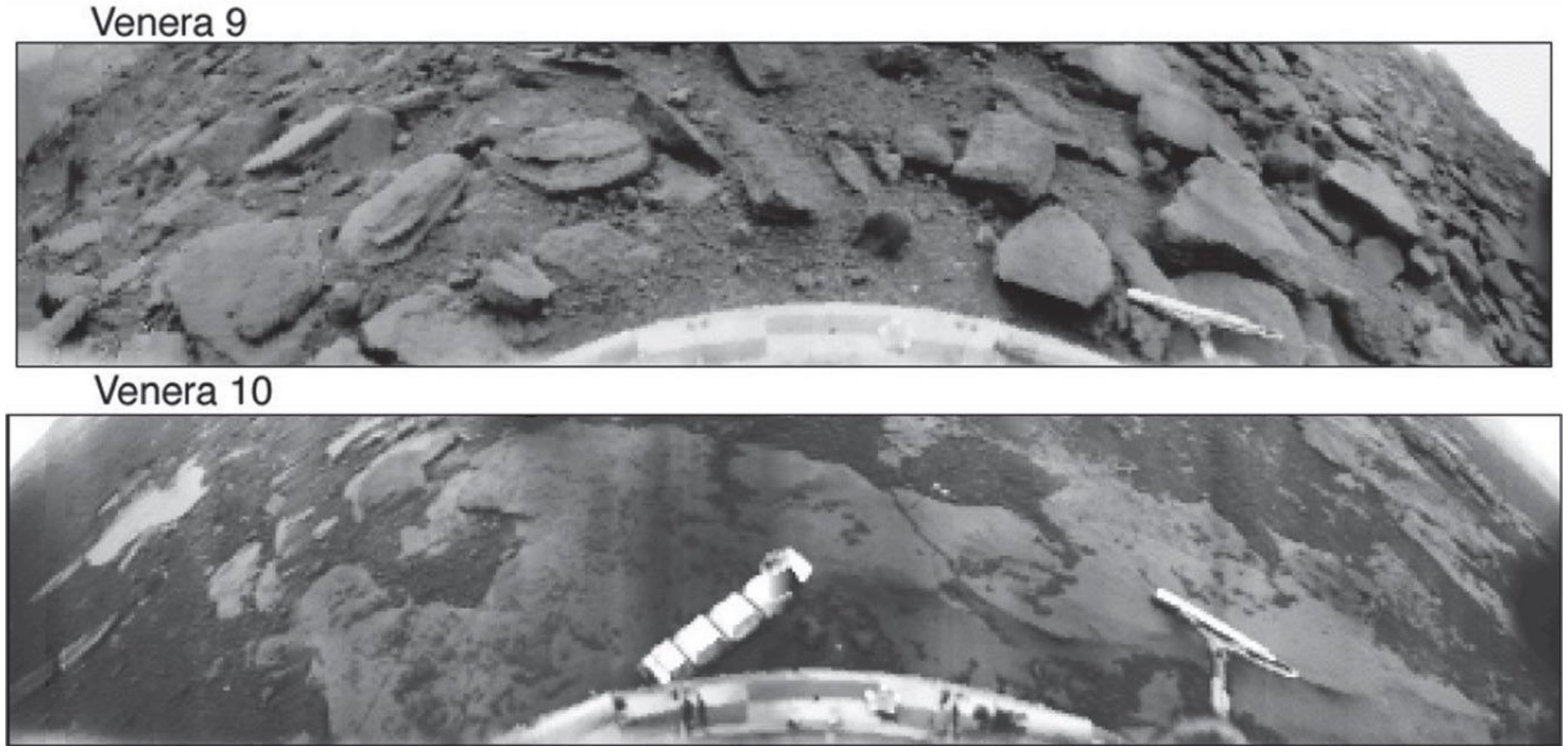

**Fig. 16.7** Images of the surface at the Venera 9 and 10 landing sites

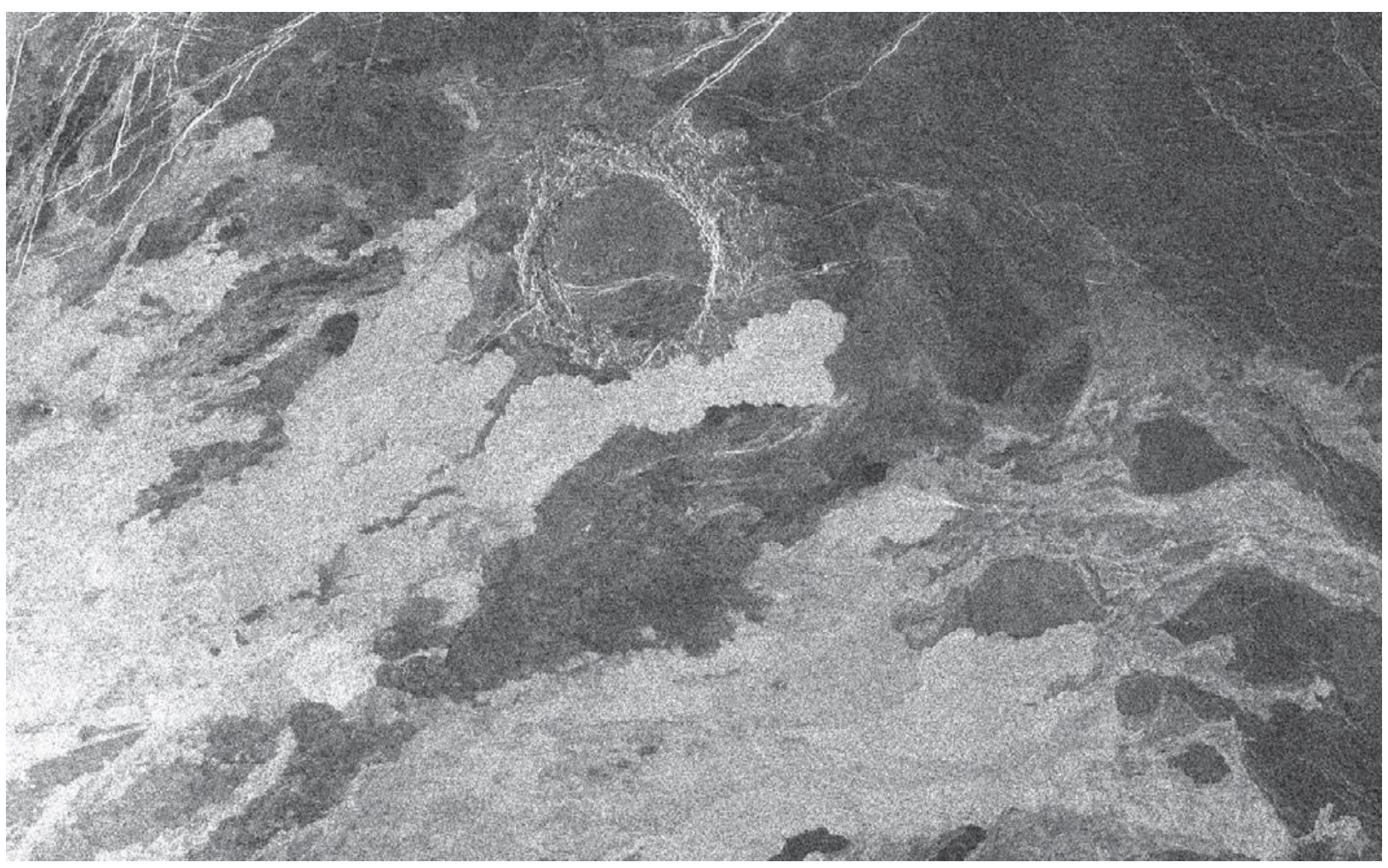

**Fig. 16.8** Magellan radar image of Venus' volcanic plains (image by NASA/JPL). The lava flows appear bright due to their rough surface. The imaged area is 140 km wide and is in the eastern part of Sapas Mons

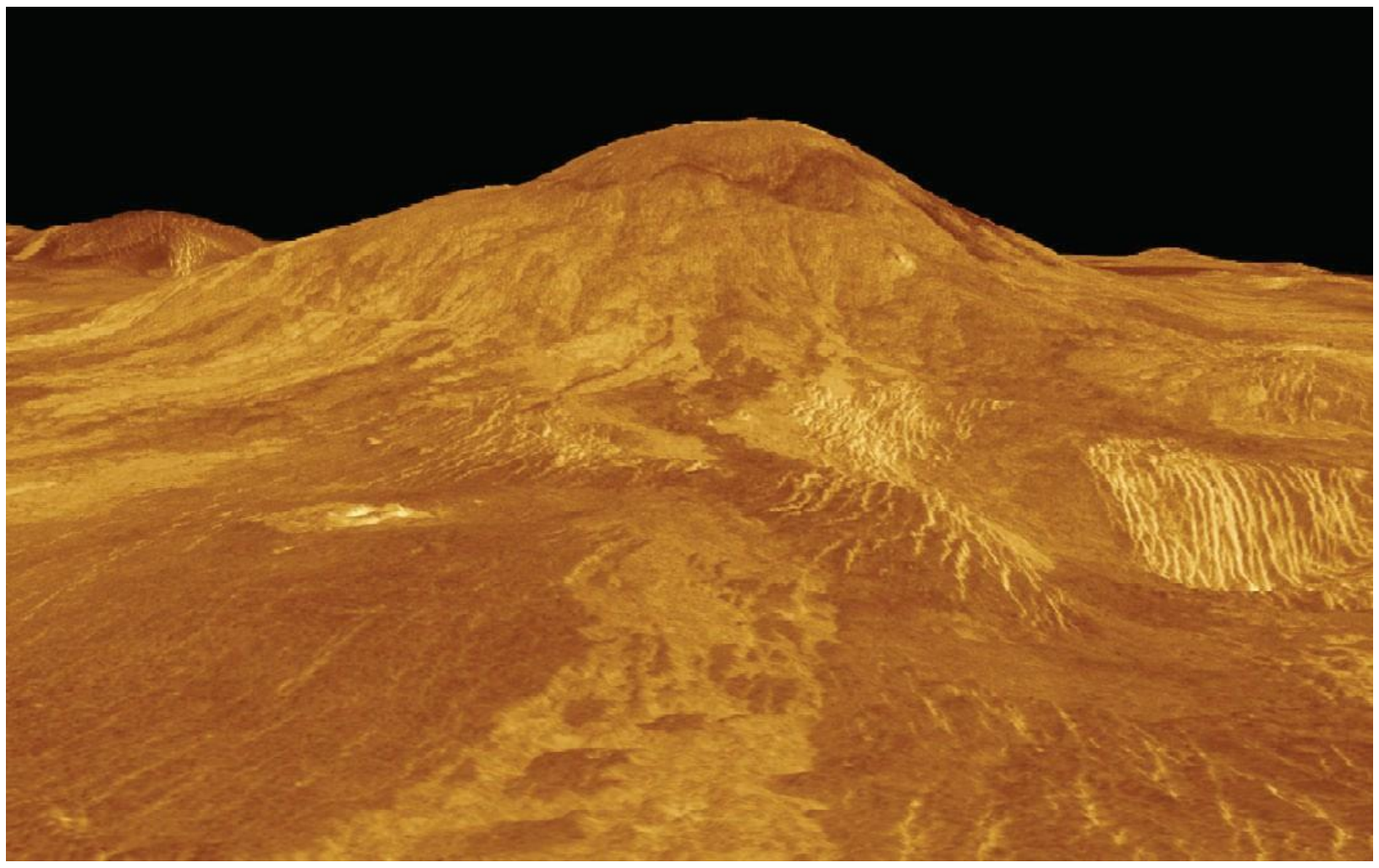

**Fig. 16.9** Computer-simulated radar image of the volcanic center, Sif Mons (NASA/JPL). Sif Mons, depicted in the upper portion of the image, is a volcano that spans ~300 km in diameter and reaches an elevation of ~2 km. Here, Magellan radar data are merged with enhanced radar altimetry. The colors are arbitrary and reflect those recorded by the Venera landers (Fig. 16.1). Similar to Earth's basaltic rocks, the actual surface appears dark in the visible spectrum

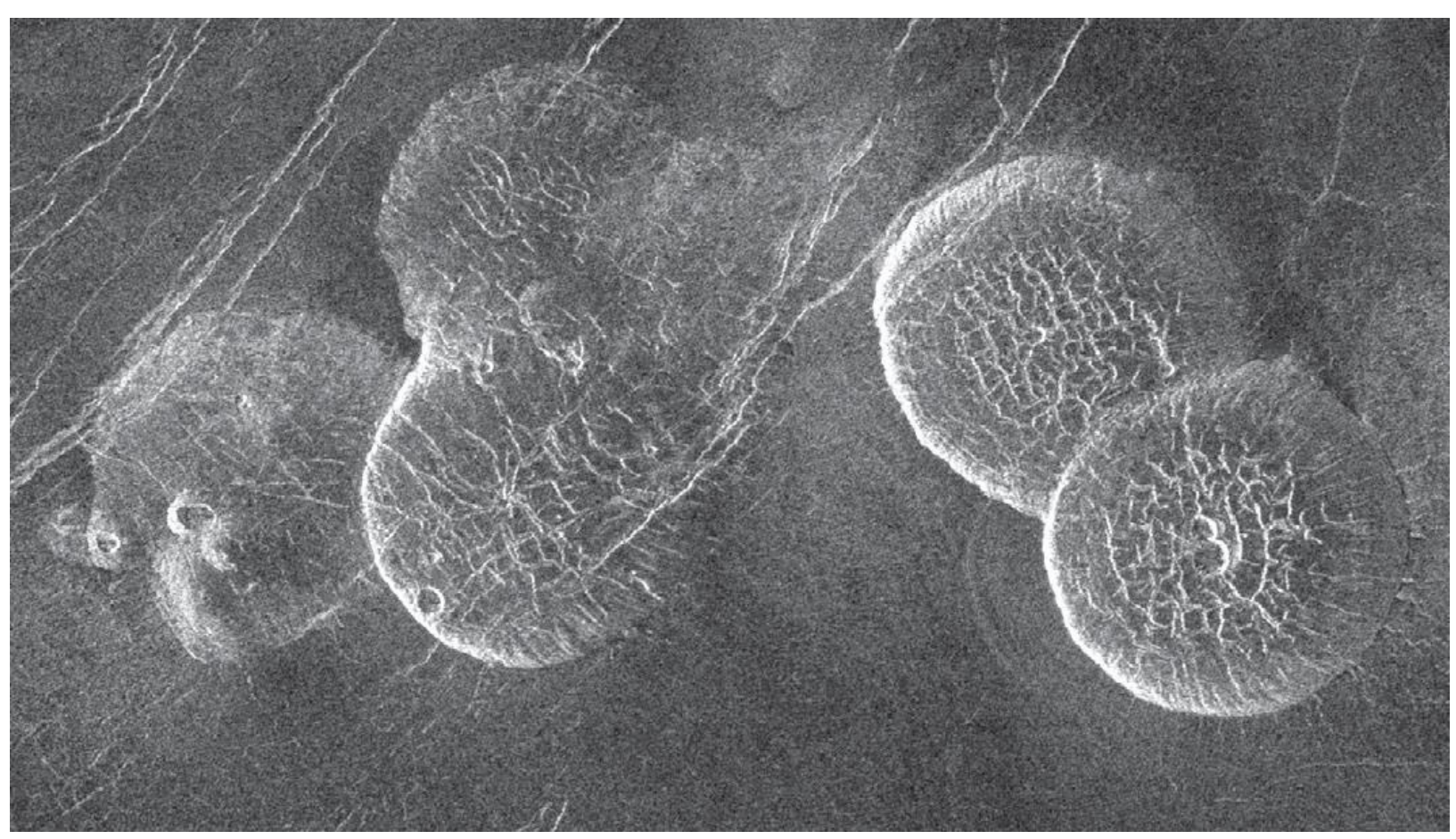

**Fig. 16.10** Magellan radar image of steep-sided domes on Venus' volcanic plains (NASA/JPL). The domes are ~25 km in diameter, stand about 2 km tall, and are located east of Alpha Regio (~29.5° S, 12° E). They may have formed through the eruption of viscous silicate magma

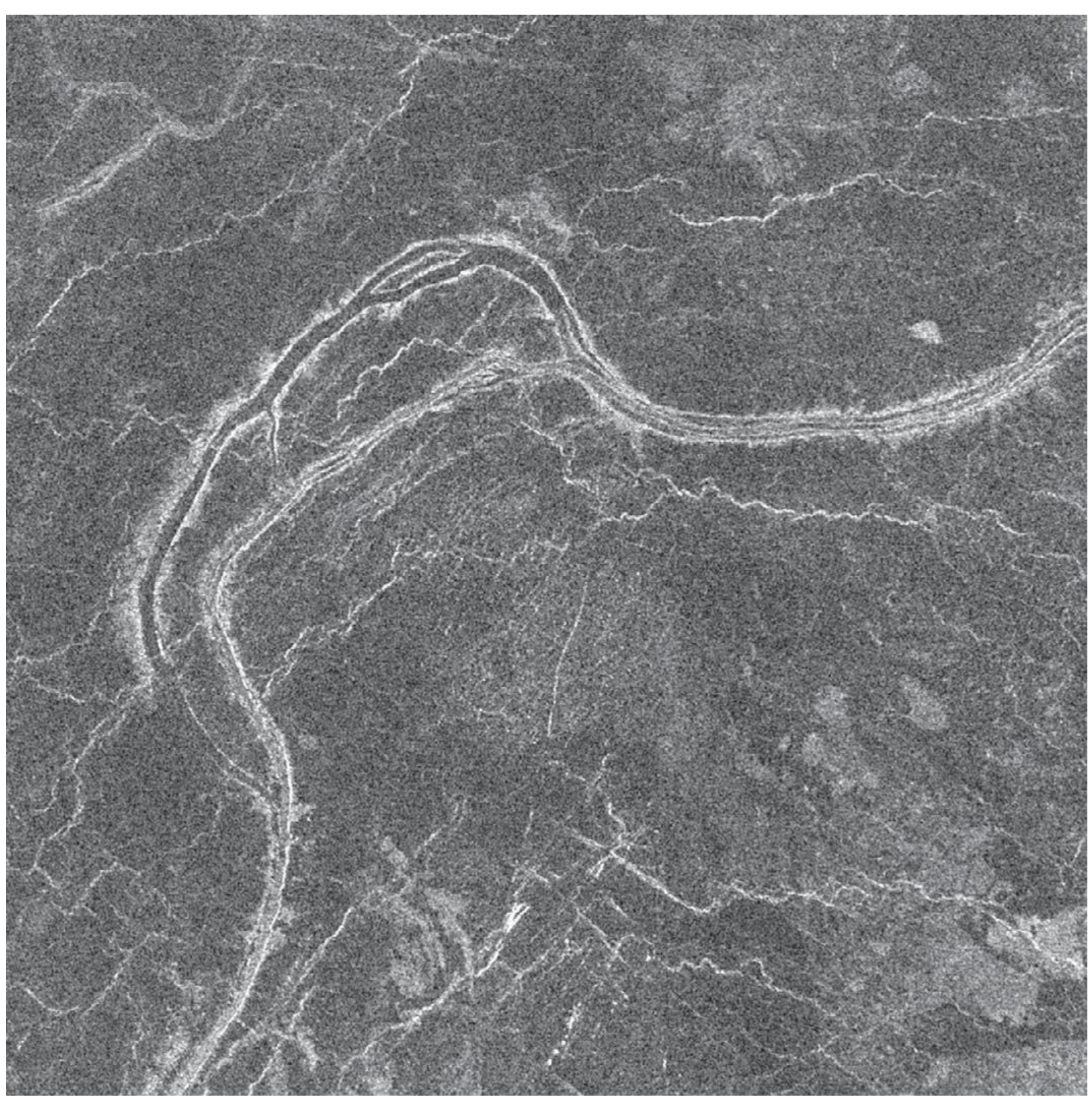

**Fig. 16.11** Magellan radar image of the Baltic Vallis lava channel on Venus (NASA/JPL photo). The channel is approximately 2 km wide and features various branches and islands. The image spans about 50 km in width in Sedna Planitia, south of Ishtar Terra (42.7° N, 340.7° E)

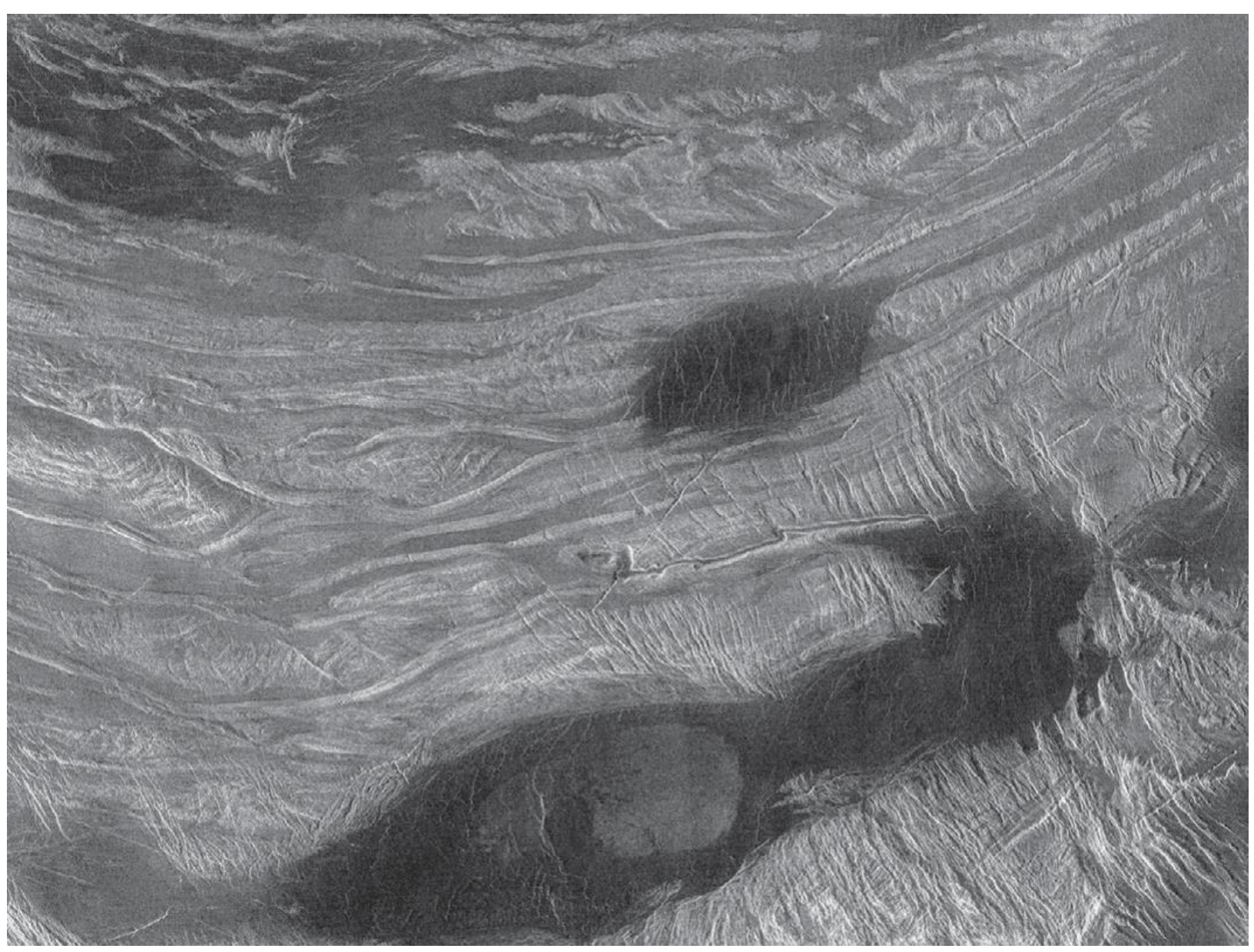

**Fig. 16.12** Magellan radar image of a part of the northern boundary of Ovda Regio on Venus (NASA/JPL photo). The image spans 300 km across. It reveals rounded linear ridges measuring 8–15 km in width and 30–60 km in length. Radar-dark material (lava and/or aeolian deposits) fills the region between the ridges. The apparent layered and deformed structures may represent oceanic sediments altered by metamorphic and/or igneous processes after the end of an aqueous period in Venus' history. Note a canali-type structure in the central part of the image that may have formed from non-silicate magma, such as chloride melt

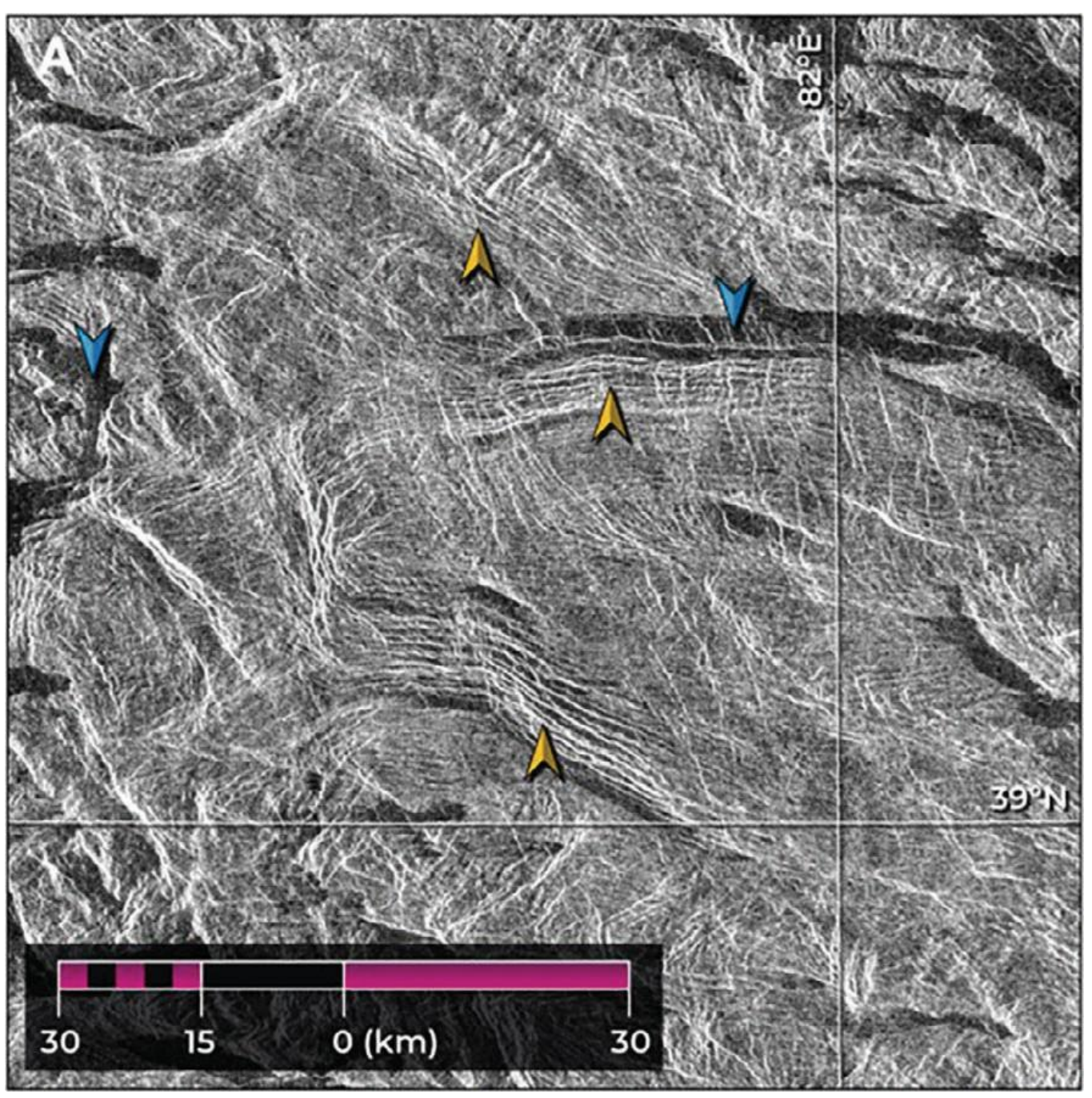

**Fig. 16.13** Magellan radar image of Tellus Tessera on Venus (NASA/JPL). The arcuate lines of high radar backscatter are indicated by gold arrows, while intratessera radar-dark material is highlighted with blue arrows (Byrne et al. 2021)

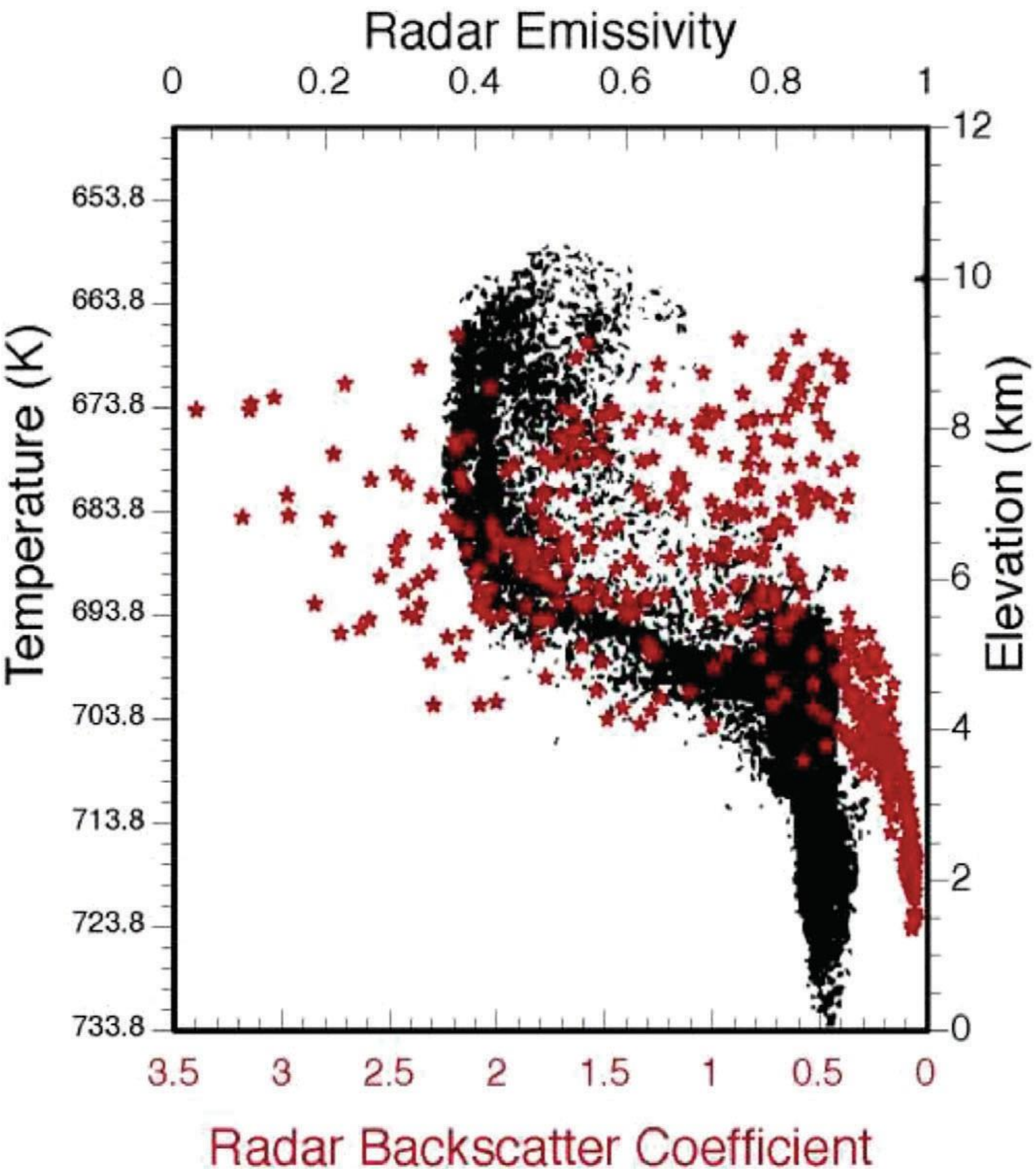

**Fig. 16.14** Radar properties of Maxwell Montes, the highest region on Venus (Treiman et al. 2016). Black symbols represent Magellan radar emissivity for all Maxwell Montes from Klose et al. (1992). Red symbols indicate Magellan SAR radar reflectance coefficients and elevations for the three areas studied by Treiman et al. (2016). The range of radar backscatter values above ∼4 km may reflect the roughness of the topography or volume scattering

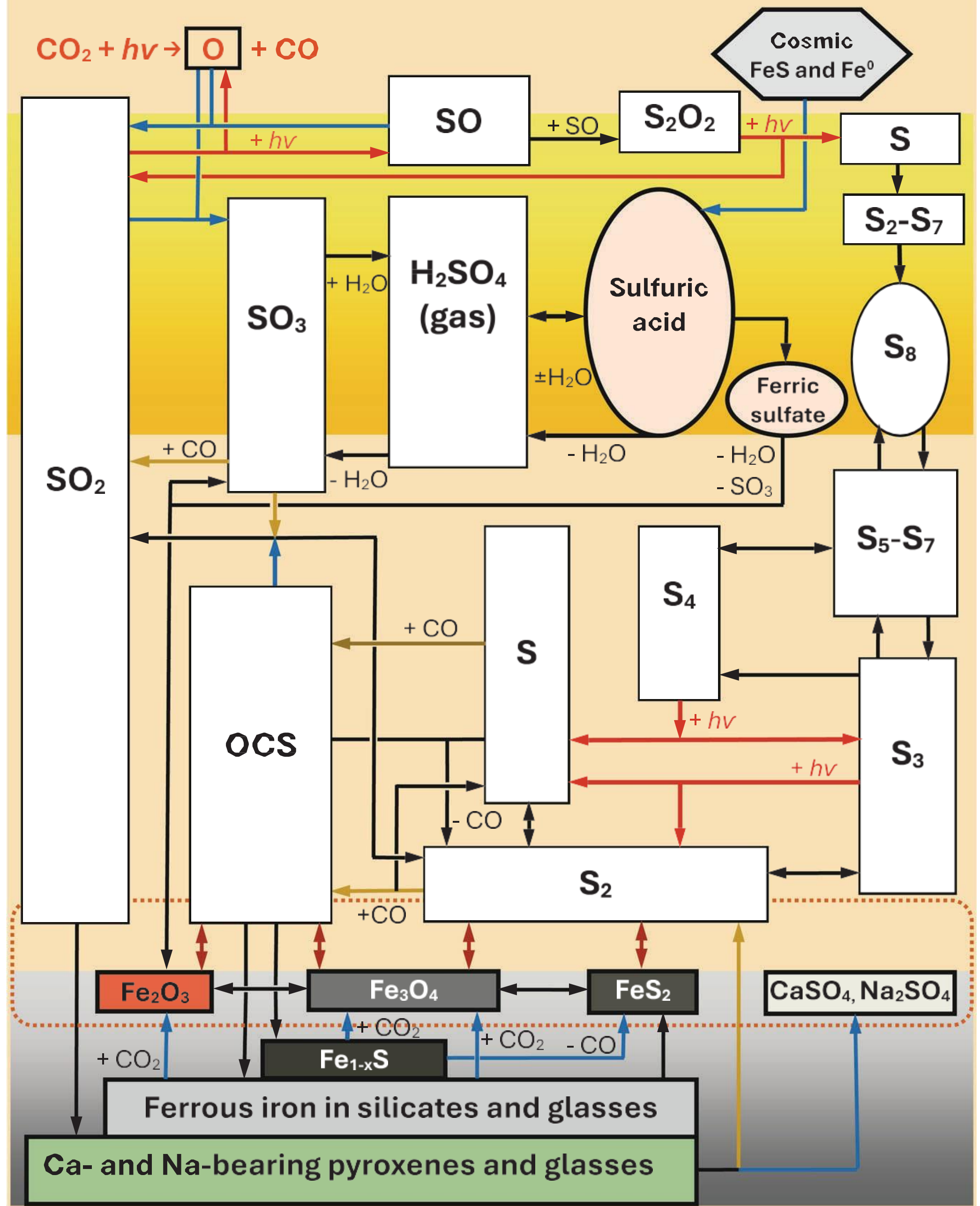

**Fig. 16.15** Chemical reactions involving sulfur-bearing compounds in the middle and lower atmosphere and in a permeable surface layer on Venus. The orange rectangle represents clouds, while the gray box designates surface materials. The dotted oval signifies near-surface conditions where gas phase and gas-solid type chemical equilibria may occur. White rectangles denote gases, and ovals illustrate aerosol components. Red arrows indicate photochemical reactions. Blue and dark yellow arrows represent oxidation and reduction reactions, respectively. Double arrows illustrate thermochemical equilibria between compounds. Only the important species and reaction pathways are shown, with details provided in Tables 16.6 and 16.7. $H_2S$, HS, $S_2O$, sulfur-carbon, and sulfur-oxygen-Cl gases are omitted. Secondary minerals, such as pyrite, anhydrite, thenardite, magnetite, and hematite, which form through gas-solid reactions, are presented; other secondary phases are discussed in Sect. 16.3.2. Some pathways, including the fates of OCS and $S_n$, and gas-solid reactions involving Fe sulfides and oxides, are hypothetical

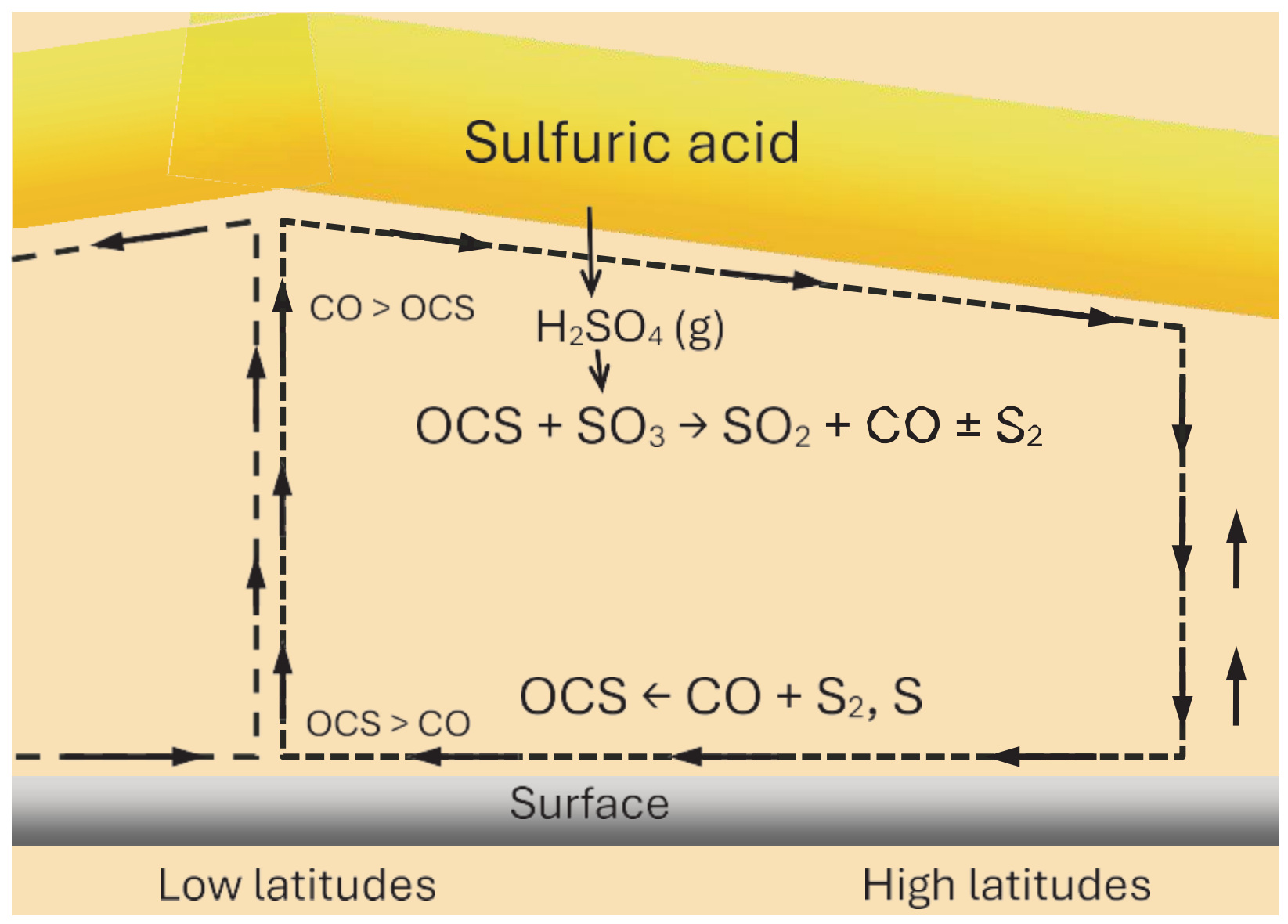

**Fig. 16.16** Fate of major sulfur-bearing species in the Hadley-cell type atmospheric circulation on Venus. Global calculation mainly involves the high-density sub-cloud atmosphere, but it also affects the clouds and the upper atmosphere. Polar upwellings are suspected (Oschlisniok et al. 2021; Marcq et al. 2023). OCS forms through reactions 16.29 and 16.30 involving CO and sulfur gases at altitudes below ~10 km. Low-latitude upwelling transports $SO_2$ and OCS toward the sulfuric acid-rich clouds. Within the clouds, $SO_2$ oxidizes to sulfuric acid (Fig. 16.15; reactions 16.7 and 16.8). OCS reacts with $SO_3$ in the mid-30 km (reactions 16.19, 16.20, and 16.21), resulting in the formation of CO, $SO_2$, and possibly $S_2$. Consequently, concentrations of CO, $SO_2$, and potentially $S_2$ increase with latitude in the middle 30 km. High-latitude downwelling brings CO-, $SO_2$-, and $S_n$-enriched gas to the lower sub-cloud atmosphere. Aside from the possible formation of $S_2$ in reaction 16.20, $S_n$ gases arise from the thermal dissociation of $S_8$ near the cloud deck (Fig. 16.15). The cycle concludes with the formation of OCS in a deep equatorward airflow that may attain chemical equilibrium in a thin near-surface layer. While gas transport through eddy diffusion enables chemical disequilibria that drive reactions throughout the atmosphere (e.g., Krasnopolsky 2007; Bierson and Zhang 2020; Yung et al. 2009), the global circulation leads to latitudinal gradients of OCS, CO, and $SO_2$ that may not be explained solely by eddy diffusion

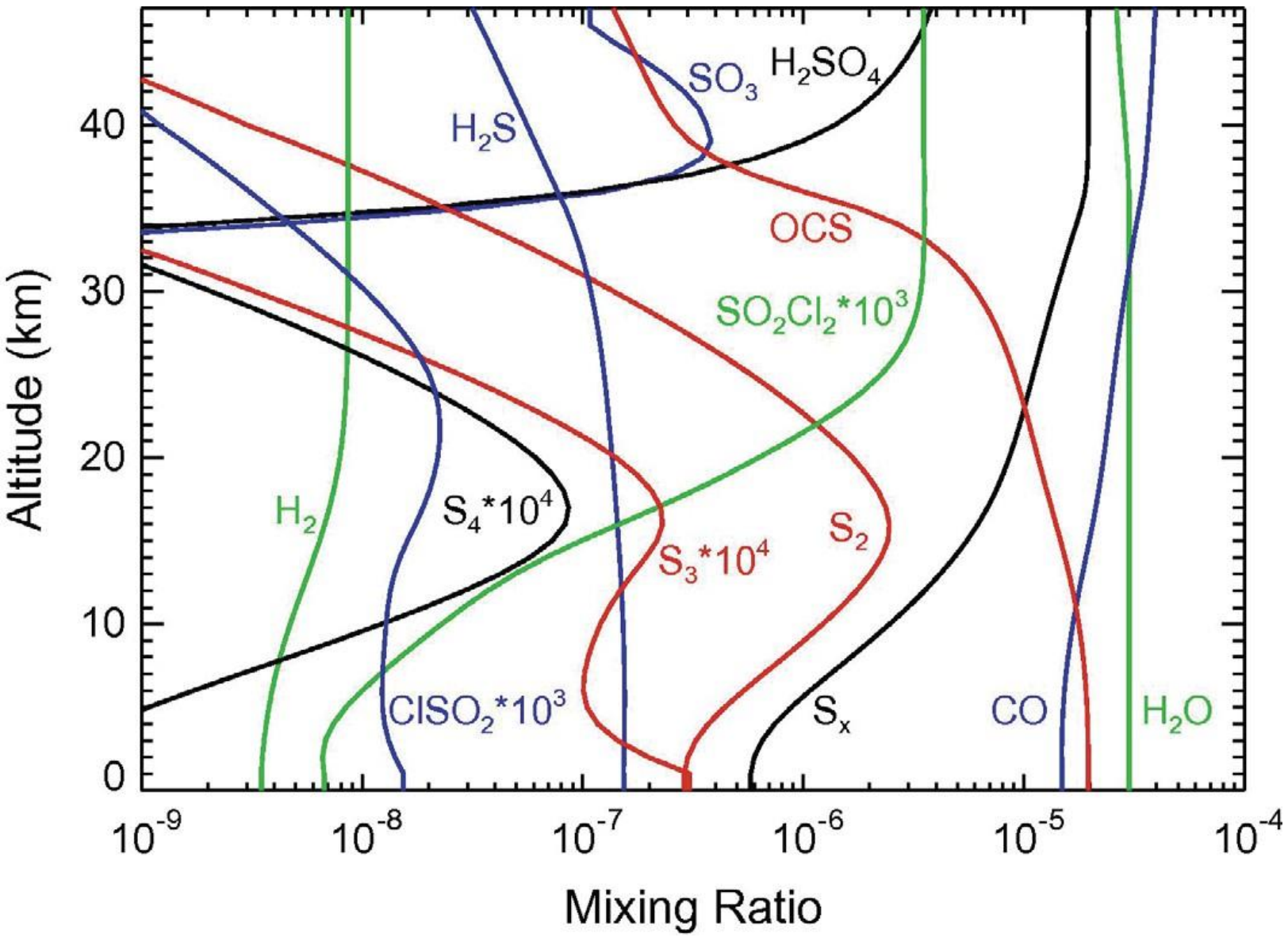

**Fig. 16.17** The modeled chemical composition of the lower atmosphere of Venus (Krasnopolsky 2013). Mixing ratios of $SO_2$ (130 ppmv) and HCl (0.5 ppmv) are constant and not shown. $S_x$ stands for the sum of $S_n$ (n = 1–8)

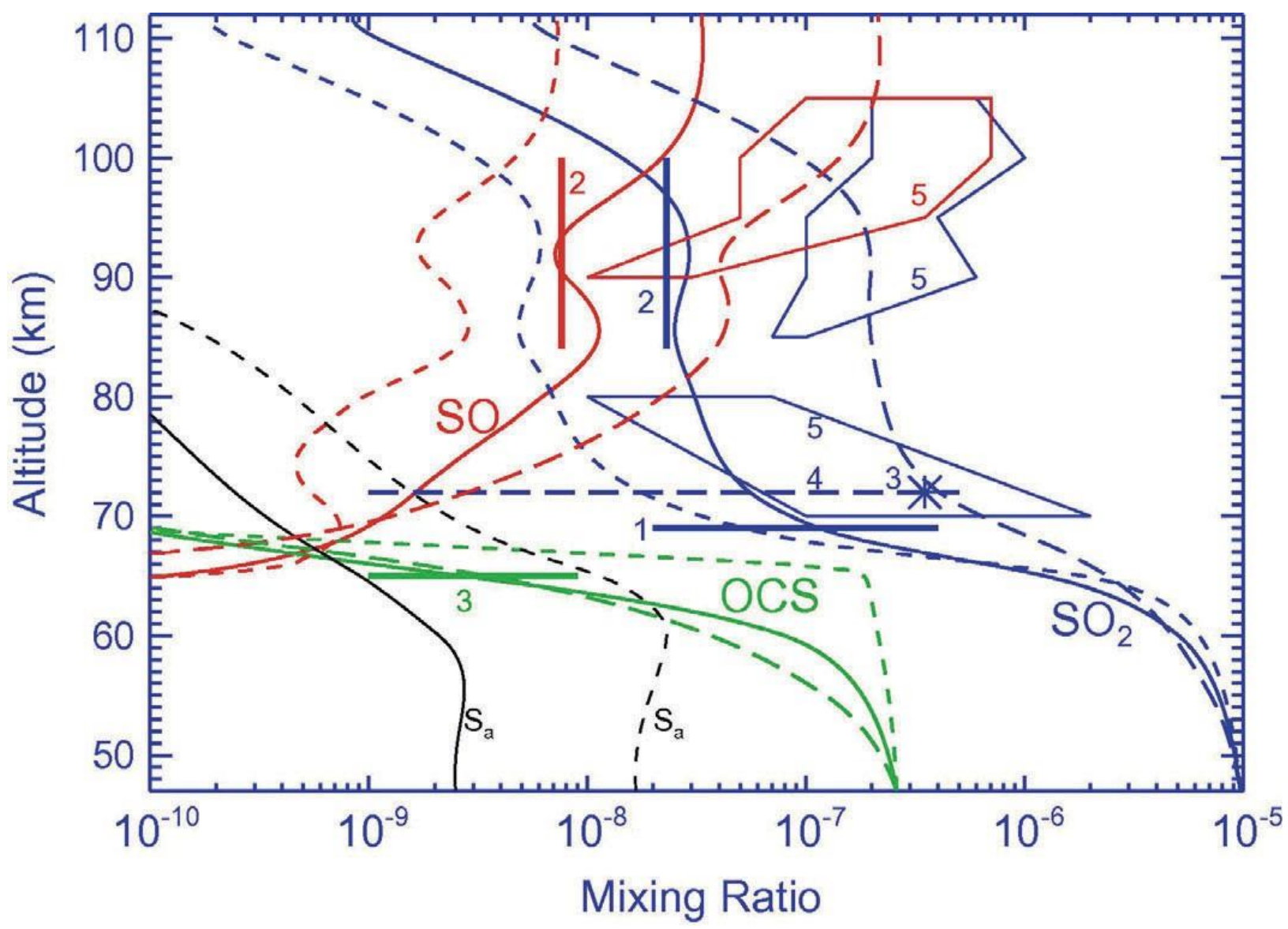

**Fig. 16.18** Sulfur-bearing gases in the middle and upper atmosphere of Venus based on models and observations (Krasnopolsky 2012). Observations include: (1) Pioneer Venus, Venera 15, HST, and rocket data (Esposito et al. 1997); (2) the mean results of Sandor et al. (2010), where the observed $xSO_2$ varies from 0 to 76 ppbv and SO from 0 to 31 ppbv; (3) Krasnopolsky (2010b); (4) Marcq et al. (2011); (5) Belyaev et al. (2012). See Table 16.1 for the observational methods used

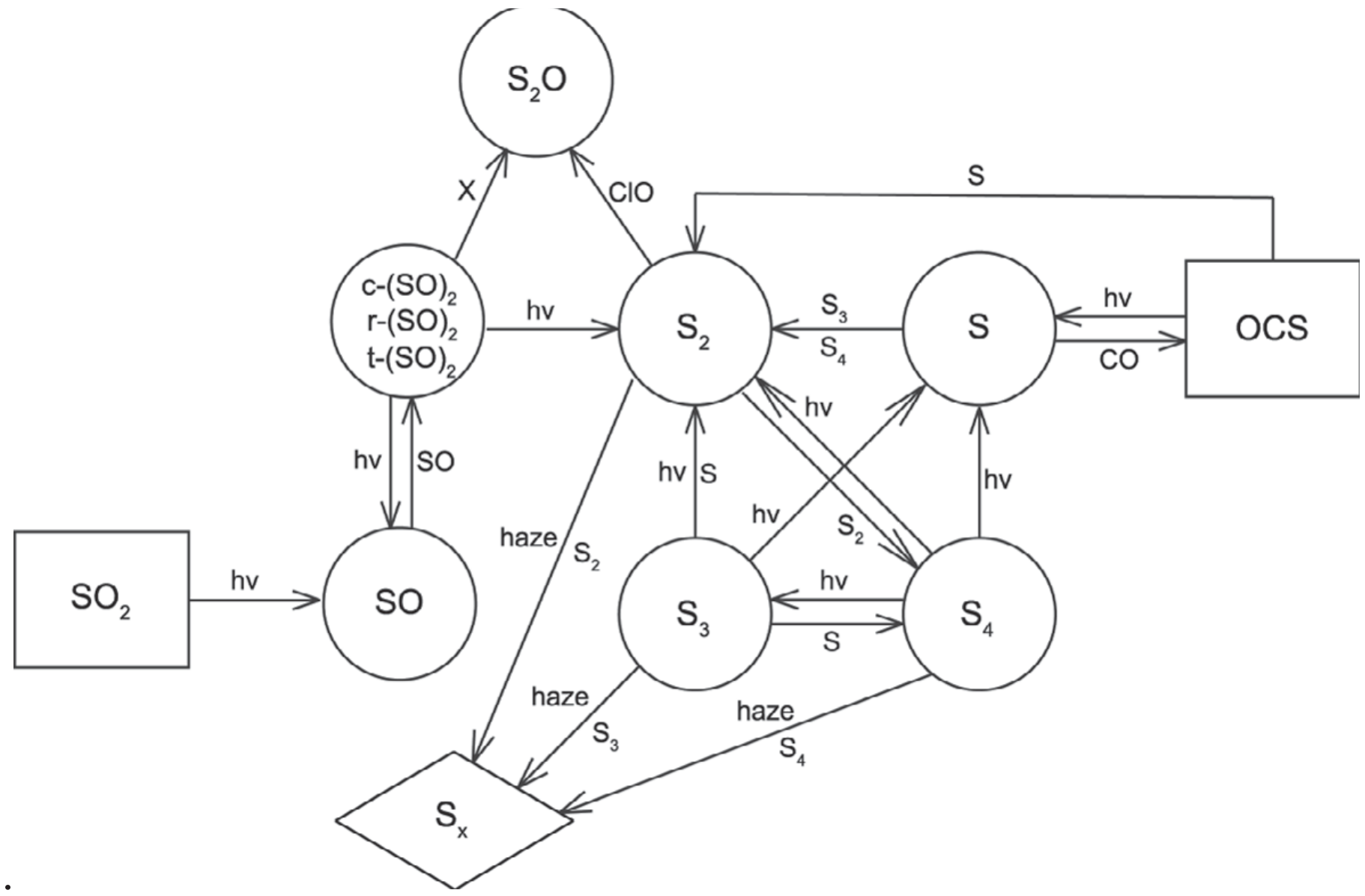

**Fig. 16.19** Schematic diagram showing possible formation pathways of condensed sulfur ($S_x$) in the clouds of Venus (Pinto et al. 2021). The parent molecules are $SO_2$ and OCS, while the intermediate species are indicated in the circles. The species required for the reactions are positioned next to the arrows. X can be O, H, NO, S, SO, or $S_2$

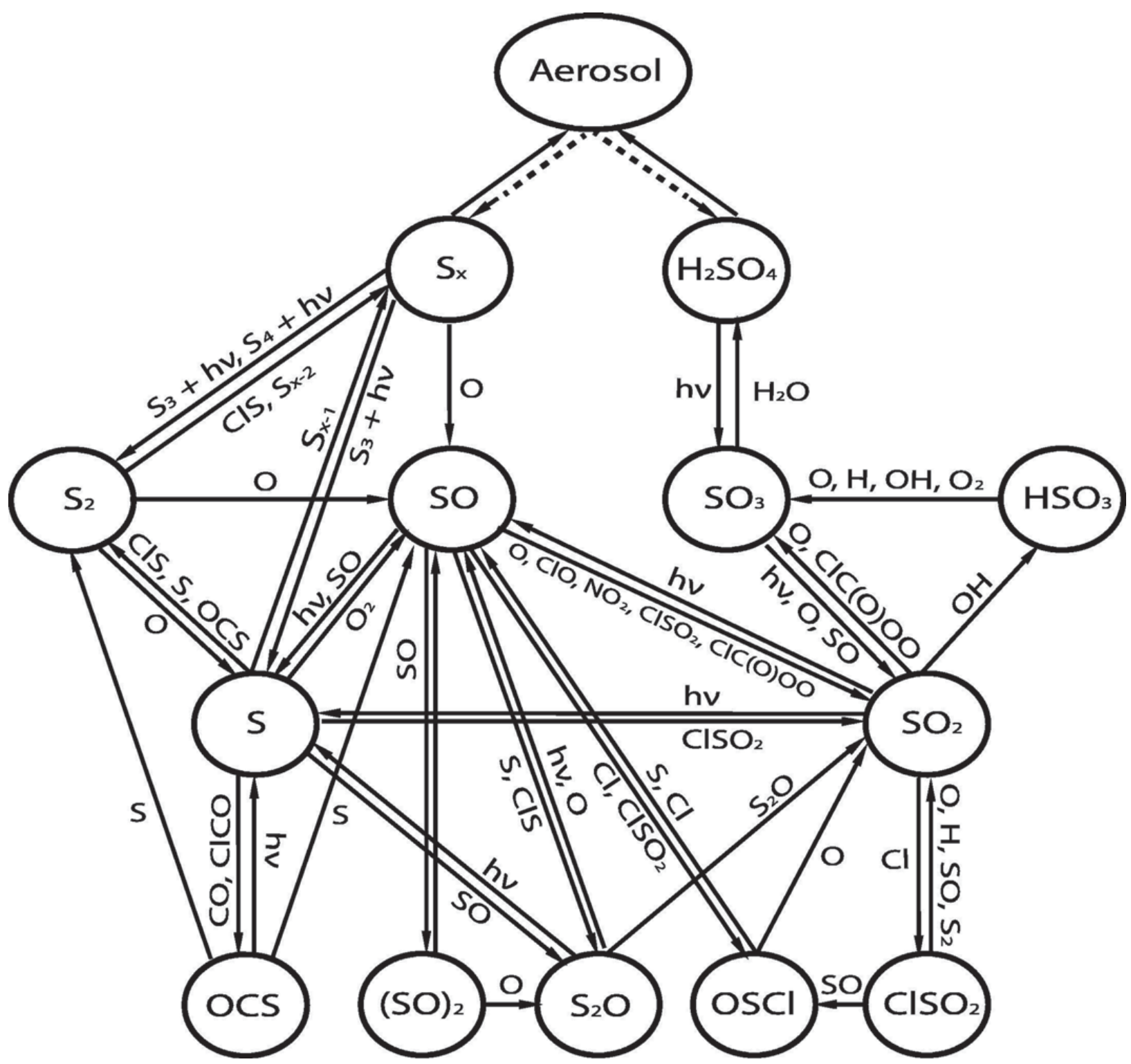

**Fig. 16.20** Significant chemical pathways for sulfur-bearing species in the middle and upper atmosphere of Venus (Zhang et al. 2012)

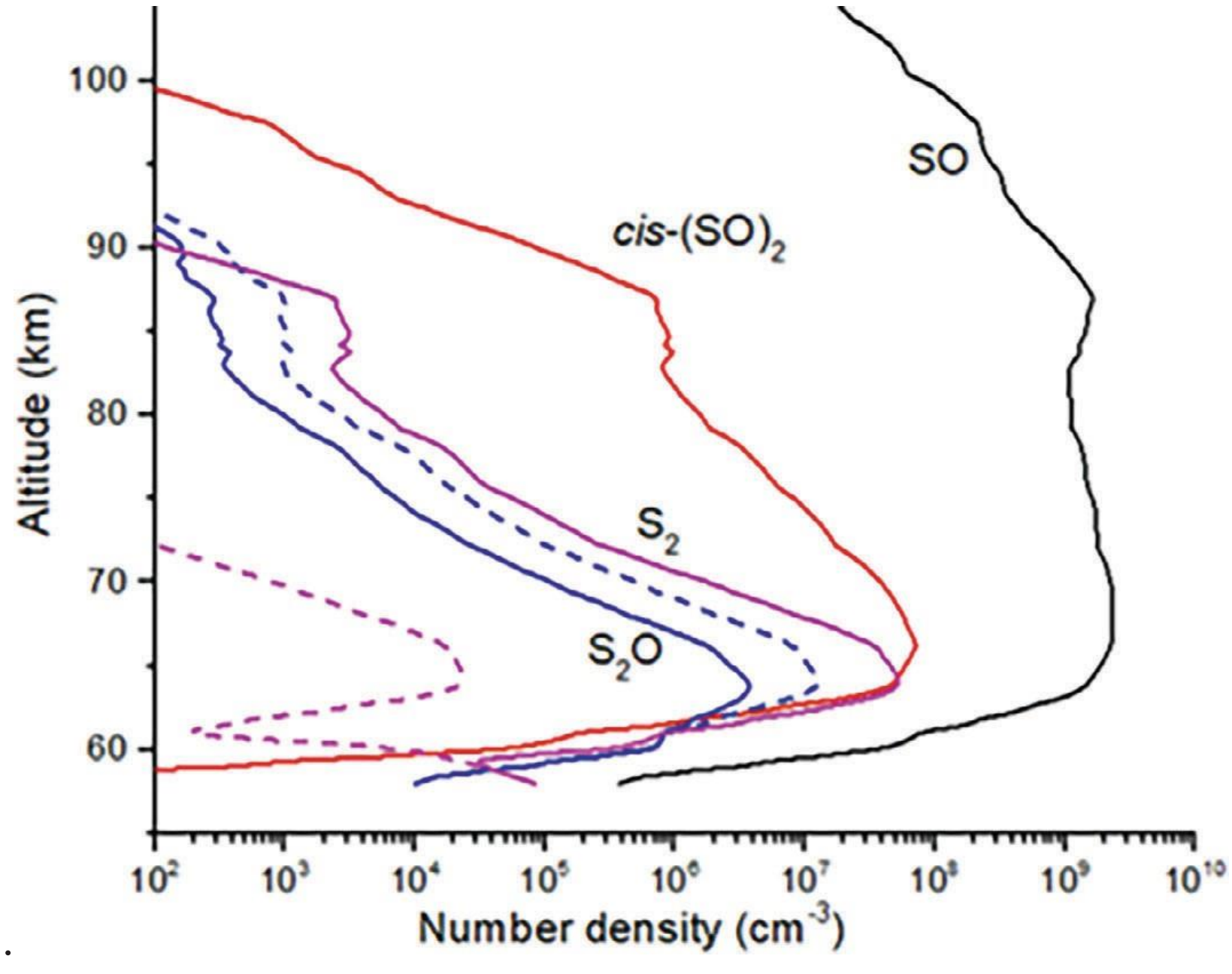

**Fig. 16.21** Estimated steady-state profiles for sulfur-bearing species (Frances-Morrenis et al. 2022). Pinto et al. (2021) $S_2$ formation mechanism is off, and the reaction of SO with $S_2O$ (reaction 16.48) becomes the dominant pathway for $S_2$. Solid curves correspond to a higher reaction rate constant, consistent with the ab initio calculation of Frances-Morrenis et al. (2022), and yield substantial $S_2$

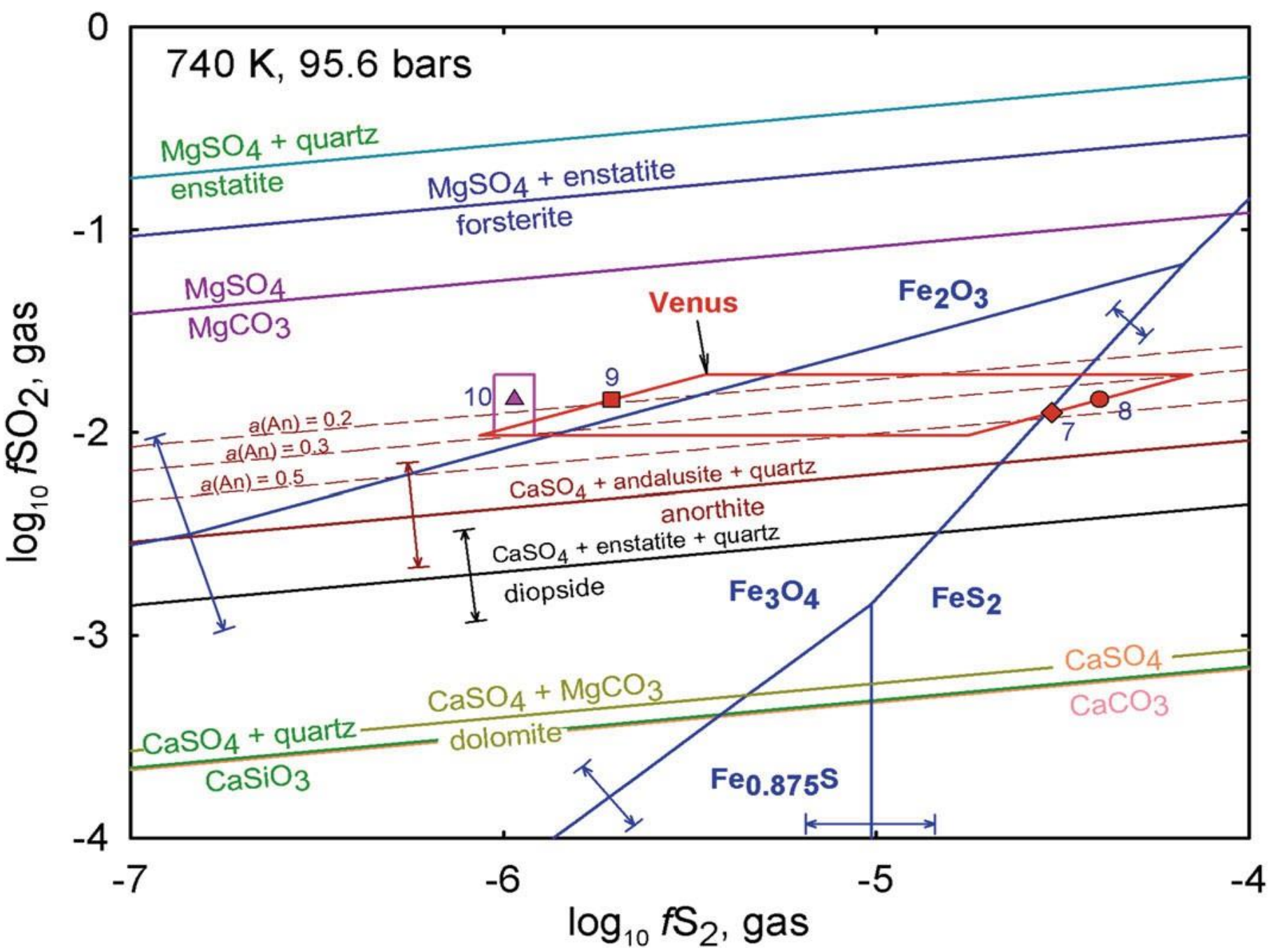

**Fig. 16.22** Stability of rock-forming silicates, carbonates, and sulfates of Ca and Mg, as well as solids in the oxygen-sulfur-Fe system, as functions of the fugacities of $SO_2$ and $S_2$ (bars) under the conditions of Venus' modal radius (740 K, 95.6 bars, 0.6 km below the 6052 km level). The equilibrium lines correspond to reactions analogous to 16.71 and 16.74. The dashed lines represent equilibrium 16.71 at the specified activity of anorthite, *a*(An); the activities of other solids are assumed to be unity. The equilibrium $fS_2$ value for the Py-Pyh reaction 16.76 is from Hong and Fegley (1998). The error bars indicate the uncertainties of selected equilibria due to the thermodynamic data of compounds. The Venus' parallelogram corresponds to the measured $xSO_2$ of 100–200 ppmv (Table 16.1) and $fS_2$ corresponding to the $S_2$-$CO_2$-$SO_2$-CO gas equilibrium at 8 and 17 ppmv CO (Table 16.2). Symbols 7 to 10 represent the corresponding models in Table 16.2. The box at symbol 10 indicates $xS_3$ of 9 to 13 pptv (Krasnopolsky 2013), $S_3$-$S_2$ gas equilibrium, and 100 to 200 ppmv $SO_2$. The diagram illustrates the instability of Ca pyroxenes and carbonates and the stability of Mg silicates and magnesite in relation to sulfatization. The stability field of anorthite is near the conditions on Venus, and plagioclase with an intermediate composition could be in equilibrium with atmospheric $SO_2$ and $S_2$ in the lowlands. The environments of Venus are similar to conditions where magnetite and pyrite or magnetite and hematite coexist. This similarity suggests oxidation and pyritization of exposed ferrous silicates, pyrrhotite, and mafic glasses at the atmosphere-surface interface in the lowlands (Fig. 16.15)

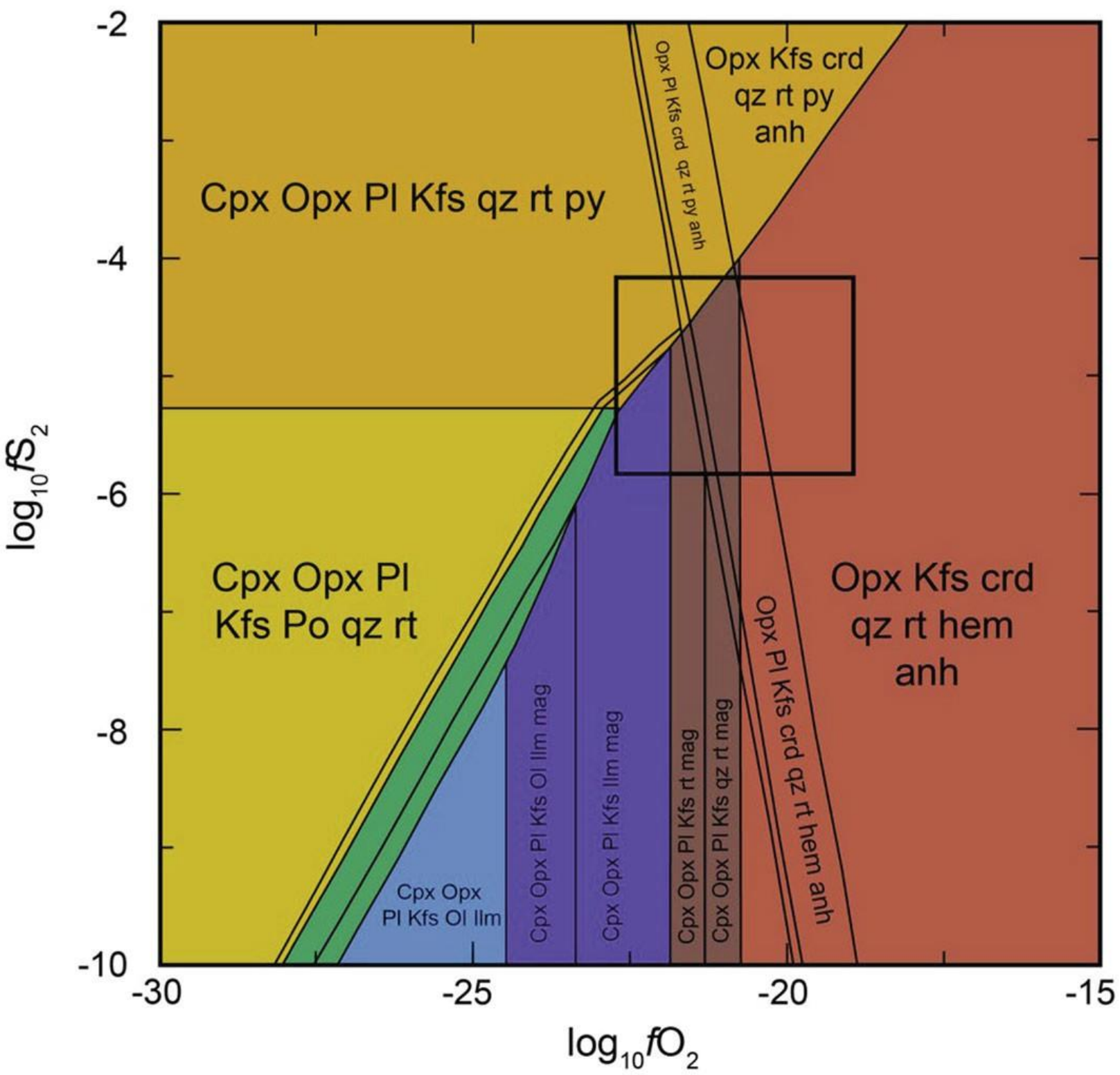

**Fig. 16.23** Mineral phase equilibria for basalt composition as a function of oxygen and sulfur fugacities at the conditions of Venus' modal radius (740 K and 95.6 bars) (Semprich et al. 2020). The black box represents a wide range of fugacities of interest at the surface. Hematite—red, magnetite—brown, pyrite—dark yellow, pyrrhotite—bright yellow, ilmenite + magnetite—purple, ilmenite—blue, ilmenite + pyrrhotite—green. Solid solutions: Cpx—clinopyroxene, Opx—orthopyroxene, Ol—olivine, Pl—plagioclase, Kfs—K-feldspar, Po—pyrrhotite, Ilm—ilmenite. Other minerals: py—pyrite, anh—anhydrite, crd—cordierite, hem—hematite, mag—magnetite, qz—quartz, rt—rutile

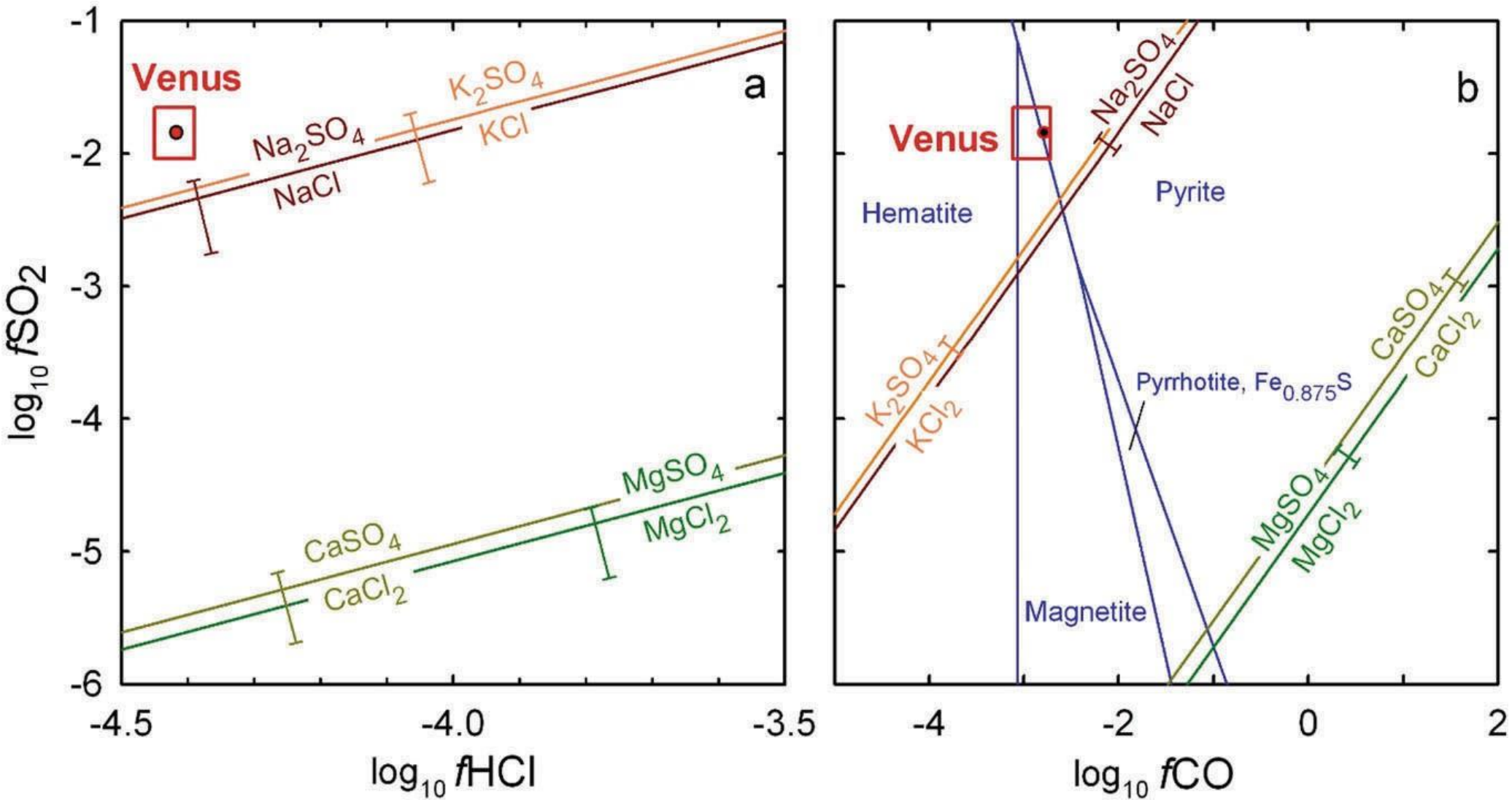

**Fig. 16.24** Stability fields of sulfates and chlorides on the surface of Venus under conditions of modal radius (740 K and 95.6 bars). The atmospheric conditions of Venus (red boxes) correspond to 95–230 ppmv $SO_2$ (Table 16.1), 0.4 ± 0.03 ppmv HCl (Krasnopolsky 2010a) in (**a**), and 8–20 ppmv CO in (**b**). The circle symbols represent 150 ppmv $SO_2$, 0.4 ppmv HCl (**a**), and 17 ppmv CO (**b**). In (**a**), phase equilibrium lines correspond to reactions such as 16.73 at 30 ppmv $H_2O$ and 0.4 ppmv $S_2$, aligning with 150 ppmv $SO_2$ at $fO_2$ of $10^{-21.36}$ (Model 8 of Table 16.2). The upper and lower ends of the error bars represent 20 ppbv and 0.8 ppbv $S_2$, respectively, corresponding to $fO_2$ values of $10^{-21.7}$ and $10^{-20.0}$ from Fegley et al. (1997b). In (**b**), the equilibrium lines are for reactions like 16.72, and phase equilibria in the oxygen-sulfur-Fe system are shown for comparison. The panels indicate that lower $fSO_2$ and $fO_2$ levels, along with higher $fCO$ and $fS_2$ levels, enhance the stability of chlorides. Chlorides of Ca and Mg are highly unstable in relation to sulfurization. Under the modal radius conditions, chlorides of K and Na can only remain stable if the atmospheric $xSO_2$ is below ~60 ppmv. The figure is modified after (Zolotov 2025)

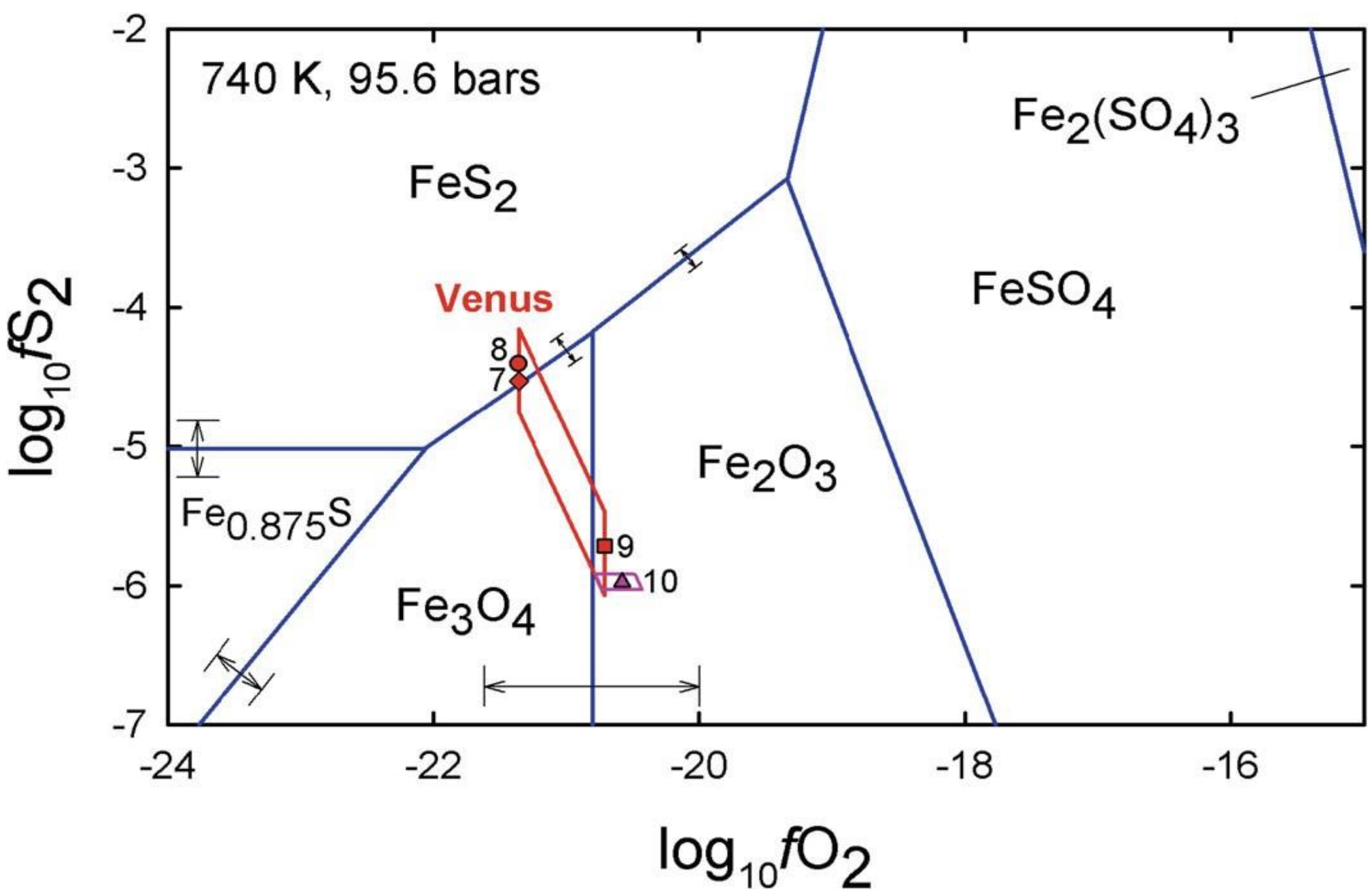

**Fig. 16.25** Stability fields of solids in the oxygen-sulfur-Fe system as functions of the fugacities of $O_2$ and $S_2$ (bars) under conditions of Venus' modal radius. The equilibrium lines correspond to reactions analogous to Eqs. 16.66 and 16.67. The value of $fS_2$ for the Py-Pyh equilibrium 16.76 is from Hong and Fegley (1998). Venus' atmospheric conditions are described in the caption of Fig. 16.22. If the composition of Venus' near-surface atmosphere corresponds to Models 7 and 8 of Table 16.2, the primary rock's pyrrhotite will oxidize to a magnetite-pyrite assemblage. Pyrrhotite could oxidize to a magnetite-hematite assemblage if the composition aligns with Models 9 and 10 of Table 16.2. The diagram illustrates that ferric and ferrous sulfates are unstable at the surface. Ferric sulfate grains that sink from the clouds decompose into hematite

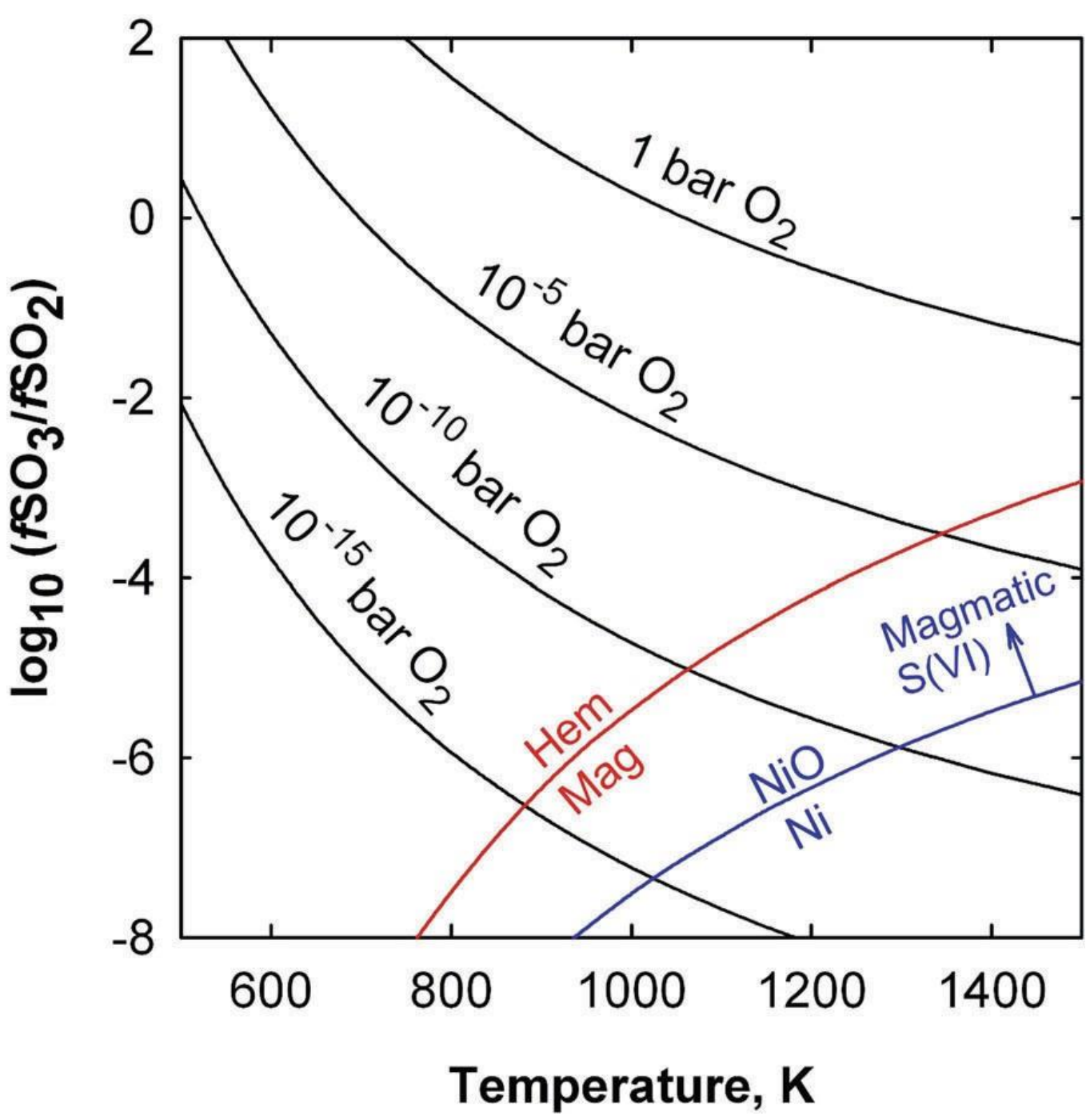

**Fig. 16.26** Relative abundance of $SO_3$ and $SO_2$ as a function of temperature and $fO_2$ in a steam atmosphere on early Venus. Other curves display the conditions of magnetite-hematite and Ni-NiO buffers. The arrow illustrates conditions under which sulfur exists in the sulfate form in silicate magmas that may have formed at the surface due to greenhouse heating. The plot indicates that $SO_3$ is more abundant than $SO_2$ only in a high-temperature, $O_2$-rich atmosphere. If $fO_2$ did not exceed values determined by the Mag-Hem buffer (equilibrium 16.66) in surface magmas or altered rocks, $SO_2$ dominated over $SO_3$

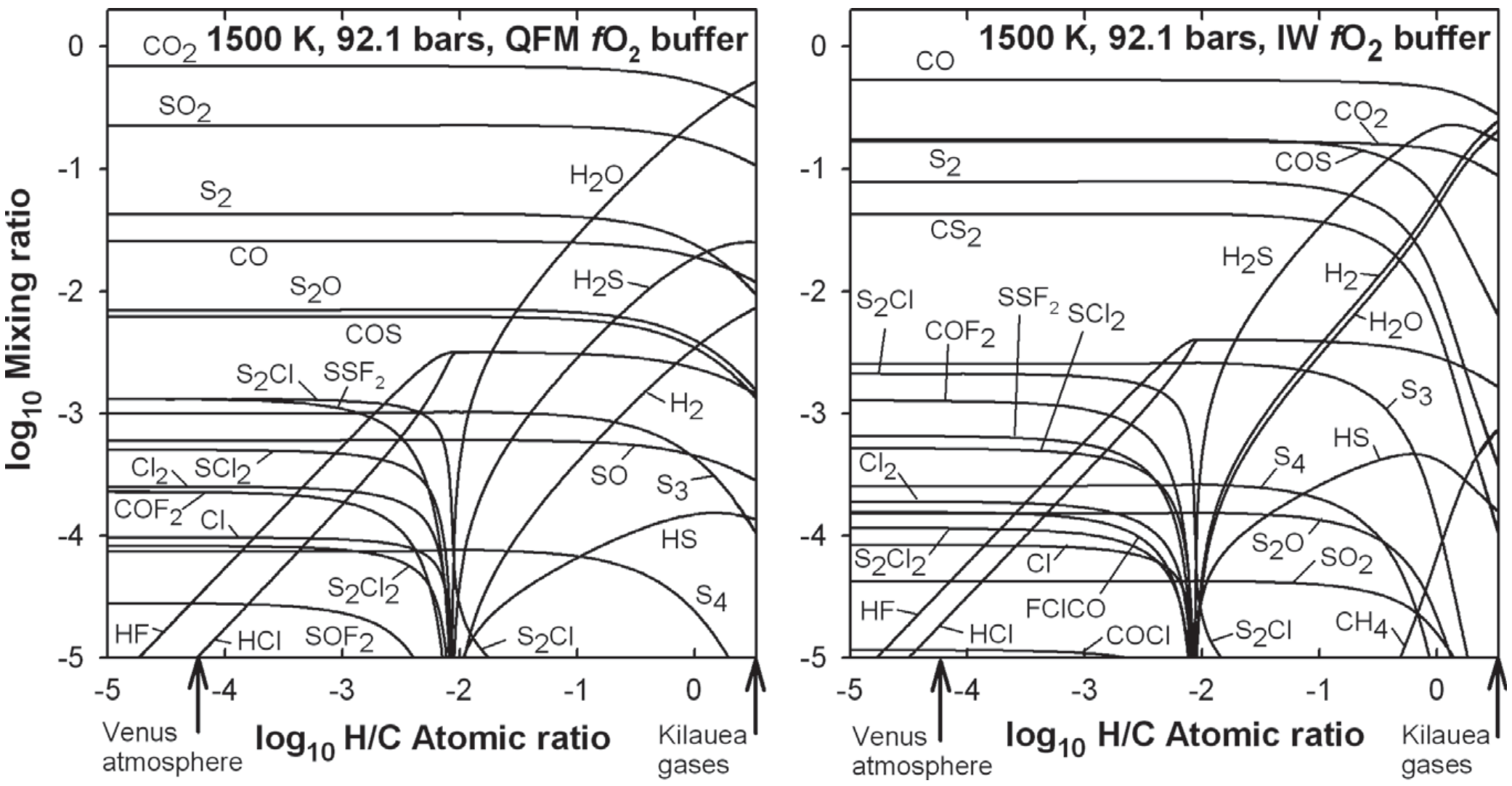

**Fig. 16.27** Speciation of Venus' volcanic gases as a function of the hydrogen/carbon ratio at 92.1 bars and 1500 K (Zolotov and Matsui 2002). The elemental composition of gases aligns with the reconstructed 1918 analysis of Kilauea magma lake gases (Gerlach 1980) and exhibits a variable hydrogen/carbon ratio. Kilauea emissions are among the most hydrogen-depleted volcanic gases on Earth. Venus' counterparts could be more hydrogen-depleted than Kilauea gases due to probable hydrogen deficiency in the interior and suppressed degassing of $H_2O$ at ambient pressure. The oxidized (QFM) and reduced (IW) oxidation states serve as $fO_2$ endmembers for Venus' mafic melts, as suggested by the MnO/FeO ratio in the Venera 13, 14, and Vega 2 probes (Sect. 16.2.2.1)

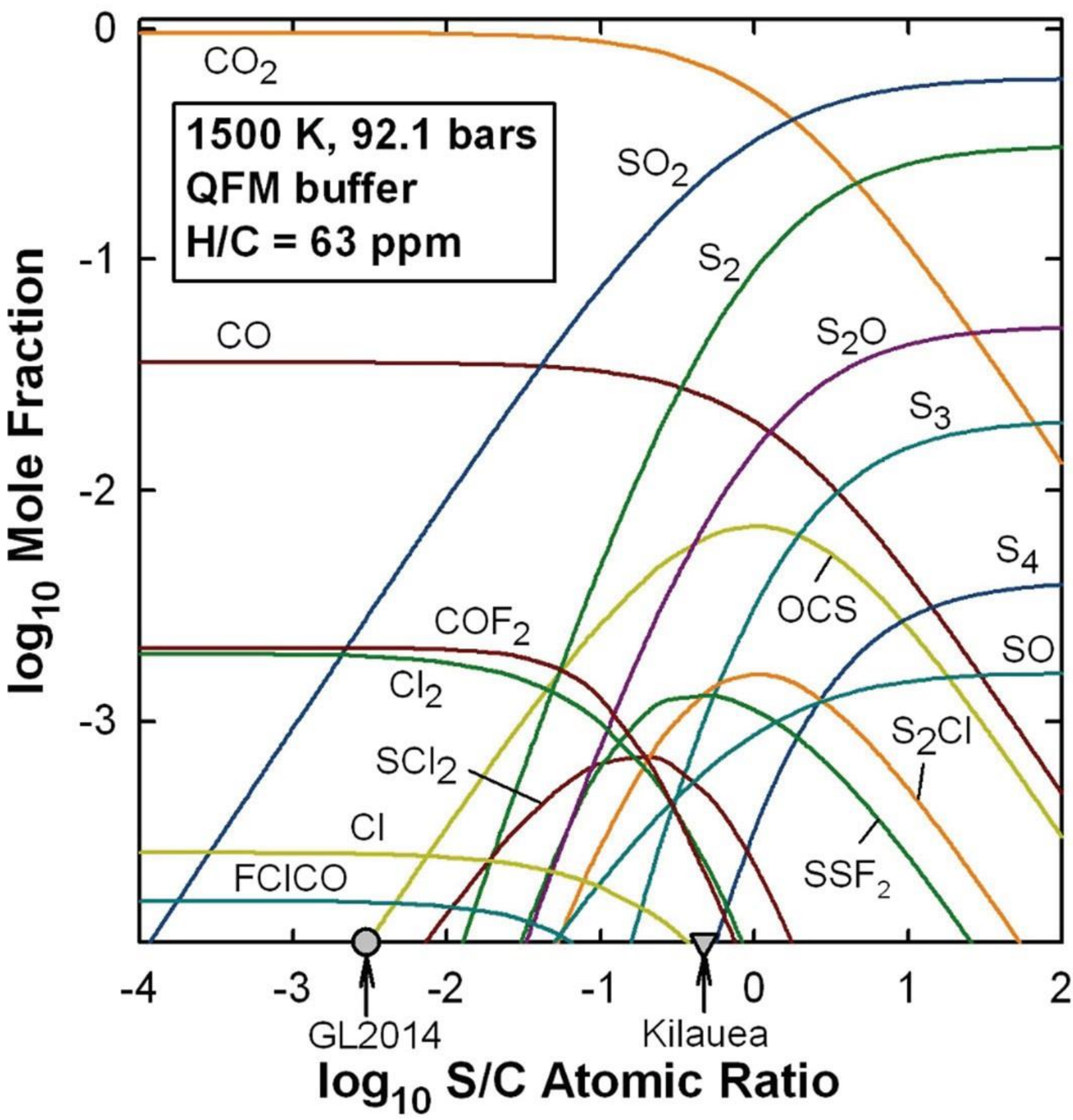

**Fig. 16.28** Speciation of possible Venus' volcanic gases as a function of sulfur/carbon ratio. The elemental composition of gases matches that in Fig. 16.27, and the hydrogen/carbon atomic ratio corresponds to that found in the atmosphere. The GL2014 arrow at a sulfur/carbon ratio of $3 \times 10^{-3}$ relates to the degassing model of Gaillard and Scaillet (2014) for ~100 bars. As mentioned in the text, comparable sulfur and carbon contents are likely for Venus, possibly resembling Kilauea lava lake samples. Higher sulfur/carbon ratios on volcanic gases could represent a significant early $CO_2$ release (e.g., from a magma ocean) and its minor volcanic degassing during the geological evolution

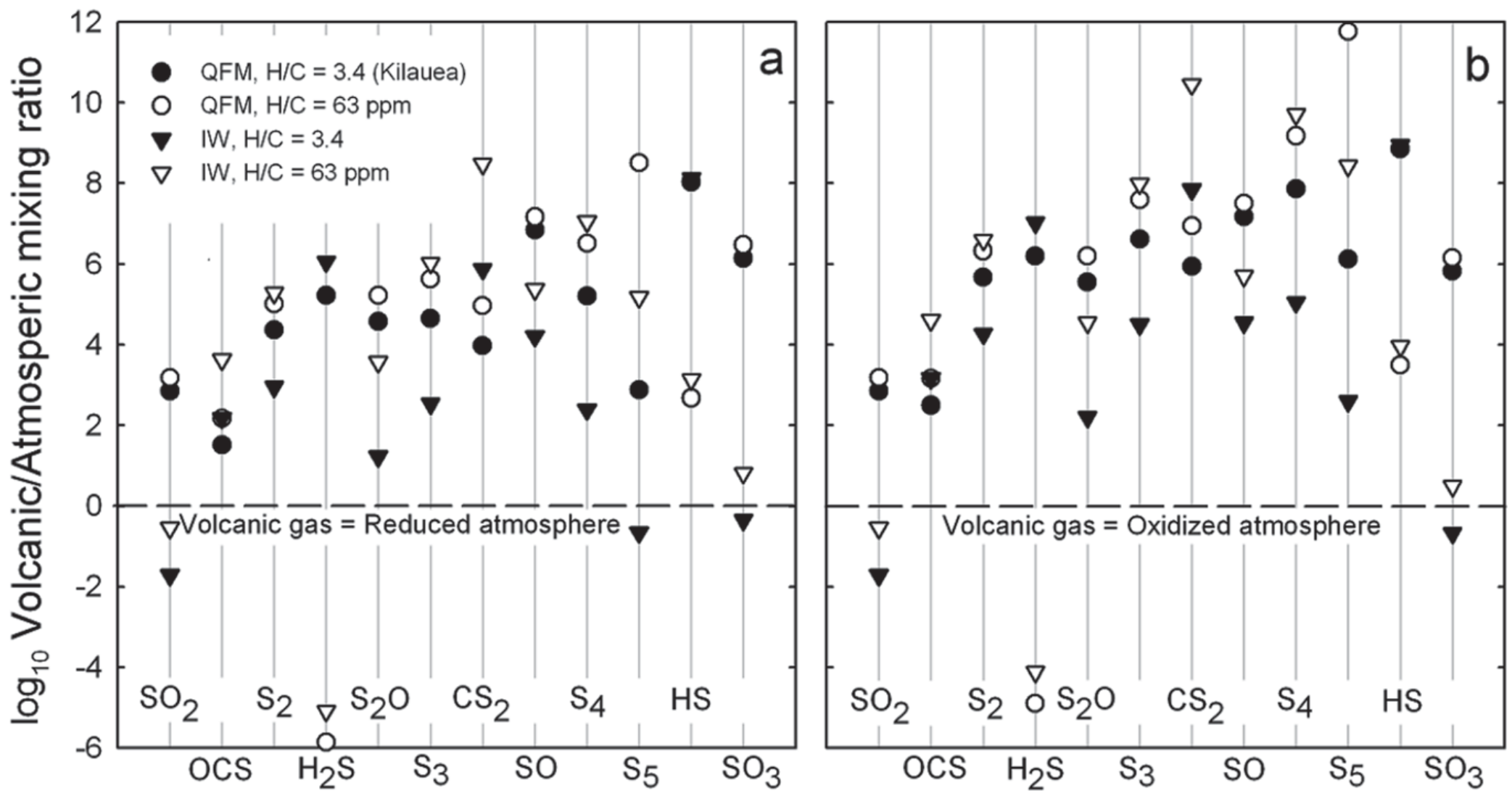

**Fig. 16.29** Venus' volcanic-to-atmospheric ratios of sulfur-bearing gases. In *a*, volcanic gas composition represents four end members regarding $fO_2$ and hydrogen/carbon atomic ratio from Fig. 16.27. Hydrogen-rich gases correspond to Kilauea-like compositions, while hydrogen-depleted gases possess Venus' atmospheric hydrogen/carbon ratio. The atmospheric gas corresponds to Model 8 (a) in Model 9 (b), which represents more reduced and oxidized compositions, respectively (Table 16.2). The dashed lines represent equal compositions of volcanic and atmospheric gases. The figure illustrates that volcanic gases are compositionally different from the near-surface atmosphere and are often more abundant than their atmospheric counterparts. This difference indicates that volcanic gases undergo reactions with atmospheric gases and with each other, moving toward gas-phase chemical equilibria in the near-surface atmosphere

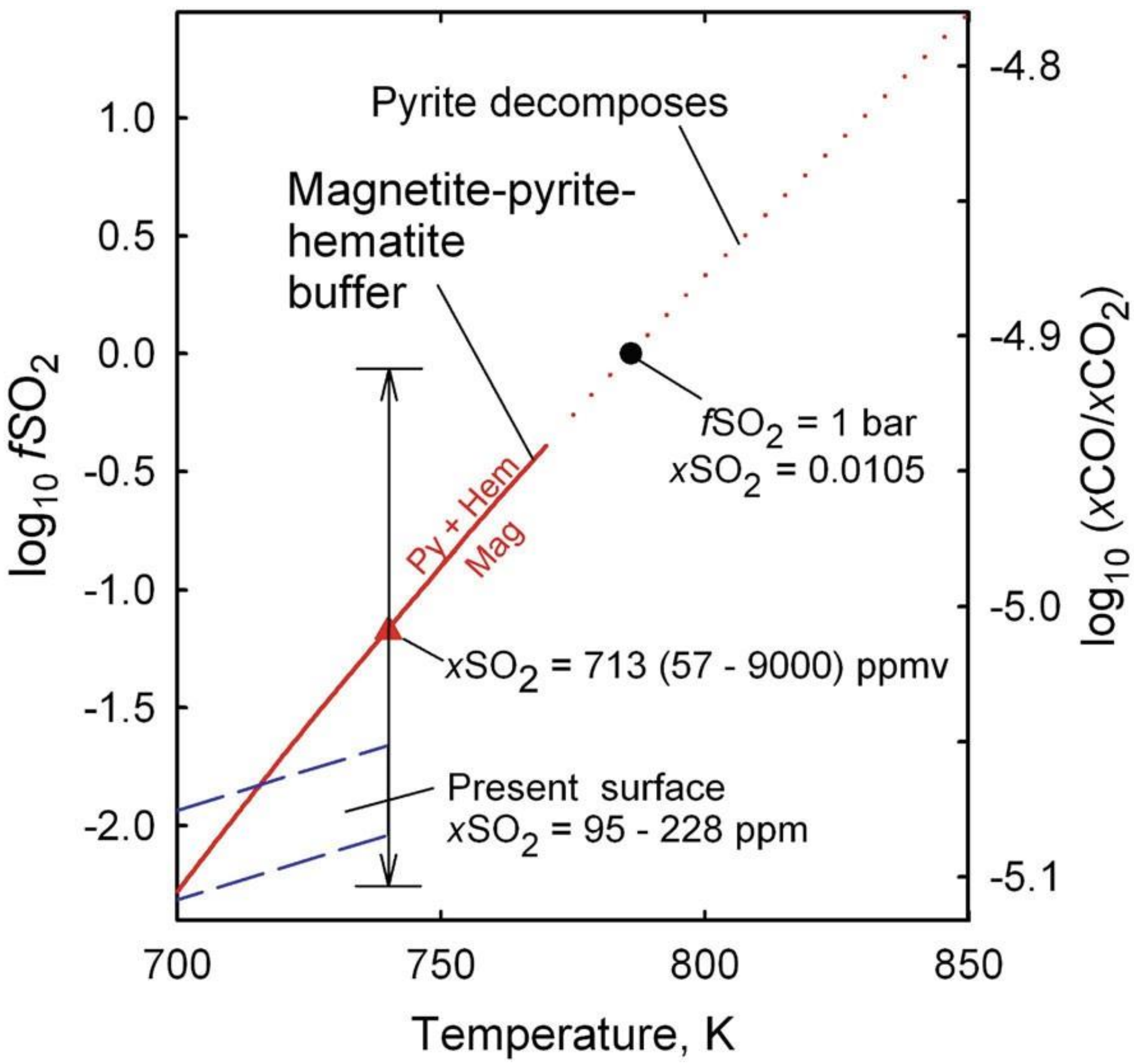

**Fig. 16.30** Possible future changes in the concentrations of atmospheric $SO_2$ and the $CO/CO_2$ ratio, as influenced by the magnetite-pyrite-hematite equilibrium (Equations 16.85 and 16.86), in surface materials at 95.6 bars. The fugacity of $SO_2$ in the current near-surface atmosphere reflects a range of measured $SO_2$ concentrations (Oyama et al. 1980; Gel'man et al. 1980; Table 16.1) and the associated uncertainties. The triangle symbol represents the nominal conditions of mineral equilibrium at 740 K (see Table 16.11 for other gases). The double arrow illustrates the potential $fSO_2$ error bar due to the thermodynamic data of minerals in Eq. 16.86. The circle symbol corresponds to a partial pressure of $SO_2$ of 1 bar at 786 K. The dotted curve illustrates potential conditions arising from the thermal decomposition of pyrite in a permeable surface layer. The figure is modified after Gillmann et al. (2022)